\documentclass{cornell}
\usepackage{amsmath,amssymb,graphicx,euscript,epsfig} 

\title{Charmless inclusive B decays and the extraction of V(ub)}

\author{Gil Paz}

\conferraldate{August}{2006}

\degreefield{Doctorate of Science}

\def\delslash{\rlap{\hspace{0.02cm}/}{\partial}}
\def\Dslash{\rlap{\hspace{0.07cm}/}{D}}
\def\nbslash{\rlap{\hspace{0.02cm}/}{\bar n}}
\def\nslash{\rlap{\hspace{0.02cm}/}{n}}
\def\pslash{\rlap{\hspace{0.02cm}/}{p}}
\def\vslash{\rlap{\hspace{0.02cm}/}{v}}
\def\F{{\EuScript F}}
\def\H{{\EuScript H}}
\def\X{{\EuScript X}}
\def\calAslash{\rlap{\hspace{0.08cm}/}{{\EuScript A}}}
\def\calDslash{\rlap{\hspace{0.1cm}/}{{\EuScript D}}}
\def\A{{\EuScript A}}
\def\D{{\EuScript D}}
\def\bm#1{\mbox{\boldmath$#1$\unboldmath}}
\def\S{{\cal S}}

\begin{document}
\setcounter{page}{0}
\maketitle

\iffinal

\makecopyright

\begin{abstract}
This work discusses charmless inclusive $B$ decays and its application
to the extraction of $|V_{ub}|$. Starting from first principles we
relate the differential decay rate to the hadronic tensor in terms of
optimal choice of kinematical variables. We review the traditional
methods of calculating the hadronic tensor, expansion in $\alpha_s$
and HQET, and discuss their shortcomings. In the kinematical region
relevant for experiment ("shape function" region), the hadronic tensor
can be factorized, at each order in $1/m_b$, as a product of
calculable hard functions and a convolution of calculable jet
functions with non perturbative shape functions.

Using SCET, we calculate the leading order hard and jet function to
first order in $\alpha_s$. Large logarithms are resummed in RGE
improved perturbation theory. Local OPE is used to relate moments of
the renormalized shape function to HQET parameters, defined in the
"shape function scheme". Beyond leading order in $1/m_b$, several
subleading shape functions arise. We derive them at tree level, where
they can be expressed as forward matrix elements of bi-local
light-cone operators. 

Based on these theoretical calculations we present two
applications. In the first, we present the "state-of-the-art "
expressions for the triple differential $\bar B\to X_u\,l\bar\nu$
decay rate and the $\bar B\to X_s\gamma$ photon spectrum. These
expressions include all known contributions and smoothly interpolate
between the "shape-function" and "OPE region". Based on these an event
generator can be constructed, from which the theoretical prediction
for any experimental cut can be extracted. In the second, a weight
function is constructed that relates the $P_+$ spectrum in $\bar B\to
X_u\,l\bar\nu$ to the normalized $\bar B\to X_s\gamma$ photon
spectrum. This weight function is independent of the leading order
shape function. At leading power the weight function contains two loop
corrections at the scale $\sqrt{m_b\Lambda_{\rm QCD}}$. Power
corrections from phase space are included exactly, while the remaining
subleading contributions are included at first order in $\Lambda_{\rm
QCD}/m_b$.

\end{abstract}

\begin{biosketch}
Gil Paz was born in Haifa, Israel. In 1999, he completed his
undergraduate study at the Technion - Israel Institute of Technology,
receiving a B.A. in Physics and a B.A. in Mathematics both summa cum
laude. In 2001 he received a M.Sc. in Physics, also from the Technion,
under the supervision of Professor Michael Gronau. The same year he
came to Cornell university. He then completed a M.Sc. in 2004 and a
Ph.D. in 2006 under the supervision of Professor Matthias Neubert.
\end{biosketch}

\begin{acknowledgements}
First, I would like to thank my collaborators: Stefan Bosch, Bj\"orn
Lange and Matthias Neubert. The work presented here was done in
collaboration with them over the last few years.

I would like to thank my advisor Matthias Neubert for his help,
encouragement and for teaching me to always try to improve the work
that we do.

I would like to thank the members of my special committee Csaba Csaki
and Lawrence Gibbons for their help and advice. Special thanks to
Maxim Perelstein for being a proxy on my B exam and for his comments
on the dissertation.

I would like to thank the Institute for Advanced Study, Princeton, NJ,
and the Institute of Nuclear Theory at the University of Washington,
where part of this research was performed.

This research was supported by the National Science Foundation under 
Grant PHY-0098631 and Grant PHY-0355005.

\end{acknowledgements}

\contentspage
\tablelistpage
\figurelistpage

\singlespacing
\setcounter{page}{1}
\pagenumbering{arabic}
\pagestyle{plain}

\chapter{Introduction}
\label{chapter_intro}
\section{Preface}

The standard model is a theory that describes our knowledge of three
of the four fundamental forces in nature: the strong, the
electromagnetic and the weak. The strong and the weak
forces especially pose interesting challenges to theoretical physics.

The strong interaction can be described at high energies by the QCD
Lagrangian, where the quarks couple to each other via an exchange of
gluons.  The coupling strength is universal, i.e. independent of the
flavor of the quarks.  At energies above roughly 1 GeV we can use
perturbation theory to calculate the effects of the strong force. At
lower energies the perturbative description, based on the use of the
interaction Lagrangian, breaks down. As a result, at low energies one
has to use either non perturbative methods such as lattice calculations 
and chiral perturbation theory, or extract information from
experiment. The separation of scales and the perturbative from the
non perturbative physics will be a major theme in this work.

The weak interaction is one of the richest parts
of the standard model. As opposed to the strong interaction, the weak
interaction is not universal and is different for different
flavors. More specifically the strength of the interaction between
quarks of different flavors is described by a unitary matrix known as
the Cabibbo-Kobayashi-Maskawa (CKM) matrix. The CKM matrix elements
are fundamental parameters of the standard model and therefore it is
important to measure them as accurately as possible. Indeed, Measuring
the CKM matrix elements has been one of the main goals of research
in particle physics, both for theory and experiment.

This work aims at precision measurements of $V_{ub}$, one of the
smallest CKM matrix elements . Currently the highest precision on
$|V_{ub}|$ can be achieved through inclusive measurement of charmless
decays of $B$ mesons. By charmless we mean both semileptonic ($\bar
B\to X_u\,l\bar\nu$) and radiative ($\bar B\to X_s\gamma$) decays. The
former is used to actually measure $|V_{ub}|$, while the latter is
used to extract, directly or indirectly, the main non perturbative
object which is the shape function. Of course, better understanding of
radiative decays is important by itself, as it has important
implications for constraining physics beyond the standard model.

The outline of this work is as follows. We begin by an introduction,
which takes the reader from ``first principles" to the starting point
of chapters \ref{chapter_pert} and \ref{chapter_sub}. These two
chapters are more theoretical, where in \ref{chapter_pert} and
\ref{chapter_sub} the perturbative and non perturbative calculations
are presented, respectively. The last two chapters are concerned with
applying these calculations. Chapter \ref{chapter_evegen} gives the
necessary ingredients for a construction of an event generator, which
allows experimenters to calculate any semileptonic spectrum they would
like to measure. In chapter \ref{chapter_weight} a weight function is
constructed, that directly relates the $P_+$ spectrum in $\bar B\to
X_u\,l\bar\nu$ to the photon spectrum in $\bar B\to X_s\gamma$. We
present our conclusions in chapter \ref{chapter_conc}.

\section{Chapter outline}
In this chapter we begin our study of inclusive charmless $B$
decays. We start with the description of the kinematics of these
decays and discuss the optimal choice of kinematical variables. We
then turn to the dynamics. The basic quantity we are trying to
calculate is the hadronic tensor, which cannot be calculated
exactly. We review the ``traditional" approximation methods, namely,
expansion in $\alpha_s$ and Heavy Quark Effective Theory (HQET) and
discuss their shortcomings. In order to overcome these problems we
introduce the appropriate effective field theory suitable for
inclusive charmless $B$ decays, the Soft Collinear Effective Theory
(SCET). This effective field theory will be used in the following
chapters to analyze both the perturbative (chapter \ref{chapter_pert})
and non perturbative (chapter \ref{chapter_sub}) corrections to the
hadronic tensor. This introduction is meant to be pedagogical, so we
present most of the derivations explicitly.

\section{Kinematics made easy}
\label{section_kin} 
In this work we focus on $\bar B\to X_u\,l\bar\nu$ decays, but also
discusses $\bar B\to X_s\gamma$ decays, both by itself and in relation
to $\bar B\to X_u\,l\bar\nu$.  These decays are inclusive in a sense
that one of the final states is something that contains a specific
quark ($u$ or $s$) but is otherwise unrestricted and can even be a
group of particles.

We begin analyzing these decays by looking at the simple question,
``How many kinematical variables are needed to describe an inclusive
event?". We are looking at a decay of a $B$ meson into $n$ particles,
where $n-1$ of them have known mass. The ``particle'' $X$ has
unspecified invariant mass. In general for a decay into $n$ particles
there are $3n-4$ kinematical variables, but not all of them are
determined from the dynamics. Since we do not have information about
the spin of the hadronic state $X$ or the spin/polarization of the
other decay products, the only four vectors we have at our disposal
are $P_B$, the four momentum of the $B$ meson, $p_1\equiv P_X$ the
four momentum of $X$, and $p_2,\,\dots,\,p_n$. Lorentz invariance
implies then that the possible variables are the $(n+1)(n+2)/2$ scalar
products of these $n+1$ vectors. Since the masses of all the particles
apart from $p_1$ are known, we have $n$ constraints of the form
$p_i^2=m_i^2,\, (i=B,2,\dots, n)$. Conservation of momentum would
allow us to eliminate all the pairs that contain $P_B$ (since
$P_B=\sum_{i=1}^{n} p_i$), i.e eliminating $n+1$ variables. That
leaves us with $n(n-1)/2$ independent variables. The rest of the
variables are undetermined from the dynamics and can be integrated
over. This argument holds for $n\leq 3$, since at 4 dimensions at most
4 vectors (corresponding to $n=3$) can be linearly independent. For
$n>3$ the argument needs to be modified.

The decay mode $\bar B\to X_s\gamma$ has $n=2$ and therefore only one
relevant variable. This variable is usually taken to be $E_\gamma$,
the energy of the photon. As we shall see, a related variable,
$P_+\equiv M_B-2 E_\gamma$ will turn out be useful. For $\bar B\to
X_u\,l\bar\nu$ we have $n=3$ and three relevant variables (see table
\ref{table_var}). There are various choices in the literature for
these variables, and in order to understand them we have to say a few
words about the dynamics of these decays.
\begin{table}[h]
\caption{Number of kinematic variables for inclusive $B$ decays. Using
the constraints on the 
 right column we can eliminate some of the
variables on the left column.}
\begin{tabular}{cccc|c}
\multicolumn{5}{c}{$\bar B\to X_s\gamma$}\\
\multicolumn{4}{c|}{Scalar Products}&Constraints\\\hline
&\rlap{\hspace{0.02cm}$\diagdown$}{$P_B^2$}&
\rlap{\hspace{0.3cm}$\diagdown$}{$P_BP_X$} &
\rlap{\hspace{0.3cm}$\diagdown$}{$P_BP_\gamma$}&$P_B=P+P_\gamma$\\
&& $P_X^2$ & \rlap{\hspace{0.3cm}$\diagdown$}{$P_XP_\gamma$}&$(P_X+P_\gamma)^2=M_B^2$\\
&& & \rlap{\hspace{0.02cm}$\diagdown$}{$P_\gamma^2$}&$P_\gamma^2=0$\\\\
\multicolumn{5}{c}{$\bar B\to X_u\,l\bar\nu$}\\
\multicolumn{4}{c|}{Scalar Products}&Constraints\\\hline
\rlap{\hspace{0.02cm}$\diagdown$}{$P_B^2$}&
\rlap{\hspace{0.3cm}$\diagdown$}{$P_BP_X$} &
\rlap{\hspace{0.3cm}$\diagdown$}{$P_BP_l$}&\rlap{\hspace{0.3cm}$\diagdown$}{$P_BP_\nu$}&$P_B=P+P_l+P_\nu$\\
&$P_X^2$& $P_XP_l$ & $P_XP_\nu$&\\
&& \rlap{\hspace{0.02cm}$\diagdown$}{$P_l^2$} & \rlap{\hspace{0.3cm}$\diagdown$}{$P_lP_\nu$}&$(P_X+P_l+P_\nu)^2=M_B^2$, $P_l^2=0$\\
&&& \rlap{\hspace{0.02cm}$\diagdown$}{$P_\nu^2$}&$P_\nu^2=0$\\
\end{tabular}\label{table_var}
\end{table}

First we need to distinguish between the hadronic level, which looks
at the decay as a decay of hadrons, and the partonic level, which
looks at the decay as a decay of quarks. At the hadronic level we have
a $B$ meson carrying momentum $P_B\equiv M_B v$ decaying into a
leptonic pair (the lepton and the anti-neutrino) carrying momentum
total $q$, and a hadronic jet carrying momentum $P_X$. Conservation of
momentum implies $M_B v-q=P_X$. At the partonic level we look at this
decay as a decay of a $b$ quark, carrying momentum $m_b v$, into a $u$
quark, carrying momentum $p$ (at tree level), and a virtual $W$
carrying momentum $q$.  The $W$ in turn decays into a lepton $l$ and
an anti-neutrino $\bar\nu_l$. (More accurately we write the momentum
of the $b$ quark as $m_b v+k$ where $k$ is $O(\Lambda_{\rm QCD})$.
and expand in powers of $k$). Conservation of momentum implies $m_b
v-q=p$. If we define $\bar\Lambda\equiv M_B-m_b$ we find that
$P_X=p+\bar\Lambda v$. Beyond tree level $p$ would be the momentum of
the jet of light partons created in the decay. We will elaborate on
the exact definition of $\bar\Lambda$ in chapter \ref{chapter_pert}.

There are generically two common choices of variables:
\begin{itemize}
\item {\bf Leptonic}: The energy of the lepton $E_l$, the energy of
the neutrino $E_\nu$ and the invariant mass of virtual $W$ boson $q^2$,
which is also the invariant mass of the lepton pair. This choice
focuses on the leptons created in the decay and was used for example
in \cite{Manohar:1993qn}
\item {\bf Partonic}: the energy of the lepton $E_l$, the energy of
the partonic jet $v\cdot p$, and the invariant mass of the partonic jet
$p^2$. This choice focuses on the partons created in the decay and was
used for example in \cite{DeFazio:1999sv}.
\end{itemize}

We will use a different choice of variables, which in some sense is
the optimal one. We start by noting the fact that any measurement of
$\bar B\to X_u\,l\bar\nu$ suffers from a large background from the
decay $\bar B\to X_c\,l\bar\nu$. The reason for the large background
is that $|V_{cb}|\gg |V_{ub}|$, meaning that the $b$ quark ``likes" to
decay to a $c$ quark much more than to a $u$ quark. In order to
eliminate the background we have to look at regions of phase space
where particles containing charm cannot be produced. For example we
can only look at events for which $P_X^2<M_D^2$. This fact implies
that the ``typical" $X_u$ state will have large energy of order $m_b$,
since it originates from a decay of a heavy quark and intermediate
invariant mass, because of the experimental cut. Since $M_D \sim
\sqrt{m_b\Lambda_{\rm QCD}}$ we can write $P_X^2 \sim
m_b\Lambda_{\rm QCD}$. This implies that some of the components of $P_X$
are larger than others. We would like the choice of our variables to
reflect that. In other words we look for two variables, one of which
scales like $m_b$ and one of which scales like $\Lambda_{\rm QCD}$. A pair of
variables which satisfy this condition is:
\begin{equation}
P_- = E_X + |\vec P_X| \,, \qquad P_+ = E_X - |\vec P_X|\,,
\end{equation}   
where $E_X$ and $P_X$ are the energy and momentum of the hadronic jet,
respectively. Note that $P_+P_-=P^2,\, P_{+}+P_-=2E_X$ and $P_+\leq
P_-$. The scaling of these variables will therefore be $P_-\sim m_b$
and $P_+\sim \Lambda_{\rm QCD}$. In order to complete our choice we should
add another variable. Specifying $P_+$ and $P_-$ would determine the
energy and the invariant mass of the lepton pair, but not the
individual energies of the lepton and the neutrino. We therefore have
to add another variable that would distinguish between the two. We
will choose:
\begin{equation}
P_l = M_B - 2 E_l\,, 
\end{equation}  
where $E_l$ is the energy of the lepton. These three variables 
allow us to specify any $\bar B\to X_u\,l\bar\nu$ event.

Having chosen the variables, the next task is to determine to phase
space in terms of $P_+$, $P_l$, and $P_-$. In the rest frame of the
$B$ meson, conservation of energy and momentum gives us the following
equations (the notation is self explanatory):
\begin{eqnarray}
E_l+E_\nu+E_X&=&M_B\\
\label{equation_momenta}\overrightarrow{P_l}+\overrightarrow{P_\nu}+\overrightarrow{P}_X&=&0.
\end{eqnarray}
We take the leptons to be massless, implying
$E_l=|\overrightarrow{P_l}|$ and $E_\nu=|\overrightarrow{P_\nu}|$. The
limits of phase space are determined by the extremal values of the
angles between the momenta. Because of conservation of momentum
(\ref{equation_momenta}) we have only two independent angels.

Let $\theta$ be the angle between $\overrightarrow{P_l}$ and
$\overrightarrow{P_\nu}$. Equation (\ref{equation_momenta}) then
implies $\overrightarrow{P}_X^2=E_l^2+E_\nu^2+2E_lE_\nu \cos
\theta$. Since $-1\leq\cos\theta\leq 1$, we have
$(E_l-E_\nu)^2\leq|\overrightarrow{P}_X|^2\leq (E_l+E_\nu)^2 $. Using
$P_+\leq P_-$, the upper limit gives us $P_-\leq M_B$, and the lower
limit $P_+\leq P_l\leq P_-$.

Let $\alpha$ be the angle between $\overrightarrow{P_l}$ and
$\overrightarrow{P}_X$. Equation (\ref{equation_momenta}) then implies
$E_\nu^2=(M_B-E_l-E_X)^2=E_l^2+|\overrightarrow{P}_X|^2+2E_l|\overrightarrow{P}_X|
\cos \alpha$. Since $-1\leq\cos\alpha\leq 1$, the upper limit gives
$(M_B-P_+)(P_- -P_l)\geq 0$, and the lower limit
$(M_B-P_-)(P_l-P_+)\geq 0$.

We have also an additional constraint from the QCD spectrum. The
lightest state containing a $u$ quark is the pion, so we must have
$M_\pi^2\leq P_X^2=P_+P_-$. (Without this constraint the lower limit
of $P_+$ would be 0). Combining all of these constraints we finally
have:
\begin{equation}
\label{equation_psl}
\frac{M_\pi^2}{P_-}\leq P_+ \leq P_l \leq P_- \leq M_B.
\end{equation}

One of the benefits of this choice of variables (apart from the easy
derivation) is that phase space has an extremely simple form, probably
the simplest form possible (compare for example the phase space limits in 
\cite{DeFazio:1999sv}). This has practical implications: since we
ultimately integrate over phase space, some of the integrations can be
done analytically. We note that a similar choice of
variables would be useful also for $\bar B\to X_c\,l\bar\nu$.

What are the phase space limits for $\bar B\to X_s\gamma$?
Conservation of momentum and energy imply that $M_B-E_X=E_\gamma$ and
$E_\gamma=|\overrightarrow{P}_X|$, or $P_-=M_B$ and
$P_+=M_B-2E_\gamma$. The limit of the phase space for $P_+$ are
determined from $E_\gamma\geq 0$, and the QCD spectrum constraint
$M_{K^*}^2\leq P_X^2=P_+P_-=M_B P_+$, where $K^*(892)$ is the lightest
strange meson that can be produced in this decay. All together we
have:
\begin{equation}
\label{equation_psg}
\frac{M_{K^*}^2}{M_B}\leq P_+ \leq M_B
\end{equation}
Notice the similarities between equation (\ref{equation_psl}) and
(\ref{equation_psg}), especially if we take the heavy quark limit,
which sets the lower limit of $P_+$ to be 0. In the following we will 
explore these similarities extensively. 

Having discussed the kinematics let us say a few words on experimental
cuts and the regions of phase space they probe. For measurement of the
photon spectrum $\bar B\to X_s\gamma$, experiments must impose a cut
on the photon energy of about 1.8 GeV (measured in the rest frame of
the B meson) \cite{Chen:2001fj,Koppenburg:2004fz,Aubert:2005cu}. This
means that the typical values of $P_+=M_B-2E_\gamma$ are of order
$\Lambda_{\rm QCD}$.

For $\bar B\to X_u\,l\bar\nu$, as we mentioned earlier, experimental
cuts are imposed to eliminate the charm background (figure
\ref{fig:PpPm}). Some of the cuts are:
\begin{itemize}
\item Cut on the charged lepton energy $E_l$. If $E_l\geq
(M_B^2-M_D^2)/2M_B\approx 2.31$ GeV, the final hadronic state will
have invariant mass smaller than $M_D$. For this cut $P_l\leq 0.66$
GeV, which implies that $P_+$ is of order $\Lambda_{\rm QCD}$
\item Cut on the hadronic invariant mass $M_X^2$. To eliminate the
charm background we need $M_X^2\leq M_D^2\,$. The cut $M_X^2=M_D^2$ is
depicted as a solid black hyperbola in figure \ref{fig:PpPm}.
\item Cut on the leptonic invariant mass $q^2$. Any cut of the form
$q^2\leq (M_B-M_D)^2 $ would not contain charm events. The cut
$q^2=(M_B-P_+)(M_B-P_-)=(M_B-M_D)^2$ is depicted as a solid blue line
in figure \ref{fig:PpPm}
\item Cut on $P_+$ will be discussed in chapter
\ref{chapter_pert}. The cut $P_+=M_D^2/M_B$ is depicted as a red
dashed line in figure \ref{fig:PpPm}
\end{itemize}
  
\begin{figure}
\begin{center}
\epsfig{file=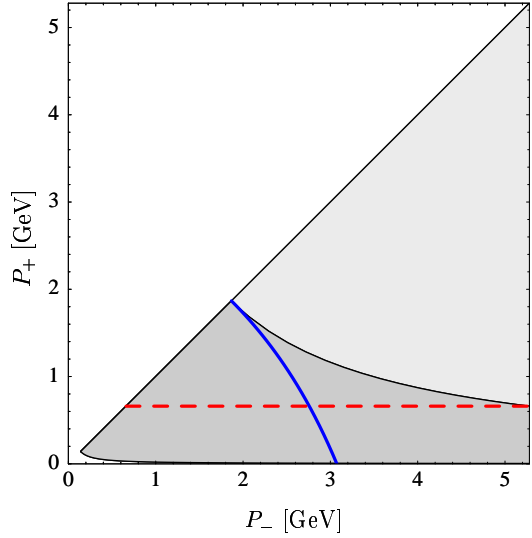,width=7cm}
\caption{The $(P+,P_-)$ phase space for $\bar B\to X_u\,l\bar\nu$. The
charm free region is the dark grey region below the black hyperbola,
which correspond to $M_X^2=P_+P_-=M_D^2$. The solid blue line is
$q^2=(M_B-P_+)(M_B-P_-)=(M_B-M_D)^2$. The red dashed line is
$P_+=M_D^2/M_B$.}
\label{fig:PpPm}
\end{center}
\end{figure}

\section{From the optical theorem to the hadronic tensor}
The basis for the calculation of an inclusive rate is the optical
theorem, which relates the decay rate to the imaginary part of the forward
scattering amplitude. In terms of the $\hat T$ matrix
\cite{Peskin:1995ev}
\begin{equation}
\label{equation_M}
\langle f|\hat T| i\rangle=(2\pi)^4\delta^4(P_i-P_f) {\cal M}(i\to f),
\end{equation} 
the optical theorem states that
\begin{equation}
2\, {\rm Im}\,{\cal M}(i\to i)=\sum_{f} \int\frac{d^3
P_f}{2E_f(2\pi)^3 }\,|{\cal M}(i\to f)|^2\,(2\pi)^4\delta^4(P_i-P_f).
\end{equation} 
Using the optical theorem we obtain the following expression for the
total rate
\begin{eqnarray}
\Gamma&=&\frac 1{2 M_B} \sum_{f}\int \frac{d^3 P_f}{2E_f(2\pi)^3 }|{\cal
M}(i\to f)|^2(2\pi)^4\delta^4(P_i-P_f)\nonumber\\
&=&\frac 1{M_B} {\rm Im}\,{\cal M}(i\to i)
\end{eqnarray}
This formula holds both for $\bar B\to X_u\,l\bar\nu$ and $\bar B\to
X_s\gamma$, but ${\cal M}$ is different in each case.

In order to proceed we need to find the matrix element of the $\hat T$
matrix, which is determined by the Hamiltonian. The relevant general
Hamiltonian contains both weak interactions and strong interactions
(unless otherwise noted, we neglect contributions from 
electromagnetic interactions). For the weak interactions we can use
perturbation theory. At the lowest order in the weak interaction
coupling $g_w$ (or alternatively $G_F$) we have for $\bar B\to X_u\,l\bar\nu$ decays:
\begin{equation}
H_W=\frac{G_F}{\sqrt{2}}V_{ub}\,\bar{u}\gamma^\mu(1-\gamma_5)b\,\bar{l}\gamma_\mu(1-\gamma_5)\nu_l
+{\rm h.c.}
\end{equation}
This Hamiltonian can be thought of as an effective Hamiltonian, obtained (at tree level) after
integrating out the heavy $W$ bosons \cite{Buras:1998ra}. For $\bar B\to X_s \gamma$ we have (see \cite{Buras:1998ra} for derivation)
\begin{equation}
H_W=\frac{G_F}{\sqrt{2}}C^{\rm\,
eff}_{7\gamma}V_{tb}V_{ts}^*\overline{m_b}\left(\frac{-e}{8\pi^2}\right)
\bar{s}\sigma^{\mu\nu}(1+\gamma_5)F_{\mu\nu}b +{\rm h.c.},
\end{equation}
There are other contributions from the weak Hamiltonian to $\bar B\to X_s \gamma$, but their contribution to the rate is suppressed by $g_s^2$ and we will ignore them.
    
Using $\hat S=T\{e^{ -i\int d^4 x{\cal H}}\}$ ($T$ is the time
ordering symbol) and $\hat S=1+i\hat T$, the non vanishing lowest-
order contribution to the $\hat T$ matrix element is given by:
\begin{equation}
\langle \bar B|\hat T|\bar B\rangle=\langle \bar B|\frac i 2 \int
d^4x_1 d^4x_2\, T\{H_W (x_1),\,H_W(x_2)\}|\bar B\rangle.
\end{equation}
If we write $H_W=O+O^\dagger$, the only non vanishing contribution to
the matrix element would come from $T\{O^\dagger,\,O\}$. Using the
delta function in equation (\ref{equation_M}) we can eliminate one of
the $x_i$ integrals to obtain the result:
\begin{equation}
{\cal M}=i\int d^4 x \langle \bar B| T\{O^\dagger(0),\,O(x)\}|\bar
B\rangle.
\end{equation}           
  
This can simplified further, by noting that up to a constant
$O$ can be written as a product of two ``currents":
\begin{equation}
O=C\,\bar q \Gamma b \,J_{NH}\equiv C\,J \,J_{NH}.  
\end{equation} 
For $\bar B\to X_u\,l\bar\nu$ transitions $J=\bar u\gamma^\nu(1-\gamma_5)b$ and
$J_{NH}=\bar{l}\gamma_\nu(1-\gamma_5)\nu_l$, and for $\bar B\to X_s\gamma$ we
have $J=\bar s \sigma^{\beta\nu}(1+\gamma_5)b$ and
$J_{NH}=2\partial_{\beta}A_\nu$. In both cases the fields in $J_{NH}$
do not have any strong interactions, which would allow us to factorize
$\cal M$ to a product of hadronic and non hadronic tensors.
We insert a complete set of states in the form of
\begin{equation}
\sum_{X} |X\rangle\langle X|\sum_{NH}|NH\rangle\langle
NH|\int\frac{d^3 P_{NH}}{2E_{NH}(2\pi)^3 }.
\end{equation}
(The notation is such that the sum over the hadronic states includes the integration over the hadronic momenta)
Thus we get:
\begin{eqnarray}
\label{equation_compset}
{\cal M}&=&i\int d^4 x |C|^2\bigg(\nonumber\\ &&\theta(-x^0) \sum_{X_q,\,
NH}\int\frac{d^3 P_{NH}}{2E_{NH}(2\pi)^3 } \cdot\nonumber\\ &&\langle
\bar B|J^\dagger(0)|X_q\rangle\langle 0|J^\dagger_{NH}(0)
|NH\rangle\langle X_q |J(x)|\bar B\rangle \langle
NH|J_{NH}(x)|0\rangle\nonumber\\ &+&\theta(x^0)\sum_{X_{bb\bar
{q}},\,NH} \int\frac{d^3 P_{NH}}{2E_{NH}(2\pi)^3 }\cdot\nonumber\\
&&\langle \bar B|J(x)|X_{bb\bar{q}}\rangle\langle 0|J_{NH}(x)
|NH\rangle\langle X_{bb\bar{q}}|J^\dagger(0)|\bar B\rangle\langle
NH|J_{NH}^\dagger(0)|0\rangle\bigg)\nonumber\\
\end{eqnarray}
where $X_{q}$ ($X_{bb\bar{q}}$) is a complete set of hadronic states
containing a $q$ quark (two $b$ quarks and a $\bar{q}$). $NH$ is a non
hadronic state, i.e., a photon for $\bar B\to X_s\gamma$ and a lepton
pair for $\bar B\to X_u\,l\bar\nu$.  The second term in equation
(\ref{equation_compset}) would not contribute to $B$ decays, since
after taking the imaginary part one gets a factor of
$\delta^4(P_B-P_{X_{bb\bar{q}}}-q)$ (see appendix
\ref{apx:hadten}). For the first term we distinguish between $\bar
B\to X_s \gamma$ and $\bar B\to X_u\,l\bar\nu$. For $\bar B\to
X_s\gamma$ we have:
\begin{eqnarray}
\sum_{NH}\langle 0|J^\dagger_{NH}(0) |NH\rangle\langle
NH|J_{NH}(x)|0\rangle&=&4\sum_{\lambda}\langle 0|\partial_\alpha
A_\mu(0) |\gamma_\lambda(q)\rangle\langle
\gamma_\lambda(q)|\partial_\beta A_\nu (x)|0\rangle\nonumber\\ &=&
-4q_\alpha q_\beta g_{\mu\nu}e^{iqx},
\end{eqnarray}
where $q$ is the photon momentum. For $\bar B\to X_u\,l\bar\nu$ we have:
\begin{eqnarray}
&&\sum_{NH}\langle 0|J^\dagger_{NH}(0) |NH\rangle\langle
NH|J_{NH}(x)|0\rangle=\nonumber\\ 
&&\sum_{\rm spins}\langle 0
|\bar{v}_{\nu_l}\gamma_\mu(1-\gamma_5)u_l|\bar{\nu}_l(p^{\nu_l})l(p^l)\rangle\langle\bar{\nu}_
l(p^{\nu_l})l(p^l)|\bar{u}_l\gamma_\nu(1-\gamma_5)v_{\nu_l}|0\rangle\nonumber\\
&&8\left(P^l_\mu P^{\nu_{l}}_\nu+P^l_\nu
P^{\nu_{l}}_\mu-g_{\mu\nu}P^l\cdot
P^{\nu_l}-i\epsilon_{\mu\nu\alpha\beta} P_l^\alpha
P_{\nu_l}^\beta\right)e^{iqx},
\end{eqnarray}
where $q=P^{\,l}+P^{\nu_l}$ and we take $\epsilon_{0123}=1$.

Using this we write $\cal M$ as:
\begin{equation}
{\cal M}=\sum_{NH}\int\frac{d^3
P_{NH}}{2E_{NH}(2\pi)^3 } A L_{\mu\nu}\langle \bar B|T^{\mu\nu}|\bar B\rangle.
\end{equation}
The current correlator $T^{\mu\nu}$ is defined as
\begin{equation}
T^{\mu\nu}=i\int d^4 x \, e^{iqx} \,T\{J^{\dagger\mu}(0),J^\nu(x)\},
\end{equation}
where the currents are:
\begin{eqnarray}
J^\mu_{b\to u}&=&\bar u \gamma^\mu(1-\gamma_5) b\nonumber\\
J^\mu_{b\to s}&=&\bar s \,\frac 1 4 [\nbslash,\gamma^\mu](1+\gamma_5) b,
\end{eqnarray}
and we have written the photon momentum as $q^\mu=n\cdot q\frac{\bar{n}^\mu}{2} $ (see below).
We have also defined:
\begin{eqnarray}
A^{b\to u}&=&\frac{G^2_F}{2}|V_{ub}|^2\nonumber\\
A^{b\to s}&=&\frac{G^2_F}{2}|C^{\rm\,
eff}_{7\gamma}|^2|V_{tb}V_{ts}^*|^2\overline{m_b}^2\frac{\alpha_{\rm em}}{16\pi^3},
\end{eqnarray}  
and
\begin{eqnarray}
L^{b\to u}_{\mu\nu}&=&8\left(P^l_\mu P^{\nu_{l}}_\nu+P^l_\nu
P^{\nu_{l}}_\mu-g_{\mu\nu}P^l\cdot
P^{\nu_l}-i\epsilon_{\nu\mu\alpha\beta} P_l^\alpha
P_{\nu_l}^\beta\right)\nonumber\\ 
L^{b\to u}_{\mu\nu}&=&-4g_{\mu\nu}(n\cdot q)^2.
\end{eqnarray}
We should note that in writing $\cal M$ in this form we have used the
fact that the second term in equation (\ref{equation_compset}) does
not contribute to $B$ decays, so we have changed its phase factor to
$e^{iqx}$. 

It is customary to define the hadronic tensor $W^{\mu\nu}$
as:
\begin{equation}
\label{equation_Wmn}
W^{\mu\nu}=\frac 1 \pi \frac{{\rm Im} \langle \bar B |T^{\mu\nu}|\bar B\rangle}{2M_B},
\end{equation}
which allows us to write the differential rate as:
\begin{equation}
\label{equation_dgsemi}
d\Gamma=\sum_{NH}\frac{d^3
P_{NH}}{2E_{NH}(2\pi)^3 }\,2\pi A L_{\mu\nu}W^{\mu\nu}.
\end{equation}
In order to continue we separate the discussion of $\bar B\to
X_s\gamma$ and $\bar B\to X_u\,l\bar\nu$.

For $\bar B\to X_s\gamma$ decays, we have a real photon ($q^2=0$)
recoiling against the hadronic jet. In the rest frame of the $B$ meson
we can take the jet to be moving in the $+z$ direction and the photon
to be moving in the $-z$ direction. The momentum of the $B$ meson in its
rest frame is then $P_B=M_B v$, where $v^\mu=(1,0,0,0)$. We also
introduce two conjugate light-like vectors $n^\mu=(1,0,0,1)$ and
$\bar{n}^\mu=(1,0,0,-1)$, which satisfy $n \cdot \bar n=2,\,n\cdot
v=\bar n\cdot v=1,\,n+\bar n=2v$. Any 4-vector $a^\mu$ can be
decomposed as:
\begin{equation}
a^\mu=\bar n \cdot a \frac{n^\mu}{2}+n \cdot a \frac{\bar{n}^\mu}{2}+a_\perp^\mu\equiv a_-^\mu+a_+^\mu+a_\perp^\mu ,
\end{equation}
where $a_\perp^\mu$ are the $x,y$ components of $a^\mu$ (for our
choice of $n$ and $\bar n$). In terms of these vectors we can write
the photon 4-momentum as $q^\mu=n\cdot
q\frac{\bar{n}^\mu}{2}=2E_\gamma\frac{\bar{n}^\mu}{2}$. We can also
express $P_+$ and $P_-$ as $P_+=n\cdot P_X,\,P_-=\bar{n}\cdot P_X$.

For $\bar B\to X_s\gamma$ the $P_{NH}$ momentum in
(\ref{equation_dgsemi}) is $q$. Integrating over the angles, 
we find that the photon spectrum in $\bar B\to
X_s\gamma$ is:
\begin{equation}
\label{equation_dgs}
d\Gamma=\frac{d^3 q}{2E_\gamma(2\pi)^3 }\,2\pi A L_{\mu\nu}W^{\mu\nu}=
\frac{-E_\gamma^3 dE_\gamma}{4\pi^4}G^2_F |C^{\rm\,
eff}_{7\gamma}|^2|V_{tb}V_{ts}^*|^2\overline{m_b}^2\alpha_{\rm em}W^\mu_\mu,
\end{equation}

We now turn to $\bar B\to X_u\,l\bar\nu$ decays. The $P_{NH}$ momenta
in (\ref{equation_dgsemi}) are $P_l$ and $P_{\nu_l}$. As we saw
before, of the 6 variables only 3 are relevant, which can be taken to
be $E_l=(M_B-P_l)/2$, $E_\nu=(M_B+P_l-P_+-P_-)/2$ and $\cos\theta=
(\overrightarrow{P}_X^2-E_l^2- E_\nu^2)/(2E_lE_\nu)$ ($\theta$ is the
angle between $\overrightarrow{P_l}$ and
$\overrightarrow{P_\nu}$). Integrating over the three irrelevant
angles and changing variables to $P_+,P_l,P_-$. We find that
\begin{equation}
d\Gamma=dP_+dP_l\,dP_-\frac{G^2_F|V_{ub}|^2(P_--P_+)}{256\pi^3} L_{\mu\nu}W^{\mu\nu}.
\end{equation}    
The hadronic tensor has the property (proved in appendix \ref{apx:hadten}) that
$W^{\mu\nu}=(W^{\nu\mu})^*$. Using this identity we can decompose the
tensor as a sum of 5 possible Lorentz structures. Given an independent pair of
4-vectors $a$ and $b$, the 5 structures are $a^\mu a^\nu,b^\mu b^\nu,(a^\mu
b^\nu+a^\nu b^\mu),g^{\mu\nu}$ and
$-i\epsilon^{\mu\nu\alpha\beta}a_\alpha b_\beta$. The coefficients of
these structures are called structure functions, denoted as $W_i$. 

Typical choices for $a$ and $b$ that appear in the literature are $v$
and $q$ \cite{Manohar:1993qn}, and $v$ and $p$
\cite{DeFazio:1999sv}. We will choose $v$ and $n$, which has the
benefit that the coefficients of the structure functions $W_i$ in the
triple decay rate are independent of $m_b$. Decomposing the hadronic
tensor in this basis we write:
\begin{eqnarray}
W^{\mu\nu}&=&(n^\mu v^\nu+n^\nu v^\mu-g^{\mu\nu}-i\epsilon^{\mu\nu\alpha\beta} n_\alpha v_\beta)\tilde{W}_1\nonumber\\
&&-g^{\mu\nu}\tilde{W}_2+v^\mu v^\nu \tilde{W}_3+(n^\mu v^\nu+n^\nu v^\mu)\tilde{W}_4
+n^\mu n^\nu\tilde{W}_5
\end{eqnarray}
The reason for defining the coefficient of $\tilde{W}_1$ in this way
is that then at tree level only $\tilde{W}_1$ is non zero. The
structure functions $\tilde{W}_i$ are (generalized) functions of $P_+$
and $P_-$ (or $n\cdot p$ and $\bar{n}\cdot p$). Having decomposed the
hadronic tensor, we can write the triple rate in terms of the
structure functions $\tilde{W_i}$. In the expression
$L_{\mu\nu}W^{\mu\nu}$ we have several scalar products: $v\cdot P_l,
v\cdot P_{\nu_l}, n\cdot P_l, n\cdot P_{\nu_l},P_l \cdot
P_{\nu_l}$. Using $P_{\nu_l}=M_Bv-P_X-P_l$ we can eliminate all the
scalar products that contain $P_{\nu_l}$. Of the remaining scalar
products $v\cdot P_l=E_l$ and $n\cdot P_l$ can be obtained from
$P_X\cdot P_l$, via $P_X\cdot P_l=[P_-(n\cdot P_l)+P_+(2E_l-n\cdot
P_l)]/2$ (since $P_X$ has no $\perp$ components). After a little bit
of algebra we find the following compact result:
\begin{eqnarray}
\label{equation_dgu}
   \frac{d^3\Gamma}{dP_+\,dP_-\,dP_l}
   &=& \frac{G_F^2|V_{ub}|^2}{16\pi^3}\,(M_B-P_+)\,
    \Big[ \nonumber\\
    &&(P_- -P_l)(M_B-P_- +P_l-P_+)\,\tilde{W}_1 
   + (M_B-P_-)(P_- -P_+)\,\frac{\tilde{W}_2}{2} \nonumber\\
   &+& (P_- -P_l)(P_l-P_+)\left(\frac y 4 \tilde{W}_3+\tilde{W}_4 +\frac 1 y \tilde{W}_5\right) 
   \Big] \,,
\end{eqnarray}
where we have defined
\begin{equation}
   y = \frac{P_- -P_+}{M_B-P_+}\,.
\end{equation}  
Notice that the triple rate only depends on one linear combination of
$\tilde{W}_3,\,\tilde{W}_4$ and $\tilde{W}_5$. The reason is that in
the limit $M_l \to 0$ (justified for electrons and muons) we have the
two constraints $q^\mu L_{\mu\nu}=q^\nu L_{\mu\nu}=0$.

Equations (\ref{equation_dgs}) and (\ref{equation_dgu}) are the main
results of this section. Of course they are not useful without the
knowledge of $W^{\mu\nu}$. The (approximate) calculation of the hadronic
tensor will be one of the main topics of this work. In the rest of the
chapter we discuss mostly $\bar B\to X_u\,l\bar\nu$ and mention in passing
what are the equivalent formulas for $\bar B\to X_s\gamma$.

\section{Free quark approximation - tree level and $\alpha_s$ corrections}
In calculating the hadronic tensor, 
the simplest approximation we can make is to look at the $b$ quark as
a free quark, ignoring the spectator quark in the $B$ meson and the
non perturbative interactions that binds them together. As we shall
see in the next section, this approximation would be the zeroth order
term in an HQET based local Operator Product Expansion (OPE).

In this approximation the $B$ meson states can be written as $|\bar
B\rangle=|b(m_b v)\rangle$. Writing the current as $J^\nu=\bar{q}\Gamma^\nu b$, the matrix element of $T^{\mu\nu}$ is:
\begin{eqnarray}
&&\langle b(m_b v) |T^{\mu\nu}|b(m_b v)\rangle=i\int d^4 x \, e^{iqx}
\,\langle b|T\{J^{\dagger\mu}(0),J^\nu(x)\}|b\rangle\nonumber\\
&=&\frac{i}{2}\sum_{b\;{\rm quark\, spin}}\int d^4 x \, e^{i(q-m_b v)\cdot x}
\,\bar{u}_b(m_b v)\bar\Gamma^\mu
T\{\,q(0),q(x)\}\Gamma^\nu u_b(m_b v)\nonumber\\
&=&-m_b\int d^4 p\,\delta^4(m_bv-q-p){\rm
Tr}\left\{\bar\Gamma^\mu\frac{\pslash}{p^2+i\epsilon}\Gamma^\nu\frac{1+\vslash}{2}\right\},
\end{eqnarray}
where in the second line we have averaged over the $b$ quark spins and  
$\bar\Gamma^\mu=\gamma^0\Gamma^{\mu\dagger}\gamma^0$.
  
Using the identity \cite{Gel'fand}
\begin{equation}
\label{equation_Impart}
{\rm Im}\, \frac 1 \pi (x\pm i\epsilon)^{-n}=\mp\frac{(-1)^{n-1}}{(n-1)!}\delta^{(n-1)}(x),
\end{equation}
we find that
\begin{equation}
\label{equation_trace}
W^{\mu\nu}=\frac{\delta(p^2)}{2} {\rm Tr}\left\{\bar\Gamma^\mu\,\pslash\,\Gamma^\nu \frac{1+\vslash}{2} \right\}.
\end{equation}
(Notice that for the free quark approximation $m_b\sim M_B$).
For $\bar B\to X_u\,l\bar\nu$ decays we can write $\delta(p^2)$ 
as $\delta(n\cdot p)/(\bar{n}\cdot p)$. Since 
\begin{equation}
\delta(n\cdot p)\,\pslash=\delta(n\cdot p)\left(\bar n \cdot p
\frac{n^\mu}{2}+n \cdot p \frac{\bar{n}^\mu}{2}\right)=\delta(n\cdot
p)\left(\bar n \cdot p \frac{n^\mu}{2}\right),
\end{equation}    
and $\Gamma^\mu=\gamma^\mu(1-\gamma^5)$ we find
\begin{equation}
W^{\mu\nu}=\delta(n\cdot p)(n^\mu v^\nu+n^\nu
v^\mu-g^{\mu\nu}-i\epsilon^{\mu\nu\alpha\beta} n_\alpha
v_\beta)\Rightarrow\tilde{W}_1=\delta(n\cdot p).
\end{equation} 
This simple result is the reason that $\tilde{W}_1$ is defined as it
is. 

As a quick check let us plug this result into
(\ref{equation_dgu}). Since $P_+=n\cdot p+\bar{\Lambda}$, we can
integrate over the total phase space to find the well known result,
that total rate at tree level is $G_F^2|V_{ub}|^2m_b^5/(192\pi^3)$.

For $\bar B\to X_s\gamma$ decays we have $\bar\Gamma^\mu=\frac14[\gamma^\mu,\nbslash](1-\gamma_5)$ and 
$\Gamma^\nu=\frac14[\nbslash,\gamma^\nu](1+\gamma_5)$. Calculating the trace in equation 
(\ref{equation_trace}) we find that 
\begin{equation}
W^\mu_\mu=-2\,\delta(n\cdot p).
\end{equation}
Using $n\cdot p=n\cdot (m_bv-q)=m_b-2E_\gamma$, we can plug this
result in equation (\ref{equation_dgs}) and find that the total $\bar
B\to X_s\gamma$ rate at tree level is $G_F^2|V_{tb}V^*_{ts}|^2
|C^{\rm\, eff}_{7\gamma}|^2 \alpha_{\rm em}\, m_b^5/(32\pi^4)$

A much more complicated calculation is to find the ${\cal O}
(\alpha_s)$ corrections to the structure functions $\tilde{W}_i$, in
the free quark approximation. Such a calculation was preformed by De
Fazio and Neubert in \cite{DeFazio:1999sv}. The hadronic tensor in
\cite{DeFazio:1999sv} was decomposed using the vectors $p$ and $v$:
\begin{eqnarray}
W^{\mu\nu}&=&(p^\mu v^\nu+p^\nu v^\mu-g^{\mu\nu}v\cdot p-i\epsilon^{\mu\nu\alpha\beta} p_\alpha v_\beta)W_1\nonumber\\
&&-g^{\mu\nu}W_2+v^\mu v^\nu W_3+(p^\mu v^\nu+p^\nu v^\mu)W_4
+p^\mu p^\nu W_5
\end{eqnarray}
The relation to our basis is:
\begin{eqnarray}
&&\tilde{W}_1=\frac{\bar n \cdot p-n\cdot p}{2}\,W_1 \quad\quad\quad\quad\quad\quad\quad\; \tilde{W}_2=W_2+n\cdot p\, W_1\nonumber\\
&&\tilde{W}_3=W_3+2n\cdot p \left(W_1+W_4+n\cdot p\, W_5\right)\\
&&\tilde{W}_4=\frac{\bar n \cdot p-n\cdot p}{2}\left(W_4+n\cdot p\, W_5\right)\quad\quad
\tilde{W}_5=\left(\frac{\bar n \cdot p-n\cdot p}{2}\right)^2 W_5\nonumber
\end{eqnarray}
The kinematical variables $z,t,\hat{p}^2$ and $x$ in
\cite{DeFazio:1999sv} are related to ours via:\\ $z=(\bar n \cdot
p+n\cdot p)/m_b,\,t=(\bar n \cdot p-n\cdot p)/(\bar n \cdot p+n\cdot
p),\,\hat{p}^2=(\bar n \cdot p)( n\cdot p)/m_b^2$, and
$x=(M_B-P_l)/m_b$, where again $P_+=n\cdot
p+\bar{\Lambda},\,P_-=\bar{n}\cdot p+\bar{\Lambda}$.

We will not transform the complete ${\cal O} (\alpha_s)$ result from
\cite{DeFazio:1999sv} to our basis. Instead, we would like to
illustrate the well known problem with the QCD calculation, namely the
appearance of double and single logarithms that might spoil the convergence
of the perturbative expansion. Let us look on one of the terms of
$W_1$:
\begin{equation}
W_1=\frac 2
{m_b^2}\left\{\delta(\hat{p}^2)+\frac{C_F\alpha_s}{4\pi}\left[-4\left(\frac{\ln\hat{p}^2}{\hat{p}^2}\right)^{[1]}_*+\dots\right]\right\},
\end{equation}
where the ``star" distribution is defined in chapter \ref{chapter_pert}
and the ``$\dots$" denote other ${\cal} O(\alpha_s)$ terms. In terms
of our basis we have:
\begin{equation}
\label{equation_Wlog2}
\tilde{W_1}=\delta(n \cdot
p)+\frac{C_F\alpha_s}{4\pi}\left\{-4\,\frac{\bar n \cdot p-n\cdot
p}{\bar n \cdot p}\left[\frac{\ln(n \cdot p/m_b)}{n \cdot
p}\right]^{[m_b]}_*+\dots\right\},
\end{equation}
The star distribution has two parts. The second part, which is
proportional to a delta function, is the source of the double
log. Inserting equation (\ref{equation_Wlog2}) into
(\ref{equation_dgu}) we have:
\begin{eqnarray}
\label{equation_d3log2}
\frac{d^3\Gamma}{dP_+\,dP_-\,dP_l}
   &=& \frac{G_F^2|V_{ub}|^2}{16\pi^3}\,(M_B-P_+)\,
     (P_- -P_l)(M_B-P_- +P_l-P_+)\nonumber\\
     &&\left\{\delta(n \cdot p)-\frac{C_F\alpha_s}{4\pi}\left[\frac{\delta(n \cdot p)}{2}\ln^2\frac {u}{m_b} +\dots\right]\right\}.
\end{eqnarray} 
In a somewhat implicit notation, $u$ is the upper limit of $n\cdot
p$ in the $P_+$ integration and depends on the specific experimental cut. As an
example, let us find the $P_l$ spectrum. The integral over $P_-$ is
automatic. For the $P_+$ integral we note that the upper limit of $n \cdot p$
is $u=P_l-\bar{\Lambda}$. Changing variables to $x=(M_B-P_l)/m_b$ we
find that:
\begin{equation}
\frac{d\Gamma}{dx}=\frac{G_F^2|V_{ub}|^2 m_b^5}{192\pi^3}2x^2(3-2x)\left\{1-\frac{C_F\alpha_s}{2\pi}\left[\ln^2(1-x)+\dots\right]\right\}
\end{equation}
which was first derived in \cite{Jezabek:1988ja}. It means that for high
values of $E_l$, i.e. $x\to 1$, the so called ``endpoint region", the double log gets
large. Notice that equation (\ref{equation_d3log2}) implies that this
is a generic problem. Anytime the value of $u$ is not of order $m_b$,
the double log would not be small. But, because of the experimental
cuts the values of $u$ are never of order $m_b$! A similar situation
arises for the single logarithms. Since these logarithms are multiplied with
$\alpha_s$, we might question the convergence of the perturbative
series.

The same double logarithms appear for $\bar B\to X_s\gamma$. From
\cite{Chetyrkin:1996vx} and earlier references cited there, we find
that the decay rate for $\bar B\to X_s\gamma$ with a lower cut on the
photon spectrum $E_\gamma>(1-\delta)m_b/2$ is:
\begin{equation}
\Gamma\left[\bar B\to X_s\gamma\right]_{E_\gamma>(1-\delta)m_b/2}=\frac{G_F^2\alpha_{\rm em}}{32\pi^4}|V_{tb}V^*_{ts}|^2 |C^{\rm\,
eff}_{7\gamma}|^2 \, m_b^5\left\{1-\frac{C_F\alpha_s}{2\pi}\left[\ln^2\delta+\dots\right]\right\}.
\end{equation}
Because of the experimental cut $\delta\sim\Lambda_{\rm QCD}/m_b$, so we encounter large logarithms again.
(The universality of the double logarithms is not a coincidence and has its origin in the properties of soft Wilson loops, see below).

Having large logarithms is of course a well known problem, encountered, for example, when
one takes into account the QCD corrections to weak decays
\cite{Buras:1998ra}. The solution to the problem is to ``resum" these
logarithms. This is done by matching to an appropriate effective field
theory and then solving the RG equations. As we shall see,
the appropriate effective field theory is SCET.

One more comment before we conclude this section. Usually when one
talks about ``large logarithms" the question arises ``how large are the logarithms?
do we need resummation?"
In order to answer this question we should note that the scale
dependence of $\alpha_s(\mu)$ is unspecified at this order. Some
prefer to set $\mu\sim m_b$, but as we shall see, the more appropriate
scale for these logarithms is $\mu\sim \sqrt{m_b \Lambda_{\rm QCD}}\sim 1.5$
GeV. Recall that because of QCD running
$\alpha_s(M_Z)=0.112,\alpha_s(m_b)=0.22,\alpha_s(1.5 {\rm GeV})\sim
0.375$. So we have the effect of  logarithms multiplied by large value of
$\alpha_s$. This further motivates the need for resummation.

\section{Effective field theories I: HQET}
Effective field theory (EFT) is a very useful tool in general and in 
particular for $B$ physics. It allows us to extract as much
perturbative physics as possible from the matrix elements, and to break down the problem of ``solving QCD" into manageable pieces,
that we can try to model or extract from experiment.

For our specific goal,  the calculation of the hadronic tensor,
given an effective field theory the procedure is as follows:
\begin{itemize}
\item A QCD current $J$ is matched onto a series of the form $J_{\rm
QCD}=C_0J_0+C_1J_1+ \nolinebreak[4]\dots\-$, where the Wilson
coefficients $C_i$ are calculable as an expansion in $\alpha_s$ and each $J_i$ is suppressed by $i$-th powers of a large scale.
\item The QCD Lagrangian is matched onto a series of the form ${\cal L}={\cal L}_0+{\cal L}_1+\dots$.
\item The $|\bar B\rangle$ meson states are matched onto eigenstates of ${\cal L}_0$, called $|B\rangle_0$ .
\item $T^{\mu\nu}$ can then be written as a series 
\begin{eqnarray*}
&&T\Big\{\sum_{i} C_iJ^\dagger_i,\sum_i C_iJ_i,e^{i\int d^4y {\cal L}_1+\dots}\Big\}=T\Big\{C_0J^\dagger_0,C_0J_0\Big\}+\\
&&+T\Big\{C_1J^\dagger_1,C_0J_0\Big\}+T\Big\{C_0J^\dagger_0,C_1J_1\Big\}+T\Big\{C_0J^\dagger_0,C_0J_0,i\int d^4y {\cal L}_1\Big\}+\dots  
\end{eqnarray*}
\item $W^{\mu\nu}$ is then written as a series of non perturbative matrix elements with calculable coefficients. 
\end{itemize} 
The effective field theories that we will use HQET and SCET, are
based on an expansion in inverse powers of the $b$ quark mass. Since
the hadronic matrix elements scale like $\Lambda_{\rm QCD}$ and
$\Lambda_{\rm QCD}<<m_{b}$, our expansion parameter will be
$\Lambda_{\rm QCD}/m_b\sim 0.1$.

We start by looking at HQET and review its predictions for the hadronic
tensor. In order to do so we go over some of the elements HQET that are
important for our purpose. For an extensive review of HQET see
\cite{Neubert:1993mb,Manohar:2000dt}. We start by writing the $b$
field as:
\begin{equation}
b(x)=e^{-im_b v\cdot x}\left[h_v(x)+H_v(x)\right], 
\end{equation}
where 
\begin{equation}
h_v(x)=e^{im_b v\cdot x}\frac{1+\vslash}{2}b(x),\quad H_v(x)=e^{im_b v\cdot x}\frac{1-\vslash}{2}b(x).
\end{equation} 
In terms of these fields the QCD Lagrangian for the $b$ quark is 
\begin{eqnarray}
\label{equation_QCD2HQET}
&&{\cal L}=\bar{b}(i\Dslash-m_b)b\nonumber\\
&&=\bar{h}_v iv\cdot D h_v-\bar{H}_v(iv\cdot D+2m_b)H_v+\bar{h}_vi\vec{\Dslash}H_v+\bar{H}_vi\vec{\Dslash}h_v,
\end{eqnarray}
where $\vec{D}^\mu=D^\mu-v^\mu v\cdot D$. From equation
(\ref{equation_QCD2HQET}) it is clear that the field $H_v$ is to be
integrated out. In zeroth order we find that
\begin{equation}
{\cal L}_0=\bar{h}_v iv\cdot D h_v.
\end{equation}
For the light quarks and gluons there is no
change, so we can write the leading order current as:
\begin{equation}
\bar{q}\,\Gamma^\nu b(x)\to J^\nu_0=e^{-im_b v\cdot x}\,\bar{q}\,\Gamma^\nu h_v(x).
\end{equation}
Armed with this information we can calculate $W^{\mu\nu}$ at tree level:
\begin{eqnarray}
T^{\mu\nu}&=&i\int d^4 x \, e^{iqx}
\,T\{J^{\dagger\mu}(0),J^\nu(x)\}\nonumber\\\nonumber &=&i\int d^4 x \,
e^{i(q-m_b v)\cdot x}
\,T\{\,\bar{h}_v\bar\Gamma^\mu q(0),\bar{q}\,\Gamma^\nu h_v(x)\}\\
&=&-\int d^4
p\,\delta^4(m_bv-q-p)\bar{h}_v\bar\Gamma^\mu\frac{\pslash}{p^2+i\epsilon}\Gamma^\nu
h_v
\end{eqnarray}
In order to find the matrix element of $T^{\mu\nu}$ we need \cite{Neubert:1993mb}:
\begin{equation}  
\langle \bar{B}|\,\bar{h}_v\Gamma h_v \,|\bar{B}\rangle=M_B \,{\rm Tr}\left\{\Gamma\frac{1+\vslash}{2} \right\}.
\end{equation}
Proceeding as before we find that the zeroth order HQET result is identical to the free quark result, namely:
\begin{equation}
\tilde{W}_1=\delta(n\cdot p),\quad W^\mu_\mu=-2\,\delta(n\cdot p)
\end{equation} 
The $\alpha_s$ corrections to this zeroth order result would give us the De Fazio-Neubert hadronic tensor.

The reason for using HQET is to calculate the $1/m_b^n$ corrections to
the hadronic tensor. Such a calculation was performed in
\cite{Manohar:1993qn,Blok:1993va} for $\bar B\to X_c\,l\bar\nu$. We
will present the results as they appear in \cite{Manohar:2000dt} after
taking the limit $m_c\to 0$. These corrections are known only at
${\cal O}(\alpha_s^0)$.

First let us look at the $1/m_b$ term in the HQET Lagrangian:
\begin{equation}
{\cal L}_1=-\bar{h}_v \frac{\vec{D}^2}{2m_b}h_v\,-g\bar{h}_v\frac{\sigma_{\mu\nu}G^{\mu\nu}}{4m_b}h_v
\end{equation}
The matrix elements of the operators in ${\cal L}_1$ define two HQET
parameters:
\begin{eqnarray}
\langle \bar{B}|\bar{h}_v \vec{D}^2 h_v|\bar{B}\rangle&=&-2 M_B\lambda_1\nonumber\\
\langle \bar{B}|\bar{h}_v \sigma_{\mu\nu}G^{\mu\nu} h_v|\bar{B}\rangle&=&-12 M_B\lambda_2
\end{eqnarray}
Intuitively we can think of $\lambda_1$ as measuring the kinetic
energy of heavy quark inside the $B$ meson and of $\lambda_2$
measuring its chromomagnetic interaction. Both parameters have mass
dimension 2 and a priori we expect $\lambda_i\sim\Lambda_{\rm QCD}^2$.

Instead of matching the current $J^\mu$ onto HQET current, the current
correlator $T^{\mu\nu}$ itself can be matched onto HQET. Doing so one
finds that the zeroth order term is identical to the free quark
result. The $1/m_b$ correction to $T^{\mu\nu}$ vanishes. The reason is
that the only operator that can appear at this order is $v^\alpha
\bar{h}_v iD^\beta h_v$. At order ${\cal O}(\alpha_s^0)$ the matrix
element of this operator would vanish because of HQET equations of
motion \cite{Chay:1990da}. This implies that the first non zero
correction appears at order $1/m_b^2$.

The hadronic tensor in \cite{Manohar:2000dt} is defined slightly
different from ours:
\begin{equation}  
\bar{W}_{\mu\nu}=-\frac1\pi\frac1{2M_B} {\rm Im} \int d^4x (-i )e^{-iqx} 
\langle \bar B |\, T\{\bar{J}^{\dagger}_\mu(x),\bar{J}_\nu (0)\}|\bar B\rangle
\end{equation} 
where $\bar{J}_\mu=\bar{u}\gamma_\mu(1-\gamma_5)b/2$. This hadronic
tensor is decomposed using the vectors $q$ and $v$:
\begin{equation}
\bar{W}_{\mu\nu}=-g_{\mu\nu}\bar W_1+v_\mu v_\nu \bar
W_2-i\epsilon_{\mu\nu\alpha\beta}v^\alpha q^\beta \bar{W}_3+q_\mu
q_\nu \bar W_4+\left(v_\mu q_\nu+v_\nu q_\mu\right)\bar W_5
\end{equation}
The relation to our basis is:
\begin{eqnarray}
&&\tilde{W}_1=4\frac{\bar{n}\cdot p -n\cdot p}{2} \bar W_3 \quad \quad
\tilde{W}_2=4\left(\bar{W}_1-\frac{\bar{n}\cdot p -n\cdot p}{2} \bar
W_3\right) \nonumber\\ 
&&\frac y 4 \tilde{W}_3+\tilde{W}_4 +\frac 1 y
\tilde{W}_5=\left(\bar{n}\cdot p -n\cdot p \right)\left(\frac{\bar
W_2}{m_b-n\cdot p}-2\bar {W}_3\right)
\end{eqnarray}
After the change of basis we find, including the zeroth order result
$\tilde{W}_1=\delta(n\cdot p)$,
\begin{eqnarray}\label{equation_OPE}
   \tilde W_1
   &=& \delta(n\cdot p) \left( 1 + \frac{2\lambda_1-3\lambda_2}{3(\bar{n}\cdot p)^2} \right) 
    + \delta'(n\cdot p) \left( \frac{2\lambda_1-3\lambda_2}{3\bar{n}\cdot p} 
    - \frac{5\lambda_1+15\lambda_2}{6m_b} \right)\nonumber \\  
    &-& \delta''(n\cdot p)\,\frac{\lambda_1}{6} \,, \nonumber \\ 
   \tilde W_2   &=& \delta(n\cdot p) \left( - \frac{4\lambda_1-6\lambda_2}{3(\bar{n}\cdot p)^2} \right) , \nonumber\\
   \frac{y}{4}\,\tilde W_3 &+& \tilde W_4
    + \frac{1}{y}\,\tilde W_5
   = \frac{\delta(n\cdot p)}{\bar{n}\cdot p} \left( \frac{2\lambda_1+12\lambda_2}{3\bar{n}\cdot p}
    - \frac{4\lambda_1+9\lambda_2}{3m_b} \right) \nonumber \\
    &+& \frac{\delta'(n\cdot p)}{\bar{n}\cdot p} \left( \frac{2\lambda_1}{3} 
    + 4\lambda_2 \right) . 
\end{eqnarray}

As we now show this calculation poses a problem when
applied for charmless semileptonic decays. Define
$\lambda\sim\Lambda_{\rm QCD}/m_b$; using this small parameter we can
define a systematic power counting. For example $m_b$ would be an
order 1 quantity, while the HQET parameters $\lambda_1,\lambda_2$ are
order $\lambda^2$.

For $\bar B\to X_c\,l\bar\nu$ both $n\cdot p$ and $\bar{n}\cdot p$
scale like $m_b$. This means that $\delta(n\cdot p)$ and its
derivatives are order 1 quantities. Considering the scaling of the HQET
parameters, we have an expansion of the form $W_i \sim
1+\lambda^2+\dots$. We can imagine extending this calculation to
higher order in $\lambda$, by introducing more HQET parameters. Each
term in the expansion would be suppressed by more powers of $\lambda$.

The situation is fundamentally different for $\bar B\to
X_u\,l\bar\nu$. We recall that because of the kinematical cuts
enforced on us due to the large charm background, $n \cdot p$ never
really scales like $m_b$. In fact, it scales like $\Lambda_{\rm
QCD}$. The appropriate power counting for $\bar B\to X_u\,l\bar\nu$ is
therefore $n \cdot p \sim \lambda$ and $\bar n \cdot p \sim 1$. Since
the n-th derivative of $\delta (x)$ scales like $1/x^{n+1}$ we
have three types of terms in (\ref{equation_OPE}): terms which scale
like $\lambda^{-1}$, e.g.  $\delta(n\cdot p), \delta''(n\cdot
p)\lambda_i$; terms which scale like $\lambda^{0}$, e.g.
$\delta'(n\cdot p)\lambda_i$; and terms which scale like
$\lambda^{1}$, e.g.  $\delta(n\cdot p)\lambda_1$. It follows that the
expansion in (\ref{equation_OPE}) must be rearranged as:
\begin{eqnarray}
\label{equation_OPEre}
  \tilde W_1 &=& \left[\delta(n\cdot p) - \delta''(n\cdot
   p)\,\frac{\lambda_1}{6} +\dots\right]\nonumber\\ &+&\left[\delta'(n\cdot p)
   \left( \frac{2\lambda_1-3\lambda_2}{3\bar{n}\cdot p} -
   \frac{5\lambda_1+15\lambda_2}{6m_b}
   \right)+\dots\right]\nonumber\\ &+&\left[ \delta(n\cdot p) \left(
   \frac{2\lambda_1-3\lambda_2}{3(\bar{n}\cdot p)^2} \right)+\dots
   \right] , \nonumber \\ \tilde W_2 &=&
   0+0+\left[\delta(n\cdot p) \left( -
   \frac{4\lambda_1-6\lambda_2}{3(\bar{n}\cdot p)^2} \right)+\dots
   \right], \nonumber\\ \frac{y}{4}\,\tilde W_3 + \tilde W_4 +
   \frac{1}{y}\,\tilde W_5 &=& 0+\left[\frac{\delta'(n\cdot
   p)}{\bar{n}\cdot p} \left( \frac{2\lambda_1}{3} + 4\lambda_2
   \right)\dots\right]\nonumber\\ &+&\left[\frac{\delta(n\cdot
   p)}{\bar{n}\cdot p} \left(
   \frac{2\lambda_1+12\lambda_2}{3\bar{n}\cdot p} -
   \frac{4\lambda_1+9\lambda_2}{3m_b} \right)+\dots \right]
   .
\end{eqnarray}
Our notation is such that in each square bracket we group terms which
scale the same way. (Notice that in $\tilde{W}_2$, for example, the
lowest order terms are of order $\lambda$). Even after this
rearranging we have a problem. By matching to higher orders in the
HQET expansion and by introducing more HQET parameters, some of the
terms might not be suppressed at all and will contribute to lower
orders. In fact, in each order in $\lambda $ we have to sum {\em
infinite} number of terms that contribute at that order. (We have
included $...$ in each square bracket to reflect this fact). This
problem was first observed in
\cite{Neubert:1993ch,Neubert:1993um,Bigi:1993ex} and is often referred
to as ``breaking down" of the OPE.  A more accurate description would
be to say that we have tried to use an HQET based OPE in a kinematical
region where it is not valid. This OPE assumes that all the components
of $p$ are large compared to $\Lambda_{\rm QCD}$, which is true for
$\bar B\to X_c\,l\bar\nu$ but not for $\bar B\to X_u\,l\bar\nu$.

Let us look at the lowest order in $\lambda$, i.e. the order
$\lambda^{-1}$ in equation (\ref{equation_OPEre}). The infinite number
of terms that are summed give us a new non perturbative object which
is no longer just a matrix element of an HQET operator, but a function
of the variable $n\cdot p$. This function is known as the leading
order shape function, denoted as $S$. Beyond leading order there are
more than one shape function and one would talk about several
subleading shape functions. This analysis is the subject of chapter
\ref{chapter_sub}.

The fact that we had to do this resummation implies that HQET is not
the most appropriate effective field theory to describe charmless
inclusive $B$ decays. We will introduce the more appropriate theory (SCET) in
the next section. As chapters \ref{chapter_pert} and
\ref{chapter_sub} will show, in this effective field theory the
various shape functions arise naturally as non local matrix elements
and there is no need to resum the HQET OPE. For the rest of this
section let us talk a little more about the leading order calculation. 

At ${\cal O}(\alpha_s^0)$ the leading order shape function can be defined as the sum of a series of distributions: 
\begin{equation}
\label{equation_sumdelta}
S(\omega)=\delta(\omega)-\frac{\lambda_1}{6}\delta''(\omega)+\dots
\end{equation}
(When no confusion can arise we will refer to the leading
order shape function simply as the ``shape function").
Moments of this function would then be related to HQET parameters. In
order to find the relation, let us write a function $f(x)$ as a sum
of infinite series of the form $f(x)=\sum_i a_i\delta^{(i)}(x)$. Using
``integration by parts" \cite{Gel'fand} we find that the n-th moment is
$\int dx \,x^n f(x)=(-1)^n n!\, a_n$. Equation
(\ref{equation_sumdelta}) implies that the first three moments of the
leading order shape function are $1,0$, and $-\lambda_1/3$.

In \cite{Neubert:1993um} it was shown that we can write the leading
order shape function as:
\begin{equation}
\frac{\langle \bar{B}|\bar{h}_v \delta(\omega-in\cdot D) h_v|\bar{B}\rangle}{2M_B}=S(\omega).
\end{equation}
Notice that the shape function is defined as a matrix element of a non
local operator.  Another important property derived in
\cite{Neubert:1993um} is the support of the shape function. In the
absence of radiative corrections (see chapter \ref{chapter_pert}) we
can think of the shape function as a probability distribution for the
$b$ quark to have residual light cone momentum $\omega\equiv n\cdot
k$. For the total $b$ quark light cone momentum fraction we have
\begin{equation}
\frac{n\cdot p_b}{n\cdot p_B}=\frac{n\cdot (m_b v+k)}{M_B}=\frac{m_b+n\cdot k}{M_B}.
\end{equation}
Since $0\leq\frac{n\cdot p_b}{n\cdot p_B}\leq 1$, (the limits corresponding to
the $b$ quark carrying none or all of the $B$ meson light cone
momentum), we have $-m_b\leq k_+\leq M_B-m_b$. In the $m_b\to \infty$
limit this correspond to $-\infty\leq n\cdot k\leq \bar{\Lambda}$,
where at tree level $\bar\Lambda=\lim_{m_b\to \infty}(M_B-m_b)$. In
chapter \ref{chapter_pert} we will define $\bar\Lambda$ beyond tree
level.

So far we have ignored the $\alpha_s$ corrections. How does the
picture change when we include them? The answer was given by
Korchemsky and Sterman in \cite{Korchemsky:1994jb} who argued that at
leading order a factorization formula holds. The factorization formula
states that the differential rate can be written as a product of
``hard" function and a convolution of a ``jet" function with the leading
order shape function, where both the hard and the jet function are calculable 
in perturbation theory. We will prove this factorization formula using
SCET in chapter \ref{chapter_pert}. At this stage we can illustrate
this formula by using, yet again, the free quark approximation. Let us
write $\tilde{W}_1$ as:
\begin{equation}
\tilde{W}_1^{(0)}=\delta(n\cdot p)=1\cdot \int d\omega \,\bar n\cdot p \, \delta [\,\bar n\cdot p\,(n\cdot p +\omega)]\,\delta (\omega)  
\end{equation} 
We now identify the term $\delta(\omega)$ with the first term in the
moment expansion of $S(\omega)$. We also define a hard function as
$H_{u1}(\bar{n} \cdot p)=1+{\cal O}(\alpha_s)$ and a jet function as
$J(u)=\delta(u)+{\cal O}(\alpha_s)$. The factorization formula is
then:
\begin{equation}
\label{equation_factu}
\tilde{W}_1=H_{u1}\cdot \int d\omega \,\bar n\cdot p \, J[\bar n\cdot p\,(n\cdot p +\omega)]\,S(\omega),
\end{equation} 
or symbolically $W_{i}\sim H_u\cdot J\otimes S$.

We can motivate this procedure in the following way. At tree level we
can think of the jet function as related to the imaginary part of the
$u$ quark propagator. Not neglecting the residual momentum $k$ of the
$b$ quark, the light quark propagator is:
\begin{equation}
i\bar{n} \cdot p\frac {\nslash}{2}\cdot\frac{1}{\bar n\cdot p\,(n\cdot p +n\cdot k)+i\epsilon}
\end{equation}      
Expanding this propagator as a power series in $n\cdot k$ would generate the leading order shape function $S(n\cdot k)$. Following this procedure we can ``implement" the shape function in any parton level calculation as:
\begin{equation}
\tilde{W}_i=f(\bar n \cdot p, n\cdot p)\to \int d\omega\,f(\bar n \cdot p, n\cdot p+\omega)S(\omega)
\end{equation}
Such procedure was used for example in \cite{DeFazio:1999sv}. We should stress though, that this procedure is justified at ${\cal O}(\alpha_s^0)$ only. A rigorous implementation of the shape function at ${\cal O}(\alpha_s)$ appears in chapter \ref{chapter_pert}.

For $\bar B\to X_s\gamma$ we find similar features. The HQET calculation preformed 
in \cite{Falk:1993dh} gives:
\begin{equation}
W^\mu_\mu=
-2\left\{\left[\delta(n\cdot p)-\frac{\lambda_1} 6 \delta''(n\cdot p) \right]+
\left[\frac{\lambda_1-3\lambda_2}{2m_b}\delta'(n\cdot p)\right]\right\}.
\end{equation}
Because of the experimental cuts $n\cdot p$ scales like $\lambda$ (see section \ref{section_kin}).
We therefore have to group the terms according to their scaling in an analogous way to 
$\bar B\to X_u\,l\bar\nu$. We also see that at leading order the hadronic tensor is proportional to the leading order shape function $S$. Including radiative effects the $W^\mu_\mu$ (and therefore the photon spectrum) can be factorized at leading order as 
\begin{equation}
\label{equation_facts}
W_\mu^\mu=H_{\gamma}\cdot \int d\omega \,m_b \, J[m_b\,(n\cdot p +\omega)]\,S(\omega).
\end{equation} 
Comparing equation (\ref{equation_factu}) to (\ref{equation_facts}), we see that the factorization formulas are very similar. The shape and jet function are the identical, but the argument of the jet function is different, since for $\bar B\to X_s\gamma$, $\bar n\cdot p = m_b$ (this is a result of the fact the kinematics is simpler for radiative decays). The hard functions are  different, with $H_{\gamma}$ turning out to be more complicated. We will utilize these similarities in chapters \ref{chapter_evegen} and \ref{chapter_weight}.    

\section{Effective field theories II: SCET}

In the previous sections we have encountered two problems in trying to
calculate the hadronic tensor. On the perturbative side we had large
double and single logarithms that threaten the convergence of the
perturbative expansion. On the non perturbative side we had to sum
infinite numbers of HQET matrix elements into non perturbative
functions, called shape functions, which are matrix elements of non
local operators. This summation procedure is defined at tree level and
it is unclear how to include radiative correction in a systematic
fashion. Both of these problems can be solved by using SCET, which
we now introduce.

We will present a brief review of SCET, focusing again on its
application for charmless $B$ decays. For a more complete discussion
see \cite{Bauer:2000ew,Bauer:2000yr,Bauer:2001yt}. The relevant effective field
theory is known in the literature as SCET-I. Our presentation of
SCET-I is based on the ``position space" formalism first presented in
\cite{Beneke:2002ph,Beneke:2002ni}. In SCET-I we distinguish between three type of
modes according to the scaling of their momenta: hard, hard-collinear, and soft. Two of these modes, namely hard
and soft, are familiar from HQET. The new mode, hard-collinear, is
designed to describe the light quark produced in the $b$ quark
decay. The various modes are listed in table \ref{table_modes}.

\begin{table}
\begin{center}
\caption{Different modes in SCET-I. We define $\lambda=\Lambda_{\rm
QCD}/m_b$ and write the light cone components of vector $a^\mu$ as
$(n\cdot a,\bar n\cdot a, a_\perp)$.}
\vspace{1em}
\begin{tabular}{c|c|c}
Mode name& Momentum scaling & Examples\\\hline
Hard&$(1,1,1)$& hard gluons\\&&\\
Hard-Collinear&$(\lambda,1,\sqrt{\lambda})$& hard-collinear quarks ($\xi$)\\
&& hard-collinear gluons ($A_{hc}$) \\
Soft&$(\lambda,\lambda,\lambda)$& heavy quarks ($h_v$), spectator quarks ($q_s$)\\ 
&&soft gluons ($A_s$)\\
\end{tabular}\label{table_modes}
\end{center}
\end{table}

In constructing SCET, we first integrate out the hard modes, which
leaves us with hard-collinear and soft modes only. For the soft quark
modes the Lagrangian is the regular QCD Lagrangian for the soft quarks
and the HQET Lagrangian for the heavy quarks. (The Yang-Mills
Lagrangian for the soft and hard-collinear modes can be found in
\cite{Beneke:2002ph,Beneke:2002ni}). We will therefore focus on the
Lagrangian for hard-collinear quarks. In order to construct it we
decompose the hard-collinear fields as:
\begin{equation}
\psi(x)=\frac{\nslash\nbslash}{4}\psi+\frac{\nbslash\nslash}{4}\psi\equiv\xi+\eta,
\end{equation}  
where $(\nslash\nbslash)/4$ and $(\nslash\nbslash)/4$ are projection
operators and $n,\bar n$ are the light like vectors introduced before. In terms of $\xi$ and $\eta$ the QCD Lagrangian for
massless hard-collinear quarks is
\begin{equation}
\label{equation_Lxy}
{\cal L}_{hc}=\bar\psi i\Dslash \psi=\bar \xi\,
\frac{\nbslash}{2}\,in\cdot D\,\xi+\bar\eta\, \frac{\nslash}{2}\,
i\bar n\cdot D\, \eta+\bar\xi\, i\Dslash_\perp\, \eta+\bar\eta \, i
\Dslash_\perp\, \xi.
\end{equation}
(The covariant derivative in (\ref{equation_Lxy}) contains both hard-collinear and soft gluons).
    
In order to decide which field is to be integrated out, let us look at
the two point function for $\xi$:
\begin{equation}
\langle 0|T\{\xi(x),\bar\xi(0)\}|0 \rangle=\int \frac{d^4p}{(2\pi)^4}e^{-ip\cdot x}\frac{i\bar{n}\cdot p}{p^2+i\epsilon}\frac{\nslash}{2}.
\end{equation}
Since $p$ is hard-collinear momentum we have $d^4p\sim\lambda^2$,
$p^2\sim \lambda$, and $\bar{n}\cdot p\sim 1$, which implies that
$\xi$ scales like $\lambda^{1/2}$. In a similar way we can show that
$\eta$ scales like $\lambda$ and it should be ``integrated out". Once
this is done equation (\ref{equation_Lxy}) becomes:
\begin{equation}
\label{equation_Lxx}
{\cal L}_{hc}=\bar \xi\, \frac{\nbslash}{2}\left(\,in\cdot D+
i\Dslash_\perp\frac{1}{i\bar{n}\cdot D}i\Dslash_\perp\right)\xi.
\end{equation}
This Lagrangian contains terms of different orders in $\lambda$ and we will expand it to the lowest order, shortly.

What about the currents? Our naive expectation is that the current
$\bar q(x)\Gamma b(x)$ would be matched, at tree level and leading
order in $\lambda$, onto $e^{-im_bv\cdot x}\bar \xi(x)\Gamma
\,h_v(x)$. The problem is that when a soft field interacts with a hard
collinear field only one component of the momentum is conserved,
namely $n\cdot p$. As a result, in position space, the soft fields need to be "multipole
expanded" \cite{Beneke:2002ph}. At leading order this amounts to placing $\xi$ at $x$ while $h_v$ is at
$x_-=\frac12\bar{n}\cdot x \,n$. As soon as we put the two fields at
different points the current is not gauge invariant. In order to make
it gauge invariant we introduce a Wilson line:
\begin{equation}
W=P\, \exp\left(ig \int_{-\infty}^0 dt \,\bar n \cdot A_{hc}(x+t\bar n)\right)
\end{equation}
The Wilson line has the right transformation properties such that the
current $e^{-im_bv\cdot x}\left(\bar\xi W\right)(x)\Gamma \,h_v(x_-)$
is gauge invariant. We can think of this Wilson line as summing
infinite number of hard-collinear gluons that can interact with the
hard-collinear quark at leading order. (Of course each interaction
``costs" us one power of $g$ which is perturbative at the hard
collinear scale). Apart from this hard-collinear Wilson line it is
also convenient to introduce a soft Wilson line:
\begin{equation}
S=P\, \exp\left(ig \int_{-\infty}^0 dt \,n \cdot A_{s}(x+tn)\right)
\end{equation}
Using this Wilson line we redefine the hard-collinear fields according to 
\begin{equation}
\xi(x)=S(x_-)\xi^{(0)}(x),\quad\quad A^\mu_{\rm hc}=S(x_-)\,A^{\mu(0)}_{\rm hc}(x)\,S^\dagger(x_-).
\end{equation}
The fields $\xi^{(0)}$ and $A^{\mu(0)}_{\rm hc}$ are ``sterile" in the
sense that they do not interact with soft gluons. In term of the
$A^{\mu(0)}_{\rm hc}$, the hard-collinear Wilson line can be written
as $W(x)=S(x_-)W^{(0)}(x)S^\dagger(x_-)$, where
\begin{equation}
W^{(0)}=P\, \exp\left(ig \int_{-\infty}^0 dt \,\bar n \cdot A^{(0)}_{hc}(x+t\bar n)\right).
\end{equation}
Thus the QCD current $\bar q(x)\Gamma b(x)$ is matched onto  $e^{-im_bv\cdot x}J^{(0)}$, where
\begin{equation}
J^{(0)}=\bar\xi W\Gamma \,h_v= \bar\xi^{(0)}S^\dagger W\Gamma
\,h_v=\bar\xi^{(0)} W^{(0)}\Gamma S^\dagger\,h_v\equiv \bar\X\Gamma
S^\dagger h_v=\bar\X(x)\Gamma \H(x_-),
\end{equation}        
and $\bar\X=\bar{\xi}^{(0)} W^{(0)}=\bar \xi W S$. Beyond tree level
and at leading order in $\lambda$ the current would be matched onto
\cite{Beneke:2002ph}
\begin{equation}
\label{equation_current}
\bar q(x)\Gamma b(x)\to e^{-im_bv\cdot x} \sum_i\int ds\,
\tilde{C}_i(s)\bar{\X}(x+s\bar n)\Gamma_i\H(x_-).
\end{equation}  
Notice that non local currents arise naturally in SCET.

We can now go back to the hard-collinear Lagrangian, equation
(\ref{equation_Lxx}), which we want to expand to the lowest order in
$\lambda$. By looking at the two point function for the hard-collinear
and soft gluons, one can show that they scale like their
momentum. That is, $A^\mu_{\rm hc}\sim(\lambda,1,\sqrt{\lambda})$ and
$A^\mu_{\rm s}\sim(\lambda,\lambda,\lambda)$. Since
$iD^\mu=i\partial^\mu+gA^\mu_{\rm hc}+gA^\mu_{\rm s}$. We find that to
the lowest order in $\lambda$
\begin{eqnarray}
i\bar{n}\cdot D&\to& i\bar{n}\cdot D_{\rm
hc}=i\bar{n}\cdot\partial+g\bar{n}\cdot A_{\rm hc}\nonumber \\
iD^\mu_\perp&\to& iD^\mu_{\perp\rm
hc}=i\partial^\mu_\perp+gA^\mu_{\perp\rm hc}\nonumber\\ in\cdot D&\to&
in\cdot D_{\rm hc}+gn\cdot A_{\rm s}=in\cdot\partial+g n\cdot A_{\rm
hc}+gn\cdot A_{\rm s}.
\end{eqnarray}
The leading order collinear Lagrangian is therefore:  
\begin{equation}
{\cal L}_{hc}^{(0)}=\bar \xi\, \frac{\nbslash}{2}\left(\,in\cdot
D_{\rm hc}+gn\cdot A_{\rm s}(x_-)+ i\Dslash_{\perp\rm
hc}\frac{1}{i\bar{n}\cdot D_{\rm hc}}i\Dslash_{\perp\rm hc}\right)\xi.
\end{equation}
Note that we have multipole expanded the soft gluon since it interacts
with collinear quarks. We now introduce the ``sterile" fields into the
leading order Lagrangian. Using the identity
\begin{equation}
S^\dagger (in\cdot \partial+gn\cdot A_{\rm s})\,S=in\cdot \partial
\end{equation} 
the Lagrangian becomes:
\begin{equation}
{\cal L}_{hc}^{(0)}=\bar \xi^{(0)}\, \frac{\nbslash}{2}\left(\,in\cdot
D^{(0)}_{\rm hc}+ i\Dslash^{(0)}_{\perp\rm hc}\frac{1}{i\bar{n}\cdot
D^{(0)}_{\rm hc}}i\Dslash^{(0)}_{\perp\rm hc}\right)\xi^{(0)}.
\end{equation}
Thus at leading order there are no interactions between soft gluons and
hard-collinear quarks.

This concludes the leading order ``derivation" of the SCET-I
Lagrangian and currents. Equation (\ref{equation_current}) will be
used in chapter \ref{chapter_pert} to derive the factorization formula
$W_i\sim H\cdot J\otimes S$. Renormalization group equations will be
used to resum the large logarithms, solving our first problem. In chapter
\ref{chapter_sub} we will use higher order corrections to the currents
and Lagrangians to calculate the subleading shape function
contributions. The shape functions will be described as matrix
elements of non local operators that arise naturally in the process of
integrating out hard-collinear modes, so we do not need to sum
infinite number of HQET parameters. Since SCET is an effective field
theory it is clear how to carry on this procedure to any order and
include radiative corrections, leading to a well defined double
expansion in $\alpha_s$ and $m_b$.

\chapter{Perturbative Corrections}
\label{chapter_pert}
\section{Chapter outline}

In this chapter we analyze the perturbative corrections to $\bar B\to
X_u\,l\bar\nu$ hadronic tensor at leading order in $1/m_b$.  We begin
by presenting a derivation of the factorization formula for inclusive
rates in the shape-function region using the position-space
formulation of SCET. We then perform a two step matching calculation
QCD\,$\to$\,SCET\,$\to$\,HQET at next-to-leading order in perturbation
theory, extracting the hard ($H$) and jet ($J$) functions.  We then
derive renormalization-group equations governing the dependence of the
functions $H$, $J$, and the shape function $S$ on the renormalization
scale, and solve these equations analytically in momentum space. Next,
we derive several model-independent properties of the shape function
$S$. In particular, we present the precise form of the relations
between renormalized shape-function moments and HQET parameters such
as $\bar\Lambda$ and $\lambda_1$, and we give an analytical formula
for the asymptotic behavior of the shape function. An unexpected
outcome of this analysis is the finding that the shape function is not
positive definite, but acquires a negative radiative tail at large
values of $|\omega|$.

\section{Factorization theorem for inclusive decay rates}
\label{sec:fact}
Recall that 
\begin{equation}\label{WandTdef}
   W^{\mu\nu} = \frac{1}{\pi}\,\mbox{Im}\,
    \frac{\langle\bar B|\,T^{\mu\nu}\,|\bar B\rangle}{2 M_B} \,,
    \qquad
   T^{\mu\nu} = i \int d^4x\,e^{iq\cdot x}\,\,
    \mbox{T}\,\{ J^{\dagger\mu}(0), J^\nu(x) \} \,,
\end{equation}
where for $\bar B\to X_u\,l^-\bar\nu$ decays $J^\mu=\bar
u\gamma^\mu(1-\gamma_5) b$. 

Below a matching scale $\mu_h\sim m_b$, the semileptonic current can be 
expanded as
\begin{equation}\label{current}
   \bar u(x)\gamma^\mu(1-\gamma_5) b(x)
   = e^{-im_bv\cdot x}\sum_{i=1}^3 \int ds\,\widetilde C_i(s)\,
   \bar\X(x+s\bar n)\,\Gamma_i^\mu\,\H(x_-) + \dots \,,
\end{equation}
where the dots denote higher-order terms in the SCET expansion, which can 
be neglected at leading power in $\Lambda_{\rm QCD}/m_b$. The position-space Wilson 
coefficient functions $\widetilde C_i(s)$ depend on the variable $s$ 
defining the position of the hard-collinear field. Here and below we 
denote functions in position space with a tilde, which is omitted from 
the corresponding Fourier-transformed functions in momentum space. A 
convenient basis of Dirac structures in (\ref{current}) is 
\begin{equation}
   \Gamma_1^\mu = \gamma^\mu(1-\gamma_5) \,, \qquad
   \Gamma_2^\mu = v^\mu(1+\gamma_5) \,, \qquad
   \Gamma_3^\mu = n^\mu\,(1+\gamma_5) \,.
\end{equation}
The current correlator in (\ref{WandTdef}) then becomes
\begin{eqnarray}\label{T0}
   T^{\mu\nu} &=& i \int d^4x\,e^{-ip\cdot x}
   \sum_{i,j=1}^3 \int ds\,dt\,\widetilde C_j^*(t)\,\widetilde C_i(s)\nonumber\\
    &&\times\mbox{T} \left\{ \bar\H(0)\,\bar\Gamma_j^\mu\,\X(t\bar n),
    \bar\X(x+s\bar n)\,\Gamma_i^\nu\,\H(x_-) \right\} + \dots \,.
\end{eqnarray}
In a second step, the hard-collinear fluctuations associated with the
light-quark jet can be integrated out by matching SCET onto HQET at an 
intermediate scale $\mu_i\sim\sqrt{m_b\Lambda_{\rm QCD}}$. At leading 
order the SCET Lagrangian (when written in terms of the gauge-invariant 
fields such as $\X$) does not contain interactions between hard-collinear 
and soft fields. Since the external $B$-meson states only contain soft 
constituents, we can take the vacuum matrix element over the 
hard-collinear fields, defining a jet function
\begin{equation}\label{Jdef}
   \langle\,\Omega|\,\mbox{T} \left\{ \X_k(t\bar n),\bar\X_l(x+s\bar n)
   \right\} |\Omega\rangle\equiv \delta_{kl}\,
   \widetilde{\cal J}(x+(s-t)\bar n) + \dots \,,
\end{equation}
where $k,l$ are color indices, and we have used translational invariance 
to determine the dependence on the coordinate vectors. At higher orders
in SCET power counting, additional jet functions would arise, but their 
contributions can be neglected at leading power. Shifting the integration 
variable from $x$ to $z=x+(s-t)\bar n$, with $z_-=x_-$, and introducing 
the Fourier-transformed Wilson coefficient functions
\begin{equation}
   C_i(\bar n\cdot p) = \int ds\,e^{is\bar n\cdot p}\,\widetilde C_i(s)
   \,,
\end{equation}
we then obtain
\begin{equation}\label{T1}
   T^{\mu\nu} = i \sum_{i,j=1}^3 C_j^*(\bar n\cdot p)\,C_i(\bar n\cdot p)
   \int d^4z\,e^{-ip\cdot z}\,\bar\H(0)\,\bar\Gamma_j^\mu\,
   \widetilde{\cal J}(z)\,\Gamma_i^\nu\,\H(z_-) + \dots \,.
\end{equation}
In the next step, we rewrite the bilocal heavy-quark operator as 
\cite{Neubert:1993ch,Neubert:1993um}
\begin{eqnarray}\label{SFops}
   \bar\H(0)\,\Gamma\,\H(z_-)
   &=& (\bar h\,S_s)(0)\,\Gamma\,e^{z_-\cdot\partial_+}\,
    (S_s^\dagger\,h)(0) 
    = \bar h(0)\,\Gamma\,e^{z_-\cdot D_+}\,h(0) \nonumber\\
   &=& \int d\omega\,e^{-\frac{i}{2}\omega\bar n\cdot z}\,
    \bar h(0)\,\Gamma\,\delta(\omega-in\cdot D)\,h(0) \,,
\end{eqnarray}
where $\Gamma$ may be an arbitrary (even $z$-dependent) Dirac structure,
and we have used the property $in\cdot D\,S_s=S_s\,in\cdot\partial$ of 
the soft Wilson line $S_s$, where $iD^\mu=i\partial^\mu+g_s A_s^\mu$ is 
the covariant derivative with respect to soft gauge transformations. When 
this expression is used in (\ref{T1}), the resulting formula for the 
correlator involves the Fourier transform of the jet function,
\begin{equation}
   \int d^4z\,e^{-ip\cdot z}\,\widetilde{\cal J}(z)
   = \pslash_-\,{\cal J}(p^2) \,,
\end{equation}
however with $p^\mu$ replaced by the combination 
$p_\omega^\mu\equiv p^\mu+\frac{1}{2}\omega\bar n^\mu$. In defining the 
momentum-space jet function ${\cal J}(p^2)$ we have taken into account 
that the matrix element in (\ref{Jdef}) vanishes when multiplied by 
$\nslash$ from either side, since $\nslash\,\X=0$. Using that 
$p_{\omega-}=p_-$, we now obtain
\begin{equation}\label{Tres}
   T^{\mu\nu} = i\sum_{i,j=1}^3 H_{ij}(\bar n\cdot p)
   \int d\omega\,{\cal J}(p_\omega^2)\,
   \bar h\,\bar\Gamma_j^\mu\,\pslash_-\Gamma_i^\nu\,
   \delta(\omega-in\cdot D)\,h + \dots \,,
\end{equation}
where $H_{ij}(\bar n\cdot p)=C_j^*(\bar n\cdot p)\,C_i(\bar n\cdot p)$ 
are called the hard functions.

In order to compute the hadronic tensor we take the discontinuity of the 
jet function,
\begin{equation}\label{jetdef}
   J(p^2) = \frac{1}{\pi}\,\mbox{Im}\,[i{\cal J}(p^2)] \,,
\end{equation}
and evaluate the $B$-meson matrix element of the soft operator using the 
HQET trace formalism, which allows us to write \cite{Neubert:1993mb}
\begin{equation}\label{Sdef}
   \frac{\langle\bar B|\,\bar h\,\Gamma\,\delta(\omega-in\cdot D)\,
   h\,|\bar B\rangle}{2M_B}
   = S(\omega)\,\frac12\,\mbox{tr}\left( \Gamma\,\frac{1+\vslash}{2}
   \right) + \dots 
\end{equation}
at leading power in the heavy-quark expansion. The soft function 
$S(\omega)$ coincides with the shape function $f(k_+)$ introduced in
\cite{Neubert:1993ch,Neubert:1993um}. This gives the factorization formula
\begin{equation}\label{Wres}
   W^{\mu\nu} = \sum_{i,j=1}^3 H_{ij}(\bar n\cdot p)\,
   \mbox{tr}\left( \bar\Gamma_j^\mu\,\frac{\pslash_-}{2}\,\Gamma_i^\nu\,
   \frac{1+\vslash}{2} \right)
   \int d\omega\,J(p_\omega^2)\,S(\omega) + \dots \,.
\end{equation}
At leading power one could replace $\pslash_-\to\pslash$ and 
$\bar n\cdot p=2v\cdot p_-\to 2v\cdot p$ in this result.

In the final expressions (\ref{Tres}) and (\ref{Wres}) the dependence on 
the three scales $\bar n\cdot p\sim m_b$, 
$p_\omega^2\sim m_b\Lambda_{\rm QCD}$ and $\omega\sim\Lambda_{\rm QCD}$ 
has been factorized into the hard, jet, and shape functions, 
respectively. Large logarithms associated with ratios of these scales can 
be resummed by solving renormalization-group equations for the scale 
dependence of these component functions. The factorization formula 
(\ref{Wres}) was derived at tree level in 
\cite{Neubert:1993ch,Neubert:1993um,Bigi:1993ex}, and was generalized to all orders in 
perturbation theory in \cite{Korchemsky:1994jb,Akhoury:1995fp}. The 
derivation presented above is equivalent to a proof of this formula 
presented in \cite{Bauer:2001yt} (see also \cite{Bauer:2000ew}). The 
limits of integration in the convolution integral are determined by the 
facts that the jet function defined in (\ref{jetdef}) has support for 
$p_\omega^2\ge 0$, and the shape function defined in (\ref{Sdef}) has 
support for $-\infty<\omega\le\bar\Lambda$ with $\bar\Lambda=M_B-m_b$, 
where $m_b$ is the heavy-quark pole mass. The argument $p_\omega^2$ of 
the jet function can be rewritten as
\begin{equation}\label{pw2}
   p_\omega^2 = p^2 + \bar n\cdot p\,\omega
   = \bar n\cdot p\,(P_+ - (\bar\Lambda-\omega))
   \equiv \bar n\cdot p\,(P_+ - \hat\omega) \,,
\end{equation}
where the variable $\hat\omega=\bar\Lambda-\omega\ge 0$. 
We shall see below 
that expressing the convolution integral in terms of the new variable 
$\hat\omega$ eliminates any spurious dependence of the decay spectra on 
the $b$-quark pole mass. 

Using the fact that the Wilson coefficients $C_i$ are real and hence 
$H_{ij}$ is symmetric in its indices, we find
\begin{eqnarray}
   \sum_{i,j=1}^3 H_{ij}\,\mbox{tr}\left( \bar\Gamma_j^\mu\,
   \frac{\pslash_-}{2}\,\Gamma_i^\nu\,\frac{1+\vslash}{2} \right)
   &=& \bar n\cdot p \bigg[ H_{11} \left( n^\mu v^\nu + n^\nu v^\mu
    - g^{\mu\nu}
    - i\epsilon^{\mu\nu\alpha\beta} n_{\alpha} v_\beta \right) \nonumber\\
    &&+ H_{22}\,v^\mu v^\nu
    + (H_{12}+H_{23})\,(n^\mu v^\nu + n^\nu v^\mu)\nonumber\\
    &&+ (2H_{13}+H_{33})\,n^\mu n^\nu\bigg] \,.
\end{eqnarray}
This result may be compared with the general Lorentz decomposition of the 
hadronic tensor:
\begin{eqnarray}\label{equation_wtdecomp}
W^{\mu\nu}&=&(n^\mu v^\nu+n^\nu v^\mu-g^{\mu\nu}-i\epsilon^{\mu\nu\alpha\beta} n_\alpha v_\beta)\tilde{W}_1\nonumber\\
&&-g^{\mu\nu}\tilde{W}_2+v^\mu v^\nu \tilde{W}_3+(n^\mu v^\nu+n^\nu v^\mu)\tilde{W}_4
+n^\mu n^\nu\tilde{W}_5
\end{eqnarray}
We see that the structure function $\tilde W_2$ is not generated at leading 
order in the SCET expansion. Since only the Wilson coefficient $C_1$ 
is non-zero at tree-level, the structure function $\tilde W_1$ receives 
leading-power contributions at tree level, whereas $\tilde W_4$ and $\tilde W_5$ 
receive leading-power contributions at $O(\alpha_s(m_b))$. The function 
$\tilde W_3$ receives leading-power contributions only at $O(\alpha_s^2(m_b))$, 
which is beyond the accuracy of a next-to-leading order calculation.

In the next section we will calculate the $H_{ij}$ from \cite{DeFazio:1999sv}. 
This paper uses the decomposition 
\begin{eqnarray}\label{Wdecomp}
W^{\mu\nu}&=&(p^\mu v^\nu+p^\nu v^\mu-g^{\mu\nu}v\cdot p-i\epsilon^{\mu\nu\alpha\beta} p_\alpha v_\beta)W_1\nonumber\\
&&-g^{\mu\nu}W_2+v^\mu v^\nu W_3+(p^\mu v^\nu+p^\nu v^\mu)W_4
+p^\mu p^\nu W_5.
\end{eqnarray} 
Equation (\ref{Wres}) then implies that:
\begin{eqnarray}
W_1&=&2H_{11}\int d\omega\,J(p_\omega^2)\,S(\omega)+ \dots\nonumber\\
W_4&=&2H_{12}\int d\omega\,J(p_\omega^2)\,S(\omega)+ \dots\nonumber\\
W_5&=&\frac{8H_{11}}{\bar n\cdot p}\int d\omega\,J(p_\omega^2)\,S(\omega)+ \dots
\end{eqnarray}

\section{Matching calculations}
\label{sec:match}

In this section we derive perturbative expressions for the hard functions 
$H_{ij}(\bar n\cdot p)$ and the jet function $J(p_\omega^2)$ in 
(\ref{Wres}) at next-to-leading order in $\alpha_s$. To this end, we 
match expressions for the hadronic tensor obtained in full QCD, SCET, and 
HQET, using for simplicity on-shell external $b$-quark states. We also 
present results for the renormalized shape function in the parton model, 
which are needed in the matching calculation. Throughout this chapter we 
use the $\overline{\rm MS}$ subtraction scheme and work in 
$d=4-2\epsilon$ space-time dimensions.

\subsection{Hard functions}

Perturbative expressions at $O(\alpha_s)$ for the structure functions 
$W_i$ in the decomposition (\ref{Wdecomp}) have been obtained in 
\cite{DeFazio:1999sv} by evaluating one-loop Feynman graphs for the 
current correlator $T^{\mu\nu}$ using on-shell external quark states with 
residual momentum $k$ (satisfying $v\cdot k=0$) in full QCD. The leading 
terms in the region of hard-collinear jet momenta are
\begin{equation}\label{Fulvia}
\begin{aligned}
   \frac12\,W_1 &= \delta(p_k^2) \left[ 1 - \frac{C_F\alpha_s}{4\pi}
    \left( 8\ln^2 y - 10\ln y + \frac{2\ln y}{1-y} + 4 L_2(1-y)
    + \frac{4\pi^2}{3} + 5 \right) \right] \\
   &\quad\mbox{}+ \frac{C_F\alpha_s}{4\pi} \left[ -4 
    \left( \frac{\ln(p_k^2/m_b^2)}{p_k^2} \right)_{\!*}^{\![m_b^2]}
    + (8\ln y-7) \left( \frac{1}{p_k^2} \right)_{\!*}^{\![m_b^2]}
    \right] + \dots \,, \\
   \frac12\,W_4 &= \delta(p_k^2)\,\frac{C_F\alpha_s}{4\pi}\,
    \frac{2}{1-y} \left( \frac{y\ln y}{1-y} + 1 \right) + \dots \,, \\
   \frac{m_b}{4}\,W_5 &= \delta(p_k^2)\,\frac{C_F\alpha_s}{4\pi}\,
    \frac{2}{1-y} \left( \frac{1-2y}{1-y}\,\ln y - 1 \right) + \dots \,, 
\end{aligned}
\end{equation}
whereas $W_2$ and $W_3$ do not receive leading-power contributions at 
this order, in accordance with our general observations made above. Here
$\alpha_s\equiv\alpha_s(\mu)$,  
$p_k^2=p^2+\bar n\cdot p\,n\cdot k$, and in the shape function region, where $P_+\ll P_-$, we have $y\approx\bar n\cdot p/m_b$. We have used that 
$2v\cdot p/m_b=\bar n\cdot p/m_b+O(\lambda)$ in the hard-collinear 
region. The star distributions are generalized plus distributions defined 
as \cite{DeFazio:1999sv}
\begin{equation}\label{star}
\begin{aligned}
   \int_{\le 0}^z\!dx\,F(x) 
   \left( \frac{1}{x} \right)_{\!*}^{\![u]}
   &= \int_0^z\!dx\,\frac{F(x)-F(0)}{x} + F(0)\,\ln\frac{z}{u} \,, \\
   \int_{\le 0}^z\!dx\,F(x)
    \left( \frac{\ln(x/u)}{x} \right)_{\!*}^{\![u]}
   &= \int_0^z\!dx\,\frac{F(x)-F(0)}{x}\,\ln\frac{x}{u} 
    + \frac{F(0)}{2}\,\ln^2\frac{z}{u} \,,
\end{aligned}
\end{equation}
where $F(x)$ is a smooth test function. For later purposes, we note the 
useful identities
\begin{equation}\label{ids}
\begin{aligned}
   \lambda \left( \frac{1}{\lambda x} \right)_{\!*}^{\![u]}
   &= \left( \frac{1}{x} \right)_{\!*}^{\![u/\lambda]}
    = \left( \frac{1}{x} \right)_{\!*}^{\![u]} + \delta(x)\,\ln\lambda 
    \,, \\
   \lambda \left( \frac{\ln(\lambda x/u)}{\lambda x}
   \right)_{\!*}^{\![u]}
   &= \left( \frac{\ln(\lambda x/u)}{x} \right)_{\!*}^{\![u/\lambda]}
    = \left( \frac{\ln(x/u)}{x} \right)_{\!*}^{\![u]}
    + \left( \frac{1}{x} \right)_{\!*}^{\![u]} \ln\lambda 
    + \frac{\delta(x)}{2}\,\ln^2\lambda \,.
\end{aligned}
\end{equation}

In order to find the hard functions $H_{ij}$, we calculate the 
discontinuity of the current correlator (\ref{T0}) between on-shell 
heavy-quark states in momentum space. We work to one-loop order in SCET, 
keeping $i,j$ fixed and omitting the Wilson coefficient functions. The 
corresponding tree diagram yields
\begin{equation}
   D^{(0)} = K\,\delta(p_k^2) \,, \qquad \mbox{with} \quad
   K = \bar u_b(v)\,\bar\Gamma_j^\mu\,\pslash_-\,\Gamma_i^\nu\,u_b(v) \,,
\end{equation}
where $u_b(v)$ are on-shell HQET spinors normalized to unity, and the 
quantity $K$ corresponds to the Dirac trace in (\ref{Wres}). The 
interpretation of this result in terms of hard, jet, and soft functions 
is that, at tree level, $J^{(0)}(p_\omega^2)=\delta(p_\omega^2)$ and 
$S_{\rm parton}^{(0)}(\omega)=\delta(\omega-n\cdot k)$. (The second 
result is specific to the free-quark decay picture.) Then the convolution 
integral $\int d\omega\,J(p_\omega^2)\,S(\omega)$ in (\ref{Wres}) 
produces $\delta(p_k^2)$, and comparison with (\ref{Fulvia}) shows that 
$H_{11}^{(0)}=1$, while all other hard functions vanish at tree level.

\begin{figure}
\begin{center}
\epsfig{file=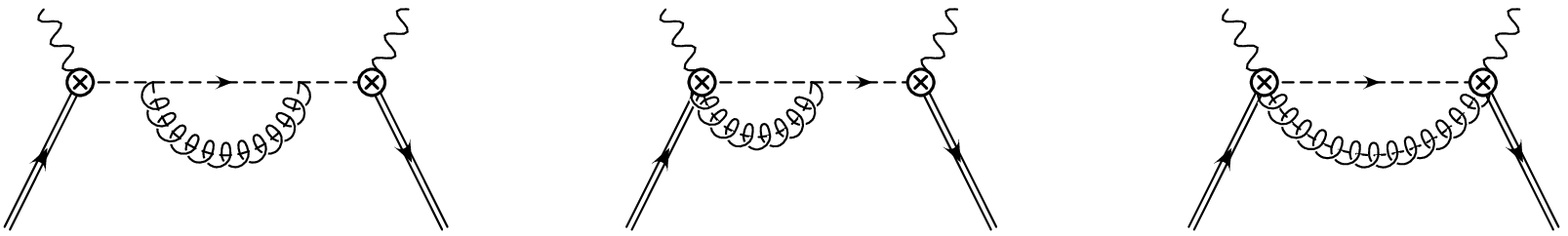,width=14cm}
\epsfig{file=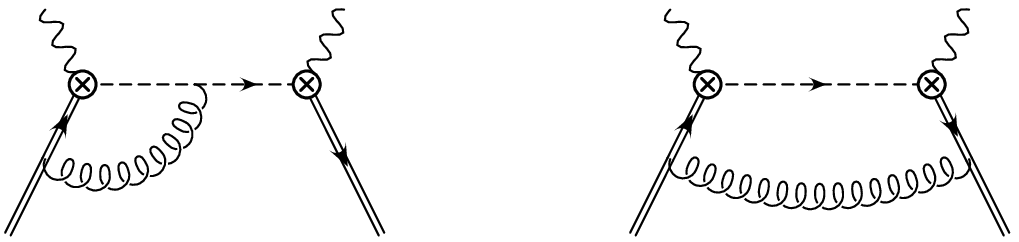,width=9.333cm}
\end{center}
\centerline{\parbox{14cm}{\caption{\label{fig:scet}
One-loop diagrams contributing to the current correlator in SCET. The
effective current operators are denoted by crossed circles, and 
hard-collinear propagators are drawn as dashed lines. Mirror graphs 
obtained by exchanging the two currents are not shown.}}}
\end{figure}

The diagrams contributing at one-loop order are shown in 
Figure~\ref{fig:scet}. They are evaluated using the Feynman rules of 
SCET. The first three graphs contain hard-collinear gluon exchanges, 
while the last two diagrams contain soft exchanges. The wave-function 
renormalization factors of the external heavy quarks equal 1 on-shell. 
For the sum of all hard-collinear exchange graphs, we find
\begin{eqnarray}\label{hcloops}
   D_{hc}^{(1)} &=& K\,\frac{C_F\alpha_s}{4\pi} \left[
   \left( \frac{4}{\epsilon^2} + \frac{3}{\epsilon} + 7 - \pi^2 \right)
   \delta(p_k^2)\right.\\\nonumber
   &+& \left. 4 \left( \frac{\ln(p_k^2/\mu^2)}{p_k^2} \right)_{\!*}^{\![\mu^2]}
   - \left( \frac{4}{\epsilon} + 3 \right)
   \left( \frac{1}{p_k^2} \right)_{\!*}^{\![\mu^2]} \right] .
\end{eqnarray}
The sum of the soft contributions is given by
\begin{eqnarray}\label{sloops}
   D_{s}^{(1)} 
   &=& K\,\frac{C_F\alpha_s}{4\pi} \Bigg[
    \left( - \frac{2}{\epsilon^2} - \frac{4}{\epsilon}\,L
    + \frac{2}{\epsilon} - 4L^2 + 4L - \frac{\pi^2}{6} \right)
    \delta(p_k^2) \nonumber\\
   &&\quad\mbox{}- 8 \left( \frac{\ln(p_k^2/\mu^2)}{p_k^2} 
    \right)_{\!*}^{\![\mu^2]}
    + \left( \frac{4}{\epsilon} + 8L - 4 \right)
   \left( \frac{1}{p_k^2} \right)_{\!*}^{\![\mu^2]} \Bigg] \,,
\end{eqnarray}
where $L=\ln(\bar n\cdot p/\mu)$. The $1/\epsilon$ poles in the sum of 
the hard-collinear and soft contributions are subtracted by a 
multiplicative renormalization factor $Z_J^2$ applied to the bare current 
correlator in (\ref{T0}), where
\begin{equation}\label{ZJres}
   Z_J = 1 + \frac{C_F\alpha_s}{4\pi} \left( - \frac{1}{\epsilon^2}
    + \frac{2}{\epsilon}\,L - \frac{5}{2\epsilon} \right)
\end{equation}
is the (momentum-space) current renormalization constant in SCET 
\cite{Bauer:2000yr}. Taking the sum of (\ref{hcloops}) and (\ref{sloops}) 
after subtraction of the pole terms, and matching it with the results in 
(\ref{Fulvia}), we find that at one-loop order
\begin{equation}\label{Hres}
\begin{aligned}
   H_{11}(\bar n\cdot p) &= 1 + \frac{C_F\alpha_s}{4\pi}
    \left( -4 L^2 + 10 L - 4\ln y - \frac{2\ln y}{1-y} - 4 L_2(1-y)
    - \frac{\pi^2}{6} -12 \right) , \\
   H_{12}(\bar n\cdot p) &= \frac{C_F\alpha_s}{4\pi}\,
    \frac{2}{1-y} \left( \frac{y\ln y}{1-y} + 1 \right) , \\
   H_{13}(\bar n\cdot p) &= \frac{C_F\alpha_s}{4\pi}\,
    \frac{y}{1-y} \left( \frac{1-2y}{1-y}\,\ln y - 1 \right) . 
\end{aligned}
\end{equation}
In deriving these results we have used the identities (\ref{ids}) to 
rearrange the various star distributions. The remaining hard functions 
start at $O(\alpha_s^2)$. Using the relation $H_{ij}=C_i\,C_j$, one can 
derive from these results expressions for the Wilson coefficients in the 
expansion of the semileptonic current in (\ref{current}). We confirm the 
expressions for these coefficients given in \cite{Bauer:2000yr}.

\subsection{Jet function}

After the hadronic tensor is matched onto HQET as shown in (\ref{Wres}), 
the SCET loop graphs in Figure~\ref{fig:scet} determine the one-loop 
contributions to the product of the jet function and the shape function 
in (\ref{Wres}). We may write this product symbolically as 
$J^{(1)}\otimes S^{(0)}+J^{(0)}\otimes S^{(1)}$, where the $\otimes$ 
symbol means a convolution in $\omega$. Whereas the jet function is a 
short-distance object that can be calculated in perturbation theory, the 
shape function is defined in terms of a hadronic matrix element and 
cannot be properly described using Feynman diagrams with on-shell 
external quark states. In a second step, we must therefore extract from 
the results (\ref{hcloops}) and (\ref{sloops}) the one-loop contribution 
$J^{(1)}$ to the jet function. To this end, we must compute the 
renormalized shape function at one-loop order in the parton model. This 
will be done in the next subsection. We may, however, already anticipate 
the result for the jet function at this point, because at one-loop order 
the graphs in Figure~\ref{fig:scet} can be separated into diagrams with 
hard-collinear (first line) or soft (second line) gluon exchange. (This 
separation would be non-trivial beyond one-loop order.) The 
hard-collinear contribution in (\ref{hcloops}) thus determines the 
convolution $J^{(1)}\otimes S^{(0)}=J^{(1)}(p_k^2)$, whereas the soft 
contribution in (\ref{sloops}) corresponds to $J^{(0)}\otimes S^{(1)}$. 
It follows that the renormalized jet function is given by the 
distribution
\begin{equation}\label{Jres}
   J(p_\omega^2) = \delta(p_\omega^2)
   + \frac{C_F\alpha_s}{4\pi} \left[ (7-\pi^2)\,\delta(p_\omega^2)
   + 4 \left( \frac{\ln(p_\omega^2/\mu^2)}{p_\omega^2} 
   \right)_{\!*}^{\![\mu^2]}
   - 3 \left( \frac{1}{p_\omega^2} \right)_{\!*}^{\![\mu^2]} \right] .
\end{equation}
This result disagrees with a corresponding expression obtained in 
\cite{Mannel:2000aj}. It will often be useful to separate the dependence 
on $\bar n\cdot p$ and $P_+$ in this result by means of the 
substitution $p_\omega^2=y\,\hat p_\omega^2$, where 
$\hat p_\omega^2=m_b(P_+-\hat\omega)$ according to (\ref{pw2}). 
Using the identities (\ref{ids}), we find
\begin{eqnarray}\label{Jrescale}
   y\,J(p_\omega^2)\equiv\hat J(\hat p_\omega^2,y)
   &=& \delta(\hat p_\omega^2) + \frac{C_F\alpha_s}{4\pi} \Bigg[
    \big( 2\ln^2 y - 3\ln y + 7 - \pi^2 \big)\,\delta(\hat p_\omega^2)
    \nonumber\\
   &+& 4 \left(
    \frac{\ln(\hat p_\omega^2/\mu^2)}{\hat p_\omega^2} 
    \right)_{\!*}^{\![\mu^2]}
    + (4\ln y - 3) 
    \left( \frac{1}{\hat p_\omega^2} \right)_{\!*}^{\![\mu^2]} \Bigg] \,.
\end{eqnarray}
The jet function is non-zero only if $y\ge 0$ and 
$P_+\ge\hat\omega$, which ensures that $p_\omega^2\ge 0$.

\subsection{Renormalized shape function}
\label{sec:SFrenorm}

Having calculated the short-distance objects $H_{ij}$ and $J$ in the 
factorization formula (\ref{Wres}) at one-loop order, we now turn to a 
study of radiative corrections to the shape function $S(\omega)$. There 
is considerable confusion in the literature about the renormalization 
properties of the shape function, and several incorrect results for its 
anomalous dimension have been published. We therefore present our 
calculation in some detail in this subsection and the following section.

\begin{figure}
\begin{center}
\epsfig{file=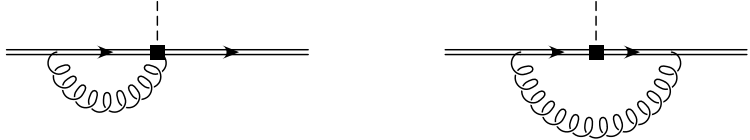,width=10cm}
\end{center}
\centerline{\parbox{14cm}{\caption{\label{fig:hqet}
Radiative corrections to the shape function. The bilocal HQET operator is
denoted by the black square. A mirror copy of the first graph is not 
shown.}}}
\end{figure}

According to (\ref{Sdef}), the shape function is defined in terms of a 
hadronic matrix element in HQET and thus cannot be calculated 
perturbatively. However, the renormalization properties of this function 
can be studied using perturbation theory. To this end, we evaluate the 
matrix element (\ref{Sdef}) in HQET using external heavy-quark states 
with residual momentum $k$. For the time being, we keep $v\cdot k$ 
non-zero to regularize infra-red singularities. The relevant one-loop 
graphs are depicted in Figure~\ref{fig:hqet}. Adding the tree 
contribution, we obtain for the matrix element of the bare shape-function 
operator $O(\omega)=\bar h\,\Gamma\,\delta(\omega-in\cdot D)\,h$ 
(expressed in terms of renormalized fields)
\begin{eqnarray}\label{Sbare}
   S_{\rm bare}(\omega)
   &=& Z_h\,\delta(\omega-n\cdot k)
    - \frac{4C_F g_s^2}{(4\pi)^{2-\epsilon}}\,\Gamma(1+\epsilon)
    \nonumber\\
   &\times& \Bigg\{ \frac{1}{\epsilon}
    \int_0^\infty\!dl\,l^{-1-2\epsilon}\,\Big[ \delta(\omega-n\cdot k)
    - \delta(\omega-n\cdot k+l) \Big] \left( 1 + \frac{\delta}{l} 
    \right)^{-\epsilon} \nonumber\\
   &&\quad\mbox{}+ \theta(n\cdot k-\omega)\,(n\cdot k-\omega)^{-\epsilon}
    (n\cdot k-\omega+\delta)^{-1-\epsilon} \Bigg\} \,,
\end{eqnarray}
where $\delta=-2v\cdot k$, and
\begin{equation}
   Z_h = 1 + \frac{4C_F g_s^2}{(4\pi)^{2-\epsilon}}\,\Gamma(2\epsilon)\,
   \Gamma(1-\epsilon)\,\delta^{-2\epsilon}
\end{equation}
is the off-shell wave-function renormalization constant of a heavy quark 
in HQET. The next step is to extract the ultra-violet poles from this 
result, which determine the anomalous dimension of the shape function. We 
define a renormalization factor through
\begin{equation}\label{Sren}
\begin{aligned}
   S_{\rm ren}(\omega) &= \int_{-\infty}^{\bar\Lambda}\!d\omega'\,
    Z_S(\omega,\omega')\,S_{\rm bare}(\omega') \,, \\
   Z_S(\omega,\omega') &= \delta(\omega-\omega')
    + \frac{C_F\alpha_s}{4\pi}\,z_S(\omega,\omega') + \dots \,.
\end{aligned}
\end{equation}
The result for $Z_S$ following from (\ref{Sbare}) must be interpreted as 
a distribution on test functions $F(\omega')$ with support on the 
interval $-\infty<\omega'\le\bar\Lambda$. We obtain
\begin{eqnarray}\label{zres}
   z_S(\omega,\omega')
   &=& \left( \frac{2}{\epsilon^2}
    + \frac{4}{\epsilon}\,\ln\frac{\mu}{\bar\Lambda-\omega}
    - \frac{2}{\epsilon} \right) \delta(\omega-\omega')
    - \frac{4}{\epsilon}\,\left( 
    \frac{\theta(\omega'-\omega)}{\omega'-\omega} \right)_{\!+}
    \nonumber\\
   &=& \left( \frac{2}{\epsilon^2} - \frac{2}{\epsilon} \right)
    \delta(\omega-\omega')
    - \frac{4}{\epsilon}\,\left( \frac{1}{\omega'-\omega}
    \right)_{\!*}^{\![\mu]} .
\end{eqnarray}
Note the peculiar dependence on the parameter $\bar\Lambda$ setting the 
upper limit on the integration over $\omega'$ in (\ref{Sren}), which 
combines with the plus distribution to form a star distribution in the 
variable $(\omega'-\omega)$. 

We can now determine the renormalized shape function from (\ref{Sren}). 
The result must once again be interpreted as a distribution, this time on 
test functions $F(\omega)$ integrated over a {\em finite\/} interval 
$-\Lambda_{\rm had}\le\omega\le\bar\Lambda$. In practice, the value of
$\Lambda_{\rm had}$ is set by kinematics or by virtue of some 
experimental cut. The result is
\begin{eqnarray}
   S_{\rm parton}(\omega) &=& \delta(\omega-n\cdot k)\,\left\{ 1
    - \frac{C_F\alpha_s}{\pi} \left[ \frac{\pi^2}{24}
    + L_2\!\left( \frac{-\delta}{\Lambda_{\rm had}+n\cdot k} \right)
    \right] \right\} \nonumber\\
   &&\mbox{}- \frac{C_F\alpha_s}{\pi}\,\Bigg\{
    \left[ \frac{\theta(n\cdot k-\omega)}{n\cdot k-\omega} \left(
    \ln\frac{n\cdot k-\omega}{\mu} + \ln\frac{n\cdot k-\omega+\delta}{\mu}
    \right) \right]_+ \nonumber\\
   &&\quad\mbox{}+ \delta(n\cdot k-\omega)\,
    \ln^2\frac{\Lambda_{\rm had}+n\cdot k}{\mu}
    + \frac{\theta(n\cdot k-\omega)}{n\cdot k-\omega+\delta}\nonumber\\
    &&+ \delta(n\cdot k-\omega)\,\ln\frac{\delta}{\mu} \Bigg\} \,. \quad
\end{eqnarray}
While it was useful to keep the heavy quark off-shell in the calculation 
of the ultra-violet renormalization factor, the limit 
$\delta=-2v\cdot k\to 0$ can be taken in the result for the renormalized 
shape functions without leading to infra-red singularities. This gives
\begin{eqnarray}\label{Sonshell}
   S_{\rm parton}(\omega) &=& \delta(\omega-n\cdot k)\,\left( 1
    - \frac{C_F\alpha_s}{\pi}\,\frac{\pi^2}{24} \right) \nonumber\\
   &&\hspace{-1cm}- \frac{C_F\alpha_s}{\pi} \left[
    2 \left( \frac{1}{n\cdot k-\omega} \ln\frac{n\cdot k-\omega}{\mu}
    \right)_{\!*}^{\![\mu]} 
    + \left( \frac{1}{n\cdot k-\omega} \right)_{\!*}^{\![\mu]}
    \right] ,
\end{eqnarray}
where the star distributions must now be understood as distributions in 
the variable $(n\cdot k-\omega)$.

We stress that these results for the renormalized shape function are 
obtained in the parton model and can in no way provide a realistic 
prediction for the functional form of $S(\omega)$. This should be 
obvious from the fact that our results depend on a single ``hadronic 
parameter'' $n\cdot k$, corresponding to a fixed residual momentum of the 
heavy quark. Only the dependence on the ultra-violet renormalization 
scale $\mu$ can be trusted. However, the one-loop result in 
(\ref{Sonshell}) is needed to complete the matching calculation of the 
jet function described in the previous subsection, which can legitimately 
be performed with on-shell external $b$-quark states. Given the 
expression for the renormalized shape function, we obtain
\begin{eqnarray}
   &&\int d\omega J^{(0)}(p_\omega^2)\,S^{(1)}(\omega)
   = \frac{1}{\bar n\cdot p}\,S^{(1)}(-p^2/\bar n\cdot p) \\
   &=& - \frac{C_F\alpha_s}{\pi} \left[
    \frac{\pi^2}{24}\,\delta(p_k^2)
    + \frac{2}{\bar n\cdot p} \left( \frac{\bar n\cdot p}{p_k^2}
    \ln\frac{p_k^2}{\bar n\cdot p\,\mu} \right)_{\!*}^{\![\mu]} 
    + \frac{1}{\bar n\cdot p} \left( \frac{\bar n\cdot p}{p_k^2}
    \right)_{\!*}^{\![\mu]} \right] . \nonumber
\end{eqnarray}
With the help of the identities (\ref{ids}) this can be shown to be equal 
to the finite part of (\ref{sloops}), as we claimed above.

\section{Renormalization-group resummation}
\label{sec:RGevol}

Equations (\ref{Hres}) and (\ref{Jres}) determine the short-distance 
objects $H_{ij}$ and $J$ in the factorization formula (\ref{Wres}) at
one-loop order in perturbation theory. However, there is no common choice 
of the renormalization scale $\mu$ that would eliminate all large 
logarithms from these results. Likewise, the shape function, being a 
hadronic matrix element, is naturally renormalized at some low scale, 
whereas the short-distance objects contain physics at higher scales. The 
problem of large logarithms arising from the presence of disparate mass 
scales can be dealt with using renormalization-group equations. 
Proceeding in three steps, our strategy will be as follows: 

i) At a high scale $\mu_h\sim m_b$ we match QCD onto SCET and extract 
matching conditions for the hard functions $H_{ij}$. The corresponding 
one-loop expressions have been given in (\ref{Hres}). At that scale, they 
are free of large logarithms and so can be reliably computed using 
perturbation theory. We then evolve the hard functions down to an 
intermediate hard-collinear scale $\mu_i\sim\sqrt{m_b\Lambda_{\rm QCD}}$ 
by solving the renormalization-group equation
\begin{equation}\label{Hevol}
   \frac{d}{d\ln\mu}\,H_{ij}(\bar n\cdot p,\mu)
   = 2\gamma_J(\bar n\cdot p,\mu)\,H_{ij}(\bar n\cdot p,\mu) \,,
\end{equation}
where $\gamma_J$ is the anomalous dimension of the semileptonic current 
in SCET.

ii) Next, we start from a model for the shape function $S(\omega,\mu_0)$ 
at some low scale $\mu_0=\mbox{few}\times\Lambda_{\rm QCD}$ large enough 
to trust perturbation theory. Such a model could be provided by a 
QCD-inspired approach such as QCD sum rules or lattice QCD, or it could 
be tuned to experimental data. We then solve the integro-differential 
evolution equation
\begin{equation}\label{Sevol}
   \frac{d}{d\ln\mu}\,S(\omega,\mu)
   = - \int d\omega'\,\gamma_S(\omega,\omega',\mu)\,S(\omega',\mu) 
\end{equation}
to obtain the shape function at the intermediate scale $\mu_i$.

iii) Finally, at the scale $\mu_i$ we combine the results for the hard 
functions and for the shape function with the jet function $J$ in 
(\ref{Jres}), which at that scale is free of large logarithms and so has 
a reliable perturbative expansion. The dependence on the matching scales 
$\mu_h$ and $\mu_i$ cancels in the final result (to the order at which we 
are working). 

We now discuss these three steps in detail.

\subsection{Evolution of the hard functions}

At one-loop order, the anomalous dimension $\gamma_J$ for the SCET 
current is twice the coefficient of the $1/\epsilon$ pole in the 
renormalization factor $Z_J$ in (\ref{ZJres}). More generally
\cite{Bauer:2000yr,Bosch:2003fc},
\begin{equation}
   \gamma_J(\bar n\cdot p,\mu)
   = - \Gamma_{\rm cusp}(\alpha_s)\,\ln\frac{\mu}{\bar n\cdot p}
    + \gamma'(\alpha_s)
   = \frac{C_F\alpha_s}{\pi}
    \left( - \ln\frac{\mu}{\bar n\cdot p} - \frac54 \right)
    + \dots \,,
\end{equation}
where $\Gamma_{\rm cusp}=C_F\alpha_s/\pi+\dots$ is the universal cusp 
anomalous dimension governing the ultra-violet singularities of Wilson 
lines with light-like segments \cite{Korchemsky:wg,Korchemskaya:1992je}. The exact solution 
to the evolution equation (\ref{Hevol}) can be written as
\begin{equation}
   H_{ij}(\bar n\cdot p,\mu_i) =  H_{ij}(\bar n\cdot p,\mu_h)\,
   U_y(\mu_h,\mu_i)\,,
\end{equation}
where
\begin{equation}
   U_y(\mu_h,\mu_i) = U(\mu_h,\mu_i)\,y^{-2a_\Gamma(\mu_h,\mu_i)} \,.
\end{equation}
The exact expression for the evolution factor reads
\begin{equation}\label{equation_U1}
   \ln U(\mu_h,\mu_i) = 2 S_\Gamma(\mu_h,\mu_i)  
   - 2a_\Gamma(\mu_h,\mu_i) \,\ln \frac{m_b}{\mu_h} 
   - 2a_{\gamma'}(\mu_h,\mu_i) \,,
\end{equation}
where the functions of the right-hand side are solutions to the 
renormalization-group equations
\begin{eqnarray}
   \frac{d}{d\ln\mu} S_\Gamma(\nu,\mu)
   &=& - \Gamma_{\rm cusp}(\alpha_s(\mu))\,\ln\frac{\mu}{\nu} \,, \nonumber\\
   \frac{d}{d\ln\mu} a_\Gamma(\nu,\mu)
   &=& - \Gamma_{\rm cusp}(\alpha_s(\mu)) \,, \qquad
   \frac{d}{d\ln\mu} a_{\gamma'}(\nu,\mu) = - \gamma'(\alpha_s(\mu)) \,,
\end{eqnarray}
with boundary conditions $S(\nu,\mu)=0$ etc.\ at $\mu=\nu$.
These equations can be integrated using that 
$d/d\ln\mu=\beta(\alpha_s)\,d/d\alpha_s$. The solutions are
\begin{equation}\label{equation_agamma}
   S_\Gamma(\nu,\mu)
   = -\int\limits_{\alpha_s(\nu)}^{\alpha_s(\mu)}\!\!d\alpha\,
    \frac{\Gamma_{\rm cusp}(\alpha)}{\beta(\alpha)} 
    \int\limits_{\alpha_s(\nu)}^{\alpha}\frac{d\alpha'}{\beta(\alpha')}
    \,, \qquad
   a_\Gamma(\nu,\mu)
   = -\int\limits_{\alpha_s(\nu)}^{\alpha_s(\mu)}\!\!d\alpha\,
    \frac{\Gamma_{\rm cusp}(\alpha)}{\beta(\alpha)} \,,
\end{equation}
and similarly for $a_{\gamma'}$. Explicit results for the Sudakov exponent $S_\Gamma$ and
the functions $a_\Gamma$ and $a_\gamma$ in (\ref{equation_U1}) at next-to-leading 
order in renormalization-group improved perturbation theory are given in Appendix~\ref{apx:Sudakovs}.

\subsection{Evolution of the shape function}

At one-loop order, the anomalous dimension for the shape function is 
twice the coefficient of the $1/\epsilon$ pole in the renormalization 
factor $Z_S$. From (\ref{zres}), we obtain
\begin{equation}\label{ourgamma}
   \gamma_S(\omega,\omega',\mu)
   = \frac{C_F\alpha_s}{\pi} \left[
   \left( 2\ln\frac{\mu}{\bar\Lambda-\omega} - 1 \right)
   \delta(\omega-\omega')
   - 2 \left( \frac{\theta(\omega'-\omega)}{\omega'-\omega}
   \right)_{\!+} \right] .
\end{equation}

The evolution equation (\ref{Sevol}) can be solved analytically using a
general method developed in \cite{Lange:2003ff}. It is convenient to 
change variables from $\omega$ to 
$\hat\omega=\bar\Lambda-\omega\in[0,\infty[$ and denote
$\hat S(\hat\omega)\equiv S(\bar\Lambda-\hat\omega)$. The 
renormalization-group equation then reads
\begin{equation}\label{oureqn}
   \frac{d}{d\ln\mu}\,\hat S(\hat\omega,\mu)
   = - \int_0^\infty\!d\hat\omega'\,
   \hat\gamma_S(\hat\omega,\hat\omega',\mu)\,
   \hat S(\hat\omega',\mu) \,,
\end{equation}
where the anomalous dimension can be written in the general form
\begin{equation}\label{hatGs}
   \hat\gamma_S(\hat\omega,\hat\omega',\mu)
   =-2\Gamma_{\rm cusp}(\alpha_s)\left(
    \frac{\theta(\hat\omega-\hat\omega')}
         {\hat\omega-\hat\omega'} \right)_{\!+} 
         +2 \left[\gamma(\alpha_s) -\,\Gamma_{\rm cusp}(\alpha_s)\,\ln\frac{\hat\omega}{\mu}
     \right] \delta(\hat\omega-\hat\omega')\,.
\end{equation}
The logarithmic term containing the cusp anomalous dimension has a 
geometric origin. Since the heavy-quark field $h(x)$ in HQET can be 
represented as a Wilson line along the $v$ direction, the field $\H(x)$ 
entering the SCET formalism contains the product of a light-like Wilson 
line (along $n$) and a time-like Wilson line (along $v$), which form a 
cusp at point $x$. The shape function contains two such cusps. According 
to the renormalization theory of Wilson lines with light-like segments, 
each cusp produces a contribution to the anomalous dimension proportional 
to $\Gamma_{\rm cusp}\,\ln\mu$ \cite{Korchemsky:wg,Korchemskaya:1992je}. 

The general solution of (\ref{oureqn}) can be obtained using the fact 
that 
\begin{equation}\label{Fdef}
   -\int_0^\infty\!d\hat\omega'\,(\hat\omega')^a\,
   \Gamma_{\rm cusp}(\alpha_s)\left(
    \frac{\theta(\hat\omega-\hat\omega')}
         {\hat\omega-\hat\omega'} \right)_{\!+} 
   = \hat\omega^a\,\Gamma_{\rm cusp}(\alpha_s)\,
   \Big[ \psi(1+a) + \gamma_E \Big]
   \end{equation}
where $\psi(z)$ is the logarithmic derivative of the Euler $\Gamma$ 
function and the integral on the left-hand side is convergent as long 
as $\mbox{Re}\,a>-1$.

Relation (\ref{Fdef}) implies that the ansatz 
\cite{Lange:2003ff}
\begin{equation}
   f(\hat\omega,\mu,\mu_0,\tau)
   = \left( \frac{\hat\omega}{\mu_0} \right)^{\tau-2a_\Gamma(\mu_0,\mu)}
   U_2(\mu,\mu_0)U_3(\tau,\mu,\mu_0)
\end{equation}
with
\begin{equation}\label{myeqs}
\begin{aligned}
\ln U_2(\mu,\mu_0) &= 2S(\mu_0,\mu)+2a_\gamma(\mu_0,\mu) \,, \\
\ln U_3(\tau,\mu,\mu_0)
   &= - 2\!\int\limits_{\alpha_s(\mu_0)}^{\alpha_s(\mu)}\!
    \frac{d\alpha}{\beta(\alpha)}\,
    \Gamma_{\rm cusp}(\alpha_s)\,
   \Big\{ \psi\left[1+\tau-2a_\Gamma(\mu_0,\mu_\alpha)\right] + \gamma_E \Big\} \,,
\end{aligned}
\end{equation}
provides a solution to the evolution equation (\ref{oureqn}) with initial 
condition, \\$f(\hat\omega,\mu_0,\mu_0,\tau)=(\hat\omega/\mu_0)^\tau$ at 
some scale $\mu_0$. Here $\mu_\alpha$ is defined such that 
$\alpha_s(\mu_\alpha)=\alpha$, $a_\gamma$ is defined as in equation (\ref{equation_agamma}) with $\Gamma_{\rm cusp}$ replaced by $\gamma$, and $\tau$ can be an arbitrary complex 
parameter. Note that $a_\Gamma(\mu,\mu_0)<0$ if $\mu>\mu_0$. 

We now assume that 
the shape function $\hat S(\hat\omega,\mu_0)$ is given at the low scale 
$\mu_0$ and define its Fourier transform with respect to 
$\ln(\hat\omega/\mu_0)$ through
\begin{equation}\label{S0def}
   \hat S(\hat\omega,\mu_0)
   = \frac{1}{2\pi} \int_{-\infty}^\infty\!dt\,{\cal S}_0(t)
   \left( \frac{\hat\omega}{\mu_0} \right)^{it} \,.
\end{equation}
The exact result for the shape function at a different scale $\mu$ is 
then given by
\begin{equation}\label{Sexact}
   \hat S(\hat\omega,\mu)
   = \frac{1}{2\pi} \int_{-\infty}^\infty\!dt\,{\cal S}_0(t)\,
   f(\hat\omega,\mu,\mu_0,it) \,. 
\end{equation}

With the help of this formula, it is straightforward to derive explicit 
expressions for the evolution of the shape function from the hadronic 
scale $\mu_0$ up to the intermediate scale $\mu_i$ at any order in 
renormalization-group improved perturbation theory: 
\begin{equation}\label{NLOapprox}
   f(\hat\omega,\mu_i,\mu_0,it)
   = U_2(\mu_i,\mu_0) e^{2a_{\Gamma}(\mu_0,\mu_i)\gamma_E }\left( \frac{\hat\omega}{\mu_0}
   \right)^{it-2a_\Gamma(\mu_0,\mu_i)}
   \frac{\Gamma(1+it)}{\Gamma\left[1+it-2a_\Gamma(\mu_0,\mu_i)\right]}\,.
\end{equation}

The leading-order result presented above can be simplified further. When 
(\ref{NLOapprox}) is inserted into (\ref{Sexact}), the integration over 
$t$ can be performed analytically. Setting 
$\eta=-2a_{\Gamma}(\mu_0,\mu_i)$, the relevant integral is
\begin{equation}
   I = \frac{1}{2\pi} \int_{-\infty}^\infty\!dt\,{\cal S}_0(t)
    \left( \frac{\hat\omega}{\mu_0} \right)^{it}
    \frac{\Gamma(1+it)}{\Gamma(1+it+\eta)} \,,
\end{equation}
where
\begin{equation}
   {\cal S}_0(t) = \int_0^\infty\!\frac{d\hat\omega'}{\hat\omega'}\,
    \hat S(\hat\omega',\mu_0)
    \left( \frac{\hat\omega'}{\mu_0} \right)^{-it}
\end{equation}
is the Fourier transform of the shape function as defined in 
(\ref{S0def}). The integrand of the $t$-integral has poles on the 
positive imaginary axis located at $t=in$ with $n\ge 1$ an integer. For 
$\hat\omega<\hat\omega'$ the integration contour can be closed in the 
lower half-plane avoiding all poles, hence yielding zero. For
$\hat\omega>\hat\omega'$ we use the theorem of residues to obtain
\begin{equation}
   I = \int_0^{\hat\omega}\!d\hat\omega'\,R(\hat\omega,\hat\omega')\,
   \hat S(\hat\omega',\mu_0) \,,
\end{equation}
where
\begin{equation}
   R(\hat\omega,\hat\omega')
   = \frac{1}{\hat\omega} \sum_{j=0}^\infty
    \left( - \frac{\hat\omega'}{\hat\omega} \right)^j
    \frac{1}{\Gamma(j+1)\,\Gamma(\eta-j)}
   = \frac{1}{\Gamma(\eta)}\,
    \frac{1}{\hat\omega^{\eta}\,(\hat\omega-\hat\omega')^{1-\eta}} \,.
\end{equation}
Note that $R(\hat\omega,\hat\omega')\to\delta(\hat\omega-\hat\omega')$ in 
the limit $\eta\to 0$, corresponding to $\mu_i\to\mu_0$, as it should be. 
Our final result for the shape function at the intermediate 
hard-collinear scale, valid at leading order in renormalization-group 
improved perturbation theory, can now be written in the simple form 
(valid for $\mu_i>\mu_0$, so that $\eta>0$)
\begin{equation}\label{wow}
   \hat S(\hat\omega,\mu_i) = U_2(\mu_i,\mu_0)\,
   \frac{e^{-\gamma_E\eta}}{\Gamma(\eta)} \int_0^{\hat\omega}\!d\hat\omega'\,
   \frac{\hat S(\hat\omega',\mu_0)}
        {\mu_0^{\eta}\,(\hat\omega-\hat\omega')^{1-\eta}} \,.
\end{equation}

From the above equation one can derive scaling relations for the 
asymptotic behavior of the shape function for $\hat\omega\to 0$ and
$\hat\omega\to\infty$ (corresponding to $\omega\to\bar\Lambda$ and 
$\omega\to-\infty$). If the function $\hat S(\hat\omega,\mu_0)$ at the 
low scale $\mu_0$ vanishes proportional to $\hat\omega^\zeta$ near the 
endpoint, the shape function at a higher scale $\mu_i>\mu$ vanishes 
faster, proportional to $\hat\omega^{\zeta+\eta}$. Similarly, if 
$\hat S(\hat\omega,\mu_0)$ falls off like $\hat\omega^{-\xi}$ for 
$\hat\omega\to\infty$, the shape function renormalized at a higher scale 
vanishes like $\hat\omega^{-\min(1,\xi)+\eta}$. Irrespective of the 
initial behavior of the shape function, evolution effects generate a 
radiative tail that falls off slower than $1/\hat\omega$. This fact 
implies that the normalization integral of $\hat S(\hat\omega,\mu)$ as 
well as all positive moments are ultra-violet divergent. The 
field-theoretic reason is that the bilocal shape-function operator in 
(\ref{SFops}) contains ultra-violet singularities as $z_-\to 0$, which 
are not subtracted in the renormalization of the shape function. The 
situation is analogous to the case of the $B$-meson light-cone 
distribution amplitude discussed in \cite{Lange:2003ff,Grozin:1996pq}. 
These divergences are never an obstacle in practice. Convolution 
integrals with the shape function are always cut off at some finite value 
of $\hat\omega$ by virtue of phase-space or some experimental cut.

\section{Properties of the shape function}
\label{sec:properties}

In this section we discuss how moments of the shape function are related
with HQET parameters. This will lead us to propose a new, physical 
scheme for defining a running heavy-quark mass, which is most appropriate 
for the study of inclusive spectra in the shape-function region. We will
also present a model-independent result for the asymptotic behavior of 
the renormalized shape function (defined in the $\overline{\rm MS}$ 
scheme), finding that it is {\em not\/} positive definite. 

Most of our discussion in this section is phrased in terms of the 
original (unhatted) shape function $S(\omega,\mu)$. At the end, we 
formulate the resulting constraints on the function 
$\hat S(\hat\omega,\mu)$.

\subsection{Shape-function moments in the pole scheme}

Naively, ignoring renormalization effects, the moments 
$M_N=\int_{-\infty}^{\bar\Lambda} d\omega\,\omega^N S(\omega)$ are given 
by hadronic parameters defined in terms of $B$-meson matrix elements of 
local HQET operators \cite{Neubert:1993ch}. In particular, $M_0=1$ fixes 
the normalization of the shape function, $M_1=0$ vanishes by the HQET 
equation of motion, and $M_2=-\lambda_1/3$ is determined by the matrix 
element of the kinetic-energy operator. The vanishing of the first moment 
is connected with the implicit definition of the heavy-quark pole mass 
built into the HQET Lagrangian via the equation of motion 
$iv\cdot D\,h=0$. These moment constraints have been implemented in 
various model parameterizations for the shape function suggested in the 
literature \cite{Mannel:1994pm,Kagan:1998ym,Bigi:2002qq}. Typically, one 
makes an ansatz for the shape function depending on a few HQET 
parameters such as $\bar\Lambda$ and $\lambda_1$, and determines the 
values of these parameters from a fit to experimental data.

Beyond tree level, all moments $M_N$ with $N\ge 0$ receive ultra-violet 
divergences from the region $\omega\to-\infty$ (or 
$\hat\omega\to\infty$). However, as we have mentioned above, the values 
of $\omega$ needed for the description of physical decay rates are always 
restricted to a finite interval. It is thus sufficient for all purposes 
to define the moments of the renormalized shape function as
\begin{equation}\label{MNdef}
   M_N(\Lambda_{\rm UV},\mu) = \int_{-\Lambda_{\rm UV}}^{\bar\Lambda}\!
   d\omega\,\omega^N S(\omega,\mu) \,.
\end{equation}
The dependence of these moments on the renormalization scale $\mu$ is
controlled by the evolution equation (\ref{Sevol}). In addition, the 
moments depend on the lower cutoff on the $\omega$ integral. The choice 
of $\Lambda_{\rm UV}$ is a matter of convenience, and so we are free to 
pick a value that is numerically (if not parametrically) large compared 
with $\Lambda_{\rm QCD}$. In this case, as we will now show, the 
dependence on $\Lambda_{\rm UV}$ can also be controlled using 
short-distance methods. 

For sufficiently large values of $\Lambda_{\rm UV}$ it is possible to 
expand the moments\\$M_N(\Lambda_{\rm UV},\mu)$ in a series of $B$-meson 
matrix elements of local HQET operators. If for simplicity we set 
$\Gamma=1$ in the shape-function operator (which is legitimate, since the 
Dirac structure is unaltered in HQET), the operators in question are 
Lorentz-scalar, ``leading-twist'' operators containing $\bar h\dots h$ 
\cite{Neubert:1993ch,Bigi:1993ex}. These are the operators that mix with 
$\bar h\,(in\cdot D)^N h$ under renormalization. It is straightforward to 
find the corresponding operators of a given dimension. The unique 
dimension-3 operator is $\bar h h$. The two operators of dimension~4 are 
$\bar h\,in\cdot D\,h$ (class-1) and $\bar h\,iv\cdot D\,h$ (class-2). 
The class-2 operator vanishes by the HQET equation of motion. The 
possible dimension-5 operators are
\begin{equation}
\begin{aligned}
   \mbox{class-1:} \qquad
   &\bar h\,(in\cdot D)^2 h \,, \quad
    \bar h\,(iD_\perp)^2 h \,, \\
   \mbox{class-2:} \qquad
   &\bar h\,(iv\cdot D)^2 h \,, \quad
    \bar h\,in\cdot D\,iv\cdot D\,h \,, \quad
    \bar h\,iv\cdot D\,in\cdot D\,h \,,
\end{aligned}
\end{equation}
where again the class-2 operators vanish by the equation of motion. 
Moreover, it follows from the Feynman rules of HQET that the two class-1 
operators do not mix under renormalization, so the operator 
$\bar h\,(iD_\perp)^2 h$ can be ignored. From dimension~6 on the 
situation is more complicated, because several class-1 operators exist 
that can mix with $\bar h\,(in\cdot D)^N h$. For $N=3$ these are of the 
form $\bar h\,iD\,G\,h$ or $\sum_q \bar h\dots q\,\bar q\dots h$, where 
we omit Lorentz and color indices. We will restrict our discussion to 
operators of dimension less than 6.

For the operator product expansion of the moments in (\ref{MNdef}) we 
need the forward matrix elements 
\begin{equation}
   \langle O\rangle
   = \frac{\langle\bar B(v)|\,O\,|\bar B(v)\rangle}{2M_B}
\end{equation}
of the leading-twist operators between $B$-meson states in HQET. Using 
the equation of motion, it can be shown that $\langle\bar hh\rangle=1$, 
$\langle\bar h\,in\cdot D\,h\rangle=0$, and 
$\langle\bar h\,(in\cdot D)^2 h\rangle=-\lambda_1/3$ 
\cite{Neubert:1993ch}. We can thus write an expansion of the form
\begin{equation}\label{beauty}
   M_N(\Lambda_{\rm UV},\mu) = \Lambda_{\rm UV}^N
   \left\{ K_0^{(N)}(\Lambda_{\rm UV},\mu)
   + K_2^{(N)}(\Lambda_{\rm UV},\mu)\,\cdot
    \frac{(-\lambda_1)}{3\Lambda_{\rm UV}^2}
   + O\bigg[ \left( \frac{\Lambda_{\rm QCD}}{\Lambda_{\rm UV}} \right)^3
   \bigg] \right\} .
\end{equation}
This expansion is useful as long as the cutoff $\Lambda_{\rm UV}$ is 
chosen much larger than the typical hadronic scale characterizing the 
matrix elements of the local operators. The matching coefficients 
$K_n^{(N)}$ in this relation can be calculated using on-shell external 
$b$-quark states with residual momentum $k$. For operators of dimension 
up to 5 it suffices to calculate two-point functions. (Three and 
four-point functions would have to be considered at dimension~6.) We 
first evaluate the moments of the renormalized shape function in 
(\ref{Sonshell}), finding at one-loop order
\begin{eqnarray}\label{mom1}
   &&M_N^{\rm parton}(\Lambda_{\rm UV},\mu)\\\nonumber
   &&= (n\cdot k)^N\,\Bigg\{ 1 - \frac{C_F\alpha_s}{\pi} \left( 
    \ln^2\frac{\Lambda_{\rm UV}+n\cdot k}{\mu}
    + \ln\frac{\Lambda_{\rm UV}+n\cdot k}{\mu} + \frac{\pi^2}{24}
    \right) \nonumber\\\nonumber
   &&\mbox{}- \frac{C_F\alpha_s}{\pi}
    \sum_{j=1}^N \frac{1}{j} \left( 1
    + 2\ln\frac{\Lambda_{\rm UV}+n\cdot k}{\mu}
    - \sum_{l=j}^N \frac{2}{l} \right)
    \left[ \left( - \frac{\Lambda_{\rm UV}}{n\cdot k} \right)^j - 1
    \right] \Bigg\}. 
\end{eqnarray}
We then expand this result in powers of $n\cdot k/\Lambda_{\rm UV}$. 
Keeping the first three terms in the expansion, we obtain
\begin{equation}\label{mom2}
\begin{aligned}
   &M_0^{\rm parton}(\Lambda_{\rm UV},\mu) = 1 - \frac{C_F\alpha_s}{\pi}
    \left( \ln^2\frac{\Lambda_{\rm UV}}{\mu}
    + \ln\frac{\Lambda_{\rm UV}}{\mu} + \frac{\pi^2}{24} \right) \\
   &\quad\mbox{}- \frac{C_F\alpha_s}{\pi} \left[
    \frac{n\cdot k}{\Lambda_{\rm UV}}
    \left( 2\ln\frac{\Lambda_{\rm UV}}{\mu} + 1 \right)
    + \frac{(n\cdot k)^2}{\Lambda_{\rm UV}^2}
    \left( - \ln\frac{\Lambda_{\rm UV}}{\mu} + \frac12 \right) + \dots
    \right] , \\
   &M_1^{\rm parton}(\Lambda_{\rm UV},\mu) = n\cdot k \left[ 1
    - \frac{C_F\alpha_s}{\pi}
    \left( \ln^2\frac{\Lambda_{\rm UV}}{\mu}
    - \ln\frac{\Lambda_{\rm UV}}{\mu} + \frac{\pi^2}{24} - 1 \right) 
    \right] \\
   &\quad\mbox{}- \frac{C_F\alpha_s}{\pi} \left[
    \Lambda_{\rm UV}
    \left( - 2\ln\frac{\Lambda_{\rm UV}}{\mu} + 1 \right)
    + \frac{(n\cdot k)^2}{\Lambda_{\rm UV}}\,
    2\ln\frac{\Lambda_{\rm UV}}{\mu} + \dots \right] , \\
   &M_2^{\rm parton}(\Lambda_{\rm UV},\mu) = (n\cdot k)^2 \left[ 1
    - \frac{C_F\alpha_s}{\pi}
    \left( \ln^2\frac{\Lambda_{\rm UV}}{\mu}
    - 2\ln\frac{\Lambda_{\rm UV}}{\mu} + \frac{\pi^2}{24} - \frac12
    \right) \right] \\
   &\quad\mbox{}- \frac{C_F\alpha_s}{\pi} \left[
    \Lambda_{\rm UV}^2\,\ln\frac{\Lambda_{\rm UV}}{\mu}
    + n\cdot k\,\Lambda_{\rm UV}
    \left( - 2\ln\frac{\Lambda_{\rm UV}}{\mu} + 3\right) + \dots
    \right] .
\end{aligned}
\end{equation}
In the next step, we calculate the one-loop matrix elements of the local
operators $\bar h\,(in\cdot D)^N h$ between heavy-quark states with 
residual momentum $k$. The relevant diagrams are the same as in 
Figure~\ref{fig:hqet}, where now the black square represents the local 
operators. Keeping $v\cdot k$ non-zero to regularize infra-red 
singularities, we obtain for the bare matrix elements
\begin{eqnarray}
   \langle\bar h\,(in\cdot D)^N h\rangle
   &=& (n\cdot k)^N \Bigg\{ 1 - \frac{4C_F g_s^2}{(4\pi)^{2-\epsilon}}\,
    (-2v\cdot k)^{-2\epsilon} \sum_{j=1}^N
    \bigg( \begin{array}{c} N \\ j \end{array} \bigg)
    \left( \frac{2v\cdot k}{n\cdot k} \right)^j \nonumber\\
   &&\hspace{2.5cm} \times 
    (j-1-\epsilon)\,\Gamma(j-\epsilon)\,\Gamma(2\epsilon-j) \Bigg\} \,.
\end{eqnarray}
While the individual diagrams are infra-red divergent, taking the limit 
$v\cdot k\to 0$ in the sum of all contributions is possible without
encountering singularities. Then the one-loop contributions vanish, and 
the matrix elements simply reduce to their tree-level values. In other
words, the one-loop contributions correspond to a mixing with class-2
operators, whose hadronic matrix elements vanish by the equations of 
motions. It follows that in (\ref{mom2}) we must identify 
$(n\cdot k)^n\to\langle\bar h\,(in\cdot D)^n h\rangle$. Substituting the 
results for the HQET matrix elements given earlier, we obtain for the 
Wilson coefficients of the first three moments
\begin{equation}\label{Dresults}
\begin{aligned}
   &K_0^{(0)} = 1 - \frac{C_F\alpha_s}{\pi}
    \left( \ln^2\frac{\Lambda_{\rm UV}}{\mu}
    + \ln\frac{\Lambda_{\rm UV}}{\mu} + \frac{\pi^2}{24} \right) ,\quad
     K_2^{(0)} = \frac{C_F\alpha_s}{\pi} 
    \left( \ln\frac{\Lambda_{\rm UV}}{\mu} - \frac12 \right) , \\
   &K_0^{(1)} = \frac{C_F\alpha_s}{\pi} 
    \left( 2\ln\frac{\Lambda_{\rm UV}}{\mu} - 1 \right) ,\quad \quad
    K_2^{(1)} = -2\,\frac{C_F\alpha_s}{\pi}\, 
    \ln\frac{\Lambda_{\rm UV}}{\mu},  \\
   &K_0^{(2)} = - \frac{C_F\alpha_s}{\pi}\,
    \ln\frac{\Lambda_{\rm UV}}{\mu}, \quad \quad
    K_2^{(2)} = 1 - \frac{C_F\alpha_s}{\pi}
    \left( \ln^2\frac{\Lambda_{\rm UV}}{\mu}
    - 2\ln\frac{\Lambda_{\rm UV}}{\mu} + \frac{\pi^2}{24} - \frac12 
    \right) .
\end{aligned}
\end{equation}
At tree level, this reproduces the naive moment relations mentioned at 
the beginning of this section. Beyond tree level, the moments get 
corrected by calculable short-distance effects, which can be controlled 
using fixed-order perturbation theory as long as the ratio 
$\Lambda_{\rm UV}/\mu$ is of $O(1)$. In particular, the renormalized 
first moment no longer vanishes, but is proportional to the cutoff 
$\Lambda_{\rm UV}$ up to small power corrections. 

As mentioned earlier, the value of the first moment is connected with the 
definition of the heavy-quark mass (see also 
\cite{Mannel:1999gs,Mannel:2000aj}). The first moment of the renormalized 
shape function can be made to vanish to all orders in perturbation theory 
by choosing an appropriate scheme for the definition of $m_b$. So far our 
calculations have assumed the definition of the heavy-quark mass as a 
pole mass, $m_b^{\rm pole}$, which is implied by the HQET equation of 
motion $iv\cdot D\,h=0$. Results such as (\ref{Dresults}) are valid in 
this particular scheme. A more general choice is to allow for a residual 
mass term $\delta m$ in HQET, such that $iv\cdot D\,h=\delta m\,h$ with 
$\delta m=O(\Lambda_{\rm QCD})$ \cite{Falk:1992fm}. It is well known that 
the pole mass is an ill-defined concept, which suffers from infra-red 
renormalon ambiguities \cite{Bigi:1994em,Beneke:1994sw}. The parameter 
$\bar\Lambda_{\rm pole}=M_B-m_b^{\rm pole}$, which determines the support 
of the shape function in the pole-mass scheme, inherits the same 
ambiguities. It is therefore advantageous to eliminate the pole mass in 
favor of some short-distance mass. For the analysis of inclusive 
$B$-meson decays, a proper choice is to use a so-called low-scale 
subtracted heavy-quark mass $m_b(\mu_f)$ \cite{Bigi:1996si}, which is 
obtained from the pole mass by removing a long-distance contribution 
proportional to a subtraction scale 
$\mu_f=\mbox{few}\times\Lambda_{\rm QCD}$,
\begin{equation}\label{mdef}
   m_b^{\rm pole} = m_b(\mu_f)
   + \mu_f\,\,g\Big( \alpha_s(\mu), \frac{\mu_f}{\mu} \Big)
   \equiv m_b(\mu_f) + \delta m \,.
\end{equation}
As long as $m_b(\mu_f)$ is defined in a physical way, the resulting
perturbative expressions after elimination of the pole mass are 
well-behaved and not plagued by renormalon ambiguities. Replacing the 
pole mass by the physical mass shifts the values of $n\cdot k$ and 
$\omega$ by an amount $\delta m$, since 
$n\cdot(m_b^{\rm pole} v+k)=m_b(\mu_f)+(n\cdot k+\delta m)$, and because 
the covariant derivative in the definition of the shape function in 
(\ref{Sdef}) must be replaced by $in\cdot D-\delta m$ \cite{Falk:1992fm}. 
At the same time, $\bar\Lambda_{\rm pole}=\bar\Lambda(\mu_f)-\delta m$,
where $\bar\Lambda(\mu_f)=M_B-m_b(\mu_f)$ is a physical parameter. Note 
that this leaves the parameter $\hat\omega=\bar\Lambda-\omega$ and hence 
the shape function $\hat S(\hat\omega,\mu)$ invariant. This follows 
since $\bar\Lambda_{\rm pole}-\omega_{\rm pole}=\bar\Lambda(\mu_f)%
-(\omega_{\rm pole}+\delta m)$, where $\omega_{\rm pole}$ denotes the 
value in the pole-mass scheme used so far.

\subsection{Shape-function moments in a physical scheme}

From now on we will adopt a mass scheme defined by some specific choice 
of $\delta m$. Let us denote by $\omega=\omega_{\rm pole}+\delta m$ the 
value of the light-cone momentum variable in that scheme and define 
``physical'' moments $M_N^{\rm phys}$ as in (\ref{MNdef}), but with all 
parameters replaced by their values in the new scheme, in particular 
$\bar\Lambda=\bar\Lambda(\mu_f)$. Then the expressions for the moments 
in (\ref{mom1}) and (\ref{mom2}) change according to the replacements 
$n\cdot k\to n\cdot k+\delta m$ and 
$\Lambda_{\rm UV}\to\Lambda_{\rm UV}-\delta m$ everywhere. We now 
{\em choose\/} $\delta m$ such that the first moment vanishes, thereby 
defining a low-scale subtracted heavy-quark mass (with 
$\mu_f=\Lambda_{\rm UV}$) to all orders in perturbation theory. We will 
refer to this mass as the ``shape-function mass'' $m_b^{\rm SF}$. This is 
a ``physical'', short-distance mass in the sense that it is free of 
renormalon ambiguities. (However, the definition of the shape-function 
mass depends on the renormalization scheme used to define the shape 
function.) From (\ref{mom2}) and (\ref{mdef}), it follows that at 
one-loop order
\begin{equation}\label{mSFmpole}
   m_b^{\rm pole} = m_b^{\rm SF}(\mu_f,\mu) 
   + \mu_f\,\frac{C_F\alpha_s(\mu)}{\pi} \left[
   \left( 1 - 2\ln\frac{\mu_f}{\mu} \right) 
   + \frac23\,\frac{(-\lambda_1)}{\mu_f^2}\,\ln\frac{\mu_f}{\mu}
   + \dots \right] .
\end{equation}
Note that after introduction of the shape-function mass the coefficients
$K_n^{(1)}$ in the operator-product expansion for the moments in 
(\ref{beauty}) vanish by definition. However, to first order in 
$\alpha_s$ the values for the coefficients $K_n^{(0)}$ and $K_n^{(2)}$ of 
the zeroth and second moments given in (\ref{Dresults}) remain unchanged, 
since $\delta m=O(\alpha_s)$. This would no longer be true for the 
coefficients of higher moments.

The shape-function mass can be related to any other short-distance mass 
using perturbation theory. For instance, at one-loop order its relations 
to the potential-subtracted mass introduced in \cite{Beneke:1998rk} and 
to the kinetic mass defined in \cite{Bigi:1997fj,Benson:2003kp} read
\begin{equation}\label{mSFmPS}
   m_b^{\rm SF}(\mu_f,\mu_f) = m_b^{\rm PS}(\mu_f)
   = m_b^{\rm kin}(\mu_f) + \mu_f\,\frac{C_F\alpha_s(\mu_f)}{3\pi} \,.
\end{equation}
Note that, in addition to the dependence on the subtraction scale 
$\mu_f$, the shape-function mass depends on the scale $\mu$ at which the 
shape function is renormalized. While it is natural to set $\mu=\mu_f$, 
as we did here, this is not necessary. Given a value for the 
shape-function mass for some choice of scales, we can solve 
(\ref{mSFmpole}) to obtain its value for any other choice, using the fact 
that the pole mass is scale independent.

Proceeding in an analogous way, we can use the second moment to define a
physical kinetic-energy parameter, commonly called $\mu_\pi^2$. This 
quantity can be used to replace the HQET parameter $\lambda_1$, which 
like the pole mass suffers from infra-red renormalon ambiguities 
\cite{Martinelli:1995zw,Neubert:1996zy}. At one-loop order, we obtain
\begin{eqnarray}\label{mupi2pole}
   &&\frac{\mu_\pi^2(\Lambda_{\rm UV},\mu)}{3}
   \equiv \frac{M_2^{\rm phys}(\Lambda_{\rm UV},\mu)}
                 {M_0^{\rm phys}(\Lambda_{\rm UV},\mu)} \\
   &=& - \frac{C_F\alpha_s(\mu)}{\pi}\,\Lambda_{\rm UV}^2\,
    \ln\frac{\Lambda_{\rm UV}}{\mu} 
    + \frac{(-\lambda_1)}{3} \left[ 1 + \frac{C_F\alpha_s(\mu)}{\pi}
    \left( 3\ln\frac{\Lambda_{\rm UV}}{\mu} + \frac12 \right) \right]
    + \dots \,. \nonumber
\end{eqnarray}
Taking the ratio of $M_2^{\rm phys}$ and $M_0^{\rm phys}$ has the 
advantage of eliminating the double logarithmic radiative corrections 
from this expression. Our definition is similar to the running parameter 
$\mu_\pi^2$ defined in the kinetic scheme 
\cite{Bigi:1997fj,Benson:2003kp}. At one-loop order, the two parameters 
are related by
\begin{equation}\label{mupi2kin}
   \mu_{\pi}^2(\mu_f,\mu_f) = - \mu_f^2\,\frac{C_F\alpha_s(\mu_f)}{\pi}
   + [\mu_{\pi}^2(\mu_f)]_{\rm kin}
   \left[ 1 + \frac{C_F\alpha_s(\mu_f)}{2\pi} \right] .
\end{equation}
Given a value for the kinetic energy in the shape-function scheme for 
some choice of scales, we can solve (\ref{mupi2pole}) to obtain its value 
for any other choice, using that $\lambda_1$ is scale independent.

Similarly, each new moment of the renormalized shape function can be used 
to define a new physical, scale-dependent parameter 
\begin{displaymath}
A_N(\Lambda_{\rm UV},\mu)\equiv M_N^{\rm phys}(\Lambda_{\rm UV},\mu)/%
M_0^{\rm phys}(\Lambda_{\rm UV},\mu), 
\end{displaymath}
which coincides with the 
corresponding HQET parameter $A_N=\langle\bar h\,(in\cdot D)^N h\rangle$ 
at tree level, and which beyond tree level is related to HQET parameters 
through well-controlled perturbative expressions. Obviously, the presence 
of power divergences implies that higher moments are progressively less 
sensitive to HQET parameters, since they are dominated by the 
perturbative terms of order $\alpha_s\Lambda_{\rm UV}^N$. 

\subsection{Moments of the scheme-independent function 
\boldmath$\hat S(\hat\omega,\mu)$\unboldmath}

It will be useful to rewrite the moment relations derived above in terms 
of the variable $\hat\omega=\bar\Lambda-\omega$, which is invariant under
redefinitions of the heavy-quark mass. Defining a new set of 
scheme-independent moments
\begin{equation}\label{hatMNdef}
   \hat M_N(\mu_f,\mu) = \int\limits_0^{\mu_f+\bar\Lambda(\mu_f,\mu)}\!\!
   d\hat\omega\,\hat\omega^N\,\hat S(\hat\omega,\mu) \,,
\end{equation}
we obtain
\begin{eqnarray}\label{M0toM2}
   \hat M_0(\mu_f,\mu)
   &=& 1 - \frac{C_F\alpha_s(\mu)}{\pi}
    \left( \ln^2\frac{\mu_f}{\mu} + \ln\frac{\mu_f}{\mu} 
    + \frac{\pi^2}{24} \right)  \\
    &+&\frac{C_F\alpha_s(\mu)}{\pi} 
    \left( \ln\frac{\mu_f}{\mu} - \frac12 \right) 
    \frac{\mu_\pi^2(\mu_f,\mu)}{3\mu_f^2} + \dots , \nonumber\\
   \frac{\hat M_1(\mu_f,\mu)}{\hat M_0(\mu_f,\mu)}
   &=& \bar\Lambda(\mu_f,\mu) \,, \qquad\qquad
   \frac{\hat M_2(\mu_f,\mu)}{\hat M_0(\mu_f,\mu)}
    = \frac{\mu_\pi^2(\mu_f,\mu)}{3} + \bar\Lambda(\mu_f,\mu)^2 \,,\nonumber
\end{eqnarray}
where the parameters $\bar\Lambda(\mu_f,\mu)=M_B-m_b^{\rm SF}(\mu_f,\mu)$ 
and $\mu_\pi^2(\mu_f,\mu)$ should be considered as known physical 
quantities. Using the relations in the previous subsection, we have
\begin{eqnarray}\label{mustar}
   m_b^{\rm SF}(\mu_f,\mu)
   &=& m_b^{\rm SF}(\mu_*,\mu_*) + \mu_*\,\frac{C_F\alpha_s(\mu_*)}{\pi}
    \nonumber\\
   &&\mbox{}- \mu_f\,\frac{C_F\alpha_s(\mu)}{\pi} \left[ 
    \left( 1 - 2\ln\frac{\mu_f}{\mu} \right) 
    + \frac23\,\frac{\mu_\pi^2(\mu_f,\mu)}{\mu_f^2}\,
    \ln\frac{\mu_f}{\mu} \right] , \nonumber\\
   \mu_\pi^2(\mu_f,\mu)
   &=& \mu_\pi^2(\mu_*,\mu_*) \left[ 1 - \frac{C_F\alpha_s(\mu_*)}{2\pi}
    + \frac{C_F\alpha_s(\mu)}{\pi}
    \left( 3\ln\frac{\mu_f}{\mu} + \frac12 \right) \right]\nonumber\\
    &-& 3\mu_f^2\,\frac{C_F\alpha_s(\mu)}{\pi}\,\ln\frac{\mu_f}{\mu} \,,
    \end{eqnarray}
where $\mu_*$ denotes the scale at which initial values for the two
parameters are obtained, for instance using relations such as 
(\ref{mSFmPS}) and (\ref{mupi2kin}). These relations are particularly 
simple if one chooses $\mu_f=\mu$.

In the relations above we have eliminated the unphysical HQET parameter 
$\lambda_1$ in favor of the physical parameter $\mu_\pi^2$ defined in the 
shape-function scheme. At first sight, this seems to threaten the 
convergence of the operator product expansion. For instance, the term 
proportional to $\mu_\pi^2$ in the expression for the zeroth moment 
$\hat M_0$ in (\ref{M0toM2}) contains a {\em leading-power\/} 
perturbative contribution of order $\alpha_s^2(\mu)$, and similar 
contributions would arise from all other terms in the expansion. These 
contributions would have to be subtracted from the Wilson coefficient of 
the first term, if this coefficient were computed to two-loop order. The 
overall convergence of the operator product expansion is unaffected by 
this reorganization of perturbative corrections.

\subsection{Asymptotic behavior of the shape function}

The fact that for sufficiently large values of the cutoff the moments of 
the shape function can be calculated using an operator-product expansion 
implies that a similar expansion can be used to obtain a 
model-independent description of the asymptotic behavior of the shape 
function. Taking the derivative of the zeroth moment $\hat M_0$ in 
(\ref{hatMNdef}) with respect to $\mu_f$, one obtains
\begin{equation}
   \hat S(\hat\omega,\mu) \Big|_{\hat\omega=\mu_f+\bar\Lambda(\mu_f,\mu)}
   = \left( 1 - \frac{dm_b^{\rm SF}(\mu_f,\mu)}{d\mu_f} \right)^{-1}
   \frac{d}{d\mu_f}\,\hat M_0(\mu_f,\mu) \,.
\end{equation}
This relation can be trusted as long as $\mu_f\gg\Lambda_{\rm QCD}$. It 
allows us to determine the behavior of the shape function for large 
values of $\hat\omega$. From (\ref{M0toM2}) we find at one-loop order
\begin{eqnarray}\label{Sasymp}
   \hat S(\hat\omega,\mu) &=& - \frac{C_F\alpha_s(\mu)}{\pi}\,
   \frac{1}{\hat\omega-\bar\Lambda}
   \left[ \left( 2\ln\frac{\hat\omega-\bar\Lambda}{\mu} + 1 \right)\right.\nonumber\\
   &+& \left.\frac23\,\frac{\mu_\pi^2}{(\hat\omega-\bar\Lambda)^2}
   \left( \ln\frac{\hat\omega-\bar\Lambda}{\mu} - 1 \right)
   + \dots \right] .
\end{eqnarray}
The precise definitions of $\bar\Lambda$ and $\mu_\pi^2$ are not 
specified at this order. (Note that the shape function cannot depend on
the value of the cutoff $\mu_f$.) We have checked that this asymptotic
behavior of the shape function is consistent with the evolution equation 
(\ref{wow}) when expanded to first order in $\alpha_s$.

Relation (\ref{Sasymp}) is a model-independent result as long as 
$\hat\omega\gg\Lambda_{\rm QCD}$. We stress the remarkable fact that this 
radiative tail of the shape function is {\em negative}, in contrast with 
the naive expectation based on a probabilistic interpretation of the 
shape function as a momentum distribution function. The point is that the 
definition of the renormalized shape function requires scheme-dependent 
ultra-violet subtractions. From (\ref{Sasymp}) it follows that the shape 
function must have a zero, which for sufficiently large $\mu$ is located 
at a value $\hat\omega_0\approx\bar\Lambda+\mu/\sqrt{e}$.

\section{Conclusions}
\label{sec:recap}
The hadronic physics governing the inclusive semileptonic decay
$\bar B\to X_u\,l^-\bar\nu$ is encoded in the structure functions $\tilde{W}_i$
appearing in the Lorentz decomposition of the hadronic tensor 
$W^{\mu\nu}$ in (\ref{equation_wtdecomp}). In the shape-function region, only two
combinations of these functions are required at leading order in 
$\Lambda_{\rm QCD}/m_b$. They follow from the factorization formula 
(\ref{Wres}) using the explicit results for the hard functions $H_{ij}$
and the jet function $J$ derived in this chapter. Explicitly, we obtain at 
next-to-leading order in renormalization-group improved perturbation 
theory
\begin{eqnarray}\label{master}
&&\tilde{W}_1=U_y(\mu_h,\mu_i)\left\{1 +\! \frac{C_F\alpha_s(\mu_h)}{4\pi} \!\left[\!
    - 4\ln^2\!\frac{y m_b}{\mu_h} + 10\ln\frac{y m_b}{\mu_h} - 4\ln y
    - \frac{2\ln y}{1-y}\right.\right.\nonumber\\ 
    &&\left.\left.- 4L_2(1\!-\!y) - \frac{\pi^2}{6} -\! 12 \right] \right\}\int_0^{P_+}\!d\hat\omega\,y m_b\,J(y m_b(P_+-\hat\omega),\mu_i)\,\hat S(\hat\omega,\mu_i)+\dots\nonumber\\\nonumber\\
&&\tilde{W}_4 + \frac1y \tilde{W}_5
   = U_y(\mu_h,\mu_i)\frac{C_F\alpha_s(\mu_h)}{4\pi}\,\frac{2\ln y}{1-y}\times\nonumber\\
 &&  \int_0^{P_+}\!d\hat\omega\,y m_b\,J(y m_b(P_+-\hat\omega),\mu_i)\,\hat S(\hat\omega,\mu_i)
    +\dots\,, 
\end{eqnarray}
where the dots represent power corrections in $\Lambda_{\rm QCD}/m_b$. 

In all our results, $m_b$ 
denotes the heavy-quark mass defined in the ``shape-function scheme'' 
introduced in Section~\ref{sec:properties}.
The jet function 
$J(y m_b(P_+-\hat\omega),\mu_i)$ at an intermediate hard-collinear scale 
$\mu_i\sim\sqrt{m_b\Lambda_{\rm QCD}}$ can be calculated in fixed-order 
perturbation theory. The relevant expression valid at one-loop order is 
given in (\ref{Jres}). The above results contain the  
renormalization-group function $U_y(\mu_h,\mu_i)$, which arises in the solution of evolution 
equations discussed in Section~\ref{sec:RGevol}. Our results are formally independent of the precise choice of 
high-energy matching scale $\mu_h\sim m_b$. The numerical effect of the residual $\mu_h$ dependence 
remaining after truncation of the perturbative expansion has been studied 
in \cite{Bosch:2003fc} and was found to be small. 

The function $\hat S(\hat\omega,\mu_i)$ in (\ref{master}) is the shape 
function after the transformation of variables from $\omega$ to 
$\hat\omega=\bar\Lambda-\omega$. The limits of integration for the 
variable $\hat\omega$ (i.e., $0\le\hat\omega\le P_+$) are set by 
hadronic kinematics and are independent of the definition of the 
heavy-quark mass. The shape function is a non-perturbative object, which 
at present cannot be predicted from first principles. It enters our 
results (\ref{master}) renormalized at the intermediate hard-collinear 
scale $\mu_i$. In (\ref{wow}), we have presented an analytic formula 
(valid at all orders in renormalization-group improved perturbation 
theory) that relates the shape function at a high scale to the shape 
function renormalized at a low hadronic scale. Many properties of the 
shape function that were so far unknown have been derived in 
Section~\ref{sec:properties}. In particular, we have given explicit 
formulae relating the moments of the shape function to HQET parameters, 
and we have proved that the shape function has a negative tail for large 
values of $\hat\omega$, whose explicit form can be calculated using an 
operator product expansion. These new insights about the shape function 
will be very helpful in constructing a realistic model for the function 
$\hat S(\hat\omega,\mu_i)$, which can then be refined by tuning it to 
experimental data such as the photon energy spectrum in inclusive 
$\bar B\to X_s\gamma$ decays.

\chapter{Non-Perturbative Corrections}
\label{chapter_sub}
\section{Introduction}

In the previous chapter we have seen how to incorporate 
leading order perturbative corrections to $\bar
B\to X_u\,l\bar\nu$. Corrections suppressed by a power of 
$\Lambda_{\rm QCD}/m_b$ are considered the second largest source of 
uncertainty. for inclusive charmless $B$ decays. The present chapter is 
devoted to a more thorough study of power 
corrections to inclusive $B$ decays distributions in the shape-function 
region, using the two-step matching procedure.  

In a first step, hard 
fluctuations are integrated out by matching the QCD currents onto 
soft-collinear effective theory (SCET) \cite{Bauer:2000yr,Beneke:2002ph}. 
The current correlator is then expanded in terms of light-cone 
operators in heavy-quark effective theory (HQET) \cite{Neubert:1993mb}, 
thereby integrating out fluctuations at the hard-collinear scale. The fact 
that the relevant HQET operators live on the light cone follows from the 
structure of the multipole expansion of soft fields in SCET. We carry out 
the matching procedure at tree level and to order 
$\Lambda_{\rm QCD}/m_b$ in the heavy-quark expansion. We indicate how our 
results would change if loop corrections were included.

\section{Short-distance expansion of the hadronic tensor}

SCET is the appropriate effective field theory for the description of the 
interactions among soft and hard-collinear degrees of freedom. Its Lagrangian 
is organized in an expansion in powers of $\sqrt{\lambda}$. The leading-order 
Lagrangian is
\begin{eqnarray}\label{L0}
   {\cal L}_{\rm SCET}^{(0)}
   &=& \bar\xi\,\frac{\nbslash}{2}
    \left( in\cdot D_{hc} + g n\cdot A_s(x_-)
    + i\Dslash_{\perp hc}\,\frac{1}{i\bar n\cdot D_{hc}}\,i\Dslash_{\perp hc}
    \right) \xi \nonumber\\
   &&\mbox{}+ \bar q\,i\Dslash_s\,q + \bar h\,iv\cdot D_s\,h
    + {\cal L}_{\rm YM}^{(0)} \,,
\end{eqnarray}
where $\xi$ is a hard-collinear quark field, $q$ is a soft, massless quark 
field, $h$ is a heavy-quark field defined in HQET, $A_s$ is a soft gluon 
field, and $iD_{hc}^\mu=i\partial^\mu+g A_{hc}^\mu$ is the covariant 
derivative containing a hard-collinear gluon field. All fields in the above 
Lagrangian are evaluated at point $x$, except for the soft gluon field in the 
first term, which is evaluated at $x_-=\frac12\,(\bar n\cdot x)\,n$. The 
explicit form of the leading-order Yang-Mills Lagrangian can be found in 
\cite{Beneke:2002ph,Beneke:2002ni}. 

The terms up to second order in the expansion in $\sqrt{\lambda}$ are
\begin{eqnarray}
   {\cal L}_{\rm SCET}^{(1)}
   &=& {\cal L}_\xi^{(1)} + {\cal L}_{\xi q}^{(1)} + {\cal L}_{\rm YM}^{(1)}
    \,, \nonumber\\
   {\cal L}_{\rm SCET}^{(2)}
   &=& {\cal L}_\xi^{(2)} + {\cal L}_{\xi q}^{(2)} + {\cal L}_h^{(2)}
    + {\cal L}_{\rm YM}^{(2)} \,,
\end{eqnarray}
where
\begin{equation}
   {\cal L}_h^{(2)} = \frac{1}{2m_b} \left[ \bar h\,(iD_s)^2 h
   + \frac{C_{\rm mag}}{2}\,\bar h\,\sigma_{\mu\nu}\,g G_s^{\mu\nu} h
   \right]
\end{equation}
is the next-to-leading term in the expansion of the HQET Lagrangian 
\cite{Neubert:1993mb}. Expressions for the remaining Lagrangian corrections 
have been presented in \cite{Beneke:2002ph}.

While it is consistent to apply a perturbative expansion at the hard and 
hard-collinear scales, the soft-gluon couplings to hard-collinear fields are 
non-perturbative and must be treated to all orders in the coupling constant. 
For instance, the hard-collinear quark propagator derived from (\ref{L0}) 
should be taken to be the propagator in the background of the soft gluon 
field, summing up arbitrarily many insertions of the field $A_s$. The most 
convenient way of achieving this summation is to decouple the leading-order 
interactions between the soft gluon field and hard-collinear fields in 
(\ref{L0}) with the help of a field redefinition, under which 
\cite{Bauer:2000yr,Becher:2003qh}
\begin{equation}
   \xi(x) = S(x_-)\,\xi^{(0)}(x) \,, \qquad
   A_{hc}^\mu(x) = S(x_-)\,A_{hc}^{(0)\mu}(x)\,S^\dagger(x_-) \,,
\end{equation}
where
\begin{equation}
   S(x) = P\exp\Bigg( ig \int\limits_{-\infty}^0\!dt\,n\cdot A_s(x+tn)
   \Bigg)
\end{equation}
is a soft Wilson line along the $n$ direction. Introducing the new fields into 
the Lagrangian yields
\begin{equation}\label{Ldecoupled}
   {\cal L}_\xi^{(0)} = \bar\xi^{(0)}\,\frac{\nbslash}{2}
   \left( in\cdot D_{hc}^{(0)} 
   + i\Dslash_{\perp hc}^{(0)}\,\frac{1}{i\bar n\cdot D_{hc}^{(0)}}\,
   i\Dslash_{\perp hc}^{(0)} \right) \xi^{(0)} \,,
\end{equation}
and similarly all interactions between soft and hard-collinear gluon fields 
are removed from the Yang-Mills Lagrangian ${\cal L}_{\rm YM}^{(0)}$. The 
propagator of the new hard-collinear quark field is now given by the simple
expression
\begin{equation}\label{xiprop}
   \Delta_\xi(x-y) 
   = \langle 0|\,T\{ \xi^{(0)}(x)\,\bar\xi^{(0)}(y) \}\,|0\rangle
   = \frac{\nslash}{2} \int\frac{d^4p}{(2\pi)^4}\,
   \frac{i\bar n\cdot p}{p^2+i\epsilon}\,e^{-ip\cdot(x-y)} \,.
\end{equation}
The effect of soft-gluon attachments is taken into account by factors of the 
Wilson line $S$ in the results below.

We now list the expressions for the subleading corrections to the SCET
Lagrangian in terms of the redefined fields, using the formalism of 
gauge-invariant building blocks \cite{Hill:2002vw}. We define
\begin{equation}
   \X = W^\dagger \xi^{(0)} \,, \qquad
   \A_{hc}^\mu = W^\dagger (iD_{hc}^{(0)\mu} W) \,,
\end{equation}
where
\begin{equation}
   W = P\exp\Bigg( ig \int\limits_{-\infty}^0\!dt\,
   \bar n\cdot A_{hc}^{(0)}(x+t\bar n) \Bigg)
\end{equation}
is a hard-collinear Wilson line. These ``calligraphic'' fields are invariant
under both hard-collinear and soft gauge transformations. Note that in the
light-cone gauge, $\bar n\cdot A_{hc}^{(0)}=0$, we simply have $\X=\xi^{(0)}$ 
and $\A_{hc}^\mu=gA_{hc}^{(0)\mu}$. In terms of these fields, the results 
compiled in \cite{Beneke:2002ph} take the form
\begin{eqnarray}
   {\cal L}_\xi^{(1)} &=& \bar\X\,\frac{\nbslash}{2}\,x_\perp^\mu n^\nu
    \left( S^\dagger g G_{\mu\nu} S \right)_{x_-}\!\X \,, 
     \nonumber
\end{eqnarray}
\begin{eqnarray}\label{L1and2}
   {\cal L}_\xi^{(2)} &=& \bar\X\,\frac{\nbslash}{2} \left( 
    \frac{n\cdot x}{2}\,\bar n^\mu n^\nu 
    \left( S^\dagger g G_{\mu\nu} S \right)_{x_-}
    + \frac{x_\perp^\mu x_\perp^\rho}{2}\,n^\nu
    \left( S^\dagger [D_\rho,g G_{\mu\nu}] S \right)_{x_-} \right)\!\X 
    \nonumber\\
   &+& \bar\X\,\frac{\nbslash}{2} \left( 
    i\calDslash_{\perp hc}\,\frac{1}{i\bar n\cdot\partial}\,
    \frac{x_\perp^\mu}{2}\,\gamma_\perp^\nu
    \left( S^\dagger g G_{\mu\nu} S \right)_{x_-}\right.\nonumber\\
    &+&\left. \frac{x_\perp^\mu}{2}\,\gamma_\perp^\nu
    \left( S^\dagger g G_{\mu\nu} S \right)_{x_-}
    \frac{1}{i\bar n\cdot\partial}\,i\calDslash_{\perp hc} \right) \X \,,
    \nonumber\\
   {\cal L}_{\xi q}^{(1)} &=& \left( \bar q S \right)_{x_-}\!
    i\calDslash_{\perp hc}\,\X + \mbox{h.c.} \,,
\end{eqnarray}
where $i\D_{hc}^\mu=i\partial^\mu+\A_{hc}^\mu$, and we have dropped the 
subscript ``$s$'' on the soft covariant derivative and field strength. The 
expression for $ {\cal L}_{\xi q}^{(2)}$ will not be needed for our analysis. 
The notation $(\dots)_{x_-}$ indicates that, in interactions with 
hard-collinear fields, soft fields are multipole expanded and live at position 
$x_-$, whereas hard-collinear fields are always evaluated at position $x$. 
Because the SCET Lagrangian is not renormalized \cite{Beneke:2002ph}, the 
above expressions are valid to all orders in perturbation theory.

Next, we need the expressions for heavy-light current operators in SCET. In 
general, a QCD current $\bar q\,\Gamma\,b$ matches onto
\begin{equation}
   \bar q(x)\,\Gamma\,b(x)
   = e^{-im_b v\cdot x} \left( J_A^{(0)} + J_A^{(1)} + J_A^{(2)}
   + J_B^{(1)} + J_B^{(2)} + \dots \right) , 
\end{equation}
where we distinguish between type-$A$ ``two-particle'' operators and type-$B$ 
``three-particle'' operators \cite{Hill:2004if}. The operators arising at tree 
level are \cite{Beneke:2002ph,Hill:2004if,Pirjol:2002km}
\begin{eqnarray}
   J_A^{(0)} &=& \bar\X\,\Gamma \left( S^\dagger h \right)_{x_-} ,
    \nonumber\\
   J_A^{(1)} &=& \bar\X\,\Gamma\,x_\perp^\mu 
    \left( S^\dagger D_\mu h \right)_{x_-}
    + \bar\X\,\frac{\nbslash}{2}\,i\!\overleftarrow{\delslash}\!_\perp\,
    \frac{1}{i\bar n\cdot\overleftarrow{\partial}}\,\Gamma
    \left( S^\dagger h \right)_{x_-} , \nonumber\\
   J_A^{(2)} &=& \bar\X\,\Gamma \left[
    \frac{n\cdot x}{2} \left( S^\dagger \bar n\cdot D h \right)_{x_-} 
    + \frac{x_\perp^\mu x_\perp^\nu}{2}
    \left( S^\dagger D_\mu D_\nu h \right)_{x_-} 
    + \left( S^\dagger \frac{i\Dslash}{2m_b}\,h \right)_{x_-} \right]
    \nonumber\\
   &+& \bar\X\,\frac{\nbslash}{2}\,
    i\!\overleftarrow{\delslash}\!_\perp\,
    \frac{1}{i\bar n\cdot\overleftarrow{\partial}}\,\Gamma\,x_\perp^\mu
    \left( S^\dagger D_\mu h \right)_{x_-} ,
\end{eqnarray}
and
\begin{eqnarray}
   J_B^{(1)} &=&\mbox{}- \bar\X\,\frac{\nbslash}{2}\,\calAslash_{\perp hc}\,
    \frac{1}{i\bar n\cdot\overleftarrow{\partial}}\,\Gamma
    \left( S^\dagger h \right)_{x_-} 
    - \bar\X\,\Gamma\,\frac{\nslash}{2m_b}\,
    \calAslash_{\perp hc} \left( S^\dagger h \right)_{x_-} , \nonumber\\
   J_B^{(2)} &=&\mbox{}- \bar\X\,\Gamma \left(
    \frac{1}{i\bar n\cdot\partial} + \frac{\nslash}{2m_b} 
    \right) n\cdot\A_{hc} \left( S^\dagger h \right)_{x_-} \nonumber\\
   &-& \bar\X \left( \frac{\nbslash}{2}\,\calAslash_{\perp hc}\,
    \frac{1}{i\bar n\cdot\overleftarrow{\partial}}\,\Gamma
    + \Gamma\,\frac{\nslash}{2m_b}\,\calAslash_{\perp hc} \right)
    x_\perp^\mu \left( S^\dagger D_\mu h \right)_{x_-} \nonumber\\
   &-& \bar\X\,\Gamma\,\frac{1}{i\bar n\cdot\partial}\,
    \frac{(i\calDslash_{\perp hc}\,\calAslash_{\perp hc})}{m_b}
    \left( S^\dagger h \right)_{x_-}\nonumber\\
    &+& \bar\X\,\frac{i\overleftarrow{\calDslash\!}_{\perp hc}}{m_b}\,
    \frac{1}{i\bar n\cdot\overleftarrow{\partial}}\,\frac{\nbslash}{2}\,
    \Gamma\,\frac{\nslash}{2}\,\calAslash_{\perp hc}
    \left( S^\dagger h \right)_{x_-} \,.
\end{eqnarray}
The expressions for the currents beyond tree level are more complicated, 
primarily because several new Dirac structures appear. The relevant formulae 
are known at leading \cite{Bauer:2000yr} and next-to-leading order 
\cite{Hill:2004if} in the power expansion. The corresponding results for the 
currents $ J_A^{(2)}$ and $J_B^{(2)}$ have not yet been derived. They would be 
needed if the analysis in this chapter should be extended beyond tree level.

If perturbative corrections at the hard-collinear scale are neglected, the
hard-collinear gluon fields can be dropped, and the above expressions for the 
effective Lagrangians and currents simplify. In this approximation 
$\X\to\xi^{(0)}$, $\A_{hc}^\mu\to 0$, $i\D_{hc}^\mu\to i\partial^\mu$, and 
$J_B^{(n)}\to 0$. While this leads to great simplifications in the 
calculation, we stress that the structures of soft fields that arise do not 
simplify. We find operators containing $S^\dagger g G_{\mu\nu} S$, 
$S^\dagger [D_\rho,g G_{\mu\nu}] S$, $S^\dagger h$, $S^\dagger D_\mu h$, and 
$S^\dagger D_\mu D_\nu h$, and the same operators would arise if the 
calculation was extended beyond the tree approximation. The only exception is 
that we no longer retain Lagrangian corrections containing the soft quark 
field, because ${\cal L}_{\xi q}^{(1)}\to 0$ in the limit where the 
hard-collinear gluon field is neglected. (The term 
$\bar q S\,i\delslash_\perp\xi^{(0)}$ is forbidden by momentum conservation.) 
In Section~\ref{sec:4quark}, we analyze the subleading shape functions 
introduced at $O(\alpha_s)$ by two insertions of ${\cal L}_{\xi q}^{(1)}$.

With all the definitions in place, we are now ready to evaluate the current
correlator $T_{ij}$ including terms of up to second order in $\sqrt{\lambda}$, 
working at lowest order in $\alpha_s$ at the hard and hard-collinear scales. 
The leading term is readily found to be
\begin{equation}\label{Tij0}
   T_{ij}^{(0)} = - \int d^4x\,e^{i(q-m_b v)\cdot x}
   \int \frac{d^4p}{(2\pi)^4}\,e^{ip\cdot x}\,
   \frac{\bar n\cdot p}{p^2+i\epsilon}\,
   \left( \bar h S \right)_{0} \Gamma_i\,\frac{\nslash}{2}\,\Gamma_j
   \left( S^\dagger h \right)_{x_-} .
\end{equation}
First-order corrections in $\sqrt{\lambda}$ vanish by rotational invariance in 
the transverse plane (provided we choose the coordinate system such that 
$v_\perp=0$ and $q_\perp=0$), i.e., $T_{ij}^{(1)}=0$. At second order in the 
expansion the correlator receives several contributions, which can be 
represented symbolically as
\begin{equation}\label{JJ}
   J^{\dagger(2)}\,J^{(0)} \,, \quad
   J^{\dagger(1)}\,J^{(1)} \,, \quad
   J^{\dagger(0)}\,J^{(2)} \,,
\end{equation}
\begin{equation}\label{JJL}
   J^{\dagger(1)}\,J^{(0)} \int d^4z\,{\cal L}_\xi^{(1)} \,,
    \quad
   J^{\dagger(0)}\,J^{(1)} \int d^4z\,{\cal L}_\xi^{(1)} \,,
    \quad
   J^{\dagger(0)}\,J^{(0)} \int d^4z \left[ {\cal L}_\xi^{(2)}
    + {\cal L}_h^{(2)} \right] ,
\end{equation}
and
\begin{equation}\label{JJLL}
   J^{\dagger(0)}\,J^{(0)} \int d^4z\,{\cal L}_\xi^{(1)}
   \int d^4w\,{\cal L}_\xi^{(1)} \,,
\end{equation}
where at tree level only the type-$A$ current operators appear. Examples of 
these time-ordered products are depicted in Figure~\ref{fig:graphs}. At 
$O(\alpha_s)$, one must include a tree-level contribution of the form
\begin{equation}\label{Lxiq2}
   J^{\dagger(0)}\,J^{(0)} \int d^4z\,{\cal L}_{\xi q}^{(1)}
   \int d^4w\,{\cal L}_{\xi q}^{(1)} \,,
\end{equation}
shown by the last diagram in the figure. Beyond tree level, one would also 
have to include the contributions from the type-$B$ current operators.

\begin{figure}
\begin{center}
\epsfig{file=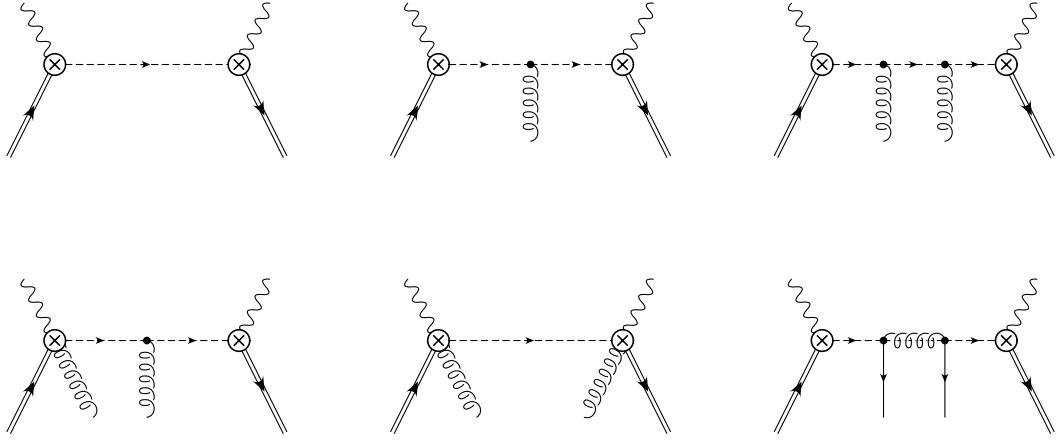,width=14cm}
\end{center}
\vspace{-0.2cm}
\centerline{\parbox{15cm}{\caption{\label{fig:graphs}
Representative examples of time-ordered products contributing to the current 
correlator $T_{ij}$ in SCET. Double lines show heavy-quark fields, dashed ones 
hard-collinear fields, and wavy lines denote the external currents. Lagrangian 
insertions and higher-order effective current operators are exemplified by 
soft gluons.}}}
\end{figure}

Because the external $B$-meson states in the definition of the hadronic tensor 
contain only soft constituents, and because the hard-collinear fields have 
been decoupled from the soft fields in the leading-order SCET Lagrangian 
(\ref{Ldecoupled}), it is possible to contract all hard-collinear fields in 
the time-ordered products. At tree level, all we need is the hard-collinear 
quark propagator (\ref{xiprop}). Derivatives acting on hard-collinear fields 
give powers of $p$ in momentum space, whereas components of $x^\mu$ appearing
in the multipole-expanded expressions for the effective Lagrangians and 
currents can be turned into derivatives $\partial/\partial p$ acting on the 
momentum-space amplitudes. Each insertion of a SCET Lagrangian correction 
introduces an integral over soft fields located along the $n$ light-cone. 
Consequently, the time-ordered products in (\ref{JJ}) lead to expressions 
involving bi-local operators as in (\ref{Tij0}), while those in (\ref{JJL}), 
(\ref{JJLL}), and (\ref{Lxiq2}) also lead to tri- and quadri-local operators.
At first sight, this would seem to require the introduction of complicated 
subleading shape functions depending on up to three momentum variables 
$\omega_i$, defined in terms of the Fourier transforms of the matrix elements 
of the non-local operators. However, the non-localities can be reduced using 
partial-fraction identities for the resulting hard-collinear quark 
propagators. At tree level, it suffices to define shape functions of a single 
variable $\omega$.

To see how this works, consider the effect of an insertion of the Lagrangian 
${\cal L}_\xi^{(1)}$ in (\ref{L1and2}). Since all hard-collinear fields must 
be contracted, we may consider without loss of generality the expression
\begin{equation}\label{typical}
   L(y_1,y_2) 
   = i\!\int d^4z\,\Delta_\xi(y_1-z)\,\frac{\nbslash}{2}\,z_\perp^\mu\,n^\nu
   \left( S^\dagger g G_{\mu\nu} S \right)_{z_-} \Delta_\xi(z-y_2) \,. 
\end{equation}
The fact that the soft fields live at position $z_-$ implies that the two 
hard-collinear quark propagators carry the same momentum components 
$\bar n\cdot p$ and $p_\perp$. In analogy with the definition of the 
hard-collinear calligraphic gluon field, we now introduce the soft field 
\cite{Hill:2002vw}
\begin{equation}\label{Asdef}
   \A_{s\mu}(x) = S^\dagger(x)\,(iD_\mu S)(x)
   = - \int\limits_{-\infty}^0\!dt\,n^\nu 
   \left( S^\dagger g G_{\mu\nu} S \right)(x+tn) \,,
\end{equation}
which allows us to write $n^\nu\,(S^\dagger g G_{\mu\nu} S)_{z_-}$ as a 
derivative, $-n\cdot\partial_z\,\A_{s\mu}(z_-)$. Integrating by parts in 
(\ref{typical}), we find
\begin{eqnarray}
   L(y_1,y_2) &=& i\!\int d^4z\,\Delta_\xi(y_1-z)\,
    \frac{\nbslash}{2}\,z_\perp^\mu\,\A_{s\mu}(z_-)\,
    n\cdot\partial_z\,\Delta_\xi(z-y_2) \nonumber\\
   &+& i\!\int d^4z\,[n\cdot\partial_z\,\Delta_\xi(y_1-z)]\,
    \frac{\nbslash}{2}\,z_\perp^\mu\,\A_{s\mu}(z_-)\,\Delta_\xi(z-y_2) . 
\end{eqnarray}
The hard-collinear propagator is a Green's function obeying the differential 
equation
\begin{equation}
   \left( n\cdot\partial + \frac{\partial_\perp^2}{\bar n\cdot\partial}
   \right) \Delta_\xi(x) = \frac{\nslash}{2}\,\delta^{(4)}(x) \,.
\end{equation}
This allows us to write
\begin{eqnarray}
   L(y_1,y_2)
   &=& i \Big[ y_{2\perp}\cdot\A_s(y_{2-}) - y_{1\perp}\cdot\A_s(y_{1-})
    \Big]\,\Delta_\xi(y_1-y_2) \nonumber\\
   &&\mbox{}- i\!\int d^4z\,\Delta_\xi(y_1-z)\,\frac{\nbslash}{2}\,
    z_\perp\cdot\A_s(z_-)\,
    \frac{\partial_{z\perp}^2}{\bar n\cdot\partial_z}\,\Delta_\xi(z-y_2)
    \nonumber\\
   &&\mbox{}- i\!\int d^4z
    \left[ \frac{\partial_{z\perp}^2}{\bar n\cdot\partial_z}\,
    \Delta_\xi(y_1-z) \right]
    \frac{\nbslash}{2}\,z_\perp\cdot\A_s(z_-)\,\Delta_\xi(z-y_2) . 
\end{eqnarray}
The terms involving transverse derivatives vanish at tree level, because they 
provide powers of transverse momenta of external lines, which are zero (recall 
that $v_\perp=0$ and $q_\perp=0$). This leaves the terms shown in the first 
line, in which the $z$ integral has been eliminated, and in which the product 
of two propagators has been reduced to a single propagator. In momentum space, 
these manipulations correspond to the partial-fraction identity
\begin{eqnarray}\label{partialfractions}
   &&\frac{1}{\bar n\cdot p\,n\cdot p+p_\perp^2}\,
   \frac{1}{\bar n\cdot p\,(n\cdot p+\omega)+p_\perp^2}=\nonumber\\
   &=& \frac{1}{\bar n\cdot p\,\omega}
   \left[ \frac{1}{\bar n\cdot p\,n\cdot p+p_\perp^2}
   -    \frac{1}{\bar n\cdot p\,(n\cdot p+\omega)+p_\perp^2} \right] ,
\end{eqnarray}
valid for two momenta that differ only in their $n\cdot p$ components.

In the sum of all terms many cancellations and simplifications take place, and
we find the rather simple result
\begin{equation}\label{Tresult}
   T_{ij}^{(2)} = - \int d^4x\,e^{i(q-m_b v)\cdot x}
   \int \frac{d^4p}{(2\pi)^4}\,e^{ip\cdot x}\,
   \frac{\bar n\cdot p}{p^2+i\epsilon}\,\sum_{n=1}^4 O_n(x_-) \,,
\end{equation}
where
\begin{eqnarray}
   O_1(x_-) &=& i\int d^4z\,T\{
    \left( \bar h S \right)_{0} \Gamma_i\,\frac{\nslash}{2}\,\Gamma_j 
    \left( S^\dagger h \right)_{x_-} {\cal L}_h^{(2)}(z) \} \,, \nonumber\\
   O_2(x_-)
   &=& \frac{1}{2m_b} \left[ 
    \left( \bar h S \right)_{0} \Gamma_i\,\frac{\nslash}{2}\,\Gamma_j 
    \left( S^\dagger i\Dslash h \right)_{x_-}
    + \left( \bar h (-i\!\overleftarrow{\Dslash}) S \right)_{0}
    \Gamma_i\,\frac{\nslash}{2}\,\Gamma_j \left( S^\dagger h \right)_{x_-} 
    \right] , \nonumber\\
   O_3(x_-)
   &=& \frac{1}{\bar n\cdot p} \left[
    \left( \bar h S \right)_{0} \Gamma_i\,\frac{\nslash\nbslash}{4}\,
    \gamma_\mu\,\Gamma_j \left( S^\dagger iD_\perp^\mu h \right)_{x_-}\right.\nonumber\\
    &+& \left.\left( \bar h iD_\perp^\mu S \right)_{0} \Gamma_i\,\gamma_\mu\,
    \frac{\nbslash\nslash}{4}\,\Gamma_j \left( S^\dagger h \right)_{x_-}
    \right] , \nonumber\\[-0.15cm]
   O_4(x_-)
   &=& \frac{i}{\bar n\cdot p} \int\limits_0^{\bar n\cdot x/2}\!dt\,
    \left( \bar h S \right)_{0} \Gamma_i\,
    \Big( S^\dagger i\Dslash_\perp \frac{\nslash}{2}\,i\Dslash_\perp S
    \Big)_{tn}\,\Gamma_j \left( S^\dagger h \right)_{x_-} .
\end{eqnarray}
In deriving these expressions, we have made use of the identity 
$iD^\mu=S (i\partial^\mu+\A_s^\mu) S^\dagger$ with $\A_s^\mu$ as defined in 
(\ref{Asdef}). Note that on the space of forward matrix elements the result 
for $T_{ij}^{(2)}$ is hermitean, because we can integrate by parts and use 
translational invariance.

In the last step, we can simplify the operator $O_2$ by noting that the HQET
equation of motion, $iv\cdot D h=0$, along with $\vslash h=h$, implies
\begin{equation}
   S^\dagger i\Dslash h = S^\dagger i\Dslash_\perp h
   + (\vslash-\nslash)\,in\cdot\partial \big( S^\dagger h \big) \,.
\end{equation}
It follows that all gauge-covariant derivatives of the heavy-quark fields are
perpendicular derivatives. This fact restricts the number of subleading shape 
functions.

\section{Definition of subleading shape functions}

The Dirac structure of the operators $O_n$ can be simplified noting that the 
heavy-quark fields $h$ of HQET are two-component spinor fields, so that
between $\bar h\dots h$ the Dirac basis collapses to a set of four basis 
matrices $(\bm{1},\bm{\sigma})$. In four-component notation, these are the
upper-left $2\times 2$ blocks of $(\bm{1},\bm{\gamma}\gamma_5)$ (in the Dirac
representation). Instead of $\bm{\gamma}\gamma_5$, we are free to take the 
matrices $\gamma_\perp^\mu\gamma_5$ and $\nslash\gamma_5$. It follows that, 
between the $P_v=\frac12(1+\vslash)$ projectors supplied by the heavy-quark 
fields, any Dirac matrix $\Gamma$ can be decomposed as
\begin{equation}\label{HQETdecomp}
   \Gamma \to \frac12\,\mbox{tr} \left( \Gamma\,P_v \right) \bm{1}
   - \frac12\,\mbox{tr} \left[ \Gamma\,P_v\,(\vslash-\nslash)\,\gamma_5
   \right] \nslash\gamma_5
   - \frac12\,\mbox{tr} \left( \Gamma\,P_v\,\gamma_{\perp\mu}\gamma_5 \right) 
   \gamma_\perp^\mu\gamma_5 \,.
\end{equation}
We denote by
\begin{equation}
   \langle\bar h\dots h\rangle
   \equiv \frac{\langle\bar B(v)|\,\bar h\dots h\,|\bar B(v)\rangle}{2m_B}
\end{equation}
the forward $B$-meson matrix element of any HQET operator. Rotational 
invariance in the transverse plane implies that transverse indices can only be 
contracted using the symmetric and anti-symmetric tensors (we set 
$\epsilon_{0123}=1$)
\begin{equation}
   g_\perp^{\mu\nu} = g^{\mu\nu} - \frac{n^\mu\bar n^\nu+n^\nu\bar n^\mu}{2}
    \,, \qquad
   \epsilon_\perp^{\mu\nu}
   = \epsilon^{\mu\nu\alpha\beta}\,v_\alpha n_\beta \,.
\end{equation}
It follows that the only non-vanishing matrix elements are
\begin{eqnarray}\label{shapes}
   \langle \left(\bar h S \right)_{0} \left( S^\dagger h \right)_{x_-} \rangle
   &=& \int d\omega\,e^{-\frac{i}{2}\omega\bar n\cdot x}\,S(\omega) \,,
    \nonumber\\
   \langle i\int d^4z\,T\{ 
   \left(\bar h S \right)_{0} \left( S^\dagger h \right)_{x_-}
   {\cal L}_h^{(2)}(z) \} \rangle
   &=& \frac{1}{m_b} \int d\omega\,e^{-\frac{i}{2}\omega\bar n\cdot x}\,
    s(\omega) \,, \nonumber\\
   \langle \left(\bar h S \right)_{0} \gamma_\perp^\rho\gamma_5
    \left( S^\dagger iD_\perp^\mu h \right)_{x_-} \rangle
   &=& - \frac{i\epsilon_\perp^{\rho\mu}}{2} 
    \int d\omega\,e^{-\frac{i}{2}\omega\bar n\cdot x}\,t(\omega) \,,
    \nonumber\\[-0.15cm]
   -i\int\limits_0^{\bar n\cdot x/2}\!dt\,\langle \left(\bar h S \right)_{0}
    \left( S^\dagger iD_\perp^\mu iD_\perp^\nu S \right)_{tn}
    \left( S^\dagger h \right)_{x_-} \rangle
   &=& \frac{g_\perp^{\mu\nu}}{2}
    \int d\omega\,e^{-\frac{i}{2}\omega\bar n\cdot x}\,u(\omega) \,,
    \nonumber\\[-0.2cm]
   -i\int\limits_0^{\bar n\cdot x/2}\!dt\,
   \langle \left(\bar h S \right)_{0} \nslash\gamma_5
    \left( S^\dagger iD_\perp^\mu iD_\perp^\nu S \right)_{tn}
    \left( S^\dagger h \right)_{x_-} \rangle
   &=& - \frac{i\epsilon_\perp^{\mu\nu}}{2}
    \int d\omega\,e^{-\frac{i}{2}\omega\bar n\cdot x}\,v(\omega) \,.\nonumber\\
\end{eqnarray}
If radiative corrections at the hard scale are included, it would be more
appropriate to split up 
$s(\omega)=s_{\rm kin}(\omega)+C_{\rm mag}\,s_{\rm mag}(\omega)$, where
$C_{\rm mag}$ is the Wilson coefficient of the chromo-magnetic operator in
the subleading HQET Lagrangian. This ensures that the shape functions remain
independent of the heavy-quark mass. The definitions of the functions $t$, 
$u$, $v$ are chosen such that
\begin{eqnarray}\label{tuvdefs}
   \langle\bar h(0)\,\nslash\,[0,x_-]\,(i\Dslash_\perp h)(x_-)\rangle
   &=& \int d\omega\,e^{-\frac{i}{2}\omega\bar n\cdot x}\,t(\omega) \,,
    \nonumber\\[-0.15cm]
   -i\int\limits_0^{\bar n\cdot x/2}\!dt\,
   \langle\bar h(0)\,[0,tn]\,(iD_\perp)^2(tn)\,[tn,x_-]\,h(x_-)\rangle
   &=& \int d\omega\,e^{-\frac{i}{2}\omega\bar n\cdot x}\,u(\omega) \,,
    \nonumber\\[-0.2cm]
   -i\int\limits_0^{\bar n\cdot x/2}\!dt\,
   \langle\bar h(0)\,\frac{\nslash}{2}\,[0,tn]\,
   \sigma_{\mu\nu}^\perp\,gG_\perp^{\mu\nu}(tn)\,[tn,x_-]\,h(x_-)\rangle 
   &=& \int d\omega\,e^{-\frac{i}{2}\omega\bar n\cdot x}\,v(\omega) \,,\nonumber\\
\end{eqnarray}
where $[x,y]\equiv S(x)\,S^\dagger(y)$ is a product of two infinite-length 
soft Wilson lines, which on the light cone (i.e., for $x,y\parallel n$) 
collapses to a straight Wilson line of finite length connecting $x$ and $y$. 

We also need a variation of the first matrix element in (\ref{tuvdefs}), in
which the derivative is located at position 0. Using hermitean conjugation, 
translational invariance, and the reality of $t(\omega)$,  which follows from
parity and time-reversal invariance of the strong interactions, we find that
\begin{equation}
   \langle(\bar h i\Dslash_\perp)(0)\,\nslash\,[0,x_-]\,h(x_-)\rangle
   = \langle\bar h(0)\,\nslash\,[0,x_-]\,(i\Dslash_\perp h)(x_-)\rangle \,,
\end{equation}
implying that all terms containing a single insertion of $D_\perp$ can be 
related to the function $t(\omega)$. From this relation, it follows that
\begin{eqnarray}\label{talt}
   &&\int d\omega\,e^{-\frac{i}{2}\omega\bar n\cdot x}\,t(\omega)
   = \langle \left( \bar h S \right)_{0} \frac{\nslash}{2}\,
    \Big[ \calAslash_{s\perp}(x_-) - \calAslash_{s\perp}(0) \Big]
    \left( S^\dagger h \right)_{x_-} \rangle \nonumber\\
   &=& - \int\limits_0^{\bar n\cdot x/2}\!dt\,
    \langle\bar h(0)\,\frac{\nslash}{2}\,[0,tn]\,\gamma_\perp^\mu n^\nu 
    g G_{\mu\nu}(tn)\,[tn,x_-]\,h(x_-)\rangle \,,
\end{eqnarray}
which defines the function $t(\omega)$ in terms of a matrix element of the 
field-strength tensor.

It is now straightforward to express the forward matrix element of the 
current correlator $T_{ij}$ in terms of shape functions. The resulting traces
of Dirac matrices can be simplified using identities for the 
$\epsilon_\perp^{\mu\nu}$ tensor derived in \cite{Lange:2003pk}. Taking the 
imaginary part, we obtain for the hadronic tensor
\begin{eqnarray}\label{final}
   W_{ij}^{(0)}
   &=& \int d\omega\,\delta(n\cdot p+\omega)\,S(\omega)\,T_1 \,, \qquad
    W_{ij}^{(1)} = 0 \,, \nonumber\\
   W_{ij}^{(2)} &=& \int d\omega\,\delta(n\cdot p+\omega) \left[
    \frac{\omega\,S(\omega) + t(\omega)}{m_b}\,T_2
    + \frac{s(\omega)}{m_b}\,T_1\right.\nonumber\\
    &+&\left. \frac{t(\omega)}{\bar n\cdot p}\,T_3
    + \frac{u(\omega)}{\bar n\cdot p}\,T_1
    - \frac{v(\omega)}{\bar n\cdot p}\,T_4 \right] ,
\end{eqnarray}
where now $p=m_b v-q$, and
\begin{eqnarray}\label{traces}
   T_1 &=& \frac14\,\mbox{tr}\left[ \Gamma_i\,\nslash\,\Gamma_j\,
    \frac{1+\vslash}{2} \right] , \hspace{1.05cm}
   T_3 = \frac14\,\mbox{tr}\left[ \Gamma_i\,\gamma_\rho^\perp\gamma_5\,
    \Gamma_j\,\frac{1+\vslash}{2}\,\gamma_\perp^\rho\gamma_5 \right] ,
    \nonumber\\
   T_2 &=& \frac18\,\mbox{tr}\,\Big[ \Gamma_i\,\nslash\,\Gamma_j\,
    (\vslash-\nslash) \Big] \,, \qquad
   T_4 = \frac14\,\mbox{tr}\left[ \Gamma_i\,\nslash\gamma_5\,\Gamma_j\,
    \frac{1+\vslash}{2}\,(\vslash-\nslash)\,\gamma_5 \right] .\nonumber\\
\end{eqnarray}
It follows from (\ref{final}) that the subleading shape functions $s(\omega)$
and $u(\omega)$ always come with the same trace as the leading shape function 
$S(\omega)$. However, $u(\omega)$ is divided by the kinematic variable 
$\bar n\cdot p$, and so it does {\em not\/} enter in a universal (i.e., 
process-independent) combination with $S(\omega)$.
 
We can now specialize our result to the case of semileptonic decay, for which 
$\Gamma_i=\gamma^\mu(1-\gamma_5)$ and $\Gamma_j=\gamma^\nu(1-\gamma_5)$. This 
yields
\begin{eqnarray}\label{Wsl}
   W^{\mu\nu}
   &=& \int d\omega\,\delta(n\cdot p+\omega)\,\Bigg\{\!
    \left( n^\mu v^\nu + n^\nu v^\mu - g^{\mu\nu}
    - i\epsilon^{\mu\nu\alpha\beta}\,n_\alpha v_\beta \right) \nonumber\\
   &&\times \left[ \left( 1 + \frac{\omega}{m_b} \right) S(\omega)
    + \frac{s(\omega) + t(\omega)}{m_b}
    + \frac{u(\omega) - v(\omega)}{\bar n\cdot p} \right] \nonumber\\
  &&\mbox{}- 2 (n^\mu v^\nu + n^\nu v^\mu)\,\frac{t(\omega)}{\bar n\cdot p}
   + 2 n^\mu n^\nu \left[ - \frac{\omega\,S(\omega)}{m_b}
   - \frac{t(\omega)}{m_b}
   + \frac{t(\omega) + v(\omega)}{\bar n\cdot p} \right] \Bigg\} \,.\nonumber\\
\end{eqnarray}
Similarly, for the contribution of the dipole operator $Q_{7\gamma}$ to 
$\bar B\to X_s\gamma$ decay, the Dirac structures are (up to prefactors) 
$\Gamma_i=\frac14[\gamma_\perp^\mu,\nbslash](1-\gamma_5)$ and
$\Gamma_j=\frac14[\nbslash,\gamma_\perp^\nu](1+\gamma_5)$, where the indices
$\mu,\nu$ are contracted with the transverse polarization vector of the photon.
In this case we obtain
\begin{eqnarray}\label{Wgamma}
   W^{\mu\nu} &=& (i\epsilon_\perp^{\mu\nu} - g_\perp^{\mu\nu})
   \int d\omega\,\delta(n\cdot p+\omega)
   \left[ \left( 1 - \frac{\omega}{m_b} \right) S(\omega)\right.\nonumber\\
   &+& \left.\frac{s(\omega) - t(\omega)}{m_b}
   + \frac{u(\omega) - v(\omega)}{\bar n\cdot p} \right] .
\end{eqnarray}
We stress, however, that while at leading power in $\Lambda_{\rm QCD}/m_b$ the 
dipole operator gives the only tree-level contribution to the 
$\bar B\to X_s\gamma$ decay rate, this is no longer the case when power 
corrections are included. For instance, interference terms of the dipole 
operator with current-current operators can lead to new subleading shape 
functions even at lowest order in perturbation theory. To derive these 
structures, it would be necessary to match the entire effective weak 
Hamiltonian for $\bar B\to X_s\gamma$ decay onto SCET operators 
\cite{Neubert:2004dd}. This task still has to be completed beyond the leading 
order in $\lambda$. Contrary to claims in \cite{Bauer:2001mh,Bauer:2002yu}, a 
complete description of tree-level subleading shape-function effects in 
$\bar B\to X_s\gamma$ decay is therefore still lacking.

\section{Moment relations and comparison with the literature}

Moments of the shape functions can be related to forward $B$-meson matrix 
elements of local HQET operators \cite{Neubert:1993ch}. In particular, setting 
$x=0$ in the defining relations (\ref{shapes}) yields expressions for the 
normalization integrals of the shape functions. They are
\begin{equation}\label{norms}
   \int d\omega\,S(\omega) = 1 \,, \qquad
   \int d\omega\,\{s(\omega), t(\omega), u(\omega), v(\omega)\} = 0 \,.
\end{equation}
The vanishing of the norm of all subleading shape functions is a consequence 
of Luke's theorem \cite{Luke:1990eg}, and it ensures that there are no 
first-order $\Lambda_{\rm QCD}/m_b$ corrections to total inclusive decay 
rates. For the functions $t$, $u$, $v$ this is an obvious consequence of the 
fact that the integration domain in (\ref{talt}) and (\ref{tuvdefs}) shrinks 
to zero in the limit $x\to 0$. The interpretation of (\ref{norms}) is that 
subleading shape functions lead to local distortions of inclusive spectra, 
which cancel out when the spectra are integrated over a sufficiently large 
region in phase space. The first moments characterize the strength of the 
distortions, while higher moments determine their shape.

Taking a derivative $in\cdot\partial_x$ in the definitions (\ref{shapes}) 
brings down a factor of $\omega$ under the integrals on the right-hand side. 
Setting then $x\to 0$ yields a set of relations for the first moments of the 
shape functions. The resulting matrix elements can be evaluated by means of 
the relations \cite{Falk:1992wt}
\begin{equation}
   \langle \bar h\,\Gamma_{\alpha\beta}\,iD^\alpha iD^\beta h\rangle
   = \frac12\,\mbox{tr} \left( \Gamma_{\alpha\beta}\,\frac{1+\vslash}{2}
   \left[ \left( g^{\alpha\beta} - v^\alpha v^\beta \right)
   \frac{\lambda_1}{3} + i\sigma^{\alpha\beta}\,\frac{\lambda_2}{2} \right]
   \frac{1+\vslash}{2} \right) , 
\end{equation} 
and \cite{Manohar:1993qn}
\begin{equation}
   \langle i\int d^4z\,T\{ (\bar h iD^\mu h)(0)\,{\cal L}_h^{(2)}(z) \}
   \rangle
   = - v^\mu\,\langle {\cal L}_h^{(2)}(0) \rangle
   =  - v^\mu\,\frac{\lambda_1 + 3C_{\rm mag}\,\lambda_2}{2m_b} \,, 
\end{equation} 
where $\lambda_1$ and $\lambda_2$ are the familiar HQET parameters arising in 
the parameterization of second-order power corrections to inclusive decay 
spectra, and $C_{\rm mag}=1$ at tree level. We obtain
\begin{eqnarray}\label{moments}
   \int d\omega\,\omega\,s(\omega)
   &=& - \frac{\lambda_1 + 3\lambda_2}{2} \,, \qquad 
   \int d\omega\,\omega\,u(\omega) = \frac{2\lambda_1}{3} \,, \nonumber\\
   \int d\omega\,\omega\,t(\omega)
   &=& - \lambda_2 \,, \hspace{2.05cm}
   \int d\omega\,\omega\,v(\omega) = \lambda_2 \,.
\end{eqnarray}
We also recall that the first two moments of the leading shape function are \\
$\int d\omega\,\omega\,S(\omega)=0$ and 
$\int d\omega\,\omega^2\,S(\omega)=-\lambda_1/3$ \cite{Neubert:1993ch}.

\section{Contributions from four-quark operators}
\label{sec:4quark}

The last diagram in Figure~\ref{fig:graphs} shows a tree-level contribution to 
the hadronic tensor involving two insertions of the subleading SCET Lagrangian 
${\cal L}_{\xi q}^{(1)}$. The exchange of a hard-collinear gluon implies that
this graph is of order $g^2$, and so it vanishes in the limit where 
$\alpha_s$ is set to zero. The fact that the suppression factor is 
$\pi\alpha_s$ instead of $\alpha_s/\pi$ reflects the phase-space enhancement 
of four-quark tree-level graphs compared with loop diagrams 
\cite{Neubert:1996we}. One might therefore expect that the four-quark 
contribution is numerically as important as the other tree-level subleading
shape-function contributions.

Applying the partial-fraction identity (\ref{partialfractions}) twice, we find 
that the resulting contribution to the correlator $T_{ij}^{(2)}$ can be 
written as in (\ref{Tresult}), adding a fifth operator to the sum. It reads
\begin{equation}\label{O5}
   O_5(x_-) = \frac{\pi\alpha_s}{\bar n\cdot p}
   \int\limits_0^{\bar n\cdot x/2}\!dt_1
   \int\limits_{t_1}^{\bar n\cdot x/2}\!dt_2
   \left( \bar h S\right)_{0} \Gamma_i\,\nslash\gamma_\rho^\perp\,t_a
   \left( S^\dagger q \right)_{t_1 n}
   \left( \bar q S\right)_{t_2 n} \gamma_\perp^\rho\nslash\,\Gamma_j\,t_a
   \left( S^\dagger h \right)_{x_-} ,
\end{equation}
where $t_a$ are the generators of color SU($N_c$). Note that the field 
insertions are ordered according to ``light-cone time'' $x_-$, just as they 
appear in the Feynman diagram in Figure~\ref{fig:graphs}. This is a general 
result. Because the minus components $\bar n\cdot p_{hc}$ of hard-collinear 
momenta are large, of order $m_b$, hard-collinear fields always propagate 
forward in light-cone time. Turning, for instance, a forward-moving 
hard-collinear quark into a backward-moving hard-collinear anti-quark would 
require a hard quantum fluctuation, which is already integrated out in SCET. 
As a result, Feynman amplitudes in SCET are ordered with respect to light-cone 
time, and that ordering is preserved in the matching onto HQET. This 
discussion explains why all our operators have the property that the 
coordinates $z_-$ of soft fields range from 0 to $x_-$ in an ordered fashion. 
The results can therefore always be expressed in terms of bi-local operators 
depending only on 0 and $x_-$. However, at present we cannot exclude the 
possibility of non-trivial weight functions under these integrals, which could 
arise at higher orders in perturbation theory. If present, they may require a 
generalization of our definitions of subleading shape functions.

Returning to the case of the four-quark operator in (\ref{O5}), we note that 
its contribution vanishes in the vacuum-insertion approximation due to the 
color-octet structure of the heavy-light quark bilinears. While this 
approximation is admittedly naive, phenomenological evidence based on studies 
of $B$-meson lifetimes \cite{Neubert:1996we,Bigi:1994wa,Bigi:1995jr} and theoretical 
arguments based on lattice calculations 
\cite{DiPierro:1998ty,Becirevic:2001fy} and QCD sum rules \cite{Baek:1998vk}
support the notion that matrix elements which vanish in the vacuum-insertion 
approximation are numerically suppressed, typically by an order of magnitude. 
It is thus very unlikely that the four-quark operator in (\ref{O5}) could give 
a larger contribution than loop-suppressed $O(\alpha_s)$ corrections, which we 
have neglected. Note also that the contribution of the corresponding local 
operator to total inclusive rates, which is what remains when the 
corresponding shape functions are integrated over a sufficiently large domain, 
is bound to be tiny. The effect is of order $(\Lambda_{\rm QCD}/m_b)^3$, and 
it is $\alpha_s$-suppressed with respect to a four-quark contribution 
discussed by Voloshin \cite{Voloshin:2001xi}, whose effect on the total decay 
rate is believed to be at most 3\%. 

Nevertheless, it is interesting to study the structure of the operator
$O_5(x_-)$ in more detail and define corresponding subleading shape functions. 
The decomposition (\ref{HQETdecomp}) implies that the Dirac structure can be
rearranged in the form (omitting factors of $S$, $S^\dagger$ and color indices 
for simplicity)
\begin{equation}\label{step1}
   \Gamma_i\,\nslash\gamma_\rho^\perp\,q\,
   \bar q\,\gamma_\perp^\rho\nslash\,\Gamma_j
   = - \frac12 \left( \begin{array}{c} \bm{1} \\ \nslash\gamma_5 \\
   - \gamma_\perp^\mu\gamma_5 \end{array} \right)
   \bar q\,\gamma_\perp^\rho\nslash\,\Gamma_j\,\frac{1+\vslash}{2}
   \left( \begin{array}{c} \bm{1} \\ (\nslash-\vslash)\gamma_5 \\
   \gamma_\mu^\perp\gamma_5 \end{array} \right)
   \Gamma_i\,\nslash\gamma_\rho^\perp\,q \,,
\end{equation}
where the notation implies a sum over the three rows in the equation. In the
next step, we use that between $\nslash\dots\nslash$ any Dirac matrix can be
decomposed as
\begin{equation}
   \Gamma \to \frac14\,\mbox{tr}(\Gamma\,\nslash)\,\frac{\nbslash}{2}
   - \frac14\,\mbox{tr}(\Gamma\,\nslash\gamma_5)\,
   \frac{\nbslash}{2}\,\gamma_5
   - \frac14\,\mbox{tr}(\Gamma\,\nslash\gamma_\perp^\alpha)\,
   \frac{\nbslash}{2}\,\gamma_\alpha^\perp \,.
\end{equation}
Only the first two terms survive after contraction of the index $\rho$ in 
(\ref{step1}). At this stage we are left with four-quark operators with Dirac 
structures
\begin{equation}
   \bar h\,h\,\bar q\,\nslash\,(\gamma_5)\,q \,, \qquad
   \bar h\,\nslash\gamma_5\,h\,\bar q\,\nslash\,(\gamma_5)\,q \,, \qquad
   \bar h\,\gamma_\perp^\mu\gamma_5\,h\,\bar q\,\nslash\,(\gamma_5)\,q \,,
\end{equation}
where $(\gamma_5)$ means $\bm{1}$ or $\gamma_5$. Lorentz and parity invariance 
imply that only the two Lorentz-scalar structures 
$\bar h\,h\,\bar q\,\nslash\,q$ and 
$\bar h\,\nslash\gamma_5\,h\,\bar q\,\nslash\gamma_5\,q$ can have non-zero 
forward matrix elements between $B$-meson states. Putting back color factors 
and soft Wilson lines, we define the corresponding subleading shape functions
\begin{eqnarray}\label{4qdef}
   2(-i)^2 &&\hspace{-0.6cm} \int\limits_0^{\bar n\cdot x/2}\!dt_1
   \int\limits_{t_1}^{\bar n\cdot x/2}\!dt_2\,
   \langle \big[ \left( \bar h S\right)_{0} t_a \big]_k 
   \big[ t_a \left( S^\dagger h \right)_{x_-} \big]_l
   \big[ \left( \bar q S\right)_{t_2 n} \big]_l\,\nslash
   \big[ \left( S^\dagger q \right)_{t_1 n} \big]_k \rangle \nonumber\\
   &=& \int d\omega\,e^{-\frac{i}{2}\omega\bar n\cdot x}\,f_u(\omega) \,,
    \nonumber\\
   2(-i)^2 &&\hspace{-0.6cm} \int\limits_0^{\bar n\cdot x/2}\!dt_1
   \int\limits_{t_1}^{\bar n\cdot x/2}\!dt_2\,
   \langle \big[ \left( \bar h S\right)_{0} t_a \big]_k\,\nslash\gamma_5
   \big[ t_a \left( S^\dagger h \right)_{x_-} \big]_l
   \big[ \left( \bar q S\right)_{t_2 n} \big]_l\,\nslash\gamma_5
   \big[ \left( S^\dagger q \right)_{t_1 n} \big]_k \rangle \nonumber\\
   &=& \int d\omega\,e^{-\frac{i}{2}\omega\bar n\cdot x}\,f_v(\omega) \,,
\end{eqnarray}
where $k$, $l$ are color indices. Using the same notation as in (\ref{final}), 
the corresponding contributions to the hadronic tensor are given by
\begin{equation}
   W_{ij}^{(2)} \Big|_{\rm 4q}
   = - \pi\alpha_s \int d\omega\,\delta(n\cdot p+\omega)
   \left[ \frac{f_u(\omega)}{\bar n\cdot p}\,T_1
   + \frac{f_v(\omega)}{\bar n\cdot p}\,T_4 \right] ,
\end{equation}
with the same traces as defined in (\ref{traces}). As far as the spin structure
is concerned, the four-quark contributions can thus be absorbed into a 
redefinition of the subleading shape functions $u$ and $v$, namely
\begin{equation}\label{tildeuv}
   \tilde u(\omega)\equiv u(\omega) - \pi\alpha_s\,f_u(\omega) \,, \qquad
   \tilde v(\omega)\equiv v(\omega) + \pi\alpha_s\,f_v(\omega) \,.
\end{equation}
Note that the definitions (\ref{4qdef}) imply that the normalization integrals 
as well as the first moments of the functions $f_u(\omega)$ and $f_v(\omega)$ 
vanish (because the integration domain is of second order in $\bar n\cdot x$), 
and therefore the new functions $\tilde u$ and $\tilde v$ have the same first 
moments as the original ones, see (\ref{moments}). 

While the shape functions $S$, $s$, $t$, $u$, $v$ are expected to be identical 
for charged and neutral $B$ mesons up to tiny isospin-breaking corrections, 
this is no longer the case for the four-quark shape functions $f_u$ and $f_v$. 
The values of these functions will depend crucially on whether the light-quark 
flavor $q$ in the four-quark operators matches that of the $B$-meson spectator 
quark. In the semileptonic decay $\bar B\to X_u\,l^-\bar\nu$, the difference 
between the subleading shape functions $f_u$ and $f_v$ for $B^-$ and 
$\bar B^0$ mesons is likely to be one of the dominant sources of 
isospin-breaking effects on the decay distributions.

\section{Applications}

We are now ready to summarize our results. We 
absorb the contributions from four-quark operators into the functions 
$\tilde u(\omega)$ and $\tilde v(\omega)$ defined in (\ref{tildeuv}). The 
moment constraints on the shape functions can be summarized by their 
expansions in distributions \cite{Neubert:1993ch}, which read
\begin{eqnarray}\label{distr}
   S(\omega) &=& \delta(\omega) - \frac{\lambda_1}{6}\,\delta''(\omega)
    + \dots \,, \quad
    s(\omega) = \frac{\lambda_1+3\lambda_2}{2}\,\delta'(\omega) + \dots \,, \\
   t(\omega) &=& \lambda_2\,\delta'(\omega) + \dots \,, \quad
    \tilde u(\omega) = - \frac{2\lambda_1}{3}\,\delta'(\omega) + \dots \,, 
    \quad
    \tilde v(\omega) = - \lambda_2\,\delta'(\omega) + \dots \,. \nonumber
\end{eqnarray}
These expressions allow us to test our results against existing predictions 
for inclusive spectra obtained using a conventional heavy-quark expansion. 

The analytic properties of the shape functions are such that they have support 
for $-\infty<\omega<\bar\Lambda_\infty$, where $\bar\Lambda_\infty$ is the 
asymptotic value of the mass difference $(m_B-m_b)_{m_b\to\infty}$ in the 
heavy-quark limit. This parameter differs from the physical value of 
$\bar\Lambda$ by power-suppressed terms, 
\begin{equation}
   \bar\Lambda\equiv m_B-m_b = \bar\Lambda_\infty
   - \frac{\lambda_1+3\lambda_2}{2m_b} + \dots \,.
\end{equation}
We would like the support in $\omega$ to extend over the physical interval
$-\infty<\omega<\bar\Lambda$, since this will ensure that the kinematic 
boundaries for decay distributions take their physical values set by the true 
$B$-meson mass. This can be achieved by shifting the arguments of all shape 
functions by a small amount $\Delta\omega=\frac12(\lambda_1+3\lambda_2)/m_b$. 
For the subleading shape functions this changes nothing to the order we are 
working, since $t(\omega+\Delta\omega)=t(\omega)+\dots$ etc., where the dots
represent terms of higher order in $1/m_b$. For the leading-order shape 
function, however, this shift produces a new $1/m_b$ correction:
$S(\omega)=S(\omega+\Delta\omega)-\Delta\omega\,S'(\omega+\Delta\omega)+\dots$,
where the prime denotes a derivative with respect to the argument. Using the 
fact that $S(\omega)$ and $s(\omega)/m_b$ always appear together, we can 
absorb the extra term into a redefinition of the subleading-shape function 
$s$, defining a new function 
\begin{equation}
   s_0(\omega)\equiv s(\omega) - \frac12(\lambda_1+3\lambda_2)\,S'(\omega) \,.
\end{equation}
From (\ref{moments}), it follows that the first moment of the function $s_0$ 
vanishes. In terms of these definitions,
\begin{equation}
   \tilde S(\omega+\Delta\omega)
   \equiv S(\omega+\Delta\omega) + \frac{s_0(\omega+\Delta\omega)}{m_b}
   = S(\omega) + \frac{s(\omega)}{m_b} + \dots \,.
\end{equation}
In other words, our expressions for the hadronic tensors in (\ref{Wsl}) and 
(\ref{Wgamma}) remain valid when in all shape functions the argument is 
shifted from $\omega$ to $\omega+\Delta\omega$, except that the subleading 
shape function $s$ must be replaced with the redefined function $s_0$, whose 
norm and first moment vanish. Once this is done, the integrals over $\omega$ 
extend from $-\infty$ up to the physical value of $\bar\Lambda$.

It is now convenient to  
express the shape functions as functions of 
$\hat\omega\equiv\bar\Lambda-\omega$ \cite{Bosch:2004th}. They have support in 
this variable for $0\le\hat\omega<\infty$. The $\delta$-functions in the 
tree-level expressions (\ref{Wsl}) and (\ref{Wgamma}) for the hadronic tensors 
set $\hat\omega=P_+$. We denote functions of $\hat\omega$ by a hat, e.g.\
$\hat S(\hat\omega)\equiv\tilde S(\bar\Lambda-\hat\omega+\Delta\omega)%
=\tilde S(\bar\Lambda_\infty-\hat\omega)$, 
$\hat t(\hat\omega)\equiv t(\bar\Lambda_\infty-\hat\omega)$, and similarly for 
the other functions. In the equations below, $\bar\Lambda$ always refers to 
the physical parameter defined with the true $B$-meson mass.

Based on equation (\ref{Wsl}), we can write the structure functions $\tilde{W}_i$ for $\bar B\to X_u\,l\bar\nu$ decays as:
\begin{eqnarray}
   \tilde{W}_1
   &=& \left(1+ \frac{\bar\Lambda-P_+}{m_b}\right)\,\hat S(P_+) + \frac{\hat t(P_+)}{m_b} 
    + \frac{\hat u(P_+) - \hat v(P_+)}{\bar n \cdot p}  , \nonumber\\
   \tilde{W}_4+\frac1y\tilde{W}_5
   &=& \frac{2}{y}
    \left[ -\frac{(\bar\Lambda-P_+)\,\hat S(P_+)}{m_b} - \frac{2\hat t(P_+)}{m_b} 
    + \frac{\hat t(P_+) + \hat v(P_+)}{\bar n\cdot p} \right] ,
\end{eqnarray}
where $y\approx\bar n\cdot p/m_b$.

For $\bar B\to X_s\gamma$ we have from equation (\ref{Wgamma}):
\begin{equation}
W^\mu_\mu=-2
   \left[ \left(1- \frac{\bar\Lambda - P_+}{m_b}\right)\,\hat S(P_+) + \frac{-\hat t(P_+)
   + \hat u(P_+) - \hat v(P_+)}{m_b} \right],
\end{equation}
where for radiative decays $\bar n\cdot p=m_b$.

\section{Conclusions}

In this chapter, we have used the formalism of SCET to perform a 
systematic study of power-suppressed effects to charmless inclusive $B$ decays. 
At tree level, the results 
can be expressed in terms of a set of subleading shape functions defined via 
the Fourier transforms of forward matrix elements of bi-local light-cone 
operators in heavy-quark effective theory. We have identified a new 
contribution arising from four-quark operators, which was not considered 
previously. We have shown that, when shape functions appearing in 
process-independent combinations are combined into single functions, then a 
total of three subleading shape functions are required to describe arbitrary 
current-induced decay distributions of $B$ mesons into light final-state 
particles. 

In the last part of this chapter, we have presented analytical expressions for 
the hadronic tensor of semileptonic and radiative charmless $B$ decays. 
While this concludes the problem of tree-level power 
corrections in semileptonic decay, we have stressed that no complete 
(tree-level) analysis of power-suppressed corrections to the 
$\bar B\to X_s\gamma$ decay exists to date. The formalism developed in this 
chapter can, however, readily be extended to this case.

\chapter{Applications: Event Generator}
\label{chapter_evegen}
\section{Introduction}\label{section_4.1}

The theoretical tools for the calculation of inclusive $B$ decays are
QCD factorization on the one hand
\cite{Neubert:1993ch,Neubert:1993um,Bigi:1993ex,Korchemsky:1994jb,
Akhoury:1995fp,Neubert:2004dd,Bauer:2003pi,Bosch:2004th,Gardi:2004ia}, and 
local operator product expansions
(OPE) on the other \cite{Manohar:1993qn,Blok:1993va}. Both approaches
perform a systematic separation of long-distance hadronic quantities from
short-distance
perturbative ones, while organizing the calculation in inverse powers
of the heavy $b$-quark mass $m_b$. The OPE is an appropriate
tool for the calculation of total inclusive rates 
(for example in $\bar B\to X_c\,l^-\bar\nu$ decay) or for
partial rates integrated over sufficiently large regions in phase space, 
where all components of the final-state hadronic momentum $P_X^\mu$ 
are large compared 
to $\Lambda_{\rm QCD}$. QCD factorization, on the other hand, is better 
suited for the calculation of partial rates and spectra near kinematical 
boundaries, where typically some components of $P_X^\mu$ are large, while 
the invariant hadronic mass $M_X=\sqrt{P_X^2}$ is small. 

It is important to note that the heavy-quark expansions
valid in these two kinematical regions are not identical, because the power 
counting rules differ in the two regimes. 
Also the nature of the non-perturbative inputs is different. 
In the OPE region, non-perturbative
physics is encoded in a few hadronic parameters, and the heavy-quark 
expansion is the usual Wilsonian expansion in local operators. In the endpoint
(or shape-function) region, the presence of multiple scales complicates the
power counting, and the interplay between soft and collinear modes gives
rise to large non-localities. As a result,
non-perturbative physics is described by hadronic structure functions 
called ``shape functions'', and the heavy-quark expansion is an expansion in 
non-local string operators defined on the light-cone. The connections 
between the two regimes is that {\em moments\/} of the shape functions can
be expressed in terms of local operators.

In this chapter we develop a formalism that smoothly 
interpolates between the two kinematical regimes (see \cite{Tackmann:2005ub}
for a related discussion, which is however restricted to the tree
approximation). This is essential for
building an event generator for inclusive $\bar B\to X_u\,l^-\bar\nu$ and
$\bar B\to X_s\gamma$ decays, which can be used to study 
partial and differential decay rates in different kinematical domains. 
In the shape-function region, our approach relies on exact QCD factorization 
theorems, which exist in every order of power counting. They allow
us to systematically disentangle short- and long-distance physics and,
in the process, resum parametrically large logarithms order by order in 
perturbation theory. This factorization can be done with high accuracy for
the terms of leading power in $1/m_b$, and with somewhat less sophistication
for the first-order power corrections. For the second-order power corrections,
we only include contributions that do not vanish when integrated
over all phase space. This is a safe approximation; the effects of the 
remaining $1/m_b^2$ terms can to a large extent 
be absorbed by a redefinition of the subleading shape functions arising at
order $1/m_b$.

Our formalism is ``optimized'' for the shape-function region in the
sense that sophisticated theoretical technology is applied in
this regime. However, when our expressions for the 
differential decay rates are integrated over sufficiently wide domains, 
they automatically reduce to the simpler results that can be derived using
the OPE approach, up to yet unknown terms of $O(\alpha_s^2)$. 
The moment relations for the shape functions are crucial
in this context. Note that local $1/m_b^2$ corrections in the OPE receive
contributions from terms of leading power ($1/m_b^0$), subleading power 
($1/m_b$), and sub-subleading power ($1/m_b^2$) in the shape-function region, 
so the transition is highly non-trivial.  
In implementing the program outlined here, we include 
all presently known information on the triple differential 
$\bar B\to X_u\,l^-\bar\nu$ decay rate and on the differential 
$\bar B\to X_s\gamma$ decay rate in a single, unified framework. 
We neglect, for simplicity, hadronic power corrections of order $1/m_b^3$ and
higher, which are known to have a negligible effect on the observables 
considered here. The only possible exception is contributions from ``weak
annihilation'', which are estimated as part of our error analysis. 
We also ignore the existing results on $O(\beta_0\alpha_s^2)$ radiative
corrections for some single-differential distributions, because the 
corresponding corrections are not known for the double or triple differential 
$\bar B\to X_u\,l^-\bar\nu$ decay spectra. 
While these $O(\beta_0\alpha_s^2)$ terms are sometimes found to be large
when naive perturbation theory in $\alpha_s(m_b)$ is used, their effects
are expected to be small in our scheme, which is based on a complete scale
separation using QCD factorization. We see no reason why the 
$\beta_0\alpha_s^2$ terms should be
enhanced compared to other, unknown corrections of $O(\alpha_s^2)$.

A technical complication in realizing the approach described here has to do 
with the treatment of phase-space factors. The heavy-quark expansion of 
the hadronic tensor for $\bar B\to X_u\,l^-\bar\nu$ decay gives rise to
expressions that are singular at certain points in phase space. One way to 
avoid these singularities is to also expand phase-space factors order by 
order in $1/m_b$ (see, e.g., the treatment in \cite{Bosch:2004cb}). However, 
since this expansion depends on the kinematical cuts of any given analysis, 
it cannot be implemented in a straightforward way in an event generator. An
alternative is to reorganize the heavy-quark expansion in such a way that 
the expansion parameter is related to {\em hadronic\/} (as opposed to 
partonic) kinematical variables, in which case kinematical singularities are
always canceled by exact phase-space factors. Following this strategy, we
obtain expressions for decay distributions and partial decay rates which
are free of explicit reference to partonic quantities such as the
$b$-quark mass. A dependence on $m_b$ enters only implicitly via the 
first moment of the leading-order shape function 
$\hat S(\hat \omega)$. The philosophy of our approach is that this 
function\footnote{More precisely, we define a new shape function 
$\hat\S(\hat\omega)$ by the combination of leading and subleading shape 
functions contributing to $\bar B\to X_s\gamma$ decay, and we will use the 
same function to make predictions for $\bar B\to X_u\,l^-\bar\nu$ decay 
distributions.}
is extracted experimentally from a fit to the $\bar B\to X_s\gamma$ photon 
spectrum, which has been measured with good precision in the region 
where $P_+ = M_B - 2 E_\gamma \sim \Lambda_{\rm QCD}$. This is analogous 
to the extraction of parton distribution functions from deep inelastic 
scattering. The
photon spectrum is experimentally accessible to energies as low as
1.8\,GeV, which corresponds to a sampling of the shape function for
values of $\hat\omega$ up to about 1.7\,GeV. Once the shape function has 
been extracted over this range, we can use it to obtain predictions for 
arbitrary partial $\bar B\to X_u\,l^-\bar\nu$ decay rates with cuts. In doing 
so, the residual hadronic uncertainties in the extraction of $|V_{ub}|$ only 
enter at the level of power corrections.

We emphasize that the program outlined above is equivalent to an approach put 
forward in \cite{Neubert:1993um} and later refined in 
\cite{Leibovich:1999xf,Leibovich:2000ey,Neubert:2001sk}, in which $|V_{ub}|$ 
is extracted with the help of shape-function independent relations between 
weighted integrals over differential decay distributions in 
$\bar B\to X_s\gamma$ and $\bar B\to X_u\,l^-\bar\nu$. The experimental 
error in the results for these weighted integrals corresponds, in our 
approach, to the error in the prediction of $\bar B\to X_u\,l^-\bar\nu$ 
partial rates resulting from the experimental uncertainty in the extraction of 
the shape function from the $\bar B\to X_s\gamma$ photon spectrum. While the 
shape-function independent relations are very elegant, it is more convenient 
for the construction of a generator to have a formulation where the shape 
function is used as an input. In this way, it is possible to impose arbitrary 
cuts on kinematical variables without having to recompute the weight 
functions in each case.

The chapter is structured as follows: In Section~\ref{sec:BsGamma} we collect 
the relevant formulae for the calculation of the $\bar B\to X_s\gamma$ 
photon spectrum. These expressions can be used to extract
the leading non-perturbative structure function from experiment. An
analogous presentation for the triple differential decay rate in 
$\bar B\to X_u\,l^-\bar\nu$ decays is presented in Section~\ref{sec:BuLnu}. In 
order to perform a numerical analysis one needs to rely on parameterizations 
of the shape functions. A collection of several useful functional forms is 
given in Section~\ref{sec:strucfunc}. In Section~\ref{sec:analysio} we present
a full error analysis of partial $\bar B\to X_u\,l^-\bar\nu$ decay rates for a 
variety of experimental cuts. We also explore the sensitivity of the results 
to the $b$-quark mass and to the functional forms adopted for the shape 
functions. Section~\ref{sec:concl} contains our conclusions.

\section{Inclusive radiative decays}
\label{sec:BsGamma}

The decay process $\bar B\to X_s \gamma$, while more complex in its
short-distance physics, is considerably simpler in its kinematics than the
semileptonic process $\bar B\to X_u\,l^-\bar\nu$. Since the
radiated photon is on-shell, the hadronic variables $P_\pm$ that
describe the momentum of the $X_s$ system are trivially related to the
photon energy $E_\gamma$ by $P_+ = M_B - 2 E_\gamma$ and 
$P_-=M_B$. In the crudest approximation, namely at tree level and leading
power, the photon-energy spectrum is directly proportional to the
leading shape function, $d\Gamma_s/dE_\gamma \propto \hat S(P_+)$. 
In this section we collect all relevant formulae needed to compute the 
$\bar B\to X_s\gamma$ photon spectrum or, equivalently, the invariant 
hadronic mass distribution. It is implicitly assumed that these spectra are 
sufficiently ``smeared'' (e.g., by experimental resolution)
to wash out any sharp hadronic structures. In cases where the 
resolution is such that the $K^{*}$ resonance peak is observed, it 
can be accounted for by combining the formulae 
in this section with the prescription for subtracting the $K^*$ peak 
proposed in \cite{Kagan:1998ym}.

The differential $\bar B\to X_s\gamma$ decay rate can be written as
\begin{equation}\label{BsG:general}
   \frac{d\Gamma_s}{d E_\gamma} = \frac{G_F^2\alpha}{2\pi^4}\,E_\gamma^3\,
   |V_{tb} V_{ts}^*|^2\,\overline{m}_b^2(\mu_h)\,
   [C_{7\gamma}^{\rm eff}(\mu_h)]^2\,U(\mu_h,\mu_i)\,\F_\gamma(P_+) \,,
\end{equation}
where the structure function $\F_\gamma$ depends on the photon energy
via $P_+=M_B-2E_\gamma$. The prefactor contains the
electromagnetic fine-structure constant $\alpha$ normalized at $q^2=0$, 
two powers of the running $b$-quark mass (defined in the $\overline{\rm MS}$ 
scheme) originating from the electromagnetic dipole
operator $Q_{7\gamma}$ in the effective weak 
Hamiltonian, and the square of the corresponding Wilson coefficient 
$C_{7\gamma}^{\rm eff}$, which is needed at next-to-leading order in 
renormalization-group improved perturbation theory \cite{Chetyrkin:1996vx}.
Renormalization-group running from the hard scale 
$\mu_h\sim m_b$ to the intermediate scale 
$\mu_i\sim\sqrt{m_b\Lambda_{\rm QCD}}$ gives rise to the evolution factor
$U(\mu_h,\mu_i)$, whose explicit form is discussed in 
Appendix~\ref{apx:Sudakovs}.
We keep $U$ and $(C_{7\gamma}^{\rm eff})^2$ outside of
the structure function $\F_\gamma$; it is understood that when
combining the various terms in (\ref{BsG:general}) all perturbative quantities
should be expanded for consistency to the required order in $\alpha_s$. 

\subsection{Leading-power factorization formula}

At leading order in $1/m_b$ the structure
function $\F_\gamma$ factorizes as \cite{Neubert:2004dd}
\begin{equation}\label{BsG:LO}
   \F_\gamma^{\rm (0)}(P_+) = |H_s(\mu_h)|^2 \int_0^{P_+}\!d\hat\omega\,
   m_b\,J(m_b(P_+ - \hat\omega),\mu_i)\,\hat S(\hat\omega,\mu_i) \,. 
\end{equation}
At this order a single non-perturbative parton distribution function arises, 
called the leading shape function \cite{Neubert:1993um} and denoted by 
$\hat S(\hat \omega,\mu_i)$. Our notation is adopted from 
\cite{Bosch:2004th,Bosch:2004cb}: hatted shape functions have support for 
$\hat\omega\ge 0$. The function $\hat S$ is
defined in terms of a non-local matrix element in heavy-quark effective
theory (HQET). Renormalization-group
running between the intermediate scale and a low hadronic scale is
avoided when using the shape functions renormalized at the intermediate 
scale $\mu_i$. Evolution effects below this scale are universal (i.e., 
process independent) and so can be absorbed into the renormalized shape 
function. Short-distance contributions from scales above $\mu_h\sim m_b$ are 
included in the hard function $H_s$, which in practice is obtained
by matching the effective weak Hamiltonian onto a current operator in 
soft-collinear effective theory (SCET). 
At next-to-leading order in perturbation theory, the result reads
\begin{eqnarray}\label{Hsresult}
   H_s(\mu_h)
   &=& 1 + \frac{C_F\alpha_s(\mu_h)}{4\pi} \left( -2\ln^2\frac{m_b}{\mu_h}
    + 7\ln\frac{m_b}{\mu_h} - 6 - \frac{\pi^2}{12} \right)
    + \varepsilon_{\rm ew} + \varepsilon_{\rm peng}\nonumber\\
   &&\mbox{}+ \frac{C_{8g}^{\rm eff}(\mu_h)}{C_{7\gamma}^{\rm eff}(\mu_h)}\,
    \frac{C_F\alpha_s(\mu_h)}{4\pi}
    \left( - \frac83 \ln\frac{m_b}{\mu_h} + \frac{11}{3} - \frac{2\pi^2}{9}
    + \frac{2\pi i}{3} \right) \nonumber\\
   &&\mbox{}+ \frac{C_1(\mu_h)}{C_{7\gamma}^{\rm eff}(\mu_h) }\,
    \frac{C_F\alpha_s(\mu_h)}{4\pi}
    \left( \frac{104}{27} \ln\frac{m_b}{\mu_h} + g(z)
    - \frac{V_{ub} V_{us}^*}{V_{tb} V_{ts}^*}\,\big[ g(0) - g(z) \big] \right)
     \,,\nonumber\\
\end{eqnarray}
where the variable $z=(m_c/m_b)^2$ denotes the ratio of quark
masses relevant to charm-loop penguin diagrams, and the ``penguin function'' 
$g(z)$ can be approximated by the first few terms of its Taylor expansion,
\begin{eqnarray}
   g(z) &=& - \frac{833}{162} - \frac{20\pi i}{27} + \frac{8\pi^2}{9}\,z^{3/2}
    \nonumber\\
   &+& \frac{2z}{9} \left[ 48 - 5\pi^2\! - 36\zeta_3
    + (30\pi - 2\pi^3) i + \!(36 - 9\pi^2 + 6\pi i) \ln z\right.\nonumber\\
   &+&\left.  \!(3 + 6\pi i) \ln^2 z + \ln^3 z \right] \nonumber\\
   &+& \frac{2z^2}{9} \left[ 18 + 2\pi^2 - 2\pi^3 i
    + (12 - 6\pi^2) \ln z + 6\pi i\ln^2 z + \ln^3 z \right] \nonumber\\
   &+& \frac{z^3}{27} \left[ -9 - 14\pi^2 + 112\pi i
    + (182 - 48\pi i) \ln z - 126\ln^2 z \right] + \dots \,.
\end{eqnarray}
The Wilson coefficients $C_1$ and $C_{8g}^{\rm eff}$ in (\ref{Hsresult}) 
multiply the current-current operators $Q_1^{u,c}$ and the chromo-magnetic 
dipole operator $Q_{8g}$ in the effective weak Hamiltonian.
The quantities $\varepsilon_{\rm ew}\approx -1.5\%$ and 
$\varepsilon_{\rm peng}\approx -0.6\%$ 
account for small electroweak corrections 
and the effects of penguin contractions of operators other than $Q_1^{u,c}$,
respectively. 
The differential decay rate (\ref{BsG:general}) is formally
independent of the matching scales $\mu_h$ and $\mu_i$. The $\mu_h$ 
dependence of the evolution factor $U(\mu_h,\mu_i)$ cancels the scale 
dependence of the product 
$\overline{m}_b^2(\mu_h)\,[C_{7\gamma}^{\rm eff}(\mu_h)]^2\,|H_s(\mu_h)|^2$, 
while its $\mu_i$ dependence compensates the scale dependence of the
convolution integral $J(\mu_i)\otimes \hat S(\mu_i)$. 

Finally let us discuss the jet function $J$, which appears as the 
hard-scattering kernel in the convolution integral in (\ref{BsG:LO}). It can 
be written in terms of distributions that act on the shape function $\hat S$. 
At one-loop order, the jet function is given by 
\cite{Bauer:2003pi,Bosch:2004th}
\begin{equation}\label{BsG:jetfunc}
   J(p^2,\mu) = \delta(p^2) \left[ 1
   + \frac{C_F \alpha_s(\mu)}{4\pi}(7-\pi^2) \right]
   + \frac{C_F \alpha_s(\mu)}{4\pi} \left[ \frac{1}{p^2}
   \left( 4\ln \frac{p^2}{\mu^2} - 3\right) \right]_*^{[\mu^2]} \,,
\end{equation}
where the star distributions have the following effect on a function
$f$ when integrated over a domain $Q^2$ \cite{DeFazio:1999sv}:
\begin{eqnarray}\label{stardistris}
   \int_{\le 0}^{Q^2} dp^2\,\left[\frac{1}{p^2} \right]_*^{[\mu^2]} f(p^2)
   &=& \int_0^{Q^2} dp^2\,\frac{f(p^2)-f(0)}{p^2} \,
    + f(0) \,\ln \frac{Q^2}{\mu^2} \,, \nonumber\\
   \int_{\le 0}^{Q^2} dp^2\,\left[\frac{1}{p^2} \ln \frac{p^2}{\mu^2}
    \right]_*^{[\mu^2]} f(p^2) 
   &=& \int_0^{Q^2} dp^2\,\frac{f(p^2)-f(0)}{p^2}\ln \frac{p^2}{\mu^2}
    + \frac{f(0)}{2}\,\ln^2 \frac{Q^2}{\mu^2} \,.  \nonumber\\
\end{eqnarray}

\subsection{Kinematical power corrections}

There exists a class of power corrections to (\ref{BsG:LO}) that do not
involve new hadronic quantities. Instead, the power suppression results from
the restriction of certain variables ($P_+$ in the present case) 
to a region where they are kinematically suppressed (here $P_+\ll M_B$).
The corresponding terms are known in fixed-order perturbation theory, without 
scale separation and renormalization-group resummation 
\cite{Greub:1996tg,Ali:1995bi} (see also 
\cite{Kagan:1998ym}). To perform a complete RG analysis of even the 
first-order terms in $1/m_b$ is beyond the scope of this work. 
Since, as we will see later, power corrections only account for small
corrections to the decay rates, an approximate treatment will suffice. 
To motivate it, we note the following two facts \cite{Neubert:2004dd}: 
First, while the 
anomalous dimensions of the relevant subleading SCET and HQET operators are 
only known for a few cases \cite{Hill:2004if}, the leading Sudakov double 
logarithms are the 
same as for the terms of leading power, because they have a geometric origin in
terms of Wilson lines \cite{Becher:2003kh}. The leading 
Sudakov double logarithms are therefore the same as those resummed into the 
function $U$ in (\ref{BsG:general}). Secondly, the kinematical power
corrections in $\bar B\to X_s\gamma$ decay  
are associated with gluon 
emission into the hadronic final state $X_s$. Because of the kinematical 
restriction to low-mass final states, i.e.\ $M_X^2\sim M_B\Lambda_{\rm QCD}$, 
we associate a coupling $\alpha_s(\bar\mu)$ with these terms, where typically
$\bar\mu\sim\mu_i$. Strictly speaking, however, the scale ambiguity associated
with the choice of $\bar\mu$ could only be resolved by computing the relevant
anomalous dimensions. 

Within this approximation, the kinematical power corrections to the structure
function $\F_\gamma$ can be extracted from \cite{Kagan:1998ym,Neubert:2004dd}. 
We find it convenient to express the result in terms of the variable
\begin{equation}\label{xdef}
   x = \frac{P_+ - \hat\omega}{M_B - P_+} \,,
\end{equation}
which in the shape-function region scales like $\Lambda_{\rm QCD}/m_b$.
We obtain
\begin{eqnarray}\label{realrads}
   \F_\gamma^{\rm kin}(P_+)
   &=& \frac{1}{M_B-P_+}\,\frac{C_F\alpha_s(\bar\mu)}{4\pi} \hspace{-0.1cm}
    \sum_{\scriptsize\begin{array}{c} i,j=1,7,8 \\[-0.1cm] i\le j \end{array}}
    \hspace{-0.1cm}
    \frac{C_i(\mu_h)\,C_j(\mu_h)}{C_{7\gamma}^{\rm eff}(\mu_h)^2}
    \int_0^{P_+}\!d\hat\omega\,\hat S(\hat\omega,\mu_i)\,h_{ij}(x)
    \nonumber\\[-0.3cm]
   &&\mbox{}- \frac{\lambda_2}{9m_c^2}\,
    \frac{C_1(\mu_h)}{C_{7\gamma}^{\rm eff}(\mu_h)}\,\hat S(P_+,\mu_i) \,.
\end{eqnarray}
The coefficient functions $h_{ij}(x)$ are 
\begin{eqnarray}
   h_{77}(x) &=& -3(5+2x) + 2(8+9x+3x^2) \ln\left( 1 + \frac{1}{x} \right) ,
    \nonumber\\
   h_{88}(x) &=& \frac29\,(1+3x+4x^2+2x^3) \left[ 2\ln\frac{m_b}{m_s}
    - \ln\left( 1 + \frac{1}{x} \right) \right]\nonumber\\
    &&- \frac19\,(3+9x+16x^2+8x^3) \,, \nonumber\\
   h_{78}(x) &=& \frac23\,(5+8x+4x^2) - \frac83\,x(1+x)^2\,
    \ln\left( 1 + \frac{1}{x} \right) , \nonumber\\
   h_{11}(x) &=& \frac{16}{9} \int_0^1\!du\,(1+x-u) 
    \left|\,\frac{z(1+x)}{u}\,
    G\!\left(\frac{u}{z(1+x)}\right) + \frac12\,\right|^2 , \nonumber\\
   h_{17}(x) &=& -3 h_{18}(x)
    = - \frac83 \int_0^1\!du\,u\,\mbox{Re}\left[\,
    \frac{z(1+x)}{u}\,G\!\left(\frac{u}{z(1+x)}\right) + \frac12 \,\right] , \nonumber\\
\end{eqnarray}
where as before $z=(m_c/m_b)^2$, and
\begin{equation}
   G(t) = \left\{ \begin{array}{ll}
    -2\arctan^2\!\sqrt{t/(4-t)} & ~;~ t<4 \,, \\[0.1cm]
    2 \left( \ln\!\Big[(\sqrt{t}+\sqrt{t-4})/2\Big]
    - \displaystyle\frac{i\pi}{2} \right)^2 & ~;~ t\ge 4 \,.
   \end{array} \right.
\end{equation}
In the shape-function region the expressions for $h_{ij}(x)$ could, if 
desired, be expanded in a power series in $x=O(\Lambda_{\rm QCD}/m_b)$, 
and this would generate a series of power-suppressed terms 
$\F_\gamma^{{\rm kin}(n)}(P_+)$ with $n\ge 1$, where the superscript ``$n$'' 
indicates the order in the $1/m_b$ expansion. Note that this expansion would
contain single logarithms $\ln x\sim\ln(\Lambda_{\rm QCD}/m_b)$. These are
precisely the logarithms that would be resummed in a more proper 
treatment using effective field-theory methods.

Outside the shape-function region the variable $x$ can take on arbitrarily
large positive values, and $\F_\gamma^{\rm kin}(P_+)$ is no longer power 
suppressed. Note that for $P_+\to M_B$ (corresponding to $x\to\infty$ and 
$E_\gamma\to 0$) most functions $h_{ij}(x)$ grow like $x^2$ or weaker, so 
that the spectrum tends to a constant. The only (well known) exception is 
$h_{88}(x)$, which grows like $x^3$, giving rise to a $1/E_\gamma$ 
soft-photon singularity \cite{Ali:1995bi}.
The main effect of the kinematical power corrections (\ref{realrads}) to the
photon spectrum is to add a radiative tail extending into the
region of small photon energies. These corrections therefore become the more 
significant the larger the integration domain over $E_\gamma$ is.

\subsection{Subleading shape-function contributions}
\label{subsec:BsG:SSF}

At order $1/m_b$ in power counting, different combinations of 
subleading shape functions enter the $\bar B\to X_s\gamma$ and 
$\bar B\to X_u\,l^-\bar\nu$ decay distributions 
\cite{Bauer:2001mh,Bauer:2002yu,Leibovich:2002ys,Neubert:2002yx}.
They provide the dominant hadronic power corrections, which must be 
combined with 
the kinematical power corrections discussed in the previous section.
We include their effects using the results of recent calculations in
\cite{Bosch:2004cb,Lee:2004ja,Beneke:2004in}. Little is known about the 
subleading shape functions apart from expressions for their first few
moments. In particular, the norms of these functions vanish at tree level, 
while their first moments are determined by the HQET parameters $\lambda_1$ 
and $\lambda_2$, which are defined via the forward $B$-meson matrix elements 
of the kinetic-energy and the chromo-magnetic operators, respectively 
\cite{Falk:1992wt}. 

For the case of $\bar B\to X_s\gamma$ decay, 
subleading shape-function contributions are currently only known for 
the matrix elements of the dipole 
operator $Q_{7\gamma}$, and the corresponding hard and jet functions
have been computed at tree level. Adopting the notations of 
\cite{Bosch:2004cb}, the relevant subleading shape functions are 
$\hat t(\hat\omega)$, $\hat u(\hat\omega)$, and $\hat v(\hat\omega)$. 
An additional function, called $s_0$, has been absorbed by a redefinition of 
the leading shape function, and it is included in our definition of 
$\hat S(\hat\omega)$. 
Roughly speaking, $\hat u(\hat\omega)$ is the ``light-cone
generalization'' of the local HQET kinetic-energy operator. The
functions $\hat v(\hat\omega)$ and $\hat t(\hat\omega)$ are both 
generalizations of the local chromo-magnetic HQET operator, but 
$\hat t(\hat\omega)$ contains also a
light-cone chromo-electric operator, which has no equivalent in the
local OPE expansion. (Such a contribution arises since there are 
two external 4-vectors in the SCET expansion, $n$ and $v$, while there is 
only $v$ in the HQET expansion.) The contribution of subleading shape
functions to the $\bar B\to X_s\gamma$ photon spectrum is
\begin{equation}\label{BsgSSF}
   \F_\gamma^{\rm hadr(1)}(P_+) = \frac{1}{M_B-P_+}
   \left[ - (\bar\Lambda - P_+)\,\hat S(P_+) - \hat t(P_+)
   + \hat u(P_+) - \hat v(P_+) \right] . 
\end{equation}
Compared to \cite{Bosch:2004cb}, we have replaced $1/m_b$ with 
$1/(M_B-P_+)$ in the prefactor, which is legitimate at this order. 
(The form of the shape 
functions restricts $P_+$ to be of order $\Lambda_{\rm QCD}$.)
The appearance of the HQET parameter $\bar\Lambda=(M_B-m_b)_{m_b\to\infty}$ 
is peculiar to the 
subleading shape-function contributions. This quantity is defined via the
first moment of the leading-order shape function \cite{Bosch:2004th}.

The formula given above can be modified to suit the purpose of
extracting the shape function from the photon spectrum better. To this
end, we absorb a linear combination of the subleading shape functions into
a redefinition of the leading shape function, in such a way that the
moment relations for this function remain unchanged to the
order we are working. This is accomplished by defining
\begin{equation}\label{def:ScriptS}
   \hat\S(\hat \omega)\equiv \hat S(\hat\omega)
   + \frac{2(\bar\Lambda-\hat\omega)\,\hat S(\hat\omega) - \hat t(\hat\omega)
           + \hat u(\hat\omega) - \hat v(\hat\omega)}{m_b} \,.
\end{equation}
When using $\hat\S$ instead of $\hat S$ in the leading-power formula 
(\ref{BsG:LO}), the subleading shape-function contribution becomes
\begin{equation}\label{BsgSSF2}
   \F_\gamma^{\rm hadr(1)}(P_+) = - \frac{3(\bar\Lambda-P_+)}{M_B-P_+}\,
   \hat\S(P_+) \,. 
\end{equation}

The hatted shape functions used in this chapter  are related to the 
original definitions in \cite{Bosch:2004cb} by 
\begin{eqnarray}
   \hat S(\hat\omega) &=& S(\bar\Lambda-\hat\omega)
    + \frac{s_0(\bar\Lambda-\hat\omega)}{m_b} \,, \nonumber\\
   \hat t(\hat\omega) &=& t(\bar\Lambda-\hat\omega) \,, \qquad 
    \hat u(\hat\omega) = u(\bar\Lambda-\hat\omega) \,, \qquad
    \hat v(\hat\omega) = v(\bar\Lambda-\hat\omega) \,,
\end{eqnarray}
where the unhatted functions have support on the interval between $-\infty$ 
and $\bar\Lambda$. It is convenient to rewrite
$\bar\Lambda-\hat\omega=\omega+\Delta\omega$, where
\begin{equation}\label{Delw}
   \Delta\omega\equiv \bar\Lambda - (M_B - m_b) 
   = \frac{\lambda_1+3\lambda_2}{2m_b} + \dots
\end{equation}
accounts for the mismatch between the HQET parameter $\bar\Lambda$ and the
difference $(M_B-m_b)$ due to power-suppressed terms in the $1/m_b$ 
expansion \cite{Neubert:1993mb}. It follows that the variable 
$\omega=(M_B-m_b)-\hat\omega$ runs from $-\infty$ to 
$(M_B-m_b)$. The moment relations for the leading and subleading shape 
functions derived in \cite{Neubert:1993um} and 
\cite{Bauer:2001mh,Bosch:2004cb} can be summarized as
\begin{eqnarray}\label{SSF:moments}
   \hat S(\hat\omega)
   &\equiv& S(\omega+\Delta\omega) + \frac{s_0(\omega+\Delta\omega)}{m_b}
    = \delta(\omega) - \frac{\lambda_1}{6}\,\delta''(\omega)
    + \frac{\lambda_1+3\lambda_2}{2m_b}\,\delta'(\omega) + \dots \,,
    \nonumber\\
   \hat t(\hat\omega)
   &\equiv& t(\omega+\Delta\omega) = \lambda_2\,\delta'(\omega) + \dots \,, 
    \nonumber\\
   \hat u(\hat\omega) 
   &\equiv& u(\omega+\Delta\omega) = - \frac{2\lambda_1}{3}\,\delta'(\omega)
    + \dots \,, \nonumber\\
   \hat v(\hat\omega) 
   &\equiv& v(\omega+\Delta\omega) = - \lambda_2\,\delta'(\omega) + \dots \,.
\end{eqnarray}
The function $\hat\S$ has the same moment expansion as $\hat S$.
The hadronic parameter $\lambda_2$ determines the leading contribution to 
the hyperfine splitting between the masses of $B$ and $B^*$ mesons through 
$m_{B^*}^2-m_B^2=4\lambda_2+O(1/m_b)$ \cite{Falk:1992wt}, from which it
follows that $\lambda_2\approx 0.12$\,GeV$^2$. The value of the parameter
$\lambda_1$ is more uncertain. In much the same way as the $b$-quark pole 
mass, it is affected by infrared renormalon ambiguities 
\cite{Martinelli:1995zw,Neubert:1996zy}.
It is therefore better to eliminate $\lambda_1$ in favor of some 
observable, for which we will choose the width of the
leading shape function. 

\subsection{Residual hadronic power corrections}
\label{subsec:BsGHQE}

At order $1/m_b^2$ a new set of sub-subleading shape
functions enter, which so far have not been classified completely in the 
literature. Since the functional form of even the subleading shape
functions is rather uncertain, there is no need
to worry too much about the precise form of sub-subleading shape functions.
Most of their effects can be absorbed into the subleading functions. An
exception, however, are terms that survive when the sub-subleading shape 
functions are integrated over a wide domain. Whereas the norms of all 
subleading ($\sim 1/m_b$) shape functions vanish, the norms of the 
sub-subleading shape functions ($\sim 1/m_b^2$) are in general 
non-zero and given in terms of the heavy-quark parameters $\lambda_1$ and 
$\lambda_2$. (At tree level, the class of functions with non-zero norm
has been studied in \cite{Tackmann:2005ub}.) Our strategy in this chapter will be as follows: We start from the well-known expressions for the 
(tree-level) second-order power corrections to the $\bar B\to X_s\gamma$ 
photon spectrum \cite{Falk:1993dh} (and similarly for the triple-differential 
$\bar B\to X_u\,l^-\bar\nu$ decay distribution 
\cite{Blok:1993va,Manohar:1993qn}, see Section~\ref{sec:MWterms}). They are of 
the form $\lambda_i/m_b^2$ times one of the singular distributions 
$\delta(p^2)$, $\delta'(p^2)$, or $\delta''(p^2)$, where 
$p^2=(m_b v-q)^2$ is the invariant partonic mass squared of the final-state 
jet. As mentioned earlier, the power counting in the shape-function region is 
different from the one used in OPE calculations, and indeed a good portion
of the $1/m_b^2$ terms in the OPE is already accounted for by the contributions
proportional to the leading and subleading shape functions in (\ref{BsG:LO}) 
and (\ref{BsgSSF}). We identify the corresponding terms
using the moment relations for the shape functions in (\ref{SSF:moments}).
In particular, this reproduces all terms at order $1/m_b^2$ in the OPE 
which contain derivatives of $\delta(p^2)$. We include the remaining terms
of the form $(\lambda_i/m_b^2)\,\delta(p^2)$ by replacing
\begin{eqnarray}
   \delta(p^2)
   &=& \delta(p_+ p_-)
    = \frac{1}{p_- - p_+}  \int d\omega\,\delta(p_+ + \omega)\,\delta(\omega)
    \nonumber\\
   &\to& \frac{1}{P_- - P_+} \int d\hat\omega\,\delta(P_+ - \hat\omega)\,
    \hat S(\hat\omega) = \frac{\hat S(P_+)}{P_- - P_+} \,.
\end{eqnarray}
Here $p_\pm$ are the light-cone projections of the partonic momentum $p^\mu$, 
which are related to the hadronic quantities $P_\pm$ by 
$P_\pm=p_\pm+(M_B-m_b)$. Similarly, $\hat\omega=(M_B-m_b)-\omega$.

The result of these manipulations is
\begin{equation}
   \F_\gamma^{\rm hadr(2)}
   = \frac{\lambda_1}{(M_B-P_+)^2}\,\hat S(P_+) \,.
\end{equation}
Together with (\ref{BsG:LO}) and (\ref{BsgSSF}) this accounts for
all known first- and second-order power corrections to the 
$\bar B\to X_s\gamma$ photon spectrum, both in the shape-function region and 
in the OPE region. The redefinition (\ref{def:ScriptS}) of the
leading shape function from $\hat S$ to $\hat\S$ leaves the form of the 
second-order power corrections unaffected. 

In Section~\ref{sec:analysio} we study the numerical impact of
second-order power corrections on various $\bar B\to X_u\,l^-\bar\nu$
partial rates and find their effects to be tiny. It is therefore a safe 
approximation to neglect hadronic power corrections of order $1/m_b^3$ or
higher. The only possible exception to this conclusion relates to the 
so-called weak annihilation terms in $\bar B\to X_u\,l^-\bar\nu$ decay, 
which will be included in our error analysis.

\section{Inclusive semileptonic decays}
\label{sec:BuLnu}
In terms of the $\tilde W_i$ functions introduced in chapter \ref{chapter_intro}, the triple
differential decay rate for $\bar B\to X_u\,l^-\bar\nu$ reads
\begin{eqnarray}\label{eq:tripleRate}
   \frac{d^3\Gamma_u}{dP_+\,dP_-\,dP_l}
   &=& \frac{G_F^2|V_{ub}|^2}{16\pi^3}\,U_y(\mu_h,\mu_i)\,(M_B-P_+)\,\nonumber\\
    &&\Big[ (P_- -P_l)(M_B-P_- +P_l-P_+)\,\F_1 \nonumber\\
   &&\mbox{}+ (M_B-P_-)(P_- -P_+)\,\F_2 + (P_- -P_l)(P_l-P_+)\,\F_3 \Big] \,,\nonumber\\
\end{eqnarray}
where we have collected the relevant combinations of $\tilde W_i$ into the
three functions
\begin{eqnarray}
   &&U_y(\mu_h,\mu_i)\,\F_1 = \tilde W_1 \,, \qquad
   U_y(\mu_h,\mu_i)\,\F_2 = \frac{\tilde W_2}{2} \,, \nonumber\\
   && U_y(\mu_h,\mu_i)\,\F_3 = \left( \frac{y}{4}\,\tilde W_3 + \tilde W_4
    + \frac{1}{y}\,\tilde W_5 \right)
\end{eqnarray}
and defined a new kinematical variable
\begin{equation}\label{ydef}
   y = \frac{P_- -P_+}{M_B-P_+} \,,
\end{equation}
which can take values $0\le y\le 1$.
The leading evolution factor $U_y(\mu_h,\mu_i)$ has been factored out in 
(\ref{eq:tripleRate}) for convenience, as we have done earlier in 
(\ref{BsG:general}). The function $U_y(\mu_h,\mu_i)$ differs from the 
corresponding function in $\bar B\to X_s\gamma$ decay by a $y$-dependent
factor, 
\begin{equation}\label{eq:SudakovFactor}
   U_y(\mu_h,\mu_i) = U(\mu_h,\mu_i)\,y^{-2a_\Gamma(\mu_h,\mu_i)} \,,
\end{equation}
where the function $a_\Gamma$ in the exponent is related to the cusp
anomalous dimension and is given in Appendix~\ref{apx:Sudakovs}.

Eq.~(\ref{eq:tripleRate}) for the triple differential rate is exact.
Note that there is
no reference to the $b$-quark mass at this point. The only dependence
on $m_b$ is through the structure functions $\F_i(P_+,y)$ (via hard matching
corrections and via the moment constraints on the shape function $\hat S$),
which are independent of the leptonic variable
$P_l$. The fact that the total decay rate $\Gamma_u$ is proportional
to $m_b^5$ is not in contradiction with (\ref{eq:tripleRate}). It is 
instructive to demonstrate how these five powers of $m_b$ are recovered in 
our approach. At tree level and leading
power the functions $\F_2$ and $\F_3$ vanish, while $\F_1=\hat S(P_+)$. 
Integrating over the full range of $P_l$ and $P_-$ builds up
five powers of $(M_B-P_+)$. For the purpose of illustration, let us
rename the $P_+$ variable to $\hat\omega$ in the last integration, so that 
the total decay rate is given as
\begin{eqnarray}\label{ex:totalRateOPErecovery}
   \Gamma_u 
   &=& \frac{G_F^2|V_{ub}|^2}{192\pi^3} \int_0^{M_B}\!d\hat\omega\,
    (M_B-\hat\omega)^5\,\hat S(\hat\omega)
   = \frac{G_F^2|V_{ub}|^2}{192\pi^3} \int_{-m_b}^{M_B-m_b}\!\! 
    d\omega\,(m_b+\omega)^5\,S(\omega) \nonumber\\
   &=& \frac{G_F^2|V_{ub}|^2}{192\pi^3} (m_b+\langle\omega\rangle)^5
    \left[ 1 + O\left(\frac{1}{m_b^2}\right) \right] .
\end{eqnarray}
At tree level, the first moment of the shape function $S(\omega)$ vanishes.
Beyond tree level this is no longer the
case, and the average $\langle\omega\rangle$ depends on the size of
the integration domain. The above observation motivates the use of the
shape-function scheme \cite{Bosch:2004th}, in which the $b$-quark mass is 
defined as $m_b^{\rm SF}=m_b^{\rm pole}+\langle\omega\rangle+O(1/m_b)$.  
After this is done, (\ref{ex:totalRateOPErecovery})
recovers the form of the conventional OPE result.

\begin{figure}
\begin{center}
\epsfig{file=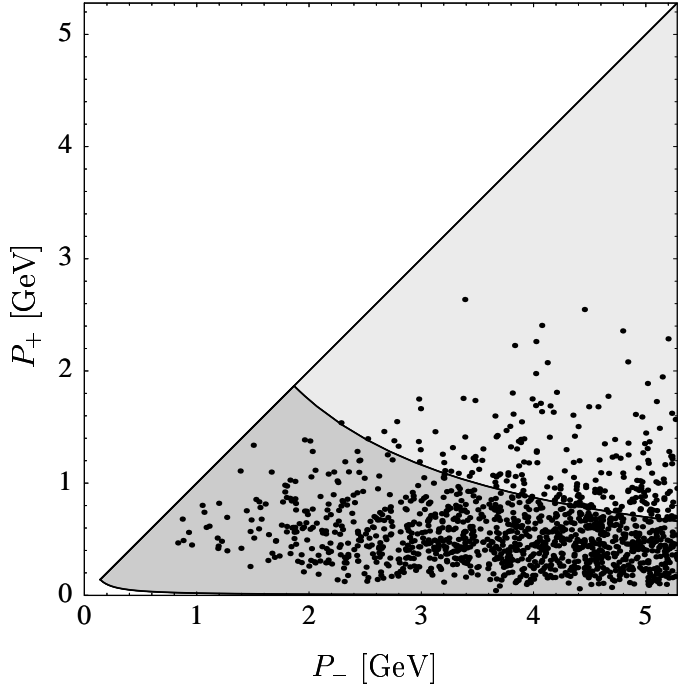, width=9cm}
\caption{\label{fig:scatterPlot}
The hadronic phase space in $P_+$ and $P_-$. The light gray region contains 
background from $\bar B\to X_c\,l^-\bar\nu$ decays, while the dark gray region 
is only populated by $\bar B\to X_u\,l^-\bar\nu$ events. The line separating 
the two regions is the contour where $M_X^2=P_+ P_-=M_D^2$.
Each point represents a $\bar B\to X_u\,l^-\bar\nu$ event 
in a Monte-Carlo simulation using the results of this chapter. While the 
shape-function region of large $P_-$ and small $P_+$ is highly populated, 
there is not a single event with $P_+$ larger than 3\,GeV out of the 1300 
events generated.}
\end{center}
\end{figure}

Eq.~(\ref{eq:tripleRate}) and the above argument tell us that the
differential rate is a priori rather insensitive to the $b$-quark mass
in the endpoint region, where $P_+$ (and therefore $\langle\hat\omega\rangle$) 
is a small quantity. Only when the rates are integrated over a sufficiently
wide domain, so that shape-function integrals can be approximated using
a moment expansion, a dependence on $m_b$ enters indirectly via the first
moment of the leading-order shape function. Likewise, a dependence on other
HQET parameters such as $\lambda_1$ enters via the sensitivity to higher
moments.

In the remainder of this section we present the various
contributions to the structure functions $\F_i$, following the same line 
of presentation as we did in the case of $\bar B\to X_s\gamma$ decay in 
Section~\ref{sec:BsGamma}. As before, while the resulting expressions 
are ``optimized'' for the shape-function region, they can
be used over the entire phase space and give the correct result for
the total decay rate up to corrections of $O(\alpha_s^2)$. In the
shape-function region, where $P_+$ is a small quantity, one may
organize each $\F_i$ as a series in inverse powers of $1/(M_B-P_+)$. No
assumption about the variable $y$ is made, which is treated as an
$O(1)$ quantity.\footnote{In the shape-function region, where $P_+\ll P_-$, we 
have $y\approx p_-/m_b$, which is the variable used in the leading-power
analysis in \cite{Bosch:2004th}.}
A preview of the results of our calculation is depicted in
Figure~\ref{fig:scatterPlot}, which shows 
an illustration of our prediction for the distribution of events in 
the plane $(P_+,P_-)$. 

\subsection{Leading-power factorization formula}
\label{sec:BuWleadPow}

The leading-power expressions for the hadronic structure functions
$W_i$ have been calculated in \cite{Bosch:2004th} at one-loop order in
renormalization-group improved perturbation theory. 
At this level $\F_2$ does not obtain a contribution,
whereas $\F_1$ and $\F_3$ do. Symbolically, they take the factorized form
$H_{ui}\,J\otimes\hat S$, consisting of hard functions $H_{ui}$ and the 
convolution of the jet function $J$ with the leading shape function $\hat S$. 
More precisely,
\begin{equation}\label{eq:BuLP}
   \F_i^{\rm (0)}(P_+,y)
   = H_{ui}(y,\mu_h) \int_0^{P_+}\!d\hat\omega\,y m_b\,
    J(y m_b(P_+-\hat\omega),\mu_i)\,\hat S(\hat\omega,\mu_i) \,,
\end{equation}
where the hard functions are given by
\begin{eqnarray}\label{eq:LeadHardFunction}
   H_{u1}(y,\mu_h)
   &=& 1 +\! \frac{C_F\alpha_s(\mu_h)}{4\pi} \!\left[\!
    - 4\ln^2\!\frac{y m_b}{\mu_h} + 10\ln\frac{y m_b}{\mu_h} \right.\nonumber\\ 
    &&\left. - 4\ln y - \frac{2\ln y}{1-y} - 4L_2(1\!-\!y) - \frac{\pi^2}{6} -\! 12 \right]\! ,
    \nonumber\\   
   H_{u3}(y,\mu_h)
   &=& \frac{C_F\alpha_s(\mu_h)}{4\pi}\,\frac{2\ln y}{1-y} \,,
\end{eqnarray}
and $H_{u2}=0$.
As before, the differential decay rate is independent of the matching scales 
$\mu_h\sim m_b$ and $\mu_i\sim\sqrt{m_b\Lambda_{\rm QCD}}$.
The jet function $J$ has already been given in (\ref{BsG:jetfunc}). 
Note that the $b$-quark mass appears
only as the argument of logarithms, where it plays the role of setting 
the renormalization scale.

\subsection{Kinematical power corrections}

As in the case of $\bar B\to X_s\gamma$ decay, there is a class of power
corrections to the $\bar B\to X_u\,l^-\bar\nu$ decay distributions which
are small only because of the restriction to certain regions in phase space,
but which are not associated with new hadronic parameters. In the present 
case, these terms can be extracted from the one-loop expressions derived 
in \cite{DeFazio:1999sv}. They are then convoluted with the leading shape 
function. As previously, the scale separation that can be achieved for these 
power-suppressed terms is only approximate, and we thus assign a coupling
$\alpha_s(\bar\mu)$ with them, where the scale $\bar\mu$ is expected to be 
of order $\mu_i\sim\sqrt{m_b\Lambda_{\rm QCD}}$.

The resulting expressions for the structure functions can be written in
a compact form in terms of the variables $x$ and $y$ defined in 
(\ref{xdef}) and (\ref{ydef}). We find
\begin{eqnarray}\label{eq:fullDFN}
   \F_1^{\rm kin}(P_+,y)
   &=& \frac{1}{M_B-P_+}\,\frac{C_F\alpha_s(\bar\mu)}{4\pi}
    \int_0^{P_+}\!d\hat\omega\,\hat S(\hat\omega,\mu_i) \nonumber\\
   &&\times \left[ \frac{f_1(x,y)}{(1+x)^2 y(x+y)}
    - \frac{2 g_1(x,y)}{x(1+x)^2 y^2(x+y)} \right.\nonumber\\
    &&\left.\times\ln\left(1+\frac{y}{x} \right)
    - \frac{4}{x} \ln\left( y+\frac{y}{x} \right) \right] , \nonumber\\
   \F_2^{\rm kin}(P_+,y)
   &=& \frac{1}{M_B-P_+}\,\frac{C_F\alpha_s(\bar\mu)}{4\pi}
    \int_0^{P_+}\!d\hat\omega\,\hat S(\hat\omega,\mu_i) \nonumber\\
   &&\times \left[ \frac{f_2(x,y)}{(1+x)^2y^2(x+y)}
    - \frac{2x\,g_2(x,y)}{(1+x)^2 y^3(x+y)} 
    \ln\left(1+\frac{y}{x}\right) \right] , \nonumber \\
   \F_3^{\rm kin}(P_+,y)
   &=& \frac{1}{M_B-P_+}\,\frac{C_F\alpha_s(\bar\mu)}{4\pi}
    \int_0^{P_+}\!d\hat\omega\,\hat S(\hat\omega,\mu_i) \nonumber \\
   &&\times \left[ \frac{f_3(x,y)}{(1+x)^2y^3(x+y)}
    + \frac{2 g_3(x,y)}{(1+x)^2y^4(x+y)} 
    \ln\left( 1+\frac{y}{x} \right) \right] \,,\nonumber\\
\end{eqnarray}
where the functions $f_i$, $g_i$ are given by
\begin{eqnarray}
   f_1(x,y) &=& -9 y + 10 y^2 + x (-16+12y+6y^2) 
   + x^2(13 y-12) \,, \nonumber\\
   g_1(x,y) &=& -2y^3 -2 x y^2(4+y) -x^2 y(12+4y+y^2) - 4x^3 (y+2)
    + 3 x^4(y-2) \,, \nonumber\\
   f_2(x,y) &=& y^2+xy(8+4y+y^2) +3x^2y(10+y)+x^3(12+19y)+10x^4 \,,
    \nonumber\\
   g_2(x,y) &=&  2y^2+4xy(1+2y)+x^2y(18+5y) + 6x^3(1+2y)+5x^4 \,, \nonumber\\
   f_3(x,y) &=& 2y^3(2y-11) +xy^2(-94+29y+2y^2) + 2x^2y(-72+18y+13y^2)
    \nonumber\\
   &&\mbox{}+ x^3(-72-42y+70y^2-3y^3) - 10x^4(6-6y+y^2) \,,\nonumber \\
   g_3(x,y) &=& 4y^4 -6x(y-5)y^3 - 4x^2y^2(-20+6y+y^2) \nonumber\\
    &&+ x^3y(90-10y-28y^2+y^3) 
   + x^4(36+36y-50y^2+4y^3) \nonumber\\
   &&+5x^5(6-6y+y^2) \,.\nonumber\\
\end{eqnarray}
The above formulae are the exact $O(\alpha_s)$ corrections to the
leading-power expression. This means that, when integrated over the entire 
phase space, they will give rise to the correct result for the total rate up 
to that order. In the shape-function region (where $P_+\ll P_-$)
the integrands in (\ref{eq:fullDFN}) can be expanded in powers of $1/m_b$ by 
counting $y=O(1)$ and $x=O(1/m_b)$.
Note that this organizes the $1/m_b$ expansion as an expansion
in powers of the hadronic variable $1/(M_B-P_+)$. The leading terms read
\begin{eqnarray}\label{eq:kinNLO}
   \F_1^{\rm kin(1)}(P_+,y)
   &=& \frac{1}{M_B-P_+}\,\frac{C_F\alpha_s(\bar\mu)}{4\pi}
    \int_0^{P_+}\!d\hat\omega\,\hat S(\hat\omega,\mu_i)
    \left[ 6 - \frac5y + \left( \frac{12}{y}-4 \right) \ln\frac{y}{x} \right]
    , \nonumber\\
   \F_2^{\rm kin(1)}(P_+,y)
   &=& \frac{1}{M_B-P_+}\,\frac{C_F\alpha_s(\bar\mu)}{4\pi}
    \int_0^{P_+}\!d\hat\omega\,\hat S(\hat\omega,\mu_i)
    \left[ \frac1y \right] , \nonumber\\
   \F_3^{\rm kin(1)}(P_+,y)
   &=& \frac{1}{M_B-P_+}\,\frac{C_F\alpha_s(\bar\mu)}{4\pi}
    \int_0^{P_+}\!d\hat\omega\,\hat S(\hat\omega,\mu_i) 
    \left[ 4 - \frac{22}{y} + \frac8y \ln\frac{y}{x} \right] .
\end{eqnarray}
Further accuracy can be achieved by adding the next-order corrections, 
for which we obtain
\begin{eqnarray}\label{eq:kinNNLO}
   \F_1^{\rm kin(2)}(P_+,y)
   &=& \frac{1}{(M_B-P_+)^2}\,\frac{C_F\alpha_s(\bar\mu)}{4\pi}
    \int_0^{P_+}\!d\hat\omega\,(P_+ -\hat\omega)\,\hat S(\hat\omega,\mu_i)
    \nonumber\\
   &&\times \left[ -12 + \frac{16}{y} + \frac{3}{y^2} 
    + \left( \frac{12}{y^2} - \frac{20}{y} + 6 \right) 
    \ln\frac{y}{x} \right] , \nonumber\\
   \F_2^{\rm kin(2)}(P_+,y)
   &=& \frac{1}{(M_B-P_+)^2}\,\frac{C_F\alpha_s(\bar\mu)}{4\pi}
    \int_0^{P_+}\!d\hat\omega\,(P_+ -\hat\omega)\,\hat S(\hat\omega,\mu_i)
    \nonumber\\
   &&\times \left[ 1 + \frac2y + \frac{7}{y^2} - \frac{4}{y^2} \ln\frac{y}{x}
    \right] , \nonumber\\
   \F_3^{\rm kin(2)}(P_+,y)
   &=& \frac{1}{(M_B-P_+)^2}\,\frac{C_F\alpha_s(\bar\mu)}{4\pi}
    \int_0^{P_+}\!d\hat\omega\,(P_+ -\hat\omega)\,\hat S(\hat\omega,\mu_i)
    \nonumber \\
   &&\times \left[ - 6 + \frac{69}{y} - \frac{64}{y^2}
    + \left( \frac{52}{y^2} - \frac{28}{y} \right) \ln\frac{y}{x} \right] .
\end{eqnarray}
In the various phase-space regions of interest to the determination of 
$|V_{ub}|$, the above terms (\ref{eq:kinNLO}) and (\ref{eq:kinNNLO}) 
approximate the full result (\ref{eq:fullDFN}) very well (see 
Section~\ref{sec:analysio} below).  

Let us comment here on a technical point already mentioned in section \ref{section_4.1}. 
When combined with the phase-space factors
in (\ref{eq:tripleRate}), the exact expressions for $\F_i^{\rm kin}$ in 
(\ref{eq:fullDFN}) are regular in the limit $P_-\to P_+$, corresponding to 
$y\to 0$. However, this feature is not automatically ensured when the
structure functions, but not the phase-space factors, are expanded about the 
heavy-quark limit. With our choice of the variables $x$ and $y$, we encounter
terms as singular as $1/y^n$ at $n$-th order in the expansion, as is obvious 
from the explicit expressions above. Phase space
scales like $y^2$ in the limit $y\to 0$ (note that $P_l\to P_+$ as 
$P_-\to P_+$ because of (\ref{equation_psl})), so that the results 
(\ref{eq:kinNLO}) and (\ref{eq:kinNNLO}) can be applied without encountering
any kinematical singularities. In order to achieve this, it was crucial to 
define the variable
$y$ in the way we did in (\ref{ydef}). We emphasize this point because
straightforward application of the technology of SCET and HQET developed 
in \cite{Bosch:2004cb,Lee:2004ja,Beneke:2004in} would give an expansion of 
the structure functions $\F_i$ in powers of $1/p_-$, whereas phase space
is proportional to $4\vec{p}^{\,2}=(p_- -p_+)^2\propto y^2$. In the 
kinematical region where $p_+<0$, which is allowed due to off-shell effects in
the $B$ meson, this leads to singularities as $p_-\to 0$. In order to
avoid these singularities, we have reorganized the SCET expansion as an 
expansion in $1/(p_- -p_+)$ instead of $1/p_-$, where $|p_+|\ll p_-$ in the 
shape-function region.

\subsection{Subleading shape-function contributions}
\label{sec:BuWsubSF}

The contributions from subleading shape functions to arbitrary 
$\bar B\to X_u\,l^-\bar\nu$ decay distributions have been derived (at tree 
level) in \cite{Bosch:2004cb,Lee:2004ja,Beneke:2004in}. 
The results involve the same set of subleading shape functions as previously 
discussed in Section~\ref{subsec:BsG:SSF}. 
Again, the structure function $\F_2$ does not obtain a contribution, while
\begin{eqnarray}
   \F_1^{\rm hadr(1)}(P_+,y)
   &=& \frac{1}{M_B-P_+}
    \left[ (\bar\Lambda-P_+)\,\hat S(P_+) + \hat t(P_+) 
    + \frac{\hat u(P_+) - \hat v(P_+)}{y} \right] , \nonumber\\
   \F_3^{\rm hadr(1)}(P_+,y)
   &=& \frac{1}{M_B-P_+}\,\frac{2}{y}
    \left[ - (\bar\Lambda-P_+)\,\hat S(P_+) - 2\hat t(P_+) 
    + \frac{\hat t(P_+) + \hat v(P_+)}{y} \right] .\nonumber\\
\end{eqnarray}
At this point we recall the discussion of Section~\ref{subsec:BsG:SSF}, where 
we have argued that the $\bar B\to X_s\gamma$ photon spectrum should be used 
to fit the function $\hat\S$ of (\ref{def:ScriptS}), which is defined to
be a linear combination of the leading shape function $\hat S$ and the
subleading shape functions $\hat t$, $\hat u$, $\hat v$. When the above 
results are rewritten in terms of the new function $\hat\S$
nothing changes in the expressions for $\F_i^{\rm (0)}$ except for
the simple replacement $\hat S\to \hat\S$, which we from now on
assume. At the level of subleading shape functions 
$\F_2^{\rm hadr(1)}=0$ and $\F_3^{\rm hadr(1)}$ remain unchanged, while
\begin{eqnarray}\label{upsy}
   \F_1^{\rm hadr(1)}(P_+,y) &=& \frac{1}{M_B-P_+}
   \left[ - (\bar\Lambda-P_+)\,\hat\S(P_+) + 2\hat t(P_+) \right.\nonumber\\
   &&\left.+ \left[ \hat u(P_+) - \hat v(P_+) \right]
   \left( \frac{1}{y} - 1 \right) \right] .
\end{eqnarray}
It follows that there reside some linear combinations of subleading shape
functions in the triple differential decay rate that cannot be
extracted from information on the photon spectrum in $\bar B\to X_s\gamma$
decays. In the end, this dependence gives rise to a theoretical
uncertainty.

\subsection{Residual hadronic power corrections}
\label{sec:MWterms}

In analogy with our treatment for the case of $\bar B\to X_s\gamma$ decay, we
start from the expressions for the $1/m_b^2$ corrections to the triple
differential $\bar B\to X_u\,l^-\bar\nu$ decay rate obtained by applying the
OPE to the hadronic tensor \cite{Manohar:1993qn,Blok:1993va}. We saw in chapter 
\ref{chapter_intro} that:
\begin{eqnarray}\label{eq:OPEres}
   \tilde W_1^{(2)}
   &=& \delta(p_+) \left( 1 + \frac{2\lambda_1-3\lambda_2}{3p_-^2} \right) 
    + \delta'(p_+) \left( \frac{2\lambda_1-3\lambda_2}{3p_-} 
    - \frac{5\lambda_1+15\lambda_2}{6m_b} \right)\nonumber \\  
    &&- \delta''(p_+)\,\frac{\lambda_1}{6} \,, \nonumber \\ 
   \tilde W_2^{(2)}
   &=& \delta(p_+) \left( - \frac{4\lambda_1-6\lambda_2}{3p_-^2} \right) , \nonumber \\
   \frac{y}{4}\,\tilde W_3^{(2)} &+& \tilde W_4^{(2)}
    + \frac{1}{y}\,\tilde W_5^{(2)}
   = \frac{\delta(p_+)}{p_-} \left( \frac{2\lambda_1+12\lambda_2}{3p_-}
    - \frac{4\lambda_1+9\lambda_2}{3m_b} \right)\nonumber \\
    &&+ \frac{\delta'(p_+)}{p_-} \left( \frac{2\lambda_1}{3} 
    + 4\lambda_2 \right) .
\end{eqnarray}
The desired $1/(M_B-P_+)^2$ corrections to the structure functions
$\F_i$ can then be extracted by expanding the leading and subleading
contributions $\F_i^{\rm (0)}$ and $\F_i^{\rm hadr(1)}$ in terms of their
moments in (\ref{SSF:moments}), and by subtracting the results from
(\ref{eq:OPEres}). Following the same procedure as in 
Section~\ref{subsec:BsGHQE} to express the remaining power corrections in 
terms of the leading shape function, we obtain
\begin{eqnarray}
   \F_1^{\rm hadr(2)}(P_+,y)
   &=& \frac{1}{(M_B-P_+)^2} \left( \frac{4\lambda_1-6\lambda_2}{3y^2} 
    - \frac{\lambda_1+3\lambda_2}{3} \right) \hat S(P_+) \,,
    \nonumber\\
   \F_2^{\rm hadr(2)}(P_+,y)
   &=& \frac{1}{(M_B-P_+)^2}
    \left( \frac{-2\lambda_1+3\lambda_2}{3y^2} \right) \hat S(P_+) \,,
    \nonumber\\
   \F_3^{\rm hadr(2)}(P_+,y)
   &=& \frac{1}{(M_B-P_+)^2} \left( \frac{4\lambda_1+24\lambda_2}{3y^2} 
    - \frac{4\lambda_1+9\lambda_2}{3y} \right) \hat S(P_+) \,.
\end{eqnarray}
These expressions remain unchanged when the shape function $\hat\S$ is used
instead of $\hat S$.

\subsection{Weak annihilation contributions}
\label{Sec:WA}

In the OPE calculation several contributions appear at third order in the
power expansion: $1/m_b$ corrections to the kinetic and chromo-magnetic
operators, the Darwin and spin-orbit terms, and weak annihilation 
contributions. The Darwin and spin-orbit terms correspond to the forward 
$B$-meson matrix elements of (light) flavor-singlet 
operators \cite{Uraltsev:1999rr}. The corresponding HQET parameters 
$\rho^3_D$ and $\rho_{LS}^3$ can in principle be extracted from moments of 
inclusive $\bar B\to X_c\,l^-\bar\nu$ decay spectra. They are insensitive 
to the flavor of the spectator quark inside the $B$ meson. The weak 
annihilation contribution, on the other hand, results from four-quark 
operators with flavor non-singlet structure.
Graphically, this contribution corresponds to a process in which
the $b$ and $\bar u$ quark annihilate into a $W^-$. Weak annihilation terms 
come with a phase-space enhancement factor of $16\pi^2$ and so are potentially 
more important than other power corrections of order $1/m_b^3$.
Because of the
flavor dependence, these contributions can effect neutral and charged
$B$ mesons differently \cite{Bigi:1993bh}. 
One choice of basis for the
corresponding four-quark operators is \cite{Neubert:1996we}
\begin{equation}
   \langle\bar B|\bar b_L\gamma_\mu u_L\,\bar u_L\gamma^\mu b_L|\bar B\rangle
    = \frac{f_B^2 M_B^2}{4}\,B_1 \,, \qquad
   \langle\bar B|\bar b_R u_L\,\bar u_L b_R|\bar B\rangle
    = \frac{f_B^2 M_B^2}{4}\,B_2 \,,
\end{equation}
where $f_B$ is the $B$-meson decay constant, and $B_i$ are hadronic
parameters. In the vacuum saturation approximation they are given by
$B_1=B_2=1$ for charged $B$ mesons and $B_1=B_2=0$ for neutral ones. 
The total semileptonic rate is proportional to the
difference $(B_2-B_1)$, which implies that the weak annihilation contribution 
would vanish in this approximation. Currently, only rough estimates are 
available for the magnitude of the deviation of this difference from zero. 
The resulting effect on the total branching ratio is 
\cite{Voloshin:2001xi}
\begin{equation}
   \delta B(\bar B\to X_u\,l^-\bar\nu)\approx
   3.9 \left( \frac{f_B}{0.2\,{\rm GeV}} \right)^2
   \left( \frac{B_2-B_1}{0.1} \right) |V_{ub}|^2 \,.
\end{equation}
Again, we expect this effect to be different for charged and neutral $B$
mesons. The most important feature of weak annihilation is that it is 
formally concentrated
at the kinematical point where all the momentum of the heavy quark is
transferred to the lepton pair \cite{Bigi:1993bh}. At the parton level
this implies that the corresponding contribution is proportional to
$\delta(q^2-m_b^2)$. It is therefore included in
every cut that includes the $q^2$ endpoint, and its effect is
independent of the specific form of the cut. 

We suggest two different strategies to control this effect. The
first is to include it in the error estimate as a constant contribution 
proportional to the total rate. A recent study \cite{TomsThesis} puts
a limit on this effect
of $\pm 1.8\%$ on the total rate (at 68\% confidence level) by analyzing 
CLEO data. The second one is to impose a cut $q^2\le q_{\rm max}^2$, 
thus avoiding the region where the weak annihilation contribution is
concentrated. The maximal value of $q^2$ is $(M_B-M_\pi)^2$, but one
must exclude a larger region of phase space, such that the {\em excluded\/} 
contribution to the decay rate at large $q^2$ 
(corresponding to a region near the origin in the 
$(P_-,P_+)$ plane) can be reliably calculated. In our
numerical analysis, we will study the effect of a cut $q^2\le(M_B-M_D)^2$,
which satisfies this criterion.

For completeness, we note that even after the weak annihilation contribution
near maximum $q^2$ has been removed, there could in principle exist other, 
flavor-specific contributions to the semileptonic decay amplitudes that are
different for charged and neutral $B$ mesons. The leading terms of this kind  
contribute at order $1/m_b$ in the shape-function region and are parameterized 
by a set of four-quark subleading shape functions 
\cite{Bosch:2004cb,Lee:2004ja,Beneke:2004in}. Model estimates of these 
contributions show 
that they are very small for all observables considered for an extraction of 
$|V_{ub}|$ \cite{Beneke:2004in,Neubert:2004cu}. If only flavor-averaged decay
rates are measured, the effects of four-quark subleading shape functions can 
be absorbed entirely by a redefinition of the functions $\hat u(\hat\omega)$ 
and $\hat v(\hat\omega)$ \cite{Bosch:2004cb}, without affecting the moment 
relations in (\ref{SSF:moments}).

\section{Shape-function parameterizations}
\label{sec:strucfunc}

Hadronic-physics effects enter the description of inclusive decay
rates via non-perturbative shape functions. Perturbation theory cannot
tell us much about the local form of these functions, but moments of
them are calculable provided that the domain of integration is much
larger than $\Lambda_{\rm QCD}$. Since the shape functions contain
information about the internal structure of the 
$B$ meson, knowledge of them relates directly to the determination
of the $b$-quark mass $m_b$, the kinetic-energy parameter $\lambda_1$, and 
in principle the matrix elements of higher-dimensional
operators. Improved measurements of the shape of the $\bar B\to X_s\gamma$
photon spectrum will therefore lead directly to a more precise
determination of HQET parameters. This argument can be 
turned around to constrain the leading shape function using
knowledge of $m_b$ and $\lambda_1$ from other physical processes such
as a $b\to c$ moment analysis \cite{Aubert:2004aw}. We emphasize,
however, that there are obviously infinitely many locally different
functions that have identical first few moments. In this section we
present a few functional forms that can be used to model the shape
functions and to fit the current experimental data.

To achieve stringent constraints on the leading shape function 
a precise definition of the HQET
parameters is required. It is a well-known fact that the pole-mass
scheme introduces uncontrollable ambiguities. To avoid these
uncertainties several short-distance definitions have been proposed,
such as the $\overline{\rm MS}$ scheme, the potential-subtraction
scheme \cite{Beneke:1998rk}, the $\Upsilon(1S)$ scheme
\cite{Hoang:1998hm}, the kinetic scheme \cite{Bigi:1996si}, or the
shape-function scheme \cite{Bosch:2004th}. While the decay rates are of 
course independent of the particular choice, it is advantageous to use a 
mass scheme that is designed for the physics problem at hand. 
In the case of inclusive
$B$ decays into light particles, this is the shape-function scheme.

\subsection{Models for the leading shape function}

Model-independent constraints on the shape function $\hat S(\hat\omega,\mu_i)$ 
can be derived by analyzing moments defined with an upper limit of 
integration $\hat\omega_0$, i.e.\
\begin{equation}\label{Def:Moments}
   M_N(\hat\omega_0,\mu_i)\equiv \int_0^{\hat\omega_0}\!d\hat\omega\,
   \hat\omega^N\,\hat S(\hat\omega,\mu_i) \,.
\end{equation}
For practical applications,
$\hat\omega_0$ should be taken of order the size of the window where
the $\bar B\to X_s\gamma$ photon spectrum is experimentally accessible,
$\hat\omega_0=M_B-2E_\gamma^{\rm min}$ with 
$E_\gamma^{\rm min}\approx 1.8$\,GeV. These moments can be expanded in
terms of matrix elements of local operators as long as $\hat\omega_0$
is large compared to $\Lambda_{\rm QCD}$. In the 
shape-function scheme, HQET parameters are defined to all orders in 
perturbation theory through ratios of such moments, e.g.\ 
\cite{Bosch:2004th}
\begin{eqnarray}\label{MomentRelations}
   \frac{M_1(\mu_f+\bar\Lambda(\mu_f,\mu_i),\mu_i)}
        {M_0(\mu_f+\bar\Lambda(\mu_f,\mu_i),\mu_i)}
   &=& \bar\Lambda(\mu_f,\mu_i) \,, \nonumber\\
   \frac{M_2(\mu_f+\bar\Lambda(\mu_f,\mu_i),\mu_i)}
        {M_0(\mu_f+\bar\Lambda(\mu_f,\mu_i),\mu_i)}
   &=& \frac{\mu_\pi^2(\mu_f,\mu_i)}{3} + \bar\Lambda^2(\mu_f,\mu_i) \,. 
\end{eqnarray}
Here, the factorization scale $\mu_f\gg\Lambda_{\rm QCD}$ is related to 
the size of the integration domain via the implicit equation 
$\hat\omega_0=\mu_f+\bar\Lambda(\mu_f,\mu_i)$. In practice $\mu_f$ is close to 
the intermediate scale $\mu_i$. At tree level, the relations between
parameters in the shape-function scheme and the pole scheme are
$\bar\Lambda(\mu_f,\mu_i)=\bar\Lambda^{\rm pole}$ and
$\mu_\pi^2(\mu_f,\mu_i)=-\lambda_1$. The corresponding relations at 
one- and two-loop order have been worked out in \cite{Bosch:2004th} and
\cite{Neubert:2004sp}, respectively. These relations allow us to obtain
precise determinations of $\bar\Lambda(\mu_f,\mu_i)$ and
$\mu_\pi^2(\mu_f,\mu_i)$ from other physical processes.

For reference purposes, it is helpful to quote values for $\bar\Lambda$
and $\mu_\pi^2$ using only a single scale $\mu_*$ instead of two
independent scales $\mu_f$ and $\mu_i$. To one-loop order, these parameters 
can be related to those determined from the moments via \cite{Bosch:2004th}
\begin{eqnarray}\label{eq:atMuStar}
   \bar\Lambda(\mu_*,\mu_*)
   &=& \bar\Lambda(\mu_f,\mu_i) 
    + \mu_* \frac{C_F\alpha_s(\mu_*)}{\pi} 
    - \mu_f \frac{C_F\alpha_s(\mu_i)}{\pi} \nonumber \\
    &&\times\left[ 1
    - 2\,\bigg( 1 - \frac{\mu_\pi^2(\mu_f,\mu_i)}{3\mu_f^2} \bigg)
    \ln\frac{\mu_f}{\mu_i} \right] ,\nonumber \\
   \mu_\pi^2(\mu_*,\mu_*)
   &=& \mu_\pi^2(\mu_f,\mu_i) 
    \left[ 1+\frac{C_F\alpha_s(\mu_*)}{2\pi} 
    - \frac{C_F\alpha_s(\mu_i)}{\pi} \left( \frac12 + 3 
    \ln\frac{\mu_f}{\mu_i} \right) \right]\nonumber \\
    &&+ 3\mu_f^2\,\frac{C_F\alpha_s(\mu_i)}{\pi}\,\ln\frac{\mu_f}{\mu_i} \,,
\end{eqnarray}
where we have neglected higher-dimensional operator matrix elements
that are suppressed by inverse powers of $\mu_f$. A typical choice for
the scale $\mu_*$ is 1.5\,GeV, which we will use as the reference scale 
throughout this chapter.
It will be convenient to connect the parameter $\bar\Lambda$
extracted from the first moment of the shape function with a low-scale
subtracted quark-mass definition referred to as the ``shape-function'' mass. 
Following \cite{Bosch:2004th}, we define
\begin{equation}
   m_b(\mu_f,\mu_i)\equiv M_B - \bar\Lambda(\mu_f,\mu_i) \,.
\end{equation}

The general procedure for modeling the leading shape function $\hat
S(\hat\omega,\mu_i)$ from a given functional form $F(\hat\omega)$ is
as follows. The shape of $F(\hat\omega)$ is assumed to be tunable so
that it can be used to fit the $\bar B\to X_s\gamma$ photon spectrum. Only
the norm of the shape function is fixed theoretically. Note that
the moment relations (\ref{MomentRelations}) are insensitive to the
norm, so that formulae for $\bar\Lambda$ and $\mu_\pi^2$ follow
directly from the functional form of $F(\hat\omega)$. Examples of such
formulae will be given below. We define moments
$M_N^{[F]}(\hat\omega_0)$ of $F$ in analogy with
(\ref{Def:Moments}). The first relation in (\ref{MomentRelations})
implies that for a given $\hat\omega_0$ the factorization scale is
\begin{equation}
   \mu_f = \hat\omega_0
   - \frac{M_1^{[F]}(\hat\omega_0)}{M_0^{[F]}(\hat\omega_0)} \,.
\end{equation}
Now that $\mu_f$ is known, the norm is determined by requiring that
the zeroth moment of the shape function is \cite{Bosch:2004th}
\begin{eqnarray} \label{SFnorm}
   M_0(\hat\omega_0,\mu_i) &=& 1 - \frac{C_F \alpha_s(\mu_i)}{\pi} 
   \left( \ln^2 \frac{\mu_f}{\mu_i} + \ln \frac{\mu_f}{\mu_i} 
    + \frac{\pi^2}{24} \right)\nonumber\\
   &&+ \frac{C_F\alpha_s(\mu_i)}{\pi} \left( \ln\frac{\mu_f}{\mu_i} 
    - \frac12 \right) \frac{\mu_\pi^2(\mu_f,\mu_i)}{3\mu_f^2} + \dots \,.
\end{eqnarray}
It follows that $[M_0(\hat\omega_0,\mu_i)/M_0^{[F]}(\hat\omega_0)]\,
F(\hat\omega)$ serves as a model of $\hat S(\hat\omega,\mu_i)$ or 
$\hat\S(\hat\omega,\mu_i)$. 

We now suggest three two-parameter models for the leading-order
shape function based on an exponential-type function $F^{\rm (exp)}$, a
gaussian-type function $F^{\rm (gauss)}$, and hyperbolic-type function
$F^{\rm (hyp)}$.
We use two parameters that can be tuned to fit the photon spectrum: a
dimensionful quantity $\Lambda$ which coincides with the position of
the average $\langle \hat\omega \rangle$, and a positive number $b$
which governs the behavior for small $\hat\omega$. The functions we propose 
are
\begin{eqnarray}\label{SF:threeModels}
   F^{\rm (exp)}(\hat\omega;\Lambda,b)
   &=& \frac{N^{\rm (exp)}}{\Lambda}
    \left( \frac{\hat\omega}{\Lambda} \right)^{b-1}
    \exp\left( - d_{\rm (exp)} \frac{\hat\omega}{\Lambda} \right) , 
    \nonumber\\
   F^{\rm (gauss)}(\hat\omega;\Lambda,b)
   &=& \frac{N^{\rm (gauss)}}{\Lambda}
    \left( \frac{\hat\omega}{\Lambda} \right)^{b-1}
    \exp\left( - d_{\rm (gauss)} \frac{\hat\omega^2}{\Lambda^2} \right) , 
    \nonumber\\
   F^{\rm (hyp)}(\hat\omega;\Lambda,b)
   &=& \frac{N^{\rm (hyp)}}{\Lambda}
    \left( \frac{\hat\omega}{\Lambda} \right)^{b-1}
    \cosh^{-1}\left( d_{\rm (hyp)} \frac{\hat\omega}{\Lambda} \right) .
\end{eqnarray}
For convenience, we normalize these functions to unity. The parameters 
$d_{(i)}$ are determined by the choice $\Lambda=\langle\hat\omega\rangle$. 
We find
\begin{equation}
\begin{aligned}
   N^{\rm (exp)} &= \frac{d_{\rm (exp)}^b}{\Gamma(b)} \,, 
    & d_{\rm (exp)} &= b \,, \\
   N^{\rm (gauss)} &= \frac{2 d_{\rm (gauss)}^{b/2}}{\Gamma(b/2)} \,, 
    & d_{\rm (gauss)}
    &= \left(\frac{\Gamma(\frac{1+b}{2})}{\Gamma(\frac{b}{2})}\right)^2 \,,
    \\
   N^{\rm (hyp)} &= \frac{[4 d_{\rm (hyp)}]^b}{2\Gamma(b) 
    \left[ \zeta(b,\frac14) - \zeta(b,\frac34) \right]} \,, \quad
    & d_{\rm (hyp)}
    &= \frac{b}{4}\, 
    \frac{\zeta(1+b,\frac14) - \zeta(1+b,\frac34)}
         {\zeta(b,\frac14) - \zeta(b,\frac34)} \,,
\end{aligned}
\end{equation}
where $\zeta(b,a)=\sum_{k=0}^\infty (k+a)^{-b}$ is the generalized
Riemann zeta function.
An illustration of the different functional forms is given on the
left-hand side in Figure~\ref{fig:SFmodels}. We show a plot with the
choice $b=2$, corresponding to a linear onset for small $\hat\omega$. 

\begin{figure}
\begin{center}
\epsfig{file=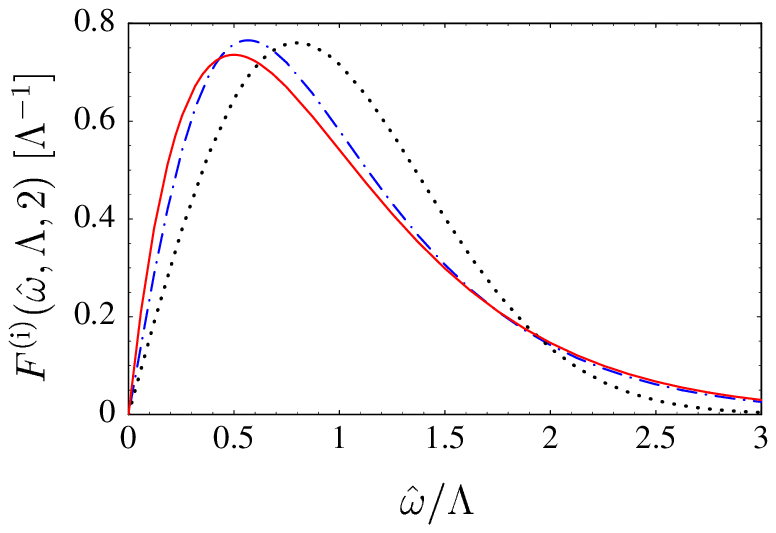, width=8cm}\hspace{3mm}%
\epsfig{file=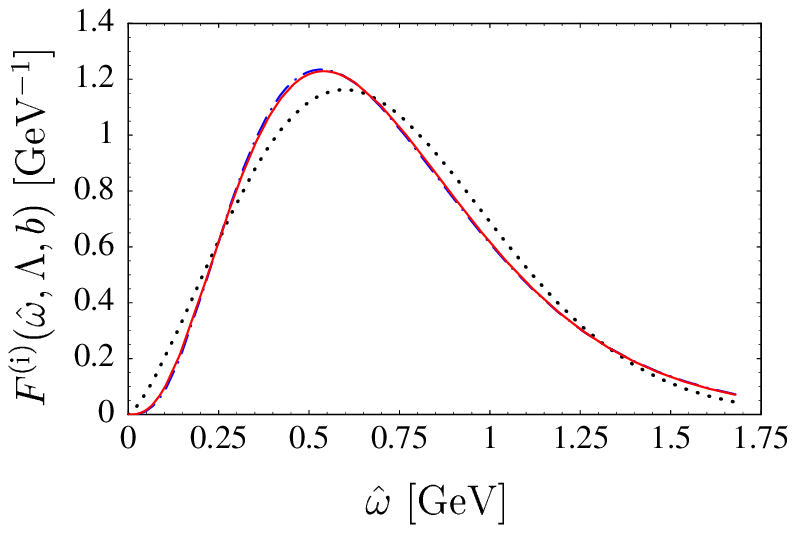, width=8cm}
\caption{\label{fig:SFmodels}
{\sc Top:} Different functional forms for the leading shape function. We 
show $F^{\rm (exp)}(\hat\omega,\Lambda,2)$ (solid), 
$F^{\rm (gauss)}(\hat\omega,\Lambda,2)$ (dotted), and 
$F^{\rm (hyp)}(\hat\omega,\Lambda,2)$ (dash-dotted) as functions of the 
ratio $\hat\omega/\Lambda$. 
{\sc Bottom:} The same functions with the parameters $\Lambda$ and $b$
tuned such that $m_b(\mu_*,\mu_*)=4.61$\,GeV and
$\mu_\pi^2(\mu_*,\mu_*) = 0.2$ GeV$^2$. See text for explanation.}
\end{center}
\end{figure}

For the first two models, analytic expressions for the HQET parameters
$\bar\Lambda$ and $\mu_\pi^2$ are available. Following the discussion
above, we compute the moments on the interval $[0,\hat\omega_0]$ and
find for the exponential form $F^{\rm (exp)}(\hat\omega;\Lambda,b)$
\begin{eqnarray}\label{parsExp}
   \bar\Lambda(\mu_f,\mu_i)
   &=& \frac{\Lambda}{b}\,\frac{\Gamma(1+b)
             - \Gamma(1+b,\frac{b\,\hat\omega_0}{\Lambda})}
            {\Gamma(b) - \Gamma(b,\frac{b\,\hat\omega_0}{\Lambda})} \,,
    \nonumber\\
   \mu_\pi^2(\mu_f,\mu_i)
   &=& 3 \left[ \frac{\Lambda^2}{b^2}\, 
    \frac{\Gamma(2+b) - \Gamma(2+b,\frac{b\,\hat\omega_0}{\Lambda})}
         {\Gamma(b) - \Gamma(b, \frac{b\,\hat\omega_0}{\Lambda})}\,
    - \bar\Lambda(\mu_f,\mu_i)^2 \right] ,
\end{eqnarray}
where $\mu_f=\hat\omega_0-\bar\Lambda(\mu_f,\mu_i)$. 

A similar calculation for the gaussian form $F^{\rm (gauss)}(\hat\omega;\Lambda,b)$ yields
\begin{eqnarray}\label{parsGauss}
   \bar\Lambda(\mu_f,\mu_i)
   &=& \frac{\Lambda}{\sqrt{d_{\rm (gauss)}}}\, 
    \frac{\Gamma(\frac{1+b}{2})
     - \Gamma(\frac{1+b}{2},\frac{d_{\rm (gauss)}\hat\omega_0^2}{\Lambda^2})}
         {\Gamma(\frac{b}{2})
     - \Gamma(\frac{b}{2},\frac{d_{\rm (gauss)}\hat\omega_0^2}{\Lambda^2})}
    \,, \nonumber\\
   \mu_\pi^2(\mu_f,\mu_i)
   &=& 3 \left[ \frac{\Lambda^2}{d_{\rm (gauss)}}\, 
    \frac{\Gamma(1+\frac{b}{2})
     - \Gamma(1+\frac{b}{2},\frac{d_{\rm (gauss)}\hat\omega_0^2}{\Lambda^2})}
    {\Gamma(\frac{b}{2})
     - \Gamma(\frac{b}{2},\frac{d_{\rm (gauss)}\hat\omega_0^2}{\Lambda^2})}
    - \bar\Lambda(\mu_f,\mu_i)^2 \right] .
\end{eqnarray}
The corresponding relations for $F^{\rm (hyp)}(\hat\omega;\Lambda,b)$ must 
be obtained numerically.

Ultimately the shape function should be fitted to the $\bar B\to X_s\gamma$
photon spectrum,
and the above equations then determine $\bar\Lambda$ and $\mu_\pi^2$. On
the other hand, these formulae can be inverted to determine
$\Lambda$ and $b$ from the current values of the HQET parameters. For
example, if we adopt the values $m_b(\mu_*,\mu_*)=4.61$\,GeV 
and $\mu_\pi^2(\mu_*,\mu_*)=0.20$\,GeV$^2$ for the parameters in 
(\ref{eq:atMuStar}) at $\mu_*=1.5$\,GeV, then we find the
parameter pair $\Lambda\approx 0.72$\,GeV, $b\approx 3.95$ for the
exponential model, $\Lambda\approx 0.71$\,GeV, $b\approx 2.36$ for
the gaussian model, and $\Lambda\approx 0.73$\,GeV, $b\approx 3.81$
for the hyperbolic model. On the right-hand side of
Figure~\ref{fig:SFmodels} we show these three different functions
plotted on the interval $[0,\hat\omega_0]$ over which the moment 
constraints are imposed. While the exponential
(solid) and hyperbolic (dash-dotted) curves are barely
distinguishable, the gaussian model has quite different
characteristics. It is broader, steeper at the onset, faster to
fall off, and the maximum is shifted toward larger $\hat\omega$. 

An important comment is that, 
once a two-parameter ansatz is employed, the shape-function parameters (i.e., 
$m_b$ and $\mu_\pi^2$) can either be determined from a fit to the entire 
photon spectrum, or to the first two moments of the spectrum. Both 
methods are equivalent and should yield consistent results. If they do not, 
it would be necessary to refine the ansatz for the functional form of the 
shape function. 

In most applications shape functions are needed for arguments
$\hat\omega$ of order $\Lambda_{\rm QCD}$. However, in some cases, like
the ideal cut on hadronic invariant mass, $\hat\omega$ is required to
be as large as $M_D$, which is much larger than $\Lambda_{\rm QCD}$. 
The large-$\hat\omega$ behavior of the shape functions can be computed
in a model-independent way using short-distance methods. 
For the leading shape function, one finds \cite{Bosch:2004th}
\begin{equation}\label{SFtail}
   \hat S(\hat\omega\gg\Lambda_{\rm QCD},\mu_i)
   = - \frac{C_F\alpha_s(\mu_i)}{\pi} \frac{1}{\hat\omega - \bar\Lambda} 
   \left( 2\ln\frac{\hat\omega-\bar\Lambda}{\mu_i} + 1 \right) 
   + \dots \,.
\end{equation}
Note that this radiative tail is negative, implying that 
the shape function must go through zero somewhere near $\hat\omega
\sim \mbox{few}\,\Lambda_{\rm QCD}$. For practical purposes, we ``glue'' the 
above expression onto models of the non-perturbative shape function starting
at $\hat\omega=\bar\Lambda+\mu_i/\sqrt{e}\approx 1.6$\,GeV, where the tail 
piece vanishes. In this way we obtain a continuous shape-function model with 
the correct asymptotic behavior. We stress that for applications with a 
maximal $P_+$ not larger than about 1.6\,GeV the radiative tail of the shape 
function is never required. This includes all methods for extracting 
$|V_{ub}|$ discussed later in this chapter, except for the case of a cut on 
hadronic invariant mass, $M_X\le M_0$, if $M_0$ is above 1.6\,GeV.

\subsection{Models for subleading shape functions}
\label{sec:modelSSF}

In the last section we have been guided by the fact that the 
$\bar B\to X_s\gamma$ photon spectrum is at leading power directly determined 
by the leading shape function. This helped in finding
models that have roughly the same shape as the photon spectrum. At the
subleading level considered here, however, no such guidance is
provided to us. The available information is limited to the tree-level
moment relations (\ref{SSF:moments}), stating that the norms of the subleading 
shape functions vanish while their first moments do not. 
In \cite{Bosch:2004cb}, two classes of models have been proposed, in which
the subleading shape functions are ``derived'' from the
leading shape function. A particularly simple choice is
\begin{equation}\label{SSF:model1}
   \hat t(\hat\omega) = - \lambda_2\,\hat S'(\hat\omega) \,, \qquad
   \hat u(\hat\omega) = \frac{2\lambda_1}{3}\,\hat S'(\hat\omega) \,, \qquad
   \hat v(\hat\omega) = \lambda_2\,\hat S'(\hat\omega) \,.
\end{equation}
Below, we will sometimes refer to this set of functions as the ``default
choice''. We choose the parameter $-\lambda_1$ in the expression for 
$\hat u(\hat\omega)$  (as well as in the expressions for the second-order 
hadronic power corrections) to coincide with the quantity 
$\mu_\pi^2(\mu_f,\mu_i)$ 
given in (\ref{parsExp}) and (\ref{parsGauss}). However, for consistency with 
the tree-level moment relations, we identity the parameter $\bar\Lambda$ in 
(\ref{BsgSSF2}) and (\ref{upsy}) with the quantity $\bar\Lambda(\mu_f,\mu_i)$ 
evaluated in the limit where $\omega_0\to\infty$. This implies 
$\bar\Lambda=\Lambda$ for all three types of functions and ensures that the 
subleading shape functions have zero norm when integrated over 
$0\le\hat\omega<\infty$. 

There are of course infinitely many possibilities to find models for
subleading shape functions that are in accordance with (\ref{SSF:moments}). 
Any function with vanishing norm and first moment can be arbitrarily added to 
any model for a subleading shape function without violating the moment 
relations. Several such functions have been proposed in recent work on 
subleading shape functions, see e.g.\
\cite{Bosch:2004cb,Neubert:2002yx,Beneke:2004in,Neubert:2004cu}.
Specifically, we define the functions
\begin{eqnarray} \label{hfuncs}
   h_1(\hat\omega)
   &=& \frac{M_2}{N\Omega_0^3}\,\frac{a^{a+1}}{2\Gamma(a)}\,z^{a-1}\,e^{-az}
    \left( \frac{a-1}{z} - a(2-z) \right) , \nonumber\\
   h_2(\hat\omega)
   &=& \frac{M_2}{N\Omega_0^3}\,\frac{a^3}{2}\,e^{-az}
    \left( 1 - 2az + \frac{a^2 z^2}{2} \right) , \nonumber \\ 
   h_3(\hat\omega)
   &=& \frac{M_2}{N\Omega_0^3} \left\{
    \left[ \frac{2\sqrt{\pi\,a}}{\pi-2}\,e^{-a z^2} 
    \left( 1 - 2z\sqrt{\frac a\pi} \right) \right]
    - 2e^{-z} + 2z\,e^{-2z}\,\mbox{Ei}\,(z) \right\} , \nonumber\\
   h_4(\hat\omega)
   &=& \frac{M_2}{N\Omega_0^3} \left\{
    \left[ \frac{\pi^2}{4}\,\frac{2\sqrt{\pi\,a}}{\pi-2}\,e^{-a z^2}
    \left( 1 - 2z\sqrt{\frac a\pi} \right) \right] \right. \nonumber\\
   &&\hspace{1.5cm} \left.
    + \frac 8 {\left(1+z^2\right)^4}\,
    \bigg[ z\ln z+ \frac z2\,(1+z^2) - \frac{\pi}{4}\,
    (1-z^2) \bigg] \right\} ,
\end{eqnarray}
where $z=\hat\omega/\Omega_0$, and the reference quantity 
$\Omega_0=O(\Lambda_{\rm QCD})$ depends on the type of function, namely 
$\Omega_0=\bar\Lambda$ for $h_1$ and $h_2$, $\Omega_0=\frac23\,\bar\Lambda$
for $h_3$, and $\Omega_0=\frac{8}{3\pi}\,\bar\Lambda$ for 
$h_4$. The quantity $a$ is a free parameter. The functions 
(\ref{hfuncs}) have by construction vanishing norm and first moment. 
Their second moments are given by the parameter $M_2$, provided the 
normalization constants are chosen as $N=1$ for $h_1$ and $h_2$, and 
\begin{equation}
   N = 1 - \frac{4-\pi}{2(\pi-2)}\,\frac{1}{a} \,, \qquad
   N = 1 - \frac{\pi^2(4-\pi)}{8(\pi-2)}\,\frac{1}{a}
\end{equation}
for  $h_3$ and  $h_4$, respectively. The values for the 
parameters $a$ and $M_2$ should be chosen such that the following 
characteristics of subleading shape functions are respected: First, they
are dimensionless functions, so that their values are naturally of
$O(1)$ for $\hat\omega\sim\Lambda_{\rm QCD}$. Secondly, when integrated 
over a sufficiently large domain, their contributions are
determined in terms of their first few moments. In particular, this 
implies that for values of $\hat\omega\gg\Lambda_{\rm QCD}$
the integrals over the subleading shape functions must approach zero.
Taking these considerations into account, we use $M_2=(0.3\,\mbox{GeV})^3$ 
in all cases and choose $a=3.5$ for $h_1$, $a=5$ for $h_2$, and $a=10$ for 
$h_3$ and $h_4$.

\begin{figure}
\begin{center}
\epsfig{file=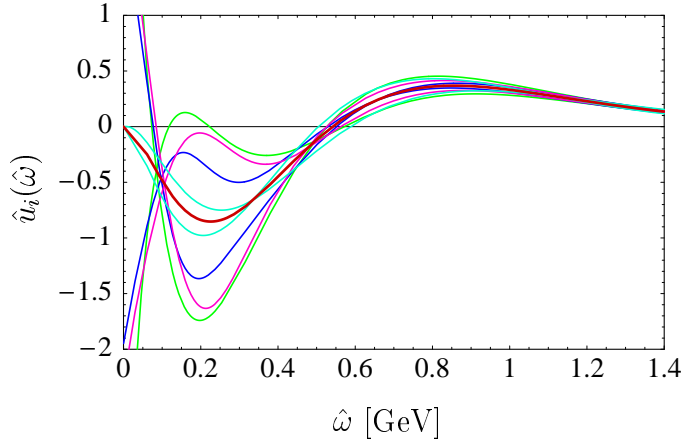, width=9cm}
\caption{\label{fig:SSFmodels}
Nine models for the subleading shape function $\hat u(\hat\omega)$ obtained 
by adding or subtracting one of the four functions $h_n(\hat\omega)$ to the 
default model in (\ref{SSF:model1}), shown as a thick line. See text for 
explanation.}
\end{center}
\end{figure}

Given the four functions (\ref{hfuncs}), we can construct several new
models for the subleading shape functions $\hat t(\hat\omega)$, 
$\hat u(\hat\omega)$, and $\hat v(\hat\omega)$. For each function, we 
construct a set of 9 models by adding or subtracting any of the functions 
$h_n(\hat\omega)$ to the default choice in (\ref{SSF:model1}). Together, this 
method yields $9^3=729$ different sets 
$\{\hat t_i(\hat\omega), \hat u_j(\hat\omega), \hat v_k(\hat\omega)\}$ with
$i,j,k=1,\dots,9$. 
This large collection of functions will be used to estimate the
hadronic uncertainties in our predictions for partial decay rates. Note 
that for most of these sets we no longer have 
$\hat t_i(\hat\omega)=-\hat v_k(\hat\omega)$, which was an ``accidental'' 
feature of the default model (\ref{SSF:model1}). 
The fact that the two functions have equal (but opposite in sign) first 
moments does not imply that their higher moments should also be related to 
each other.

For the case of $\hat u(\hat\omega)$ the resulting functions are shown in 
Figure~\ref{fig:SSFmodels}, where we have used the exponential model 
(\ref{SF:threeModels}) with parameters $\Lambda=0.72$\,GeV and $b=3.95$
for the leading shape function. In the region
$\hat\omega \sim \Lambda_{\rm QCD}$ they differ dramatically from each
other, while the large $\hat\omega$ dependence is dominated by the
moment relations (\ref{SSF:moments}). 

\subsection{Illustrative studies}

We stressed several times that the calculation of the hadronic tensor is 
``optimized'' for the shape-function region of large $P_-$ and small $P_+$, 
while it can smoothly be extended over the entire phase space. The notions
``large $P_-$'' and ``small $P_+$'' are to be understood as the sizes of 
integration domains for $P_-$ and $P_+$. Only when 
the differential distributions are integrated over a sufficiently 
large region in phase space, global quark-hadron duality ensures that the 
partonic description used in the present chapter matches the true, hadronic 
distributions with good accuracy. A more ambitious goal would be to 
calculate the differential decay rate point by point in the $(P_+,P_-)$ plane.
This can be done invoking local quark-hadron duality, as long as there is a 
sufficiently large number of hadronic final states contributing to 
the rate at any given point in phase space. 

\begin{figure}
\begin{center}
\begin{minipage}{7.8cm}
\epsfig{file=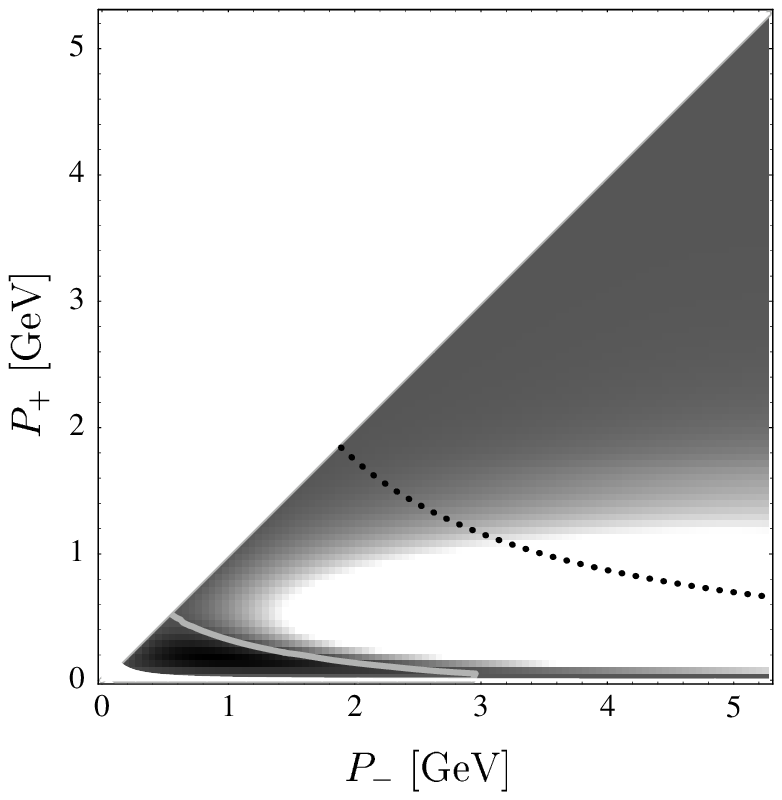, width=7.65cm}
\end{minipage}\hspace{5mm}
\begin{minipage}{8cm}
\epsfig{file=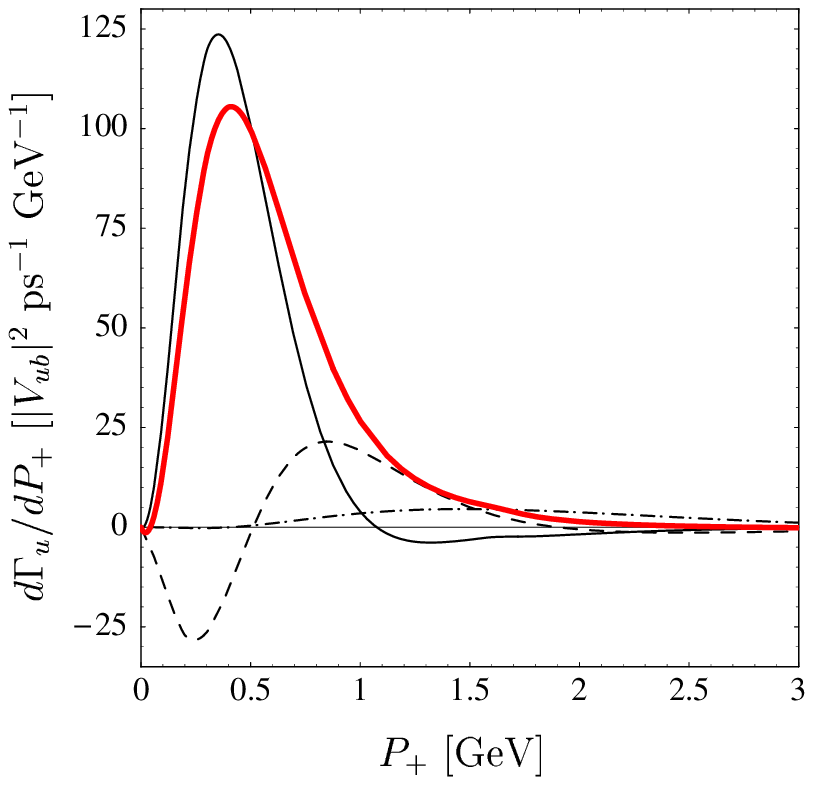, width=8.0cm}
\end{minipage}
\caption{\label{fig:d2Gamma}
{\sc Top:} Theoretical prediction for the double differential decay rate. 
The light area represents a large decay rate. Black regions
denote areas where the decay rate is close to zero. The dotted line is
given by $P_+ P_- = M_D^2$, which means that charm background is
located in the upper wedge. See text for further explanation.  
{\sc Bottom:} The $P_+$ spectrum extended to large values of $P_+$. The
thin solid line denotes the leading-power prediction, the dashed line
depicts first-order power corrections, the dash-dotted line shows 
second-order power corrections, and the thick solid line is their sum.}
\end{center}
\end{figure}

It is instructive to integrate the triple differential decay rate 
(\ref{eq:tripleRate}) over the leptonic variable $P_l$ in the range 
$P_+\le P_l\le P_-$, which yields the exact formula
\begin{eqnarray}\label{eq:doubleRate}
   \frac{d^2\Gamma_u}{dP_+\,dP_-}
   &=& \frac{G_F^2|V_{ub}|^2}{96\pi^3}\,U_y(\mu_h,\mu_i)\,
    (M_B-P_+)(P_- -P_+)^2 \nonumber\\
   &&\times \Big\{ (3M_B-2P_- -P_+)\,\F_1
    + 6 (M_B-P_-)\,\F_2 + (P_- -P_+)\,\F_3 \Big\} \,. \nonumber\\
\end{eqnarray}
Our theoretical prediction for the double differential decay rate 
(\ref{eq:doubleRate}) is shown on the left-hand side of 
Figure~\ref{fig:d2Gamma}. We use the exponential model for the leading shape
function with parameters $m_b(\mu_*,\mu_*)=4.61$\,GeV and 
$\mu_\pi^2(\mu_*,\mu_*)=0.2$\,GeV$^2$, as well as the default choice 
(\ref{SSF:model1}) for the subleading shape functions. For very small 
$P_-$ values the rate turns negative (to the left of the gray 
line in the figure), signaling a breakdown of quark-hadron duality. 
It is reassuring that the only region where this happens is the 
``resonance region'', where the hadronic invariant mass is of order 
$\Lambda_{\rm QCD}$, and local duality breaks down.

Another useful quantity to consider is the differential $P_+$ rate,
which is obtained by integrating the double differential rate over $P_-$ in 
the range $P_+\le P_-\le M_B$. The resulting $P_+$ spectrum 
is shown on the right-hand side of Figure~\ref{fig:d2Gamma}. In the plot
we also disentangle the contributions from different orders in power
counting.

\section{Predictions and error estimates for partial rates}
\label{sec:analysio}

Before discussing predictions for partial $\bar B\to X_u\,l^-\bar\nu$ rates
for various kinematical cuts, let us recapitulate the ingredients of
the calculation and general procedure. We have presented expressions
for the triple differential decay rate, which can be organized in an expansion
in inverse powers of $(M_B-P_+)$. The leading-power contribution is given
at next-to-leading order in renormalization-group 
improved perturbation theory. At first
subleading power two contributions arise. The first type involves
subleading shape functions and is included at tree level, while the
second type contributes perturbative corrections of order $\alpha_s$ 
that come with the leading shape function. Further
contributions enter at second subleading power and are again of 
the two types: perturbative corrections of order $\alpha_s$ and
non-perturbative structures at tree level. In summary, then, partial rates 
can be computed term by term in an expansion of the form
\begin{equation}\label{partialRates}
   \Gamma_u = \Gamma_u^{\rm (0)}
   + (\Gamma_u^{\rm kin(1)} + \Gamma_u^{\rm hadr(1)})
   + (\Gamma_u^{\rm kin(2)} + \Gamma_u^{\rm hadr(2)}) + \dots \,.
\end{equation}
The goal of this section is to test the convergence of this series expansion 
and to 
perform a thorough analysis of uncertainties. For the kinematical corrections
$\Gamma_u^{{\rm kin}(n)}$ the sum of all terms is known and given by the 
expressions in (\ref{eq:fullDFN}), while the first two terms in the series 
correspond to the expanded results in (\ref{eq:kinNLO}) and 
(\ref{eq:kinNNLO}). We will find that in all cases of interest the first two 
terms give an excellent approximation to the exact result for 
$\Gamma_u^{\rm kin}$.

For the purpose of illustration, we adopt the exponential model for the shape 
function and present numerical results for two sets of input parameters, which
are biased by the results deduced from fits to $\bar B\to X_c\,l^-\bar\nu$ 
moments \cite{Neubert:2004sp}. Specifically, we use
$m_b(\mu_*,\mu_*)=4.61$\,GeV, $\mu_\pi^2(\mu_*,\mu_*)=0.2$\,GeV$^2$ (set~1) 
and $m_b(\mu_*,\mu_*)=4.55$\,GeV, $\mu_\pi^2(\mu_*,\mu_*)=0.3$\,GeV$^2$ 
(set~2). The values of the $b$-quark mass coincide with those obtained at 
two-loop and one-loop order in \cite{Neubert:2004sp} (see also the discussion 
below), while the values of $\mu_\pi^2$ are close to the corresponding values
in that reference. As was mentioned before, in the future 
the leading shape function $\hat\S(\hat\omega,\mu_i)$ 
should be extracted from a fit to the $\bar B\to X_s\gamma$ photon
spectrum, in which case the uncertainty in its shape becomes an experimental 
error, which can be systematically reduced with improved data. In the process, 
the ``theoretically preferred'' parameter values used in the present work 
will be replaced with the ``true'' values extracted directly from data.
While this will change the central values for the partial rates, our 
estimates of the theoretical errors will only be affected marginally. 

The different sources of theoretical uncertainties are as follows:
First, there are uncertainties associated with the functional forms
of the subleading shape functions. To estimate them, we take the spread of
results obtained when using the large set of different models described in 
Section~\ref{sec:modelSSF}, while the central value for a partial decay rate
corresponds to the default model (\ref{SSF:model1}). Secondly, there are 
perturbative uncertainties associated with the choice of the matching scales
$\mu_h$, $\mu_i$, and $\bar\mu$. Decay rates are formally independent
of these scales, but a residual dependence remains because of the truncation 
of the perturbative series. Our error analysis is as follows:
\begin{itemize}
\item 
The hard scale $\mu_h$ is of order $m_b$. In perturbative
logarithms the scale appears in the combination $(y m_b/\mu_h)$, see
e.g.\ (\ref{eq:LeadHardFunction}). To set a
central value for $\mu_h$ we are guided by the average $\langle
y\rangle m_b$. The leading term for the double differential decay rate
$d^2\Gamma_u /d P_+ dy$ is proportional to $2y^2(3-2y)$. It follows
that the average $y$ on the interval $[0,1]$ is 0.7. 
However, in some applications $y$ is not integrated over the full domain. 
Also, there are large negative constants in the matching correction $H_{u1}$ 
in (\ref{eq:LeadHardFunction}), whose effect can be ameliorated by lowering
the scale further. In the error analysis we use the central value 
of $\mu_h=m_b/2\approx 2.3$\,GeV and vary
the scale by a factor between $1/\sqrt2$ and $\sqrt2$. For the central value 
$\alpha_s(\mu_h)\approx 0.286$.
\item 
The intermediate scale $\mu_i\sim\sqrt{m_b \Lambda_{\rm QCD}}$ serves as the 
renormalization point for the jet and shape functions. 
We fix this scale to $\mu_i=1.5$\,GeV. Variations of $\mu_i$ would
affect both the normalization and the functional form of the shape function, 
as determined by the solution to the renormalization-group equation for the
shape function discussed in \cite{Bosch:2004th,Neubert:2004dd}. In practice, 
effects on the shape are irrelevant because the shape function is fitted to
data. The only place where the intermediate scale
has a direct impact on the extraction of $|V_{ub}|$ is through the
normalization of the shape function (\ref{SFnorm}). 
In the analysis we therefore estimate the uncertainty by assigning the
value $\pm (\frac{\alpha_s(\mu_i)}{\pi})^2$ as a relative error, where
$\alpha_s(\mu_i)\approx 0.354$. 
\item 
The scale $\bar\mu$ appears as the argument of $\alpha_s$ in the
perturbative contributions $\Gamma_u^{\rm kin}$. We vary $\bar\mu$ from 
$\mu_i/\sqrt{2}$ to $\sqrt{2}\mu_i$ with the central value 
$\bar\mu=\mu_i=1.5$\,GeV.
\end{itemize}
These three errors are added in quadrature and assigned as the total 
perturbative uncertainty. Finally, we need to estimate the effects from 
higher-dimensional operators at third and higher-order in power counting.  
If the considered cut includes the region of phase space near the origin 
($P_+\sim P_-\sim\Lambda_{\rm QCD}$), then the dominant such 
contributions are weak annihilation effects, which we have discussed in 
Section~\ref{Sec:WA}. From the analysis in \cite{TomsThesis} one can derive 
a bound on the weak annihilation contribution that is $\pm$1.8\% of the total 
decay rate, for which we take $\Gamma_u\approx 70\,|V_{ub}|^2\,{\rm ps}^{-1}$ 
(see below). The resulting uncertainty 
$\delta\Gamma_u^{\rm WA}=\pm 1.3\,|V_{ub}|^2\,{\rm ps}^{-1}$ affects all 
partial rates which include the region near the origin in the $(P_+,P_-)$ 
plane. The uncertainty from weak annihilation can be avoided by
imposing a cut $q^2\le q_{\rm max}^2$ (see Section~\ref{sec:eliminateWA}).
For all observables considered in the present work, other 
power corrections of order $1/m_b^3$ 
can be safely neglected. This can be seen by multiplying the contributions
from second-order hadronic power corrections to the various 
decay rates (called $\Gamma_u^{\rm hadr(2)}$) by an additional suppression 
factor $\Lambda_{\rm QCD}/m_b\sim 0.1$.

The following subsection contains a discussion of the total decay rate. In the
remainder of this section we then present predictions for a variety of 
kinematical cuts designed to eliminate (or reduce) the charm background. 
These partial rates can be computed either numerically or, in many cases, 
semi-analytically. In Appendix~\ref{appendix:B} we discuss how to perform the
integrations over the kinematical variables $P_l$ and $P_-$ analytically.

\subsection{Total decay rate}

Before presenting our predictions for the various partial decay rates, it
is useful to have an expression for the total $\bar B\to X_u\,l^-\bar\nu$ 
decay rate expressed in terms of the heavy-quark parameters defined in the 
shape-function scheme. We start from the exact two-loop expression for the
total rate derived in \cite{vanRitbergen:1999gs}, add the second-order 
hadronic power corrections, which are known at tree level 
\cite{Blok:1993va,Manohar:1993qn}, 
and finally convert the parameters $m_b$ and $\lambda_1$ from the pole scheme
to the shape-function scheme. The relevant replacements at two-loop order
can be taken from \cite{Neubert:2004sp} and read
\begin{eqnarray}
   m_b^{\rm pole} &=& m_b + 0.424\mu_*\alpha_s(\mu) \left[
    1 + \left( 1.357 + 1.326\ln\frac{\mu}{\mu_*}
    + 0.182\,\frac{\mu_\pi^2}{\mu_*^2} \right) \alpha_s(\mu) \right]
   \nonumber\\
   &&\mbox{}+ \frac{3\lambda_2-\mu_\pi^2-0.330\mu_*^2\alpha_s^2(\mu)}{2m_b}
    + \dots \,, \nonumber\\
   -\lambda_1 &=& \mu_\pi^2 + 0.330\mu_*^2\alpha_s^2(\mu) + \dots \,,
\end{eqnarray}
where here and from now on $m_b\equiv m_b(\mu_*,\mu_*)$ and
$\mu_\pi^2\equiv\mu_\pi^2(\mu_*,\mu_*)$ are defined in the shape-function 
scheme. At a reference scale $\mu_*=1.5$\,GeV the values of these 
parameters have been determined to be $m_b=(4.61\pm 0.08)$\,GeV and 
$\mu_\pi^2=(0.15\pm 0.07)$\,GeV$^2$ \cite{Neubert:2004sp},\footnote{The values 
obtained from a one-loop analysis are $m_b=(4.55\pm 0.08)$\,GeV and 
$\mu_\pi^2=(0.34\pm 0.07)$\,GeV$^2$.} 
where we account for the small $1/m_b$ correction to 
the relation for the pole mass in the above formula (corresponding to a 
shift of about $-0.02$\,GeV in $m_b$), which was not included in that paper.

The resulting expression for the total decay rate is
\begin{eqnarray}\label{sonice}
   \Gamma_u
   &=& \frac{G_F^2 |V_{ub}|^2 m_b^5}{192\pi^3}\,\Bigg\{
    1 + \alpha_s(\mu) \left( -0.768 + 2.122\,\frac{\mu_*}{m_b} \right)
    \nonumber\\
   &&\mbox{}+ \alpha_s^2(\mu) \left[ - 2.158 + 1.019\ln\frac{m_b}{\mu}
    + \left( 1.249 + 2.814\ln\frac{\mu}{\mu_*}
    + 0.386\,\frac{\mu_\pi^2}{\mu_*^2} \right) \frac{\mu_*}{m_b}\right.\nonumber\\
    &&\left.+ 0.811\,\frac{\mu_*^2}{m_b^2} \right] 
   - \frac{3(\mu_\pi^2-\lambda_2)}{m_b^2} + \dots \Bigg\} \,.
\end{eqnarray}
We observe that for $\mu_*\approx 1.5$\,GeV and $\mu=O(m_b)$, the perturbative 
expansion coefficients are strongly reduced compared to their values in the 
pole scheme ($-0.768$ and $-2.158$, respectively), indicating a vastly 
improved convergence of the perturbative expansion. For  
$m_b=\mu=4.61$\,GeV, and $\mu_\pi^2=0.15$\,GeV$^2$ we obtain for the 
one-loop, two-loop, and power corrections inside the brackets in 
(\ref{sonice}): $\{1 - 0.017 - 0.030 - 0.004\}$. All of these are very small
corrections to the leading term.

Including the uncertainties in the values of $m_b$ and $\mu_\pi^2$ quoted
above, and varying the renormalization scale $\mu$ between $m_b$ and $m_b/2$ 
(with a central value of $m_b/\sqrt2$), we get
\begin{eqnarray}\label{totrate}
   \frac{\Gamma_u}{|V_{ub}|^2\,\mbox{ps}^{-1}}
   &=& 68.0_{\,-5.5}^{\,+5.9}\,[m_b]\,\mp 0.7\,[\mu_\pi^2]\,
   {}_{\,-0.9}^{\,+0.6}\,[\mu] \nonumber\\
   &=& \left( 68.0\mp 0.7\,[\mu_\pi^2]\,{}_{\,-0.9}^{\,+0.6}\,[\mu] \right)
   \left( \frac{m_b}{4.61\,\mbox{GeV}} \right)^{4.81} .
\end{eqnarray}
Here and below, we quote values for decay rates in units of 
$|V_{ub}|^2\,{\rm ps}^{-1}$. To convert these results to partial branching 
fractions the numbers need to be multiplied by the average $B$-meson lifetime. 
Without including the two-loop corrections, the central value in the above 
estimate increases to 70.6. For comparison, with the same set of input 
parameters our new approach based on (\ref{eq:tripleRate}) predicts a total 
decay rate of 
$\Gamma_u=(71.4_{\,-5.0}^{\,+6.2}\pm 0.5)\,|V_{ub}|^2\,\mbox{ps}^{-1}$, where 
the first error accounts for perturbative uncertainties while the second one 
refers to the modeling of subleading shape functions (to which there is
essentially no sensitivity at all in the total rate). The fact that this 
is in excellent agreement with the direct calculation using (\ref{sonice}) 
supports the notion that the formalism developed in this chapter can be used to 
describe arbitrary $\bar B\to X_u\,l^-\bar\nu$ decay distributions, both in 
the shape-function region and in the OPE region of phase space.

\subsection{Cut on charged-lepton energy}

\begin{table}
\begin{center}
\caption{\label{tab:LeptoInB}
Partial decay rate $\Gamma_u(E_0)$ for a cut on charged-lepton energy 
$E_l>E_0$ in the $B$-meson rest frame, given in units of 
$|V_{ub}|^2\,\mbox{ps}^{-1}$. Predictions are based on the shape-function 
parameter values $m_b=4.61$\,GeV, $\mu_\pi^2=0.2$\,GeV$^2$ (top) and 
$m_b=4.55$\,GeV, $\mu_\pi^2=0.3$\,GeV$^2$ (bottom).} 
\vspace{5mm}
\begin{tabular}{|c|c|c|c|c|} 
\hline
$E_0$ [GeV] & Mean & Subl.\ SF & Pert.\ & Total \\ 
\hline\hline
1.9 & 24.79 & $\pm$0.53 & $^{+1.90}_{-1.66}$ & $^{+2.34}_{-2.15}$ \\
2.0 & 18.92 & $\pm$0.60 & $^{+1.35}_{-1.20}$ & $^{+1.95}_{-1.84}$ \\
2.1 & 13.07 & $\pm$0.71 & $^{+0.82}_{-0.75}$ & $^{+1.66}_{-1.63}$ \\
2.2 &  7.59 & $\pm$0.81 & $^{+0.38}_{-0.34}$ & $^{+1.55}_{-1.54}$ \\
2.3 &  3.12 & $\pm$0.89 & $^{+0.15}_{-0.16}$ & $^{+1.55}_{-1.55}$ \\
2.4 &  0.42 & $\pm$1.05 & $^{+0.16}_{-0.22}$ & $^{+1.65}_{-1.65}$ \\
\hline
\hline
1.9 & 21.10 & $\pm$0.53 & $^{+1.57}_{-1.35}$ & $^{+2.08}_{-1.92}$ \\
2.0 & 15.83 & $\pm$0.60 & $^{+1.08}_{-0.94}$ & $^{+1.77}_{-1.68}$ \\
2.1 & 10.73 & $\pm$0.68 & $^{+0.64}_{-0.55}$ & $^{+1.57}_{-1.54}$ \\
2.2 &  6.12 & $\pm$0.74 & $^{+0.31}_{-0.23}$ & $^{+1.50}_{-1.48}$ \\
2.3 &  2.47 & $\pm$0.84 & $^{+0.17}_{-0.22}$ & $^{+1.53}_{-1.53}$ \\
2.4 &  0.29 & $\pm$0.99 & $^{+0.18}_{-0.24}$ & $^{+1.61}_{-1.62}$ \\
\hline
\end{tabular}
\end{center}
\end{table}

Traditionally, the most common variable to discriminate against the charm 
background is the charged-lepton energy $E_l$. As long as one requires that 
$E_l$ is bigger than $(M_B^2-M_D^2)/2M_B\approx 2.31$\,GeV, the final hadronic 
state cannot have an invariant mass larger than $M_D$. For this ideal cut, and 
using the default set of subleading shape functions, we find
\begin{equation}
\begin{array}{lllllll}
   & \Gamma_u^{\rm (0)}
   & +\,\, (\Gamma_u^{\rm kin(1)} & +\,\, \Gamma_u^{\rm hadr(1)}) 
   & +\,\, (\Gamma_u^{\rm kin(2)} & +\,\, \Gamma_u^{\rm hadr(2)}) & \\
   = \big[ & 6.810 & +\,\, (0.444 & -\,\, 3.967)
   & +\,\, (0.042 & -\,\, 0.555) \big] 
   & |V_{ub}|^2\,{\rm ps}^{-1} \,.
\end{array}
\end{equation}
The corrections from subleading shape functions are quite sizable, in 
accordance with the findings in
\cite{Bauer:2002yu,Leibovich:2002ys,Neubert:2002yx}. Note that 
the sum $\Gamma_u^{\rm kin(1)}+\Gamma_u^{\rm kin(2)}=0.486$ is an excellent
approximation to the exact result $\Gamma_u^{\rm kin}=0.482$ (all 
in units of $|V_{ub}|^2\,{\rm ps}^{-1}$) obtained
using (\ref{eq:fullDFN}), indicating that the expansion of the 
kinematical power corrections is converging rapidly. The same will be true
for all other observables considered below.

In practice, the cut on $E_l$ can be relaxed to some extent because the
background is well understood, thereby increasing the efficiency and reducing
the impact of theoretical uncertainties.
Our findings for different values of the cut $E_0$ are summarized in 
Table~\ref{tab:LeptoInB}. Here and below, the columns have the following 
meaning: ``Mean'' denotes the
prediction for the partial decay rate, ``Subl.\ SF'' the
uncertainty from subleading shape functions, and ``Pert.'' the total
perturbative uncertainty. In the column ``Total'' we add the
stated errors plus the uncertainty from weak annihilation in quadrature. 

\begin{table}
\begin{center}
\caption{\label{tab:LeptoInUps}
Same as Table~\ref{tab:LeptoInB}, but for the partial decay rate 
$\gamma\,\Gamma_u^{(\Upsilon)}(E_0)$ for a cut on lepton energy $E_l>E_0$ in 
the $\Upsilon(4S)$ rest frame.} 
\vspace{5mm}
\begin{tabular}{|c|c|c|c|c|} 
\hline
$E_0$ [GeV] & Mean & Subl.\ SF & Pert.\ & Total \\ 
\hline\hline
1.9 & 24.82 & $\pm$0.54 & $^{+1.91}_{-1.66}$ & $^{+2.35}_{-2.15}$ \\
2.0 & 19.00 & $\pm$0.61 & $^{+1.37}_{-1.21}$ & $^{+1.96}_{-1.85}$ \\
2.1 & 13.25 & $\pm$0.71 & $^{+0.85}_{-0.76}$ & $^{+1.68}_{-1.63}$ \\
2.2 &  7.99 & $\pm$0.78 & $^{+0.42}_{-0.37}$ & $^{+1.54}_{-1.53}$ \\
2.3 &  3.83 & $\pm$0.86 & $^{+0.18}_{-0.13}$ & $^{+1.54}_{-1.53}$ \\
2.4 &  1.31 & $\pm$0.99 & $^{+0.10}_{-0.14}$ & $^{+1.61}_{-1.61}$ \\
\hline
\hline
1.9 & 21.16 & $\pm$0.54 & $^{+1.58}_{-1.35}$ & $^{+2.09}_{-1.93}$ \\
2.0 & 15.94 & $\pm$0.60 & $^{+1.10}_{-0.95}$ & $^{+1.78}_{-1.69}$ \\
2.1 & 10.94 & $\pm$0.68 & $^{+0.66}_{-0.57}$ & $^{+1.58}_{-1.54}$ \\
2.2 &  6.49 & $\pm$0.74 & $^{+0.34}_{-0.26}$ & $^{+1.50}_{-1.48}$ \\
2.3 &  3.05 & $\pm$0.84 & $^{+0.17}_{-0.18}$ & $^{+1.53}_{-1.53}$ \\
2.4 &  0.98 & $\pm$0.92 & $^{+0.13}_{-0.18}$ & $^{+1.56}_{-1.57}$ \\
\hline
\end{tabular}
\end{center}
\end{table}

Experiments often do not measure the partial rates in the $B$-meson
rest frame, but in the rest frame of the $\Upsilon(4S)$ resonance produced 
in $e^+ e^-$ collisions. Boosting to the
$\Upsilon(4S)$ frame with $\beta=v/c\approx 0.064$ has a small
effect on the spectrum and rates. The exact formula for this boost is 
\cite{Kagan:1998ym}
\begin{equation}
   \gamma\,\Gamma_u^{(\Upsilon)}(E_0) = \frac{1}{\beta_+ -\beta_-} 
   \int_{\beta_- E_0}^{M_B/2}\!dE\,\frac{d\Gamma_u^{(B)}}{dE} 
   \left[ \beta_+ - \mbox{max}\left( \beta_-,\frac{E_0}{E} \right) 
   \right] ,
\end{equation}
where $\beta_\pm=\sqrt{1\pm\beta}/\sqrt{1\mp\beta}$, and the factor
$\gamma=1/\sqrt{1-\beta^2}\approx 1.002$ on the left-hand side takes the 
time dilation of 
the $B$-meson lifetime $\tau_B'=\gamma\,\tau_B$ into account. (In other
words, branching fractions are Lorentz invariant.) 
The above formula can be accurately
approximated by the first term in an expansion in $\beta^2$, which yields
\cite{Kagan:1998ym}
\begin{equation}
   \gamma\,\Gamma_u^{(\Upsilon)}(E_0) = \Gamma_u^{(B)}(E_0) 
   - \frac{\beta^2}{6} E_0^3 
   \left[ \frac{d}{dE}\,\frac{1}{E}\,\frac{d\Gamma_u^{(B)}}{dE}
   \right]_{E=E_0} + O(\beta^4) \,,
\end{equation}
as long as $E_0$ is not too close to the kinematical endpoint (i.e., 
$E_0\le\beta_- M_B/2\approx 2.47$\,GeV).  The
numerical results for the partial decay rate
$\gamma\,\Gamma_u^{(\Upsilon)}(E_0)$ in the rest frame of the $\Upsilon(4S)$
resonance are given in Table~\ref{tab:LeptoInUps}. 

\subsection{Cut on hadronic {\boldmath $P_+$}}

Cutting on $P_+$ samples the same hadronic phase space as a cut on the
charged-lepton energy, but with much better efficiency 
\cite{Bosch:2004th,Bosch:2004bt}. The phase space $P_+\le\Delta_P$ with the 
ideal separator $\Delta_P=M_D^2/M_B\approx 0.66$\,GeV contains well over half 
of all $\bar B\to X_u\,l^-\bar\nu$ events. Here we find with the default 
settings
\begin{equation}
\begin{array}{lllllll}
   & \Gamma_u^{\rm (0)}
   & +\,\, (\Gamma_u^{\rm kin(1)} & +\,\, \Gamma_u^{\rm hadr(1)}) 
   & +\,\, (\Gamma_u^{\rm kin(2)} & +\,\, \Gamma_u^{\rm hadr(2)}) & \\
   = \big[ & 53.225 & +\,\, (4.646 & -\,\, 11.862)
   & +\,\, (0.328 & -\,\, 0.227) \big] 
   & |V_{ub}|^2\,{\rm ps}^{-1} \,.
\end{array}
\end{equation}
We see a much better convergence of the power series than in the case of a 
cut on the charged-lepton energy, namely $53.225-7.216-0.100$ when
grouping the above numbers according to their power counting. Once again, 
the sum $\Gamma_u^{\rm kin(1)}+\Gamma_u^{\rm kin(2)}=4.973$ 
is very close to the full kinematical correction 
$\Gamma_u^{\rm kin}=4.959$ (in units of $|V_{ub}|^2\,{\rm ps}^{-1}$).

\begin{table}
\begin{center}
\caption{\label{tab:Pp}
Partial decay rate $\Gamma_u(\Delta_P, E_0)$ for a cut on the hadronic 
variable $P_+\le\Delta_P$ and lepton energy $E_l \ge E_0$, given in units of 
$|V_{ub}|^2\,\mbox{ps}^{-1}$. Predictions are based on the shape-function 
parameter values $m_b=4.61$\,GeV, $\mu_\pi^2=0.2$\,GeV$^2$ (top) and 
$m_b=4.55$\,GeV, $\mu_\pi^2=0.3$\,GeV$^2$ (bottom).} 
\vspace{5mm}
\begin{tabular}{|c|c|c|c|c|c|} 
\hline
$\Delta_P$ [GeV] & $E_0$ [GeV] & Mean & Subl.\ SF & Pert.\ & Total \\ 
\hline\hline
0.70 & 0.0 & 48.90 & $\pm$1.15 & $^{+2.83}_{-2.65}$ & $^{+3.30}_{-3.15}$ \\
0.65 & 0.0 & 45.34 & $\pm$1.46 & $^{+2.55}_{-2.41}$ & $^{+3.20}_{-3.09}$ \\
0.60 & 0.0 & 41.34 & $\pm$1.76 & $^{+2.26}_{-2.15}$ & $^{+3.13}_{-3.05}$ \\
0.55 & 0.0 & 36.91 & $\pm$2.01 & $^{+1.95}_{-1.87}$ & $^{+3.08}_{-3.02}$ \\
0.50 & 0.0 & 32.09 & $\pm$2.34 & $^{+1.64}_{-1.58}$ & $^{+3.12}_{-3.09}$ \\
\hline
0.70 & 1.0 & 43.36 & $\pm$1.02 & $^{+2.54}_{-2.39}$ & $^{+3.01}_{-2.88}$ \\
0.65 & 1.0 & 40.18 & $\pm$1.30 & $^{+2.28}_{-2.16}$ & $^{+2.92}_{-2.82}$ \\
0.60 & 1.0 & 36.59 & $\pm$1.59 & $^{+2.01}_{-1.92}$ & $^{+2.86}_{-2.80}$ \\
0.55 & 1.0 & 32.61 & $\pm$1.86 & $^{+1.73}_{-1.67}$ & $^{+2.84}_{-2.80}$ \\
0.50 & 1.0 & 28.29 & $\pm$2.19 & $^{+1.44}_{-1.40}$ & $^{+2.91}_{-2.89}$ \\
\hline
\hline
0.70 & 0.0 & 39.95 & $\pm$1.19 & $^{+2.18}_{-2.06}$ & $^{+2.79}_{-2.70}$ \\
0.65 & 0.0 & 36.94 & $\pm$1.42 & $^{+1.95}_{-1.86}$ & $^{+2.72}_{-2.66}$ \\
0.60 & 0.0 & 33.67 & $\pm$1.65 & $^{+1.71}_{-1.65}$ & $^{+2.69}_{-2.65}$ \\
0.55 & 0.0 & 30.15 & $\pm$1.88 & $^{+1.47}_{-1.43}$ & $^{+2.70}_{-2.68}$ \\
0.50 & 0.0 & 26.40 & $\pm$2.09 & $^{+1.22}_{-1.21}$ & $^{+2.73}_{-2.72}$ \\
\hline
0.70 & 1.0 & 35.42 & $\pm$1.13 & $^{+1.95}_{-1.85}$ & $^{+2.59}_{-2.51}$ \\
0.65 & 1.0 & 32.73 & $\pm$1.34 & $^{+1.74}_{-1.66}$ & $^{+2.53}_{-2.48}$ \\
0.60 & 1.0 & 29.81 & $\pm$1.55 & $^{+1.52}_{-1.47}$ & $^{+2.51}_{-2.48}$ \\
0.55 & 1.0 & 26.65 & $\pm$1.76 & $^{+1.29}_{-1.27}$ & $^{+2.52}_{-2.51}$ \\
0.50 & 1.0 & 23.29 & $\pm$1.95 & $^{+1.07}_{-1.06}$ & $^{+2.56}_{-2.55}$ \\
\hline
\end{tabular}
\end{center}
\end{table}

Often times it is required to impose an additional cut on the
charged-lepton energy, as leptons that are too soft are difficult to
detect. In Table~\ref{tab:Pp} we list results for both $E_l\ge 0$ and
$E_l\ge 1.0$ GeV. For the ideal cut we find that the prediction is
quite precise, as the total theoretical uncertainty is only about 6.8\%. For
comparison, the ideal cut for the lepton energy is uncertain by about
50\%, but rapidly improving as the energy cut is relaxed.

\subsection{Cut on hadronic invariant mass and {\boldmath $q^2$}}

\begin{table}
\begin{center}
\caption{\label{tab:MxQ2}
Partial decay rate $\Gamma_u(M_0,q_0^2)$ for combined cuts 
$M_X\le M_0$ on hadronic invariant mass, $q^2>q_0^2$ on leptonic invariant 
mass, given in units of 
$|V_{ub}|^2\,\mbox{ps}^{-1}$. Predictions are based on the shape-function 
parameter values $m_b=4.61$\,GeV, $\mu_\pi^2=0.2$\,GeV$^2$ (top) and 
$m_b=4.55$\,GeV, $\mu_\pi^2=0.3$\,GeV$^2$ (bottom).} 
\vspace{5mm}
\begin{tabular}{|c|c|c|c|c|c|c|} 
\hline
$M_0$ [GeV] & $q_0^2$ [GeV$^2$] & $E_0$ [GeV] & Mean & Subl.\ SF & Pert.\
 & Total \\
\hline\hline
$M_D$ & 0.0 & 0.0 & 59.30 & $\pm$0.36 & $^{+4.22}_{-3.73}$
 & $^{+4.42}_{-3.96}$ \\
1.70  & 0.0 & 0.0 & 53.13 & $\pm$0.73 & $^{+3.67}_{-3.31}$
 & $^{+3.95}_{-3.61}$ \\
1.55  & 0.0 & 0.0 & 45.72 & $\pm$1.16 & $^{+3.11}_{-2.84}$
 & $^{+3.55}_{-3.32}$ \\
$M_D$ & 6.0 & 0.0 & 34.37 & $\pm$0.37 & $^{+2.97}_{-2.58}$
 & $^{+3.25}_{-2.89}$ \\
1.70  & 8.0 & 0.0 & 24.80 & $\pm$0.36 & $^{+2.24}_{-1.98}$
 & $^{+2.59}_{-2.37}$ \\
$M_D$ & $(M_B-M_D)^2$ & 0.0 & 12.55 & $\pm$0.49 & $^{+1.41}_{-1.24}$
 & $^{+1.95}_{-1.83}$ \\
\hline

\hline
$M_D$ & 0.0 & 0.0 & 50.08 & $\pm$0.54 & $^{+3.52}_{-3.11}$
 & $^{+3.78}_{-3.40}$ \\
1.70  & 0.0 & 0.0 & 44.20 & $\pm$0.86 & $^{+2.98}_{-2.69}$
 & $^{+3.35}_{-3.09}$ \\
1.55  & 0.0 & 0.0 & 37.76 & $\pm$1.22 & $^{+2.46}_{-2.26}$
 & $^{+3.03}_{-2.86}$ \\
$M_D$ & 6.0 & 0.0 & 29.42 & $\pm$0.35 & $^{+2.50}_{-2.16}$
 & $^{+2.82}_{-2.52}$ \\
1.70  & 8.0 & 0.0 & 20.87 & $\pm$0.39 & $^{+1.84}_{-1.61}$
 & $^{+2.26}_{-2.08}$ \\
$M_D$ & $(M_B-M_D)^2$ & 0.0 &10.49 & $\pm$0.48 & $^{+1.16}_{-1.00}$
 & $^{+1.76}_{-1.68}$ \\
\hline
\end{tabular}
\end{center}
\end{table}

\begin{table}
\begin{center}
\caption{\label{tab:MxQ2El}
Partial decay rate $\Gamma_u(M_0,q_0^2,E_0)$ for combined cuts 
$M_X\le M_0$ on hadronic invariant mass, $q^2>q_0^2$ on leptonic invariant 
mass, and $E_l\ge E_0$ on charged-lepton energy, given in units of 
$|V_{ub}|^2\,\mbox{ps}^{-1}$. Predictions are based on the shape-function 
parameter values $m_b=4.61$\,GeV, $\mu_\pi^2=0.2$\,GeV$^2$ (top) and 
$m_b=4.55$\,GeV, $\mu_\pi^2=0.3$\,GeV$^2$ (bottom).} 
\vspace{5mm}
\begin{tabular}{|c|c|c|c|c|c|c|} 
\hline
$M_0$ [GeV] & $q_0^2$ [GeV$^2$] & $E_0$ [GeV] & Mean & Subl.\ SF & Pert.\
 & Total \\
\hline
\hline
$M_D$ & 0.0 & 1.0 & 53.49 & $\pm$0.36 & $^{+3.91}_{-3.45}$
 & $^{+4.13}_{-3.69}$ \\
1.70  & 0.0 & 1.0 & 48.25 & $\pm$0.63 & $^{+3.42}_{-3.08}$
 & $^{+3.70}_{-3.38}$ \\
1.55  & 0.0 & 1.0 & 41.81 & $\pm$1.03 & $^{+2.91}_{-2.66}$
 & $^{+3.34}_{-3.12}$ \\
$M_D$ & 6.0 & 1.0 & 33.88 & $\pm$0.37 & $^{+2.94}_{-2.55}$
 & $^{+3.22}_{-2.87}$ \\
1.70  & 8.0 & 1.0 & 24.74 & $\pm$0.36 & $^{+2.23}_{-1.97}$
 & $^{+2.59}_{-2.37}$ \\
$M_D$ & $(M_B-M_D)^2$ & 1.0 & 12.55 & $\pm$0.49 & $^{+1.41}_{-1.24}$
 & $^{+1.95}_{-1.83}$ \\
\hline

\hline
$M_D$ & 0.0 & 1.0 & 45.29 & $\pm$0.50 & $^{+3.27}_{-2.88}$
 & $^{+3.54}_{-3.18}$ \\
1.70  & 0.0 & 1.0 & 40.22 & $\pm$0.77 & $^{+2.78}_{-2.50}$
 & $^{+3.15}_{-2.90}$ \\
1.55  & 0.0 & 1.0 & 34.55 & $\pm$1.09 & $^{+2.31}_{-2.11}$
 & $^{+2.85}_{-2.69}$ \\
$M_D$ & 6.0 & 1.0 & 28.99 & $\pm$0.34 & $^{+2.48}_{-2.13}$
 & $^{+2.80}_{-2.50}$ \\
1.70  & 8.0 & 1.0 & 20.82 & $\pm$0.39 & $^{+1.83}_{-1.60}$
 & $^{+2.26}_{-2.08}$ \\
$M_D$ & $(M_B-M_D)^2$ & 1.0 &10.49 & $\pm$0.48 & $^{+1.16}_{-1.00}$ &
 $^{+1.78}_{-1.68}$ \\
\hline
\end{tabular}
\end{center}
\end{table}

The most efficient separator for the discrimination of 
$\bar B\to X_c l^-\bar\nu$ events is a cut on the invariant mass $M_X$ of the 
hadronic final state, $M_X\le M_D$ \cite{Falk:1997gj,Bigi:1997dn}.
It has also been argued \cite{Bauer:2000xf} that a cut on 
$q^2$ can reduce the shape-function sensitivity, since it avoids the 
collinear region in phase space where $P_-\gg P_+$.
In order to optimize signal efficiency and theoretical uncertainties, it was
suggested in \cite{Bauer:2001rc} to combine a $q^2$ cut with a cut on hadronic 
invariant mass. 

The theoretical predictions obtained in \cite{Bauer:2000xf,Bauer:2001rc} 
were based on a conventional OPE calculation, which was assumed to be valid 
for these cuts. The
assessment of the shape-function sensitivity was based on convolving
the tree-level decay rate with a ``tree-level shape function'', for which 
two models (a realistic model similar to the ones considered here, and an 
unrealistic $\delta$-function model) were employed. The 
shape-function sensitivity was then inferred from the comparison of the 
results obtained with the two models. The sensitivity to subleading 
shape functions was not considered, since it was assumed to be very small.
Since our formalism smoothly interpolates between the
``shape-function'' and ``OPE'' regions, and since we include radiative 
corrections as well as power corrections as far as they are known, 
we can estimate 
the sensitivity of a combined $M_X$--$q^2$ cut to the leading and subleading 
shape functions much more accurately. Contrary to \cite{Bauer:2001rc}, 
we do not find a significant reduction of the shape-function sensitivity when 
adding the $q^2$ cut to a cut on hadronic invariant mass. 

In Tables \ref{tab:MxQ2} and \ref{tab:MxQ2El} we give results for typical 
cuts on $M_X$ and $q^2$, with and without including an additional cut on 
charged-lepton energy.
Let us study the contributions for the optimal cut $M_X\le M_D$ in
detail. We find with the default settings
\begin{equation}
\begin{array}{lllllll}
   & \Gamma_u^{\rm (0)} 
   & +\,\, (\Gamma_u^{\rm kin(1)} & +\,\, \Gamma_u^{\rm hadr(1)}) 
   & +\,\, (\Gamma_u^{\rm kin(2)} & +\,\, \Gamma_u^{\rm hadr(2)}) & \\
   = \big[ & 58.541 & +\,\, (8.027 & -\,\, 9.048)
   & +\,\, (2.100 & -\,\, 0.318) \big] 
   & |V_{ub}|^2\,{\rm ps}^{-1} \,.
\end{array}
\end{equation}
Note the almost perfect (accidental) cancellation of the two terms at order 
$1/m_b$. The resulting power series,
$58.541-1.022+1.782$, again exhibits good convergence. As previously, the sum
$\Gamma_u^{\rm kin(1)}+\Gamma_u^{\rm kin(2)}=10.127$ is a good
approximation to the exact value 
$\Gamma_u^{\rm kin}=9.753$ (in units of $|V_{ub}|^2\,{\rm ps}^{-1}$). 
The analogous analysis for a combined cut $M_X\le 1.7$\,GeV and 
$q^2\ge 8.0$\,GeV$^2$ reads
\begin{equation}
\begin{array}{lllllll}
   & \Gamma_u^{\rm (0)} 
   & +\,\, (\Gamma_u^{\rm kin(1)} & +\,\, \Gamma_u^{\rm hadr(1)}) 
   & +\,\, (\Gamma_u^{\rm kin(2)} & +\,\, \Gamma_u^{\rm hadr(2)}) & \\
   = \big[ & 25.880 & +\,\, (4.049 & -\,\, 6.358)
   & +\,\, (1.399 & -\,\, 0.171) \big] 
   & |V_{ub}|^2\,{\rm ps}^{-1} \,,
\end{array}
\end{equation}
which means that the power series is $25.880-2.309+1.228$. Here we have 
$\Gamma_u^{\rm kin(1)}+\Gamma_u^{\rm kin(2)}=5.449$, which is close to
$\Gamma_u^{\rm kin}=5.160$ (in units of $|V_{ub}|^2\,{\rm ps}^{-1}$).

\subsection{Cut on {\boldmath $s_H^{\rm max}$} and {\boldmath $E_l$}}
\label{sec:babarcut}

In \cite{Aubert:2004tw}, the BaBar collaboration employed a
cut on both $E_l\ge E_0$ and a new kinematical variable 
$s_H^{\rm max}\le s_0$, where the
definition for $s_H^{\rm max}$ involves both hadronic and leptonic
variables. In the $B$-meson rest frame, it is
\begin{equation}\label{shmax}
   s_H^{\rm max} = M_B^2 + q^2 -2 M_B \left( E_l + \frac{q^2}{4E_l} \right) .
\end{equation}
Rewriting the phase space of this cut in the variables $P_+$, $P_-$, $P_l$, we 
find
\begin{eqnarray}
   0\le P_+ &\le& \mbox{min}\left( M_B-2E_0, \sqrt{s_0} \right) ,
    \nonumber \\
   P_+\le P_- &\le& \mbox{min}\left( \frac{s_0}{P_+}, M_B \right) ,
    \nonumber \\
   P_+\le P_l &\le& \mbox{min}\left( M_B-2E_0, P_- \right) ,
\end{eqnarray}
where it is understood that if $q^2=(M_B-P_+)(M_B-P_-)\le (M_B-\sqrt{s_0})^2$, 
then the interval $P_l^{\rm min}<P_l<P_l^{\rm max}$ must be {\em excluded\/} 
from the $P_l$ integration. Here
\begin{eqnarray}
   P_l^{\rm max/min}(P_+,P_-) &=& \left(\frac{P_+ + P_-}{2}
   + \frac{s_0 - P_+ P_-}{2M_B} \right)\nonumber\\
   &&\pm \sqrt{\left(\frac{P_+ + P_-}{2} + \frac{s_0 - P_+ P_-}{2M_B} \right)^2
   - s_0} \,.
\end{eqnarray}

\begin{table}
\begin{center}
\caption{\label{tab:sHmax}
Partial decay rate $\Gamma_u(s_0,E_0)$ for combined cuts 
$s_H^{\rm max}\le s_0$ and $E_l\ge E_0$, given in 
units of $|V_{ub}|^2\,\mbox{ps}^{-1}$. Predictions are based on the 
shape-function parameter values $m_b=4.61$\,GeV, $\mu_\pi^2=0.2$\,GeV$^2$ 
(top) and $m_b=4.55$\,GeV, $\mu_\pi^2=0.3$\,GeV$^2$ (bottom).} 
\vspace{5mm}
\begin{tabular}{|c|c|c|c|c|c|} 
\hline
$s_0$ [GeV$^2$] & $E_0$ [GeV] & Mean & Subl.\ SF & Pert.\ & Total \\ 
\hline\hline
3.5 & 1.8 & 17.39 & $\pm$0.62 & $^{+1.54}_{-1.36}$ & $^{+2.08}_{-1.96}$ \\ 
3.5 & 1.9 & 15.86 & $\pm$0.63 & $^{+1.33}_{-1.18}$ & $^{+1.94}_{-1.84}$ \\
3.5 & 2.0 & 13.70 & $\pm$0.66 & $^{+1.05}_{-0.94}$ & $^{+1.77}_{-1.71}$ \\
3.5 & 2.1 & 10.78 & $\pm$0.73 & $^{+0.71}_{-0.64}$ & $^{+1.62}_{-1.59}$ \\
\hline
\hline
3.5 & 1.8 & 14.57 & $\pm$0.60 & $^{+1.25}_{-1.09}$ & $^{+1.87}_{-1.77}$ \\
3.5 & 1.9 & 13.18 & $\pm$0.61 & $^{+1.06}_{-0.92}$ & $^{+1.76}_{-1.68}$ \\
3.5 & 2.0 & 11.28 & $\pm$0.64 & $^{+0.82}_{-0.71}$ & $^{+1.63}_{-1.58}$ \\
3.5 & 2.1 &  8.77 & $\pm$0.69 & $^{+0.54}_{-0.46}$ & $^{+1.54}_{-1.51}$ \\
\hline
\end{tabular}
\end{center}
\end{table}

A summary of our findings is given in Table~\ref{tab:sHmax}. When compared to 
the pure charged-lepton energy cut in Table~\ref{tab:LeptoInB}, 
the additional cut on $s_H^{\rm max}$
eliminates roughly another 20--30\% of events. However, the hope is
that this cut also reduces the sensitivity to the leading shape
function, which we expect to be sizable for the pure $E_l$ cut. The
uncertainty from subleading shape functions, however, is almost
unaffected by the $s_H^{\rm max}$ cut. 

\subsection{Eliminating weak annihilation contributions}
\label{sec:eliminateWA}

\begin{table}
\begin{center}
\caption{\label{tab:withQ2}
Examples of partial decay rates with a cut on $q^2\le(M_B-M_D)^2$ imposed to 
eliminate the weak annihilation contribution. We consider an additional cut on 
the hadronic variable $P_+\le\Delta_P$ (top), or on the hadronic invariant 
mass $M_X\le M_0$ (bottom). As before, decay rates are given in units of 
$|V_{ub}|^2\,\mbox{ps}^{-1}$. Predictions are based on the shape-function 
parameters $m_b=4.61$\,GeV and $\mu_\pi^2=0.2$\,GeV$^2$.}
\vspace{5mm}
\begin{tabular}{|c|c|c|c|c|}
\hline
$\Delta_P$ [GeV] & Mean & Subl.\ SF & Pert.\ & Total \\ 
\hline\hline
0.70 & 39.96 & $\pm$1.27 & $^{+2.16}_{-2.01}$ & $^{+2.51}_{-2.38}$ \\ 
0.65 & 37.18 & $\pm$1.50 & $^{+1.99}_{-1.85}$ & $^{+2.49}_{-2.38}$ \\
0.60 & 34.05 & $\pm$1.71 & $^{+1.82}_{-1.69}$ & $^{+2.50}_{-2.41}$ \\
0.55 & 30.61 & $\pm$1.89 & $^{+1.63}_{-1.52}$ & $^{+2.49}_{-2.42}$ \\
0.50 & 26.86 & $\pm$1.97 & $^{+1.44}_{-1.33}$ & $^{+2.44}_{-2.38}$ \\
\hline\hline
$M_0$ [GeV] & Mean & Subl.\ SF & Pert.\ & Total \\ 
\hline\hline
$M_D$ & 46.75 & $\pm$0.65 & $^{+2.82}_{-2.50}$ & $^{+2.89}_{-2.58}$ \\ 
 1.70 & 40.70 & $\pm$1.12 & $^{+2.32}_{-2.11}$ & $^{+2.58}_{-2.39}$ \\ 
 1.55 & 33.69 & $\pm$1.56 & $^{+1.88}_{-1.73}$ & $^{+2.44}_{-2.32}$ \\
\hline
\end{tabular}
\end{center}
\end{table}

In Section~\ref{Sec:WA} we have argued that a cut on {\em high\/}
$q^2$, i.e., $q^2<q_0^2$, will eliminate the effect of weak
annihilation and remove the uncertainty associated with this
contribution. The cutoff $q_0^2$ should be small enough to exclude the
region around $q^2=m_b^2$, where this contribution is concentrated. It
is instructive to assess the ``cost'' of such an additional cut in
terms of the loss of efficiency and, more importantly, the behavior of
the remaining uncertainties. In order to do this, we combine the cut
$q^2\le(M_B-M_D)^2$ with either a cut on $P_+$ or on $M_X$. 
While this particular choice for $q_0^2$ still leaves some room
to improve the efficiency by increasing $q_0^2$, it is not desirable
to raise the cut much further, since this would threaten the validity
of quark-hadron duality.

The results are summarized in Table~\ref{tab:withQ2} and can be
compared to the previous ``pure'' $P_+$ and $M_X$ cuts in
Tables~\ref{tab:Pp}, \ref{tab:MxQ2}, and \ref{tab:MxQ2El}. As an example, let us consider
the case $P_+\le 0.65$ GeV, which is close to the charm
threshold. Without the additional $q^2$ cut we found that the total
theoretical uncertainty (including the weak annihilation error) is
${}_{-6.8}^{+7.0}$\%. When cutting in addition on $q^2\le(M_B-M_D)^2$,
the efficiency decreases by about 20\% as expected. However, due to the
absence of the weak annihilation uncertainty, the overall uncertainty
decreases to ${}_{-6.4}^{+6.7}$\%. Therefore both strategies result in
comparable relative uncertainties, with a slight favor for imposing
the additional cut from the theoretical point of view.

While the small reduction of theoretical errors hardly seems worth the effort 
of imposing the $q^2$ cut, performing an analysis of the type outlined 
here and comparing its results with those obtained without the additional cut 
may help to corroborate the expectation 
that the weak annihilation contribution is 
indeed not much larger than what has been found in \cite{TomsThesis}.

\subsection{Dependence on {\boldmath $m_b$} and shape-function sensitivity}

Non-perturbative hadronic physics enters in our approach via the form of the
leading and subleading shape functions. The strongest sensitivity by far is 
to the first moment of the leading shape function, which determines the HQET
parameter $\bar\Lambda$ and with it the $b$-quark mass. Given that the 
value of $m_b\equiv m_b(\mu_*,\mu_*)$ can be determined with good precision 
from other sources (such as moments of the leptonic or hadronic invariant 
mass spectra in $\bar B\to X_c\,l^-\bar\nu$ decays), it is instructive to 
disentangle this dependence from the sensitivity to 
higher moments or, more generally, to the functional form of the shape 
functions for fixed $m_b$. 

\begin{table}
\begin{center}
\caption{\label{tab:mbdep} 
Values of the exponent $a(m_b)$ for different kinematical cuts. The parameter
$\mu_\pi^2=0.2$\,GeV$^2$ is kept fixed. Also quoted is the sensitivity of the 
partial decay rates to the functional form of the shape functions. See text 
for explanation.}
\vspace{5mm}
\begin{tabular}{|l|c|c|c|c|c|c|}
\hline
& $m_b$ [GeV] & 4.50 & 4.55 & 4.60 & 4.65 & 4.70 \\
\hline\hline
$M_X\le M_D$ & $a$ & 9.5 & 8.8 & 8.2 & 7.7 & 7.3 \\
 & Functional Form & 1.4\% & 1.1\% & 0.8\% & 0.5\% & 0.4\% \\
\hline
$M_X\le 1.7$\,GeV & $a$ & 12.5 & 11.5 & 10.5 & 9.7 & 8.9 \\
 & Functional Form & 2.9\% & 2.6\% & 2.2\% & 1.9\% & 1.6\% \\
\hline
$M_X\le 1.7$\,GeV & $a$ & 10.3 &  9.8 & 9.3 & 9.0 & 8.7 \\
$q^2\ge 8$\,GeV$^2$ & Functional Form & 2.0\% & 1.7\% & 1.5\% & 1.4\% & 1.4\%
 \\
\hline
$q^2\ge(M_B-M_D)^2$ & $a$ & 11.4 & 11.1 & 10.9 & 10.8 & 10.6 \\
 & Functional Form & 5.0\% & 4.4\% & 4.0\% & 3.6\% & 3.2\% \\
\hline
$P_+\le M_D^2/M_B$ & $a$ & 16.7 & 15.0 & 13.6 & 12.2 & 11.1 \\
 & Functional Form & 5.3\% & 4.8\% & 4.4\% & 4.0\% & 3.6\% \\
\hline
$E_l\ge 2.2$\,GeV & $a$ & 22.6 & 21.0 & 19.7 & 18.5 & 17.4 \\
 & Functional Form & 16.2\% & 13.1\% & 11.0\% & 9.3\% & 7.9\% \\
\hline
\end{tabular}
\end{center}
\end{table}

To explore the dependence on $m_b$ we define the exponent
\begin{equation}\label{eq:postulate}
   a(m_b)\equiv \frac{d\ln\Gamma_u}{d\ln m_b}
   = \left( \frac{\triangle\Gamma_u}{\Gamma_u} \right) /
   \left( \frac{\triangle m_b}{m_b} \right) ,
\end{equation}
which means that $\Gamma_u\sim (m_b)^a$. Table \ref{tab:mbdep} shows
the values of this exponent over a wide range of values of $m_b$ 
for a variety of experimental cuts. 
To estimate the sensitivity to the functional form we scan over a large set of
models for the subleading shape functions, and we also study the difference
between the results obtained using the exponential or the 
gaussian ansatz for the leading shape function. The corresponding variations
are added in quadrature and given as a relative change in the 
corresponding partial decay rates (labeled ``Functional Form'').
In all cases, $\mu_\pi^2=0.2$\,GeV$^2$ is kept fixed. 
Because we restrict ourselves to only two functional forms for the leading
shape function in this study, the resulting sensitivities should be 
interpreted with caution. 

The entries in the table are listed in roughly the order of increasing 
sensitivity to $m_b$ and to the functional form of the shape functions, with 
the hadronic invariant mass cut showing the least sensitivity and the 
lepton energy cut exhibiting the largest one. To some extent this reflects 
the different efficiencies (or ``inclusiveness'') of the various cuts.
It is reassuring that 
$a\approx 10$ for the pure $q^2$ cut, in accordance with the findings of
\cite{Neubert:2000ch,Neubert:2001ib}. Perhaps somewhat surprisingly, 
for this cut a substantial sensitivity to shape-function effects remains 
even for fixed $m_b$ and $\mu_\pi^2$. It is well known that the partial rate
with a cut $q^2\ge(M_B-M_D)^2$ can be calculated using a local OPE in powers
of $\Lambda_{\rm QCD}/m_c$ \cite{Bauer:2000xf,Neubert:2000ch}, thereby 
avoiding the notion of shape-function sensitivity. 
Differences between the functional forms of the shape functions 
in our approach correspond to effects 
that are formally of order $1/m_c^3$ and higher. It is not unreasonable that 
these effects should be of order 3--5\%.

We also checked that for much more relaxed cuts the value of $a(m_b)$
tends to $4.8$, as stated in (\ref{totrate}). For example, for a cut 
$P_+\le\Delta_P$ we find (with $m_b=4.61$\,GeV and $\mu_\pi^2=0.2$\,GeV$^2$):

\begin{center}
\begin{tabular}{|c|rrrrrrrr|} 
\hline
$\Delta_P$ [GeV] & 0.6 & 0.8 & 1.0 & 1.2 & 1.6 & 2.0 & 3.0 & $M_B$ \\ 
\hline
$a$ & 15.4 & 9.8 & 7.0 & 5.8 & 5.1 & 5.0 & 4.9 & 4.8 \\
\hline
\end{tabular}
\end{center}

\section{Conclusions}
\label{sec:concl}

A high-precision measurement of the parameters of the unitarity triangle 
is an ongoing quest, which
necessitates the close cooperation of theory and experiment. The
determination of $|V_{ub}|$ from inclusive 
$\bar B\to X_u\,l^-\bar\nu$ decay requires the measurement of partial decay
rates with kinematical cuts that eliminate the large background
from $\bar B\to X_c\,l^-\bar\nu$ decay, as well as theoretical
predictions for such quantities. To this end, it is desirable to have a
theoretical description of the triple differential decay rate, which can
be used for predicting arbitrary partial rates obtained after integrating 
over certain regions of phase space. One problem in providing such a 
description is that the power-counting rules of the heavy-quark expansion 
are different in different kinematical domains. In this chapter we have 
overcome this difficulty. 

In the shape-function region, our results are in agreement with
QCD factorization theorems, and perturbative effects have been
separated from non-perturbative shape functions. When the allowed
phase space extends over a large domain, our results smoothly
reduce to the expressions obtained from the local operator product 
expansion. We have presented a formalism in which event distributions and
partial decay rates are expressed without explicit reference to
partonic quantities such as the $b$-quark mass. The sensitivity to such
hadronic parameters enters indirectly, via the moments of shape functions.
The most important non-perturbative object, namely the leading-order shape
function, can be extracted from the photon spectrum in 
$\bar B\to X_s\gamma$ decay. 
This is analogous to extractions of parton distribution 
functions from fits to data on deep inelastic scattering.
In this way, the dominant uncertainty from our ignorance about bound-state 
effects in the $B$ meson is turned into an experimental 
uncertainty, which will reduce with increasing accuracy of the experimental 
data on the photon spectrum. Residual hadronic uncertainties are power
suppressed in the heavy-quark expansion.

One goal of this chapter was to present a detailed framework
in which this program can be carried out. We have given formulae
that can be readily used for the construction of an
event generator, as well as to estimate the remaining
theoretical uncertainties in a robust and automated fashion.

In practice the leading shape function needs to be parameterized. We have 
suggested three different functional forms, which can be used to fit the data 
of the $\bar B\to X_s\gamma$ photon spectrum. Once the data is accurately 
described by a choice of the shape functions, this function can be used in
the predictions for partial $\bar B\to X_u\,l^-\bar\nu$ rates and
spectra. Subleading shape functions give rise to
theoretical uncertainties starting at the level of $1/m_b$ power corrections. 
We have estimated these uncertainties using a large set of models, each of 
which obeys the known tree-level moment relations, but which are very
different in their functional form. A second error estimate is
determined by the residual renormalization-scale dependence. We also
considered uncertainties from weak annihilation effects, which in principle 
can be avoided by cutting away the region of phase space in which they
contribute. We have suggested a cut on high leptonic invariant mass,
which accomplishes just that.

The second half of this chapter contains detailed 
numerical predictions for a variety of partial $\bar B\to X_u\,l^-\bar\nu$ 
decay rates with different kinematical cuts, 
including cuts on the charged-lepton energy (both
in the rest frame of the $B$ meson and of the $\Upsilon(4S)$ resonance),
on the hadronic quantity $P_+=E_X-|\vec{P}_X|$, on $M_X$, on $q^2$, and on 
various combinations of these variables. Along with our predictions 
for the rates we have presented a complete analysis of theoretical
uncertainties. Once the data on
the $\bar B\to X_s\gamma$ photon spectrum are sufficiently precise to
accurately determine the leading-order shape function, a determination of
$|V_{ub}|$ with theoretical uncertainties at the 5--10\% level now seems 
feasible.

\chapter{Applications: Weight Function}
\label{chapter_weight}

\section{Introduction}\label{section_5.1}
The kinematical variable $P_+$  is common to semileptonic and radiative 
charmless inclusive $B$ decays. In $\bar B\to X_s\gamma$ decays  $P_+$
is related to the $B$-meson mass and the photon energy, 
$P_+=M_B-2E_\gamma$, and the measurement of its spectrum leads directly
to the extraction of the leading hadronic structure function, called
the shape function \cite{Neubert:1993ch,Neubert:1993um,Bigi:1993ex}. 
The $P_+$ spectrum in semileptonic $\bar B\to X_u\,l^-\bar\nu$ decays, on the 
other hand, enables us to determine $|V_{ub}|$ 
\cite{Mannel:1999gs,Bosch:2004bt,Aglietti:2002md}, but this requires a precise 
knowledge of the shape function. One approach for measuring $|V_{ub}|$ is to 
first extract the shape function from the $\bar B\to X_s\gamma$
photon spectrum, and then to use this information for predictions
of event distributions in $\bar B\to X_u\,l^-\bar\nu$. A comprehensive
description of this program has been presented in
chapter \ref{chapter_evegen}. Equivalently, it is possible to eliminate the
shape function in $\bar B\to X_u\,l^-\bar\nu$ decay rates in favor of
the $\bar B\to X_s\gamma$ photon-energy spectrum. This idea was first put
forward in \cite{Neubert:1993um} and later refined in
\cite{Leibovich:1999xf,Leibovich:2000ey,Neubert:2001sk,Hoang:2005pj}. Partial
$\bar B\to X_u\,l^-\bar\nu$ decay rates are then given as weighted
integrals over the $\bar B\to X_s\gamma$ photon-energy spectrum,
\begin{equation}\label{eq:relation}
   \Gamma_u(\Delta)
   = \underbrace{\int_0^\Delta\!dP_+\,
    \frac{d\Gamma_u}{dP_+}}_{\hbox{\footnotesize exp.\ input}}
   = |V_{ub}|^2 \int_0^\Delta\!dP_+
    \underbrace{\phantom{\frac{1}{\Gamma_s(E_*)}}\hspace{-12mm}
    W(\Delta,P_+)}_{\hbox{\footnotesize theory}}\,
    \underbrace{\frac{1}{\Gamma_s(E_*)}\,
    \frac{d\Gamma_s}{dP_+}}_{\hbox{\footnotesize exp.\ input}} \,,
\end{equation}
where the weight function $W(\Delta,P_+)$ is perturbatively calculable
at leading power in $\Lambda_{\rm QCD}/m_b$. A comparison of both
sides of the equation determines the CKM matrix element $|V_{ub}|$
directly. For the measurement of the left-hand side 
to be free of charm background, $\Delta$ must be
less than $M_D^2/M_B\approx 0.66$\,GeV. However, the $P_+$ spectrum in
$\bar B\to X_u\,l^-\bar\nu$ decays displays many of the features of
the charged-lepton energy spectrum, so that it is not inconceivable that the 
cut can be further relaxed for the same reasons that experimenters are able 
to relax the lepton cut beyond the charm threshold.
We stress that for an application of relation (\ref{eq:relation}) a 
measurement of the $\bar B\to X_s\gamma$ photon spectrum is needed only for 
$E_\gamma\ge\frac12(M_B-\Delta)\approx 2.3$\,GeV (or slightly lower, if 
the cut is relaxed into the charm region). This high-energy part of the
spectrum has already been measured with good precision.

Previous authors \cite{Neubert:1993um,Leibovich:1999xf,Leibovich:2000ey,%
Neubert:2001sk,Hoang:2005pj} have considered relations such as 
(\ref{eq:relation}) in the slightly different form
\begin{equation}\label{oldrelation}
   \underbrace{\int_0^\Delta\!dP_+\,
    \frac{d\Gamma_u}{dP_+}}_{\hbox{\footnotesize exp.\ input}}
   = \frac{|V_{ub}|^2}{|V_{tb} V_{ts}^*|^2} \int_0^\Delta\!dP_+
    \underbrace{\phantom{\frac{1}{\Gamma_s(E_*)}}\hspace{-12mm}
    \widetilde W(\Delta,P_+)}_{\hbox{\footnotesize theory}}\!
    \underbrace{\frac{d\Gamma_s}{dP_+}}_{\hbox{\footnotesize exp.\ input}}
   \hspace{-0.3cm} .
\end{equation}
Normalizing the photon spectrum by the total\footnote{Due to an
unphysical soft-photon singularity, the total decay rate is commonly
defined to include all events with photon energies above 
$E_*=m_b/20$ \cite{Kagan:1998ym}.} 
rate $\Gamma_s(E_*)$ as done in (\ref{eq:relation}) has several advantages. 
Firstly, it is a known fact that event fractions in $\bar B\to X_s\gamma$ 
decay can be calculated with better accuracy than partial decay rates (see 
\cite{Neubert:2004dd} for a recent discussion), and likewise the normalized 
rate does not suffer from the relatively large experimental error on the 
total branching ratio. Secondly, relation 
(\ref{eq:relation}) is independent of the CKM factor $|V_{tb} V_{ts}^*|$. 
Thirdly, unlike the total $\bar B\to X_s\gamma$ decay rate, the shape of the 
photon spectrum is rather insensitive to possible New Physics 
contributions \cite{Kagan:1998ym}, which could distort the outcome of a 
$|V_{ub}|$ measurement via relation (\ref{oldrelation}). Lastly, as
we will see below, the weight function $W(\Delta,P_+)$ possesses a
much better perturbative expansion than the function 
$\widetilde W(\Delta,P_+)=|V_{tb} V_{ts}^*|^2\,W(\Delta,P_+)/\Gamma_s(E_*)$. 
This last point can be traced back to the fact that most of the very large
contribution from the $O_1 - O_{7\gamma}$ operator mixing in the effective 
weak Hamiltonian cancels in the theoretical expression for the
normalized photon spectrum.

In principle, any partial $\bar B\to X_u\,l^-\bar\nu$ decay rate can
be brought into the form (\ref{eq:relation}), with complicated weight
functions. The relation between the two $P_+$ spectra is particularly simple, 
because the leading-power weight function is a constant at tree level. 
Experiments typically reject semileptonic $B$-decay events with very low 
lepton energy. The effect of such an additional cut can be determined from
\cite{Lange:2005yw}. Alternatively, it is possible to modify the weight
function so as to account for a lepton cut, however at the expense 
of a significant increase in complexity. We will not pursue this option in 
this chapter.

The weight function depends on the kinematical variable $P_+$ and on the size 
$\Delta$ of the integration domain. It possesses integrable singularities of
the form $\alpha_s^n(\mu) \ln^k[m_b(\Delta-P_+)/\mu^2]$, with $k\le n$, in
perturbation theory. Different strategies can be found in the
literature concerning these logarithms. Leibovich et al.\ resummed them
by identifying $\mu^2$ with $m_b(\Delta-P_+)$
\cite{Leibovich:1999xf,Leibovich:2000ey}. The $P_+$ dependence of the
weight function then enters via the running coupling
$\alpha_s(\sqrt{m_b(\Delta-P_+)})$. This is a legitimate choice 
of scale as long as $(\Delta-P_+)$ has a generic value of order 
$\Lambda_{\rm QCD}$; however, it is {\em not\/} a valid choice in the small 
region where $m_b(\Delta-P_+)\sim\Lambda_{\rm QCD}^2$. A key result 
underlying relation (\ref{eq:relation}) is that, by construction, the weight 
function is insensitive to soft physics. Quark-hadron duality ensures that 
the region near the point $P_+=\Delta$, which is without any physical 
significance, does not require special consideration after 
integration over $P_+$. In the approach of 
\cite{Leibovich:1999xf,Leibovich:2000ey}, the attempt to resum the above 
logarithms near the endpoint of the $P_+$ integral
leads to integrals over unphysical Landau singularities of the running 
coupling in the nonperturbative domain. Hoang et al.\ chose to calculate the
weight function in fixed-order perturbation theory at the scale $\mu=m_b$ 
\cite{Hoang:2005pj}. This leads to parametrically large logarithms, 
since $\Delta-P_+\ll m_b$. In the present chapter we separate physics effects 
from two parametrically distinct scales, a hard scale $\mu_h\sim m_b$ and
an intermediate scale $\mu_i\sim \sqrt{m_b\Lambda_{\rm QCD}}$, so that
we neither encounter Landau singularities nor introduce parametrically
large logarithms. The shape of the weight function is then governed by
a perturbative expansion at the intermediate scale, $\alpha_s^n(\mu_i)
\ln^k[m_b(\Delta-P_+)/\mu_i^2]$. As will be explained later, the coefficients 
in this series, as well as the overall normalization, possess themselves an
expansion in $\alpha_s(\mu_h)$.

The calculation of the weight function starts with the
theoretical expressions for the $P_+$ spectra in $\bar B\to X_u\,l^-\bar\nu$ 
and $\bar B\to X_s\gamma$ decays, which are given as \cite{Lange:2005yw}
\begin{eqnarray}
   \frac{d\Gamma_u}{dP_+} 
   &=& \frac{G_F^2 |V_{ub}|^2}{96\pi^3}\,U(\mu_h,\mu_i)\,(M_B-P_+)^5
    \int_0^1\!dy\,y^{2-2a_\Gamma(\mu_h,\mu_i)} 
    \Big\{ (3-2y)\,\F_1(P_+,y) \label{eq:gammaU} \nonumber\\
    &&+ 6(1-y)\,\F_2(P_+,y) + y\,\F_3(P_+,y) \Big\}
    \,, \\
   \frac{d\Gamma_s}{dP_+} 
   &=& \frac{\alpha G_F^2 |V_{tb}V^*_{ts}|^2}{32\pi^4}\,U(\mu_h,\mu_i)\,
    (M_B-P_+)^3\,\overline{m}_b^2(\mu_h)\,[C_{7\gamma}^{\rm eff}(\mu_h)]^2\,
    \F_\gamma(P_+) \,, \label{eq:gammaS}\nonumber\\
\end{eqnarray}
where 
\begin{equation} \label{eq:y}
   y = \frac{P_- -P_+}{M_B-P_+}
\end{equation}
with $P_-=E_X+|\vec{P}_X|$ 
is a kinematical variable that is integrated over the available phase space. 
Expressions for the structure functions $\F_i$ valid at next-to-leading order
(NLO) in renormalization-group (RG) improved perturbation theory and including 
first- and second-order power corrections can be found in \cite{Lange:2005yw} 
(see also \cite{Neubert:2004dd,Bauer:2003pi,Bosch:2004th}). Symbolically, they 
are written as $H(\mu_h)\cdot J(\mu_i)\otimes S(\mu_i)$, where $H(\mu_h)$ 
contains matching
corrections at the hard scale $\mu_h$. The jet function $J(\mu_i)$,
which is a perturbative quantity at the intermediate scale $\mu_i$, is
convoluted with a non-perturbative shape function renormalized at that same
scale. Separation of the two scales $\mu_h$ and $\mu_i$ allows for the
logarithms in matching corrections to be small, while logarithms of
the form $\ln\mu_h/\mu_i$, which appear at every
order in perturbation theory, are resummed in a systematic fashion and
give rise to the RG evolution functions $U(\mu_h,\mu_i)$ and
$a_\Gamma(\mu_h,\mu_i)$ \cite{Bosch:2004th}.

The leading-power jet function $J(p^2,\mu_i)$ entering the expressions for 
$\F_i$ is universal and has been computed at one-loop order in
\cite{Bauer:2003pi,Bosch:2004th}. The two-loop expression for 
$J$ has been obtained apart from a single unknown 
constant in \cite{Neubert:2005nt}, which is the two-loop coefficient of 
the local $\delta(p^2)$ term. (This term was recently calculated in \cite{Becher:2006qw}). 
This constant does not enter in the two-loop result for the weight 
function $W(\Delta,P_+)$ in (\ref{eq:relation}). Due to the
universality of the leading-power jet function, it is possible to
calculate the complete $O(\alpha_s^2(\mu_i))$ corrections to the weight
function. However, the extraction of hard corrections at two-loop order 
would require multi-loop calculations for both decay
processes, which are unavailable at present. As a result, we will be able to 
predict the $\Delta$ and $P_+$ dependence of the weight function 
$W(\Delta,P_+)$ at next-to-next-to-leading order (NNLO) in RG-improved 
perturbation theory, including exact two-loop matching contributions and 
three-loop 
running effects. However, the overall normalization of the weight function 
will have an uncertainty of $O(\alpha_s^2(\mu_h))$ from yet unknown hard 
matching corrections.

The total rate $\Gamma_s(E_*)$ has been calculated in a local operator
product expansion and reads (including only the leading non-perturbative 
corrections) \cite{Neubert:2004dd}
\begin{eqnarray}\label{eq:totalBtoS}
   \Gamma_s(E_*) &=& \frac{\alpha G_F^2 |V_{tb}V^*_{ts}|^2}{32\pi^4}\,
   m_b^3\,\overline{m}_b^2(\mu_h)\,[C_{7\gamma}^{\rm eff}(\mu_h)]^2\,\times\nonumber\\
  && \times|H_s(\mu_h)|^2\,H_\Gamma(\mu_h) \left[ 1 - \frac{\lambda_2}{9m_c^2}\, 
   \frac{C_1(\mu_h)}{C_{7\gamma}^{\rm eff}(\mu_h)} \right] ,
\end{eqnarray}
where $H_s$ is the hard function of $\F_\gamma$, and $H_\Gamma$ contains the 
remaining radiative corrections. We will present an explicit expression for 
this quantity at the end of Section~\ref{sec:results} below. The hadronic 
correction proportional to $\lambda_2/m_c^2$ cancels against an identical 
term in $\F_\gamma$. Apart from two 
powers of the running $b$-quark mass defined in the $\overline{\rm MS}$ 
scheme, which is part of the electromagnetic dipole operator $O_{7\gamma}$ in 
the effective weak Hamiltonian, three more powers of $m_b$ emerge from 
phase-space integrations. To avoid the renormalon
ambiguities of the pole scheme we use a low-scale subtracted quark-mass 
definition for $m_b$. Specifically, we adopt the shape-function mass 
$m_b^{\rm SF}(\mu_*,\mu_*)$ \cite{Bosch:2004th,Neubert:2004sp} defined at a 
subtraction scale $\mu_*=1.5$\,GeV, which relates to the pole mass as
\begin{equation}\label{eq:SFmassToPole}
   m_b^{\rm pole} = m_b^{\rm SF}(\mu_*,\mu_*)
   + \mu_*\,\frac{C_F \alpha_s(\mu_h)}{\pi} + \ldots \,.
\end{equation}
Throughout this chapter we will use $m_b^{\rm SF}(\mu_*,\mu_*)$ as the
$b$-quark mass and refer to it as $m_b$ for brevity. The present value of 
this parameter is $m_b=(4.61\pm 0.06)$\,GeV \cite{Neubert:2005nt}.

\section{Calculation of the weight function}

\subsection{Leading power}

The key strategy for the calculation of the weight function is to make use of 
QCD factorization theorems for the decay distributions on both sides of
(\ref{eq:relation}) and to arrange the resulting, factorized expressions 
such that they are both given as integrals over the shape function, 
$\int d\hat\omega\,\hat S(\hat\omega)\,g_i(\hat\omega)$, with different 
functions $g_i$ for the
left-hand and right-hand sides. Relation~(\ref{eq:relation}) 
can then be enforced by matching $g_{\rm LHS}$ to $g_{\rm RHS}$. 
Following this procedure, we find for the integrated $P_+$ spectrum in
$\bar B\to X_u\,l^-\bar\nu$ decay after a series of integration interchanges
\begin{eqnarray}\label{eq:uSwitch}
   \int\limits_0^\Delta dP_+\,\frac{d\Gamma_u}{dP_+} 
   \hspace{-0.5em}&\propto&\hspace{-0.5em} \int\limits_0^1\!dy\,y^{-2a} H_u(y) 
    \int\limits_0^\Delta\!dP_+\,(M_B-P_+)^5
    \int\limits_0^{P_+}\!d\hat\omega\,y m_b\,J(ym_b(P_+ -\hat\omega))\,
    \hat S(\hat\omega) \nonumber \\
   &=& \int\limits_0^\Delta\!d\hat\omega\,\hat S(\hat\omega) 
    \int\limits_0^{M_B-\hat\omega}\!dq\,5q^4 
    \int\limits_0^1\!dy\,y^{-2a} H_u(y)\,
    j\left(\ln \frac{m_b(\Delta_q-\hat\omega)}{\mu_i^2} + \ln y \right) ,\nonumber \\
\end{eqnarray}
where $\Delta_q=\mbox{min}(\Delta,M_B-q)$, and \cite{Neubert:2005nt}
\begin{equation}\label{jdef}
   j\left(\ln \frac{Q^2}{\mu_i^2},\mu_i \right)
   = \int_0^{Q^2}\!dp^2\,J(p^2,\mu_i)
\end{equation}
is the integral over the jet function. For the sake of transparency, we 
often suppress the explicit dependence on $\mu_i$ and $\mu_h$ when it is clear 
at which scales the relevant quantities are defined. The function $H_u$
is a linear combination of the hard functions entering the structures
$\F_i$ in (\ref{eq:gammaU}), which in the notation of \cite{Lange:2005yw} is
given by
\begin{equation}
   H_u(y,\mu_h) = 2y^2(3-2y)\,H_{u1}(y,\mu_h) + 12y^2(1-y)\,H_{u2}(y,\mu_h) 
   + 2y^3\,H_{u3}(y,\mu_h) \,.
\end{equation}
RG resummation effects build up the factor $y^{-2a}$ in (\ref{eq:uSwitch}), 
where $a\equiv a_\Gamma(\mu_h,\mu_i)$ is the value of the RG-evolution 
function
\begin{equation}\label{adef}
   a_\Gamma(\mu_h,\mu_i)
   = \int_{\mu_i}^{\mu_h}\!\frac{d\mu}{\mu}\,\Gamma_{\rm cusp}(\alpha_s(\mu))
   = - \int_{\alpha_s(\mu_h)}^{\alpha_s(\mu_i)}\!d\alpha\,
    \frac{\Gamma_{\rm cusp}(\alpha)}{\beta(\alpha)} \,,
\end{equation}
which depends only on the cusp anomalous dimension 
\cite{Korchemsky:wg,Korchemskaya:1992je}. The quantity $a$ has its origin in 
the geometry of time-like and light-like Wilson lines underlying the 
kinematics of inclusive $B$ decays into light particles. Our definition is 
such that $a$ is a positive number for $\mu_h>\mu_i$ and vanishes in the limit 
$\mu_h\to\mu_i$. We find it convenient to treat the function 
$a_\Gamma(\mu_h,\mu_i)$ as a running ``physical'' 
quantity, much like $\alpha_s(\mu)$ or $\overline{m}_b(\mu)$. Since the 
cusp anomalous dimension is known to three-loop order \cite{Moch:2004pa}, the 
value of $a$ can be determined very accurately. Note that three-loop accuracy 
in $a$ (as well as in the running coupling $\alpha_s$) is required for a 
consistent calculation of the weight function at NNLO. The corresponding 
expression is
\begin{eqnarray}
   a = a_\Gamma(\mu_h,\mu_i) &=& \frac{\Gamma_0}{2\beta_0}\,\Bigg\{
    \ln\frac{\alpha_s(\mu_i)}{\alpha_s(\mu_h)} 
    + \left( \frac{\Gamma_1}{\Gamma_0} - \frac{\beta_1}{\beta_0} \right)
    \frac{\alpha_s(\mu_i) - \alpha_s(\mu_h)}{4\pi} \nonumber\\
   &&\mbox{}+ \left[ \frac{\Gamma_2}{\Gamma_0} - \frac{\beta_2}{\beta_0}
    - \frac{\beta_1}{\beta_0}
    \left( \frac{\Gamma_1}{\Gamma_0} - \frac{\beta_1}{\beta_0} \right) \right]
    \frac{\alpha_s^2(\mu_i) - \alpha_s^2(\mu_h)}{32\pi^2} + \dots \Bigg\} \,,\nonumber\\
\end{eqnarray}
where the expansion coefficients $\Gamma_n$ and $\beta_n$ of the cusp anomalous
dimension and $\beta$-function can be found in appendix \ref{apx:Sudakovs}.

Instead of the jet function $J$ itself, we need its integral 
$j(\ln Q^2/\mu_i^2,\mu_i)$ in the second line of (\ref{eq:uSwitch}).
Since the jet function has a
perturbative expansion in terms of ``star distributions'', which are
logarithmically sensitive to the upper limit of integration
\cite{DeFazio:1999sv}, it follows that $j(L,\mu_i)$ is a simple polynomial in
$L$ at each order in perturbation theory. The two-loop result for this
quantity has recently been computed by solving the integro-differential 
evolution equation for the jet function \cite{Neubert:2005nt}. An unknown 
integration constant of $O(\alpha_s^2 L^0)$ does not enter the expression for 
the weight function.

We now turn to the right-hand side of (\ref{eq:relation}) and follow
the same steps that lead to (\ref{eq:uSwitch}). It is helpful to make
an ansatz for the leading-power contribution to the weight function, 
$W^{(0)}(\Delta,P_+)$, where the 
dependence on $\Delta$ is solely given via an upper limit of integration. To 
this end, we define a function $f(k)$ through
\begin{equation}\label{eq:fdef}
   W^{(0)}(\Delta,P_+)\propto \frac{1}{(M_B-P_+)^3} 
   \int_0^{\Delta-P_+}\!dk\,f(k)\,(M_B-P_+-k)^5 \,.
\end{equation}
This allows us to express the weighted integral over the $\bar B\to X_s\gamma$ 
photon spectrum as
\begin{eqnarray}\label{eq:sSwitch}
   &&\int\limits_0^\Delta\!dP_+\,\frac{d\Gamma_s}{dP_+}\,W^{(0)}(\Delta,P_+)
   \propto \nonumber \\
   &\propto&\int\limits_0^\Delta\!dP_+\,(M_B-P_+)^3\,W^{(0)}(\Delta,P_+)
    \int\limits_0^{P_+}\!d\hat\omega\,m_b\,J(m_b(P_+ -\hat\omega))\,
    \hat S(\hat\omega) \nonumber \\ 
   &\propto& \int\limits_0^\Delta\!d\hat\omega\,\hat S(\hat\omega) 
    \int\limits_0^{M_B-\hat\omega}\!dq\,5q^4 
    \int\limits_0^{\Delta_q-\hat\omega}\!dk\,f(k)\, 
    j\left(\ln\frac{m_b(\Delta_q-\hat\omega-k)}{\mu_i^2} \right) . \qquad
\end{eqnarray}
Note that the jet function $J$ (and with it $j$) is the same in semileptonic 
and radiative decays. The difference is that the argument of the jet 
function in (\ref{eq:uSwitch}) contains an extra factor of $y$, which is 
absent in (\ref{eq:sSwitch}). Comparing these two relations leads us to the 
matching condition
\begin{equation} \label{eq:fmatch}
   \int_0^1\!dy\, y^{-2a}\,H_u(y)\,
   j\left(\ln\frac{m_b\Omega}{\mu_i^2}+\ln y \right)
   \stackrel{!}{=} \int_0^\Omega\!dk\,f(k)\,
   j\left(\ln\frac{m_b(\Omega-k)}{\mu_i^2} \right) ,
\end{equation}
which holds to all orders in perturbation theory and allows for the
calculation of $W^{(0)}(\Delta,P_+)$ via (\ref{eq:fdef}). The main feature
of this important relation is that the particular value of
$\Omega$ is irrelevant for the determination of $f(k)$. It follows
that, as was the case for the jet function $J$, the perturbative
expansion of $f(k)$ in $\alpha_s(\mu_i)$ at the intermediate scale
involves star distributions, and $W^{(0)}(\Delta,P_+)$ depends
logarithmically on $(\Delta-P_+)$. At two-loop order it suffices to
make the ansatz
\begin{eqnarray}\label{eq:fexpansion}
  && f(k) \quad\propto\quad \delta(k) + C_F\frac{\alpha_s(\mu_i)}{4\pi}
    \left[ c_0^{(1)}\,\delta(k) 
    + c_1^{(1)} \left( \frac{1}{k} \right)_*^{[\mu_i^2/m_b]} \right]  \\
   &&\mbox{}+ C_F\left( \frac{\alpha_s(\mu_i)}{4\pi} \right)^2 
    \left[ c_0^{(2)}\,\delta(k)
    + c_1^{(2)} \left( \frac{1}{k} \right)_*^{[\mu_i^2/m_b]}
    \hspace{-0.5em} + 2c_2^{(2)} \left( \frac{1}{k}\ln\frac{m_b k}{\mu_i^2} 
    \right)_*^{[\mu_i^2/m_b]} \right] + \dots \,, \nonumber
\end{eqnarray}
where the star distributions have the following effect when integrated
with some smooth function $\phi(k)$ over an interval $\Omega$:
\begin{eqnarray}\label{stardistris5}
   \int_0^\Omega\!dk \left( \frac{1}{k} \right)_*^{[\mu_i^2/m_b]} \phi(k)
   &=& \int_0^{\Omega}\!dk\,\frac{\phi(k)-\phi(0)}{k}
    + \phi(0)\,\ln\frac{m_b\Omega}{\mu_i^2} \,, \nonumber\\
   \int_0^\Omega\!dk\,\left( \frac{1}{k}\ln\frac{m_b k}{\mu_i^2} 
    \right)_*^{[\mu_i^2/m_b]} \phi(k)
   &=& \int_0^{\Omega}\!dk\,\frac{\phi(k)-\phi(0)}{k}\,
    \ln\frac{m_b k}{\mu_i^2} \,
    + \frac{\phi(0)}{2}\,\ln^2\frac{m_b\Omega}{\mu_i^2} \,.\nonumber\\
\end{eqnarray}

A sensitivity to the hard scale $\mu_h$ enters into $f(k)$ via the
appearance of $H_u(y,\mu_h)$ in (\ref{eq:fmatch}). Because of the
polynomial nature of $j(L,\mu_i)$, all we ever need are moments of the hard 
function with respect to $\ln y$. We thus define the master integrals
\begin{equation}\label{eq:Tn}
   T_n(a,\mu_h)\equiv \int_0^1\!dy\,y^{-2a}\,H_u(y,\mu_h)\,\ln^n y \,; \qquad 
   h_n(a,\mu_h) = \frac{T_n(a,\mu_h)}{T_0(a,\mu_h)} \,,
\end{equation}
which can be calculated order by order in $\alpha_s(\mu_h)$. Therefore,
the coefficients $c_k^{(n)}$ of the perturbative expansion in
(\ref{eq:fexpansion}) at the intermediate scale have the (somewhat
unusual) feature that they possess themselves an expansion in
$\alpha_s(\mu_h)$. This is a consequence of the fact that, unlike the
differential decay rates (\ref{eq:gammaU}) and (\ref{eq:gammaS}), the
weight function itself does not obey a simple factorization
formula, in which the hard correction can be factored out. Rather, as can be 
seen from (\ref{eq:fmatch}), it is a convolution of the type 
$W=H(\mu_h)\otimes J(\mu_i)$. 
To one-loop accuracy, the hard function $H_u$ reads
\begin{eqnarray}\label{eq:Hu}
   H_u(y,\mu_h) 
   &=& 2y^2(3-2y)\,\Bigg[ 1 + \frac{C_F\alpha_s(\mu_h)}{4\pi}\,
    \bigg( -4\ln^2\frac{ym_b}{\mu_h} + 10\ln\frac{ym_b}{\mu_h} - 4\ln y 
    \nonumber\\
   &&\hspace{2.3cm}\mbox{}- 4 L_2(1-y) - \frac{\pi^2}{6} - 12 \bigg) \Bigg] 
    - \frac{C_F\alpha_s(\mu_h)}{\pi}\,3y^2\ln y \,.\nonumber\\
\end{eqnarray}
Explicit expressions for the quantities $T_0$, $c_k^{(n)}$, and $h_n$
entering the distribution function $f(k)$ will be given below.

\subsection{Subleading power}
\label{sec:corr}

Power corrections to the weight function can be 
extracted from the corresponding contributions to the two $P_+$ spectra in 
(\ref{eq:gammaU}) and (\ref{eq:gammaS}). There exists a class of power 
corrections associated with the phase-space prefactors $(M_B-P_+)^n$ in these 
relations, whose effects are treated exactly in our approach, see e.g.\ 
(\ref{eq:fdef}). This is 
important, because these phase-space corrections increase in magnitude as the 
kinematical range $\Delta$ over which the two spectra are integrated is 
enlarged. One wants to make $\Delta$ as large as 
experimentally possible so as to increase statistics and justify the 
assumption of quark-hadron duality, which underlies the theory of inclusive
$B$ decays. 

The remaining power corrections fall into two distinct classes:
kinematical corrections that start at order $\alpha_s$ and come with  
the leading shape function \cite{Kagan:1998ym,DeFazio:1999sv}, and 
hadronic power corrections that start at tree level and involve new, 
subleading shape functions 
\cite{Bauer:2001mh,Bauer:2002yu,Bosch:2004cb,Leibovich:2002ys,Neubert:2002yx,Lee:2004ja,
Beneke:2004in,Burrell:2003cf}. Because different 
combinations of these hadronic functions enter in $\bar B\to X_u\,l^-\bar\nu$ 
and $\bar B\to X_s\gamma$ decays, it is impossible to eliminate their 
contributions in relations such as (\ref{eq:relation}). As a result, at 
$O(\Lambda_{\rm QCD}/m_b)$ there are non-perturbative hadronic uncertainties
in the calculation of the weight function $W(\Delta,P_+)$, which need to be
estimated before a reliable extraction of $|V_{ub}|$ can be performed. For
the case of the charged-lepton energy spectrum and the hadronic invariant mass
spectrum, this aspect has been discussed previously in 
\cite{Bauer:2002yu,Neubert:2002yx} and \cite{Burrell:2003cf}, respectively.

Below, we will include power corrections to first order in 
$\Lambda_{\rm QCD}/m_b$. Schematically, the subleading corrections to the 
right-hand side of (\ref{eq:relation}) are computed according to
$\Gamma_u^{(1)}\sim W^{(0)}\otimes d\Gamma_s^{(1)}/dP_+%
+W^{(1)}\otimes d\Gamma_s^{(0)}/dP_+$, where the superscripts indicate the
order in $1/m_b$ power counting. The power corrections to the weight function, 
denoted by $W^{(1)}$, are derived from the mismatch in the power corrections 
to the two decay spectra. The kinematical power corrections to the two spectra 
are known at $O(\alpha_s)$, without scale separation. We assign a coupling 
$\alpha_s(\bar\mu)$ to these terms, where the scale $\bar\mu$ will be chosen
of order the intermediate scale \cite{Lange:2005yw}. At first subleading power 
the leading shape function is convoluted with either a constant or a
single logarithm of the form $\ln[(P_+ -\hat\omega)/(M_B-P_+)]$, and we have
(with $n=0,1$)
\begin{eqnarray}
   \int_0^\Delta\!dP_+\,\frac{d\Gamma_u}{dP_+} 
   &\ni& \alpha_s(\bar\mu) \int_0^\Delta\!dP_+\,(M_B-P_+)^4
    \int_0^{P_+}\!d\hat\omega\,\hat S(\hat\omega)\,
    \ln^n\frac{P_+ -\hat\omega}{M_B-P_+} \nonumber \\
   &=& \alpha_s(\bar\mu) \int_0^\Delta\!d\hat\omega\,\hat S(\hat\omega) 
    \int_0^{\Delta-\hat\omega}\!dk\,(M_B-\hat\omega-k)^4 
    \ln^n\frac{k}{M_B-\hat\omega-k} \,, \nonumber \\
\end{eqnarray}
and similarly for the photon spectrum. On the other hand, the weighted
integral in (\ref{eq:relation}) also contains terms where the photon
spectrum is of leading power and the weight function of subleading
power,
\begin{equation}
   \int_0^\Delta\!dP_+\,\frac{d\Gamma_s}{dP_+}\,W^{\rm kin(1)}(\Delta,P_+)
   \ni \int_0^\Delta\!dP_+\,(M_B-P_+)^3\,\hat S(P_+)\,
   W^{\rm kin(1)}(\Delta,P_+) \,.
\end{equation}
Therefore the kinematical corrections to the weight function must have the 
form 
\begin{eqnarray}\label{eq:wkin1}
   W^{\rm kin(1)}(\Delta,P_+)&\propto& \frac{\alpha_s(\bar\mu)}{(M_B-P_+)^3} 
   \int_0^{\Delta-P_+}\!dk\,(M_B-P_+ -k)^4 \times\nonumber \\
   &&\times\left( A + B\,\ln\frac{k}{M_B-P_+ -k} \right) ,
\end{eqnarray}
and a straightforward calculation determines the coefficients $A$ and $B$.  
The \mbox{} hadronic power corrections to the weight function, $W^{\rm hadr(1)}$, 
can be expressed in terms of the subleading shape functions 
$\hat t(\hat\omega)$, $\hat u(\hat\omega)$, and $\hat v(\hat\omega)$ defined 
in \cite{Bosch:2004cb}. These terms are known at tree level only, and at
this order their contribution to the weight function can be derived using the 
results of \cite{Lange:2005yw}.

\section{Results}
\label{sec:results}

Including the first-order power corrections and the exact phase-space
factors, the weight function takes the form
\begin{eqnarray}\label{eq:masterform}
   &&W(\Delta,P_+)
   = \frac{G_F^2 m_b^3}{192\pi^3}\,T_0(a,\mu_h)\,H_\Gamma(\mu_h)\,
    (M_B-P_+)^2 \\
   &\times& \Bigg\{ 1 + \frac{C_F\alpha_s(\mu_i)}{4\pi}
    \left[ c_0^{(1)} + c_1^{(1)} \left( \ln \frac{m_b(\Delta-P_+)}{\mu_i^2} 
    - p_1(\delta) \right) \right] \nonumber\\
   &&\hspace{0.55cm}\mbox{}+ C_F \left( \frac{\alpha_s(\mu_i)}{4\pi} \right)^2
    \left[ c_0^{(2)} + c_1^{(2)} \left( \ln \frac{m_b(\Delta-P_+)}{\mu_i^2} 
    - p_1(\delta) \right) \right. \nonumber\\
   &&\hspace{1.6cm}\left.
    \mbox{}+ c_2^{(2)} \left( \ln^2\frac{m_b(\Delta-P_+)}{\mu_i^2} 
    - 2 p_1(\delta) \ln\frac{m_b(\Delta-P_+)}{\mu_i^2} 
    + 2 p_2(\delta) \right) \right] \nonumber\\
   &&\mbox{}+ \frac{\Delta-P_+}{M_B-P_+}\,\frac{C_F\alpha_s(\bar\mu)}{4\pi}
    \left[ A(a,\mu_h)\,I_A(\delta) + B(a,\mu_h)\,I_B(\delta) \right] \nonumber\\
   &&\mbox{}+ \frac{1}{M_B-P_+}\,\frac{1}{2(1-a)(3-a)}\,
    \bigg[ 4(1-a)(\bar\Lambda-P_+) + 2(4-3a)\,\frac{\hat t(P_+)}{\hat S(P_+)}
    \nonumber\\
   &&\mbox{}+ (4-a)\,\frac{\hat u(P_+)}{\hat S(P_+)}
    + (8-13a+4a^2)\,\frac{\hat v(P_+)}{\hat S(P_+)} \bigg]
    + \frac{m_s^2}{M_B-P_+}\,\frac{\hat S'(P_+)}{\hat S(P_+)} 
    + \dots \Bigg\} \,, \nonumber
\end{eqnarray}
where $\bar\Lambda=M_B-m_b$ is the familiar mass parameter of heavy-quark 
effective theory, and $\delta=(\Delta-P_+)/(M_B-P_+)$. 
The first line denotes an overall normalization, 
the next three lines contain the leading-power contributions, and
the remaining expressions enter at subleading power. The different terms in
this result will be discussed in the remainder of this section.

For the leading-power terms in the above result we have accomplished a 
complete separation of hard and intermediate (hard-collinear) contributions 
to the weight function in a way consistent with the factorization formula
$W=H\otimes J$ mentioned in the previous section. The 
universality of the shape function, which encodes the soft physics in both 
$\bar B\to X_s\gamma$ and $\bar B\to X_u\,l^-\bar\nu$ decays, implies that 
the weight function is insensitive to physics below the intermediate scale
$\mu_i\sim\sqrt{m_b\Lambda_{\rm QCD}}$. In particular, quark-hadron duality 
ensures that the small region in phase space where the argument 
$m_b(\Delta-P_+)$ of the logarithms scales as $\Lambda_{\rm QCD}^2$ or 
smaller does not need special consideration. 
At a technical level, this can be seen by noting that the jet function is the
discontinuity of the collinear quark propagator in soft-collinear effective 
theory \cite{Bauer:2001yt,Bosch:2004th}, and so the $P_+$ integrals can be 
rewritten as a contour integral in the complex $p^2$ plane along a circle of 
radius $m_b\Delta\sim\mu_i^2$.

\subsection{Leading power}

The leading-power corrections in the curly brackets in (\ref{eq:masterform})
are determined completely 
at NNLO in RG-improved perturbation theory, including three-loop running 
effects via the quantity $a$ in (\ref{adef}), and two-loop matching 
corrections at the scale $\mu_i$ as indicated above. To this end we need
expressions for the one-loop coefficients $c_n^{(1)}$ including terms of
$O(\alpha_s(\mu_h))$, while the two-loop coefficients $c_n^{(2)}$ are needed 
at leading order only. We find
\begin{equation}\label{eq:1-loopCoeffs}
   c_0^{(1)} = -3 h_1(a,\mu_h) 
    \left[ 1 - \frac{C_F\alpha_s(\mu_h)}{\pi}\,\frac{4\mu_*}{3m_b} \right]
    + 2 h_2(a,\mu_h) \,, \qquad
   c_1^{(1)} = 4 h_1(a,\mu_h) \,,
\end{equation}
and 
\begin{eqnarray}\label{eq:2-loopCoeffs}
   c_0^{(2)}
   &=& C_F\,\Bigg[ \left( -\frac32 + 2\pi^2 - 24\zeta_3 \right) h_1(a,\mu_h) 
    + \left( \frac92 - \frac{4\pi^2}{3} \right) h_2(a,\mu_h) \nonumber\\
    &&- 6 h_3(a,\mu_h) + 2 h_4(a,\mu_h) \Bigg] \nonumber\\
   &&\mbox{}+ C_A \left[ \left( -\frac{73}{9} + 40\zeta_3 \right) h_1(a,\mu_h)
    + \left( \frac83 - \frac{2\pi^2}{3} \right) h_2(a,\mu_h) \right] 
    \nonumber\\
   &&\mbox{}+ \beta_0 \left[
    \left( -\frac{247}{18} + \frac{2\pi^2}{3} \right) h_1(a,\mu_h)
    + \frac{29}{6}\,h_2(a,\mu_h) - \frac23\,h_3(a,\mu_h) \right] , \nonumber\\
   c_1^{(2)} 
   &=& C_F\,\Big[ -12 h_2(a,\mu_h) + 8 h_3(a,\mu_h) \Big]
    + C_A \left[ \left( \frac{16}{3} - \frac{4\pi^2}{3} \right) h_1(a,\mu_h)
    \right] \nonumber\\
   &&\mbox{}+ \beta_0 \left[ \frac{29}{3}\,h_1(a,\mu_h) - 2 h_2(a,\mu_h)
    \right] , \nonumber \\
   c_2^{(2)} &=& 8 C_F\,h_2(a,\mu_h) - 2\beta_0\,h_1(a,\mu_h) \,.
\end{eqnarray}
As always $C_F=4/3$, $C_A=3$, and $\beta_0=11-2 n_f/3$ is the first 
coefficient of the QCD $\beta$-function. The term proportional to $\mu_*$ in 
the expression for $c_0^{(1)}$ arises because of the elimination of the 
pole mass in favor of the shape-function mass, see
(\ref{eq:SFmassToPole}). Since the logarithms $\ln[m_b(\Delta-P_+)/\mu_i^2]$ 
in (\ref{eq:masterform}) contain $m_b$, all coefficients except $c_n^{(n)}$ 
receive such contributions. However, to two-loop order only $c_0^{(1)}$ is 
affected.

Next, the corresponding expressions for the hard matching coefficients
$h_i$ are calculated from (\ref{eq:Tn}). To the required order they read
\begin{eqnarray}
   h_1(a,\mu_h)
   &=& - \frac{15-12a+2a^2}{2(2-a)(3-a)(3-2a)} \nonumber\\
   &&\mbox{}+ \frac{C_F\alpha_s(\mu_h)}{4\pi}\,\Bigg[
    - \frac{2(189-318a+192a^2-48a^3+4a^4)}{(2-a)^2(3-a)^2(3-2a)^2}\,
    \ln\frac{m_b}{\mu_h} \nonumber\\
   &&+\frac{2331-5844a+5849a^2-2919a^3+726a^4-72a^5}{(2-a)^3(3-a)^2(3-2a)^3}\nonumber\\
    &&- 4\psi^{(2)}(3-2a) \Bigg] + \dots \,, \nonumber
\end{eqnarray}
\begin{eqnarray}
   &&h_2(a,\mu_h)
   =  \frac{69-90a+36a^2-4a^3}{2(2-a)^2(3-a)(3-2a)^2} \nonumber\\
   &&+ \frac{C_F\alpha_s(\mu_h)}{4\pi}\,\Bigg[
    \frac{2(1692-3699a+3138a^2-1272a^3+240a^4-16a^5)}{(2-a)^3(3-a)^2(3-2a)^3}
    \,\ln\frac{m_b}{\mu_h} \nonumber\\
   &&\quad\mbox{}-
    \frac{46521-140064a+175479a^2-117026a^3+43788a^4-8712a^5+720a^6}
         {2(2-a)^4(3-a)^2(3-2a)^4} \nonumber\\
   &&\quad\mbox{}+ \frac{4(15-12a+2a^2)}{(2-a)(3-a)(3-2a)}\,\psi^{(2)}(3-2a)
    - 4\psi^{(3)}(3-2a) \Bigg] + \dots \,, \nonumber\\
   &&h_3(a,\mu_h)
   = - \frac{3(303-552a+360a^2-96a^3+8a^4)}{4(2-a)^3(3-a)(3-2a)^3} + \dots
    \,, \nonumber\\
   &&h_4(a,\mu_h)
   = \frac{3(1293-3030a+2760a^2-1200a^3+240a^4-16a^5)}{2(2-a)^4(3-a)(3-2a)^4}
    + \dots \,,\nonumber\\
\end{eqnarray}
where $\psi^{(n)}(x)$ is the $n$-th derivative of the polygamma
function. Because of the exact treatment of the phase space there are
corrections to the logarithms in (\ref{eq:masterform}), which are 
finite-order polynomials in the small ratio
$\delta=(\Delta-P_+)/(M_B-P_+)$. Explicitly,
\begin{eqnarray}
   p_1(\delta) &=& 5\delta - 5\delta^2 + \frac{10}{3}\,\delta^3
    - \frac54\,\delta^4 + \frac15\,\delta^5 \,, \nonumber\\
   p_2(\delta) &=& 5\delta - \frac52\,\delta^2 + \frac{10}{9}\,\delta^3
    - \frac{5}{16}\,\delta^4 + \frac{1}{25}\,\delta^5 \,.
\end{eqnarray}
This concludes the discussion of the leading-power expression for the
weight function. 

\subsection{Subleading power}
\label{sec:power}

The procedure for obtaining the kinematical power corrections to the weight 
function has been discussed in Section~\ref{sec:corr}. For the coefficients 
$A$ and $B$ in (\ref{eq:masterform}) we find
\begin{eqnarray}
   A(a,\mu_h)
   &=& \frac{-388+702a-429a^2+123a^3-34a^4+8a^5}{2(1-a)^2(2-a)(3-a)(3-2a)} \nonumber\\
    &&+ \left(\frac13- \frac49\,\ln\frac{m_b}{m_s} \right) 
    \frac{[C_{8g}^{\rm eff}(\mu_h)]^2}{[C^{\rm eff}_{7\gamma}(\mu_h)]^2}
    \nonumber\\
   &&- \frac{10}{3}\,
    \frac{C_{8g}^{\rm eff}(\mu_h)}{C^{\rm eff}_{7\gamma}(\mu_h)}
    + \frac83 \left( 
    \frac{C_1(\mu_h)}{C^{\rm eff}_{7\gamma}(\mu_h)} 
    - \frac13\,\frac{C_1(\mu_h)\,C_{8g}^{\rm eff}(\mu_h)}
    {[C^{\rm eff}_{7\gamma}(\mu_h)]^2}\,\right) g_1(z)\nonumber\\
    &&- \frac{16}{9}\,\frac{[C_1(\mu_h)]^2}{[C^{\rm eff}_{7\gamma}(\mu_h)]^2}\, 
    g_2(z) \,, \nonumber\\
   B(a,\mu_h) 
   &=& - \frac{2(8+a)}{(1-a)(3-a)}
    - \frac29\,\frac{[C_{8g}^{\rm eff}(\mu_h)]^2}
    {[C^{\rm eff}_{7\gamma}(\mu_h)]^2} \,.
\end{eqnarray}
Here $C_i(\mu_h)$ denote the (effective) Wilson coefficients of the
relevant operators in the effective weak Hamiltonian, which are real
functions in the Standard Model. The variable $z=(m_c/m_b)^2$ enters
via charm-loop penguin contributions to the hard function of the 
$\bar B\to X_s\gamma$ photon spectrum \cite{Kagan:1998ym}, and
\begin{equation}
   g_1(z) = \int_0^1\!dx\,x\,\mbox{Re} \left[\,
    \frac{z}{x}\,G\!\left(\frac{x}{z}\right) + \frac12 \,\right] , \qquad
   g_2(z) = \int_0^1\!dx\,(1-x) \left|\,\frac{z}{x}\,
    G\!\left(\frac{x}{z}\right) + \frac12\,\right|^2 ,
\end{equation}
with
\begin{equation}
   G(t) = \left\{ \begin{array}{ll}
   -2\arctan^2\!\sqrt{t/(4-t)} & ;~ t<4 \,, \\[0.1cm]
   2 \left( \ln\!\Big[(\sqrt{t}+\sqrt{t-4})/2\Big]
   - \displaystyle\frac{i\pi}{2} \right)^2 & ;~ t\ge 4 \,.
  \end{array} \right.
\end{equation}
Furthermore we need the integrals over $k$ in
(\ref{eq:wkin1}), which encode the phase-space corrections. They give rise to 
the functions
\begin{eqnarray}
   I_A(\delta) &=& 1 - 2\delta + 2\delta^2 - \delta^3 + \frac15\,\delta^4
    \,, \nonumber\\
   I_B(\delta) &=& I_A(\delta)\,\ln\frac{\delta}{1-\delta}
    + \frac{\ln(1-\delta)}{5\delta} - \frac45 + \frac35\,\delta
    - \frac{4}{15}\,\delta^2 + \frac{1}{20}\,\delta^3 \,.
\end{eqnarray}

The hadronic power corrections come from subleading shape functions
in the theoretical expressions for the two decay rates. We give their
tree-level contributions to the weight function in the last two lines
of (\ref{eq:masterform}), where $\hat S$ denotes the leading
shape function, and $\hat t, \hat u,\hat v$ are subleading shape
functions as defined in \cite{Bosch:2004cb}. For completeness, we also include 
a contribution proportional to $m_s^2$ resulting from finite-mass effects 
in the strange-quark propagator in $\bar B\to X_s\gamma$ decays. For 
$m_s=O(\Lambda_{\rm QCD})$ these effects are formally of the same order as 
other subleading shape-function contributions \cite{Chay:2005ck}, although
numerically they are strongly suppressed. The appearance of subleading shape
functions introduces an irreducible hadronic uncertainty to a $|V_{ub}|$
determination via (\ref{eq:relation}). In practice, this uncertainty
can be estimated by adopting different models for the subleading shape
functions. This will be discussed in detail in Section~\ref{sec:SSF} below. 
Until then, let us use a ``default model'', in which we 
assume the functional forms of the subleading shape functions 
$\hat t(\hat\omega)$, $\hat u(\hat\omega)$, and $\hat v(\hat\omega)$ to be 
particular linear combinations of the functions $\hat S'(\hat\omega)$ and 
$(\bar\Lambda-\hat\omega)\,\hat S(\hat\omega)$. These combinations are chosen 
in such a way that the 
results satisfy the moment relations derived in \cite{Bosch:2004cb}, and that 
all terms involving the parameter $\bar\Lambda$ cancel in the expression 
(\ref{eq:masterform}) for the weight function for any value of $a$. These 
requirements yield
\begin{equation}\label{defaultmodel}
   \hat t\to-\frac34\,(\bar\Lambda-\hat\omega)\,\hat S
    - \left( \lambda_2 + \frac{\lambda_1}{4} \right) \hat S' \,, \quad
   \hat u\to\frac12\,(\bar\Lambda-\hat\omega)\,\hat S
    + \frac{5\lambda_1}{6}\,\hat S' \,, \quad
   \hat v\to\lambda_2\,\hat S' \,,
\end{equation}
and the last two lines inside the large bracket in the expression 
(\ref{eq:masterform}) for the weight function simplify to
\begin{equation}\label{eq:SSFdefault}
   - \frac{\Lambda_{\rm SSF}^2(a)}{M_B-P_+}\,
   \frac{\hat S'(P_+)}{\hat S(P_+)} 
   ~\widehat{=} - \frac{\Lambda_{\rm SSF}^2(a)}{M_B-P_+}\,\delta(P_+ -\Delta)
    - \frac{4\Lambda_{\rm SSF}^2(a)}{(M_B-P_+)^2} \,,
\end{equation}
where 
\begin{equation}
   \Lambda_{\rm SSF}^2(a)\equiv - \frac{(2+a)\,\lambda_1}{3(1-a)(3-a)}
   + \frac{a(7-4a)\,\lambda_2}{2(1-a)(3-a)} - m_s^2 \,.
\end{equation}
Here $\lambda_1$ and $\lambda_2=\frac14(M_{B^*}^2-M_B^2)$ are hadronic 
parameters describing certain $B$-meson matrix elements in heavy-quark 
effective theory \cite{Falk:1992wt}. The strange-quark mass is a running 
mass evaluated at a scale typical for the final-state hadronic jet, for which 
we take 1.5\,GeV.
As mentioned above, the numerical effect of the strange-quark mass correction 
is small. For typical values of the parameters, it reduces the result for
$\Lambda_{\rm SSF}^2$ by about 10\% or less.
The expression on the right-hand side in (\ref{eq:SSFdefault}) is equivalent
to that on the left-hand side after the integration with the photon spectrum
in (\ref{eq:relation}) has been performed. It has been derived using the fact 
that the normalized photon spectrum is proportional to the shape function 
$\hat S(P_+)$ at leading order. Note that the second term in the final formula 
is power suppressed with respect to the first one. It results from our exact
treatment of phase-space factors and thus is kept for consistency.

\subsection{Normalization}

Finally, let us present explicit formulae for the overall normalization factor 
in (\ref{eq:masterform}). The new ingredient here is the factor $T_0$,
which is defined in (\ref{eq:Tn}). At one-loop order we find
\begin{eqnarray}
  &&T_0(a,\mu_h)= 
   = \frac{2(3-a)}{(2-a)(3-2a)}\,\Bigg\{ 1
    - \frac{C_F\alpha_s(\mu_h)}{4\pi} \Bigg[ 4 \ln^2 \frac{m_b}{\mu_h}\nonumber \\
   && - \frac{2(120-159a+69a^2-10a^3)}{(2-a)(3-a)(3-2a)}\,\ln \frac{m_b}{\mu_h}
    \nonumber \\
   &&\mbox{}+
    \frac{1539-3845a+3842a^2-1920a^3+480a^4-48a^5}{(2-a)^2(3-a)(3-2a)^2} 
    + 4 \psi^{(1)}(3-2a) + \frac{\pi^2}{6} \Bigg] \Bigg\} \,.  \nonumber \\
\end{eqnarray}
When the product of $T_0$ with the quantity \cite{Neubert:2004dd}
\begin{eqnarray}\label{Hgamma}
   H_\Gamma(\mu_h)
   &=& 1 + \frac{C_F\alpha_s(\mu_h)}{4\pi}\,
    \Bigg[ 4\ln^2\frac{m_b}{\mu_h} - 10\ln\frac{m_b}{\mu_h} 
    + 7 - \frac{7\pi^2}{6} + \frac{12\mu_*}{m_b} \nonumber \\
   &&\mbox{}- 2\ln^2\delta_* - (7+4\delta_*-\delta_*^2)\,\ln\delta_* 
    + 10\delta_* + \delta_*^2 - \frac23\,\delta_*^3 \nonumber \\
   &&\mbox{}+ \frac{[C_1(\mu_h)]^2}{[C^{\rm eff}_{7\gamma}(\mu_h)]^2}\, 
    \hat f_{11}(\delta_*)
    + \frac{C_1(\mu_h)}{C^{\rm eff}_{7\gamma}(\mu_h)}\,\hat f_{17}(\delta_*)
    + \frac{C_1(\mu_h)\,C_{8g}^{\rm eff}(\mu_h)}%
           {[C^{\rm eff}_{7\gamma}(\mu_h)]^2}\,\hat f_{18}(\delta_*)
    \nonumber\\
   &&\mbox{}+ \frac{C_{8g}^{\rm eff}(\mu_h)}{C^{\rm eff}_{7\gamma}(\mu_h)}\,
    \hat f_{78}(\delta_*)
    + \frac{[C_{8g}^{\rm eff}(\mu_h)]^2}{[C^{\rm eff}_{7\gamma}(\mu_h)]^2}\,
    \hat f_{88}(\delta_*) \Bigg]
\end{eqnarray}
from the total $\bar B\to X_s\gamma$ decay rate is consistently expanded to 
$O(\alpha_s(\mu_h))$, the double logarithm cancels out. Here 
$\delta_*=1-2E_*/m_b=0.9$, and the functions $\hat f_{ij}(\delta_*)$ capture 
effects from operator mixing.

\section{Numerical results}
\label{sec:num}

We are now in a position to explore the phenomenological implications of our
results. We need as inputs the heavy-quark 
parameters $\lambda_2=0.12$\,GeV$^2$, $\lambda_1=(-0.25\pm 0.10)$\,GeV$^2$, 
and the quark masses $m_b=(4.61\pm 0.06)$\,GeV \cite{Neubert:2005nt}, 
$m_s=(90\pm 25)$\,MeV \cite{Gamiz:2004ar,Aubin:2004ck}, and
$m_c/m_b=0.222\pm 0.027$ \cite{Neubert:2004dd}. Here $m_b$ is defined in the 
shape-function scheme at a scale $\mu_*=1.5$\,GeV, $m_s$ is the running mass 
in the $\overline{\rm MS}$ scheme evaluated at 1.5\,GeV, and $m_c/m_b$ is a 
scale invariant ratio of running masses. Throughout, we use the 3-loop running 
coupling normalized to $\alpha_s(M_Z)=0.1187$, matched to a 4-flavor theory 
at 4.25\,GeV. For the matching scales, we pick the default values 
$\mu_h^{\rm def}=m_b/\sqrt 2$ and 
$\mu_i^{\rm def}=\bar\mu^{\rm def}=1.5$\,GeV, which are motivated by the 
underlying dynamics of inclusive processes in the shape-function region 
\cite{Bosch:2004th,Lange:2005yw}. 

In the remainder of this section we present results for the partial decay rate 
$\Gamma_u(\Delta)$ computed by evaluating the right-hand side
of relation (\ref{eq:relation}). This is 
more informative than to focus on the value of the weight function for a  
particular choice of $P_+$. For the purpose of our discussion we use a simple 
model for the normalized photon spectrum that describes the experimental data 
reasonably well, namely
\begin{equation}\label{model}
   \frac{1}{\Gamma_s}\,\frac{d\Gamma_s}{dP_+}
   = \frac{b^b}{\Gamma(b)\Lambda^b}\,(P_+)^{b-1} 
   \exp\left( -b\,\frac{P_+}{\Lambda} \right)
\end{equation}
with $\Lambda=0.77$\,GeV and $b=2.5$. 

\subsection{Studies of the perturbative expansion}
\label{sec:pert}

\begin{figure}
\begin{center}
\epsfig{file=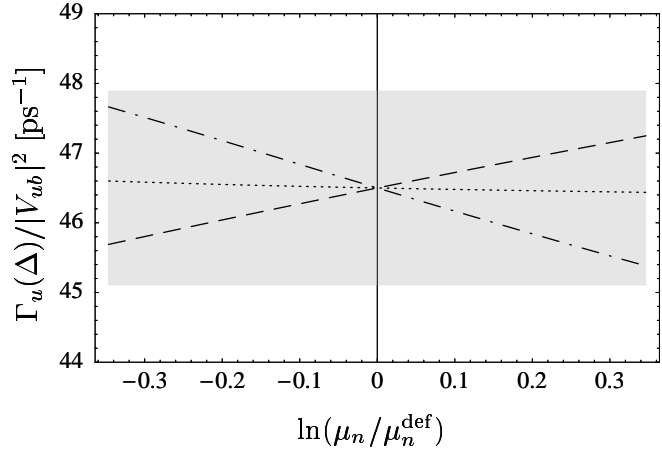,width=8.7cm}
\caption{\label{fig:scales} 
Residual scale dependence of $\Gamma_u(\Delta)$ for $\Delta=0.65$\,GeV. The 
dashed line depicts the variation of $\mu_h$ about its default value 
$m_b/\sqrt 2$, the dash-dotted line the variation of $\mu_i$ about 
$1.5$\,GeV, and the dotted line the variation of $\bar\mu$ also about 
$1.5$\,GeV. The highlighted area shows the combined perturbative uncertainty.}
\end{center}
\end{figure}

The purpose of this section is to investigate the individual
contributions to $\Gamma_u(\Delta)$ that result from the corresponding
terms in the weight function, as well as their residual dependence on
the matching scales. For $\Delta=0.65$\,GeV we find numerically
\begin{eqnarray}\label{eq:breakup}
   \frac{\Gamma_u(0.65\,{\rm GeV})}{|V_{ub}|^2\,{\rm ps}^{-1}}
   &=& 43.5\,\big( 1 + 0.158\,\hbox{\scriptsize $[\alpha_s(\mu_i)]$}  
    - 0.095\,\hbox{\scriptsize $[\alpha_s(\mu_h)]$}  
    + 0.076\,\hbox{\scriptsize $[\alpha_s^2(\mu_i)]$} \nonumber\\[-0.2cm]
   &&\hspace{1.2cm}
    \mbox{}- 0.037\,\hbox{\scriptsize $[\alpha_s(\mu_i)\alpha_s(\mu_h)]$} 
    + 0.009\,\hbox{\scriptsize [kin]}  
    - 0.043\,\hbox{\scriptsize [hadr]} \big) = 46.5 \,.\nonumber\\
\end{eqnarray}
The terms in parenthesis correspond to the contributions to the weight 
function arising at different orders in perturbation theory and in the 
$1/m_b$ expansion, as indicated by the subscripts.
Note that the perturbative contributions from the intermediate scale are 
typically twice as large as the ones from the hard scale, which is also the
naive expectation. Indeed, the two-loop $\alpha_s^2(\mu_i)$
correction is numerically of comparable size to the one-loop
$\alpha_s(\mu_h)$ contribution. This confirms the importance of
separating the scales $\mu_i$ and $\mu_h$.
The contributions from kinematical and hadronic power corrections turn out to 
be numerically small, comparable to the two-loop corrections.

The weight function (\ref{eq:masterform}) is formally independent of the 
matching scales $\mu_h$, $\mu_i$, and $\bar\mu$. In Figure~\ref{fig:scales} we 
plot the residual scale dependence resulting from the truncation of the 
perturbative series. Each of the three scales is varied independently by a 
factor between $1/\sqrt2$ and $\sqrt2$ about its default value. The scale 
variation of $\mu_i$ is still as significant as the variation of $\mu_h$, even
though the former is known at NNLO and the latter only at NLO. We have checked 
analytically that the result
(\ref{eq:masterform}) is independent of $\mu_i$ through two-loop
order, i.e.~the residual scale dependence is an $O(\alpha_s^3(\mu_i))$ effect. 
In order to obtain a conservative estimate of the perturbative uncertainty in
our predictions we add the individual scale dependencies in quadrature. This 
gives the gray band shown in the figure.

\begin{figure}
\begin{center}
\epsfig{file=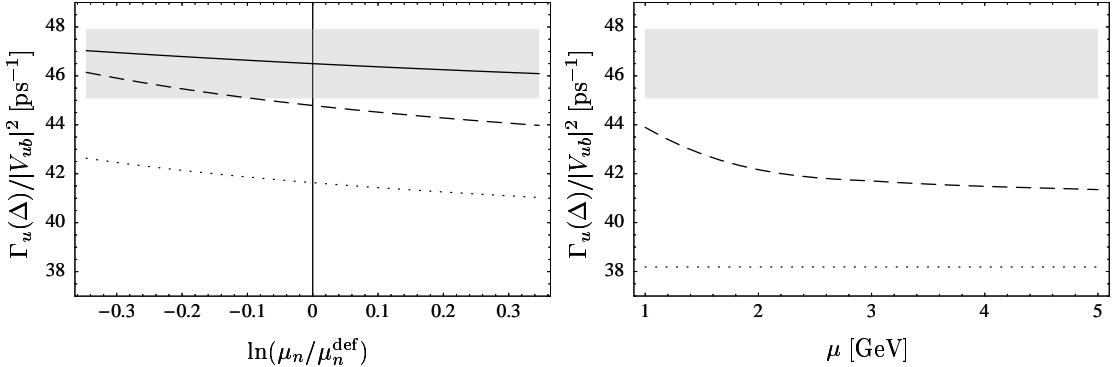,width=\textwidth}
\caption{\label{fig:onetwo} 
Convergence of the perturbative series and residual scale dependence. 
{\sc Left:} RG-improved results at LO (dotted), NLO (dashed), and NNLO (solid) 
as a function of the scales $\mu_i$, $\mu_h$, $\bar\mu$, which are varied 
simultaneously about their default values. {\sc Right:} Fixed-order results at 
tree-level (dotted) and one-loop order (dashed).}
\end{center}
\end{figure}

Figure~\ref{fig:onetwo} displays the result for $\Gamma_u(0.65\,\mbox{GeV})$ 
at different orders in RG-improved perturbation theory. At LO, we dismiss all 
$\alpha_s$ terms including the kinematical power corrections; however, leading 
logarithms are still resummed and give rise to a non-trivial dependence of 
$T_0$ on the coefficient $a$. At NLO, we include the $O(\alpha_s(\mu_h))$,
$O(\alpha_s(\mu_i))$, and $O(\alpha_s(\bar\mu))$ contributions, but drop terms 
of order $\alpha_s^2(\mu_i)$ or $\alpha_s(\mu_i)\,\alpha_s(\mu_h)$. At NNLO, 
we include all terms shown in (\ref{eq:masterform}). In studying the different 
perturbative approximations we vary the matching scales simultaneously (and 
in a correlated way) about their default values. Compared with 
Figure~\ref{fig:scales} this leads to a reduced scale variation. The gray 
bands in Figure~\ref{fig:onetwo} show the total perturbative uncertainty as
determined above. While the two-loop NNLO contributions are sizable, we 
observe a good convergence of the perturbative expansion and a reduction of 
the scale sensitivity in higher orders. 
The right-hand plot in the figure contrasts these findings with the 
corresponding results in fixed-order perturbation theory, which are obtained 
from (\ref{eq:masterform}) by setting $\mu_h=\mu_i=\bar\mu=\mu$ and truncating
the series at $O(\alpha_s(\mu))$ for consistency. We see that
the fixed-order results are also rather insensitive to the value of
$\mu$ unless this scale is chosen to be small; yet, the predicted values for
$\Gamma_u$ are significantly below those obtained in RG-improved perturbation
theory. We conclude that the small scale dependence observed in the 
fixed-order calculation does not provide a reliable estimator of the
true perturbative uncertainty. In our opinion, a fixed-order calculation at a 
high scale is not only inappropriate in terms of the underlying dynamics of 
inclusive decay processes in the shape-function region, it is also misleading 
as a basis for estimating higher-order terms in the perturbative expansion.

\subsection{Comments on the normalization of the photon spectrum}

We mentioned in section \ref{section_5.1} that the use of the normalized photon 
spectrum is advantageous because event fractions in $\bar B\to X_s\gamma$ 
decay can be calculated more reliably than partial decay rates. In this 
section we point out another important advantage, namely that the perturbative 
series for the weight function $W(\Delta,P_+)$ is much better behaved 
than that for $\widetilde W(\Delta,P_+)$. The difference of the two weight
functions lies in their normalizations, which are
\begin{equation}\label{hardfunx}
   W(\Delta,P_+)\propto m_b^3\, T_0(a,\mu_h) H_\Gamma(\mu_h) \,, \qquad
   \widetilde W(\Delta,P_+)\propto 
    \frac{T_0(a,\mu_h)}{[C_{7\gamma}^{\rm eff}(\mu_h)]^2\,%
    \overline{m}_b^2(\mu_h)\,|H_s(\mu_h)|^2} \,.
\end{equation}
Here $H_s$ is the hard function in the factorized expression for the structure 
function $\F_\gamma$ in (\ref{eq:gammaS}), which has been derived in 
\cite{Neubert:2004dd}.
Note that the two weight functions have a different dependence on the 
$b$-quark 
mass. In the case of $W$, three powers of $m_b$ enter through phase-space 
integrations in the total decay rate $\Gamma_s(E_*)$, and it is therefore 
appropriate to use a low-scale subtracted quark-mass definition, such as the
shape-function mass. In the case of $\widetilde W$, on the other hand, two 
powers of the running quark mass $\overline{m}_b(\mu_h)$ enter through the
definition of the dipole operator $O_{7\gamma}$, and it is appropriate to 
use a short-distance mass definition such as that provided by the 
$\overline{\rm MS}$ scheme. In practice, we write $\overline{m}_b(\mu_h)$ as 
$\overline{m}_b(m_b)$ times a perturbative series in $\alpha_s(\mu_h)$.

The most pronounced effect of the difference in normalization is that the 
weight function $\widetilde W$ receives very large radiative corrections 
at order $\alpha_s(\mu_h)$, which range between $-68\%$ and $-43\%$ when the 
scale $\mu_h$ is varied between $m_b$ and $m_b/2$. This contrasts the 
well-behaved perturbative expansion of the weight function $W$, for which the 
corresponding corrections vary between $-11\%$ and $-7\%$. In other words, 
the hard matching corrections for $\widetilde W$ are about six times larger
than those for $W$. Indeed, these corrections are so large that in our opinion
relation (\ref{oldrelation}) should not be used for phenomenological 
purposes.

The different perturbative behavior of the hard matching corrections to the 
weight functions is mostly due to the mixing of the dipole operator 
$O_{7\gamma}$ with other operators in the effective weak Hamiltonian for 
$\bar B\to X_s\gamma$ decay. In order to illustrate this fact, consider the
one-loop hard matching coefficients defined as
\begin{equation}
   W\propto 1 + k\,\frac{\alpha_s(\mu_h)}{\pi} + \dots \,, \qquad
   \widetilde W\propto 1 + \widetilde k\,\frac{\alpha_s(\mu_h)}{\pi}
    + \dots \,.
\end{equation}
With our default scale choices we have $k=-2.32+1.13=-1.19$, where the 
second contribution ($+1.13$) comes from operator mixing, which gives rise
to the terms in the last two lines in (\ref{Hgamma}). For the weight function
$W$, this contribution has the opposite sign than the other terms, so that 
the combined value of $k$ is rather small. For the weight function 
$\widetilde W$, on the other hand, we find $\widetilde k=-1.58-5.52=-7.10$. 
Here the contribution from operator mixing is dominant and has the same sign 
as the remaining terms, thus yielding a very large value of 
$\widetilde k$. Such a large $O(\alpha_s)$ correction was not 
observed in \cite{Hoang:2005pj}, because these authors chose to omit the 
contribution from operators mixing. Note that at a higher scale $\mu_h=m_b$, 
as was adopted in this reference, the situation is even worse. In that case we 
find $k=-2.81+1.19=-1.62$ and $\widetilde k=-0.66-9.13=-9.79$.

\begin{figure}
\begin{center}
\epsfig{file=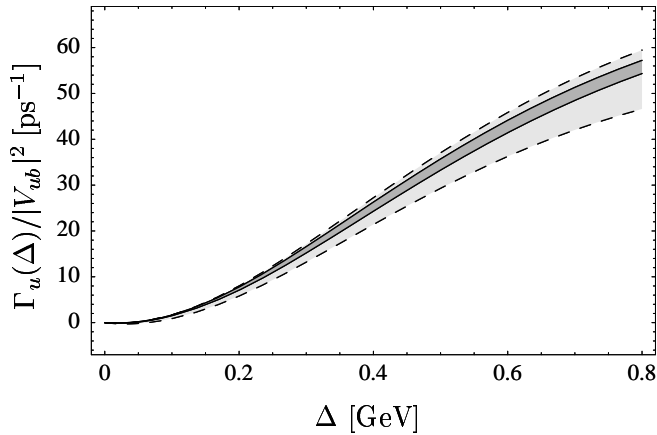,width=8.7cm}
\caption{\label{fig:band}
Perturbative uncertainties on $\Gamma_u(\Delta)$ encountered when 
using the weight function $W(\Delta,P_+)$ (dark gray) or 
$\widetilde W(\Delta,P_+)$ (light gray). In the latter case, the normalization 
of the photon spectrum is chosen such that the two predictions agree at 
$\Delta=0.65$\,GeV for central values of the matching scales.}
\end{center}
\end{figure}

A visualization of the perturbative uncertainty is depicted in
Figure~\ref{fig:band}, where predictions for $\Gamma_u(\Delta)$ are
shown using either (\ref{eq:relation}) or (\ref{oldrelation}).
In each case, the error band is obtained by 
varying the different scales about their default values, 
$\mu_n \in[\mu_n^{\rm def}/\sqrt2,\mu_n^{\rm def}\sqrt2]$, and adding the
resulting uncertainties in quadrature. The
dark-gray band bordered by solid lines denotes the perturbative
uncertainty of predictions when using the normalized photon
spectrum, as in (\ref{eq:relation}). (At the point $\Delta=0.65$\,GeV this 
uncertainty is identical to the gray band depicted in 
Figures~\ref{fig:scales} and \ref{fig:onetwo}). The light-gray band bordered
by dashed lines corresponds to the use of the absolute photon
spectrum, as in (\ref{oldrelation}). The difference in precision between the 
two methods would be even more pronounced if we used the higher default
value $\mu_h=m_b$ for the hard matching scale. Obviously, 
the use of the normalized photon spectrum will result in a more precise 
determination of $|V_{ub}|$. 

\subsection{Comments on \boldmath$\beta_0\alpha_s^2$ terms and scale 
separation\unboldmath}

The separation of different momentum scales using RG techniques, which is 
one of the key ingredients of our approach, is well motivated by the dynamics 
of charmless inclusive $B$ decays in the shape-function region. Factorized 
expressions for the $B$-decay spectra involve hard functions renormalized 
at $\mu_h$ multiplied by jet and shape functions defined at a lower scale.
While physics at or below the intermediate scale is very
similar for $\bar B\to X_u\,l^-\bar\nu$ and $\bar B\to X_s\gamma$ (as is 
manifested by the fact that the leading shape and jet functions are 
universal), the physics at the hard scale in $\bar B\to X_s\gamma$ decay is 
considerably more complicated than in semileptonic decay, and it might even 
contain effects of New Physics. Therefore it is natural to respect the
hierarchy $\mu_h\gg\mu_i$ and disentangle the various contributions, as done 
in this chapter. In fact, our ability to calculate the dominant
two-loop corrections is a direct result of this scale separation.
Nevertheless, at a technical level we can reproduce the results of a
fixed-order calculation by simply setting all matching scales equal to 
a common scale, $\mu_h=\mu_i=\bar\mu=\mu$. 
In this limit, the expressions derived in this chapter smoothly reduce to 
those obtained in conventional perturbation theory. While factorized 
expressions for the decay rates are 
superior to fixed-order results whenever there are widely separated scales in 
the problem, they remain valid in the limit where 
the different scales become of the same order.

In \cite{Hoang:2005pj}, the $O(\beta_0\alpha_s^2)$ BLM corrections 
\cite{Brodsky:1982gc} to the weight 
function $\widetilde W(\Delta,P_+)$ in (\ref{oldrelation}) were calculated in 
fixed-order perturbation theory. For simplicity, 
only the contribution of the operator $O_{7\gamma}$ to the 
$\bar B\to X_s\gamma$ decay rate was included in this work.
We note that without the contributions from other operators the expression for 
$\widetilde W$ is not renormalization-scale and -scheme invariant.
Neglecting operator mixing in the calculation of $\widetilde W$ is therefore 
not a theoretically consistent approximation. 
However, having calculated the exact NNLO corrections at the intermediate 
scale allows us to examine some of the 
terms proportional to $\beta_0\alpha_s^2(\mu_i)$ and compare them 
to the findings of \cite{Hoang:2005pj}. In this way
we confirm their results for the coefficients multiplying the logarithms
$\ln^n[m_b(\Delta-P_+)/\mu_i^2]$ with $n=1,2$ in (\ref{eq:masterform}). 
While the $\beta_0\alpha_s^2$ terms approximate the full two-loop coefficients 
of these logarithms arguably well, we stress that 
the two-loop constant at the intermediate 
scale is not dominated by terms proportional to $\beta_0$. Numerically we find
\begin{equation}
   c^{(2)}_0 = - 47.4 + 39.6\,\frac{\beta_0}{25/3}
   + a \left[ - 31.8 + 38.8\,\frac{\beta_0}{25/3} \right] + O(a^2) \,,
\end{equation}
which means that the approximation of keeping only the BLM terms would 
overestimate this coefficient by almost an order of magnitude and give the 
wrong sign. This shows the importance of a complete two-loop calculation, as 
performed in this chapter.

We believe that the perturbative approximations adopted in this chapter, 
i.e.\ working to NNLO at the intermediate scale and to NLO at the hard
scale, are sufficient for practical purposes in the sense that the residual
perturbative uncertainty is smaller than other uncertainties encountered in 
the application of relation (\ref{eq:relation}). Still, one may ask what 
calculations would be required to determine the missing 
$\alpha_s^2(\mu_h)$ terms in
the normalization of the weight function in (\ref{eq:masterform}), or at 
least the terms of order $\beta_0\alpha_s^2(\mu_h)$. For the case of 
$\bar B\to X_s\gamma$ decay, the contribution of the operator $O_{7\gamma}$ 
to the normalized photon spectrum was recently calculated at two-loop order 
\cite{Melnikov:2005bx}, while the contributions from other operators are known
to $O(\beta_0\alpha_s^2)$ \cite{Ligeti:1999ea}. What is still needed are 
the two-loop corrections to the double differential (in $P_+$ and $y$) 
$\bar B\to X_u\,l^-\bar\nu$ decay rate in (\ref{eq:gammaU}).

\subsection{Subleading corrections from hadronic structures}
\label{sec:SSF}

Due to the fact that different linear combinations of the subleading shape 
functions $\hat t(\hat\omega)$, $\hat u(\hat\omega)$, and $\hat v(\hat\omega)$
enter the theoretical description of radiative and semileptonic decays
starting at order $\Lambda_{\rm QCD}/m_b$, the weight function cannot be free
of such hadronic structure functions. Consequently, we found in
(\ref{eq:masterform}) all of the above subleading shape functions,
divided by the leading shape function $\hat S(\hat\omega)$. Our default model
(\ref{defaultmodel}) for the subleading shape functions was chosen such that 
the combined effect of all hadronic power corrections could be absorbed into
a single hadronic parameter $\Lambda_{\rm SSF}^2$. More generally, 
we define a function $\delta_{\rm hadr}(\Delta)$ via (a factor 2 is inserted
for later convenience)
\begin{equation}
   \Gamma_u(\Delta) = [\Gamma_u(\Delta)]_{\rm def}
   \left[ 1 + 2\delta_{\rm hadr}(\Delta) \right] ,
\end{equation}
where $[\Gamma_u(\Delta)]_{\rm def}$ denotes the result obtained with the 
default model for the subleading shape functions.
From (\ref{eq:masterform}), one finds that
\begin{eqnarray}\label{epsrela}
  && \delta_{\rm hadr}(\Delta)
   = \bigg\{\int_0^\Delta\!dP_+\,(M_B-P_+)^4 \big[ 2(4-3a)\,h_t(P_+)
           + (4-a)\,h_u(P_+) +\nonumber\\
           &&+ (8-13a+4a^2)\,h_v(P_+) \big]\bigg\}%
          \left\{4(1-a)(3-a)\int_0^\Delta\!dP_+\,(M_B-P_+)^5\,\hat S(P_+)\right\}^{-1} \,,
          \nonumber\\
\end{eqnarray}
where we have used that, at leading order in $\alpha_s$ and 
$\Lambda_{\rm QCD}/m_b$, the 
$\bar B\to X_s\gamma$ photon spectrum is proportional to 
$(M_B-P_+)^3\,\hat S(P_+)$. In the relation above,
$h_t(\hat\omega)\equiv\hat t(\hat\omega)-[\hat t(\hat\omega)]_{\rm def}$ etc.\ 
denote the differences between the true subleading shape functions and the 
functions adopted in our default model. By construction, these are functions
with vanishing normalization and first moment. 

The above expression for $\delta_{\rm hadr}(\Delta)$ is exact to the order we 
are working; however, in practice we do not know the precise form of the 
functions $h_i(\hat\omega)$. 
Our goal is then to find a conservative bound, 
$|\delta_{\rm hadr}(\Delta)|<\epsilon_{\rm hadr}(\Delta)$, and to interpret 
the function $\epsilon_{\rm hadr}(\Delta)$ as the relative 
hadronic uncertainty on the value of $|V_{ub}|$ extracted using relation 
(\ref{eq:relation}). To obtain the bound we scan
over a large set of realistic models for the subleading shape functions. In 
\cite{Lange:2005yw}, four different functions $h_i(\hat\omega)$ were 
suggested, which can be added or subtracted (in different combinations) to 
each of the subleading shape functions.
Together, this provides a large set of different models for these 
functions. To be conservative, we pick from this set the model 
which leads to the largest value of $|\delta_{\rm hadr}(\Delta)|$. The 
integrand in the numerator in (\ref{epsrela}) is maximized if all three 
$h_i(\hat\omega)$ functions are equal to a single function, whose choice 
depends on the value of $\Delta$. In the
denominator, we find it convenient to eliminate the shape function 
$\hat S(P_+)$ in favor of the normalized photon spectrum. Working consistently 
to leading order, we then obtain
\begin{equation}
   \epsilon_{\rm hadr}(\Delta)
   = \frac{5-5a+a^2}{(1-a)(3-a)}\,\frac{U(\mu_h,\mu_i)}{m_b^3}\,
   \max_i
   \frac{\displaystyle\left|\int_0^\Delta\!dP_+\,(M_B-P_+)^4\,h_i(P_+)\right|}%
        {\displaystyle\int_0^\Delta\!dP_+\,(M_B-P_+)^2\,
         \frac{1}{\Gamma_s(E_*)}\,\frac{d\Gamma_s}{dP_+}} \,,
\end{equation}
where as before $a=a_\Gamma(\mu_h,\mu_i)\approx 0.12$ for the default choice 
of matching scales, and $U(\mu_h,\mu_i)\approx 1.11$ \cite{Lange:2005yw}.

\begin{figure}
\begin{center}
\epsfig{file=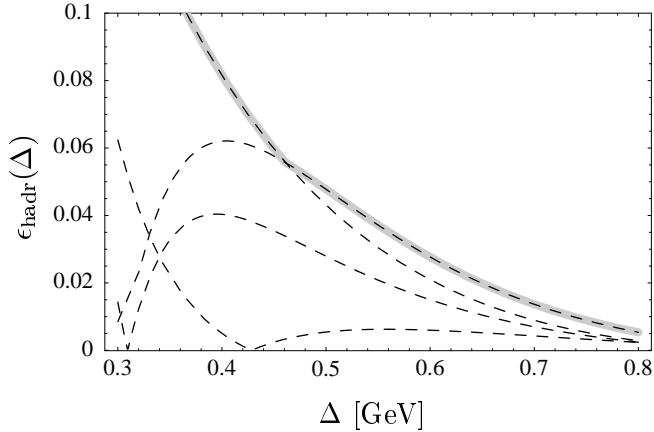,width=8.7cm}
\caption{\label{fig:sublSF}
Estimates for the hadronic uncertainty $\epsilon_{\rm hadr}(\Delta)$ obtained 
from a scan over models for the subleading shape functions. 
The dashed lines correspond to the individual results for the four 
$h_i(\hat\omega)$ functions suggested in \cite{Lange:2005yw}. The thick solid 
line, which covers one of the dashed lines, shows the maximum effect.}
\end{center}
\end{figure}

The result for the function $\epsilon_{\rm hadr}(\Delta)$ obtained this 
way is shown in Figure~\ref{fig:sublSF}. We set the matching scales to their 
default values and use the model (\ref{model}) for the photon spectrum, which 
is a good enough approximation for our purposes. 
From this estimate it is apparent that the effects of subleading shape
functions are negligible for large values $\Delta\gg\Lambda_{\rm QCD}$ and 
moderate for $\Delta\sim\Lambda_{\rm QCD}$, which is the region of interest 
for the determination of $|V_{ub}|$. In the region $\Delta<0.3$\,GeV the
accuracy of the calculation deteriorates.
For example, we find $\epsilon_{\rm hadr}=2.0\%$ for $\Delta=0.65$\,GeV, 
$\epsilon_{\rm hadr}=4.8\%$ for $\Delta=0.5$\,GeV, and
$\epsilon_{\rm hadr}=8.2\%$ for $\Delta=0.4$\,GeV.

\section{Conclusions}

Model-independent relations between weighted integrals of 
$\bar B\to X_s\gamma$ and $\bar B\to X_u\,l^-\bar\nu$ decay distributions, in 
which all reference to the leading non-perturbative shape function is avoided, 
offer one of the most promising avenues to a high-precision determination of 
the CKM matrix element $|V_{ub}|$. In order to achieve a theoretical precision 
of better than 10\%, it is necessary to include higher-order corrections in 
$\alpha_s$ and $\Lambda_{\rm QCD}/m_b$ in this approach. 

In this chapter, we have calculated the weight function $W(\Delta,P_+)$ in 
the relation between the hadronic $P_+=E_X-|\vec{P}_X|$ spectra in the two 
processes, integrated over the interval $0\le P_+\le\Delta$. 
Based on QCD factorization theorems for the differential decay 
rates, we have derived an exact formula (\ref{eq:fmatch}) that allows for the 
calculation of the leading-power weight function to any order in perturbation 
theory. We have calculated the $\Delta$- and 
$P_+$-dependent terms in the weight function exactly at 
next-to-next-to-leading order (NNLO) in renormalization-group improved 
perturbation theory, including two-loop matching corrections at 
the intermediate scale $\mu_i\sim\sqrt{m_b\Lambda_{\rm QCD}}$ and three-loop 
running between the intermediate scale and the hard scale $\mu_h\sim m_b$.
The only piece missing for a complete prediction at NNLO is the two-loop
hard matching correction to the overall normalization of the weight function.
A calculation of
the $\alpha_s^2(\mu_h)$ term would require the knowledge of both decay
spectra at two-loop order, which is currently still lacking. We also include 
various sources of power corrections. Power corrections from phase-space 
factors are treated exactly. The remaining hadronic and kinematical power
corrections are given to first order in $\Lambda_{\rm QCD}/m_b$ and to the 
order in perturbation theory to which they are known.

A dedicated study of the perturbative behavior of our result for the weight 
function has been performed for the partial $\bar B\to X_u\,l^-\bar\nu$ decay 
rate $\Gamma_u(\Delta)$ as obtained from the right-hand side of relation 
(\ref{eq:relation}). It exhibits
good convergence of the expansion and reduced scale sensitivity in higher 
orders. We find that corrections of order $\alpha_s^2(\mu_i)$ at the 
intermediate scale are typically as important as first-order $\alpha_s(\mu_h)$ 
corrections at the hard scale. We have also seen that fixed-order 
perturbation theory significantly underestimates the value of 
$\Gamma_u(\Delta)$, even though the apparent stability with respect to scale
variations would suggest a good perturbative convergence. In order to obtain 
a well-behaved expansion in powers of $\alpha_s(\mu_h)$, it is important to 
use the normalized photon spectrum in relation (\ref{eq:relation}). A similar 
relation involving the differential $\bar B\to X_s\gamma$ decay rate
receives uncontrollably large matching 
corrections at the hard scale and is thus not suitable for phenomenological
applications.
At next-to-leading order in the $1/m_b$ expansion, the weight function 
receives terms involving non-perturbative subleading shape functions, which 
cannot be eliminated. Our current ignorance about the functional forms of 
these functions leads to a hadronic uncertainty, which we have 
estimated by scanning over a large set of models. We believe that a 
reasonable estimate of the corresponding relative uncertainty 
$\epsilon_{\rm hadr}$ on $|V_{ub}|$ is given by the solid line in 
Figure~\ref{fig:sublSF}.

Let us summarize our main result for the partial $\bar B\to X_u\,l^-\bar\nu$ 
decay rate with a cut $P_+\le 0.65$\,GeV, which is close to the charm 
threshold $M_D^2/M_B$, and present a detailed list of the various sources of 
theoretical uncertainties. We find 
\begin{eqnarray}
   &&\Gamma_u(0.65\,{\rm GeV})=\nonumber\\
&=& \left( 46.5\pm 1.4\,\hbox{\scriptsize [pert]}\,
    \pm 1.8\, \hbox{\scriptsize [hadr]}\,
    \pm 1.8\, \hbox{\scriptsize [$m_b$]}\,
    \pm 0.8\, \hbox{\scriptsize [pars]}\,
    \pm 2.8\, \hbox{\scriptsize [norm]}\,
    \right) |V_{ub}|^2\,{\rm ps}^{-1} \nonumber\\
   &=& (46.5\pm 4.1)\,|V_{ub}|^2\,{\rm ps}^{-1}\,,
\end{eqnarray}
where the central value is derived assuming that the $\bar B\to X_s\gamma$ 
photon spectrum can be accurately described by the function (\ref{model}). 
The errors refer to the perturbative uncertainty as estimated in 
Section~\ref{sec:pert}, the uncertainty due to the ignorance about 
subleading shape functions as discussed in Section~\ref{sec:SSF}, the error 
in the value of the $b$-quark mass, 
other parametric uncertainties from variations of $m_c$, $m_s$, and 
$\lambda_1$, and finally a 6\% uncertainty in the calculation of the 
normalization of the photon spectrum \cite{Neubert:2004dd}. To a good
approximation the errors scale with the central value. The above numbers 
translate into a combined theoretical uncertainty of 4.4\% on $|V_{ub}|$ when 
added in quadrature.

\chapter{Conclusions}
\label{chapter_conc}
Inclusive charmless $B$ decays offer the most precise methods for
measuring $|V_{ub}|$. The physics of these decays is rich and involves
a delicate interplay of theory and experiment. In this work we have
developed and described an elaborate theoretical machinery that can be
applied to these decays.

We have started our discussion by explaining how to relate the
differential decay rates of $\bar B\to X_s\gamma$ and $\bar B\to
X_u\,l\bar\nu$ to the hadronic tensor. We have reviewed the traditional 
(approximate) methods of
calculating the hadronic tensor, namely, an expansion in the strong
coupling constant and HQET based OPE. We have explained the problems
that arise from the use of these methods, and argued that they can be
overcome by the use of SCET, the appropriate effective field theory
for charmless inclusive $B$ decays. We therefore reviewed some of
its basic elements. Using SCET one can prove that the hadronic
tensor factorizes, at each order in $1/m_b$, as a sum of products of a
calculable hard function multiplied by a calculable jet function
convoluted with a non perturbative shape function.

At leading order in $1/m_b$ all these functions are unique. We have
calculated the leading order hard and jet function by performing a
two-step matching calculation QCD$\to$SCET$\to$HQET.  We have resuumed
large logarithms that arise in the calculation at next-to-leading
order in renormalization-group improved perturbation theory. We have
employed the operator product expansion to relate moments of the
renormalized shape function with HQET parameters such as $m_b$,
$\bar\Lambda$ and $\mu_\pi^2$ defined in a new physical subtraction
scheme, called the "shape function scheme". We have obtained an
analytic expression for the asymptotic behavior of the shape function, which reveals that it is not positive definite.

Having analyzed the leading power terms we have considered the
contributions of subleading shape functions.  At tree-level, the
results can be expressed in terms of forward matrix elements of
bi-local light-cone operators. Four-quark operators, which arise at
$O(\alpha_s)$, were also included. We have shown that at tree level
only three independent subleading shape functions are needed.

Based on these calculations we turned, in the second part of this work, 
to applications.  In the first application, we have presented ``state-of-the-art''
theoretical expressions for the triple differential $\bar B\to
X_u\,l^-\bar\nu$ decay rate and for the $\bar B\to X_s\gamma$ photon
spectrum, which incorporate all known contributions and smoothly
interpolate between the ``shape-function region'' of large hadronic
energy and small invariant mass, and the ``OPE region'' in which all
hadronic kinematical variables scale with $M_B$. The differential
rates are given in a form which has no explicit reference to the mass
of the $b$ quark, avoiding the associated uncertainties. Dependence on
$m_b$ enters indirectly through the properties of the leading shape
function, which can be determined by fitting the $\bar B\to X_s\gamma$
photon spectrum. This eliminates the dominant theoretical
uncertainties from predictions for $\bar B\to X_u\,l^-\bar\nu$ decay
distributions, allowing for a precise determination of $|V_{ub}|$. In
the shape-function region, we have factorized short-distance and
long-distance contributions at next-to-leading order in
renormalization-group improved perturbation theory. The higher-order
power corrections include effects from subleading shape functions
where they are known. When integrated over sufficiently large portions
in phase space, our results would reduce to standard OPE expressions
up to yet unknown $O(\alpha_s^2)$ terms. We have presented predictions
for partial $\bar B\to X_u\,l^-\bar\nu$ decay rates with various
experimental cuts.  The elaborate error analysis that we have
performed contains all significant theoretical uncertainties,
including weak annihilation effects. We have suggested that the latter
can be eliminated by imposing a cut on high leptonic invariant mass.

In the second application we have derived a shape-function independent
relation between the partial $\bar B\to X_u\,l^-\bar\nu$ decay rate
with a cut on $P_+=E_X-|\vec P_X|\le\Delta$ and a weighted integral
over the normalized $\bar B\to X_s\gamma$ photon-energy spectrum. We
have calculated the leading-power contribution to the weight function
at next-to-next-to-leading order in renormalization-group improved
perturbation theory, including exact two-loop matching corrections at
the scale $\mu_i\sim\sqrt{m_b\Lambda_{\rm QCD}}$.  The overall
normalization of the weight function is obtained up to yet unknown
corrections of order $\alpha_s^2(m_b)$.  We have included power
corrections from phase-space factors exactly, while the remaining
subleading contributions have been included at first order in
$\Lambda_{\rm QCD}/m_b$. At this level unavoidable hadronic
uncertainties enter, which we have estimated in a conservative
way. The combined theoretical accuracy in the extraction of $|V_{ub}|$
is at the level of 5\% if a value of $\Delta$ near the charm threshold
can be achieved experimentally.

The work presented here was already implemented by the "B-factories": Babar and Belle.
As of June 2006, Babar has measured $|V_{ub}|$ using a combined $s_H$ and $E_l$ cut \cite{Aubert:2005im}, combined $M_X$ and $q^2$ cut \cite{Aubert:2005hb}, and $E_l$ cut \cite{Aubert:2005mg}. Babar has also extracted the $M_X$ spectrum for $\bar B\to X_s\gamma$ \cite{Aubert:2005cu}. As of June 2006, Belle has measured $|V_{ub}|$ using an $E_l$ cut \cite{Limosani:2005pi}, $M_X$ cut, combined  $M_X$ and $q^2$ cut, and $P_+$ cut \cite{Bizjak:2005hn}. 

We would like to conclude this work with an outlook. The reduction of the experimental errors encourages theorists to improve the theoretical predictions. As of June 2006, what theoretical developments are feasible? In general the formalism developed in this work can be extended to higher orders in $1/m_b$ and $\alpha_s$. 

At leading power the jet function \cite{Becher:2006qw} and the partonic shape function \cite{Becher:2005pd} have already been calculated to second order in $\alpha_s$. For $\bar B\to X_s\gamma$ a next-to-next-to-leading order QCD calculation is underway, where the contribution of $O_{7\gamma}$ is already known \cite{Melnikov:2005bx,Blokland:2005uk}. Once the complete result is known, one could extract the hard function for $\bar B\to X_s\gamma$ from it. An "easier" calculation is to find the leading power hard function for $\bar B\to X_u\,l^-\bar\nu$. Having these ingredients would allow to extend the analysis of chapters \ref{chapter_evegen} and \ref{chapter_weight} to a complete two loop order. 

Beyond leading power one would like to find a way to extract the subleading shape functions from experiment, or at least to constrain them. This would help to reduce a large source of uncertainty in the extraction of $|V_{ub}|$. Another possible improvement is to find the subleading shape function beyond tree level. The analysis performed in \cite{Lee:2004ja,Beneke:2004in} indicates that beyond tree level more subleading shape functions would arise. This calculation is related to the long overdue OPE calculation of the ${\cal O}(\alpha_s)$ power corrections corrections to $\bar B\to X_c\,l^-\bar\nu$ decays. The latter calculation would allow to check the results of the former, once it is performed.

Finally, as a result of the better control of the charm background, the experimental cuts are being pushed away from the shape function region and into the OPE region. In this intermediate region a multi scale OPE calculation, like the one performed in \cite{Neubert:2004dd}, is more appropriate. This calculation is yet to be done for semileptonic decays.

To summarize, despite the impressive progress achieved in our understanding of charmless inclusive $B$ decays, much is left to be done. Hopefully the research presented here would help make this progress possible.

\appendix
\chapter{The hadronic tensor}
\label{apx:hadten}
In chapter \ref{chapter_intro} we defined the current correlator $T^{\mu\nu}$ as:
\begin{equation}
T^{\mu\nu}=i\int d^4 x \, e^{iqx} \,T\{J^{\dagger\mu}(0),J^\nu(x)\}.
\end{equation}
Reinserting the complete set of hadronic states we have
\begin{eqnarray}
\langle \bar B|T^{\mu\nu}|\bar B\rangle&=&i\int d^4 x \, e^{iqx}\bigg[
\theta(-x^0)\sum_{X_q} \langle
\bar B|J^{\dagger\mu}(0)|X_q\rangle\langle X_q |J^\nu(x)|\bar B\rangle\nonumber\\
&+&\theta(x^0)\sum_{X_{bb\bar q}} \langle
\bar B|J^\nu(x)|X_{bb\bar q}\rangle\langle X_{bb\bar q}|J^{\dagger\mu}(0)|\bar B\rangle\bigg].
\end{eqnarray}
Following \cite{Manohar:2000dt} we write the steps function as:
\begin{equation}
\theta(\pm x^0)=\mp\frac{1}{2\pi i}\int_{-\infty}^{\infty}d\omega\frac{e^{-i\omega x^0}}{\omega\pm i\epsilon}. 
\end{equation}
(these identities can be proved by contour integration, or by integrating the integral representation of the delta function).
Translation invariance allows us to write:
\begin{eqnarray}
\langle X|J^\nu(x)|\bar B\rangle&=&e^{i(P_X-P_B)\cdot x}\langle X|J^\nu(0)|\bar B\rangle\nonumber\\
\langle \bar B|J^\nu(x)|X\rangle&=&e^{i(P_B-P_X)\cdot x}\langle \bar B|J^\nu(0)|X\rangle.
\end{eqnarray}
Integrating over $x$ and $\omega$ we find that in the rest frame of the $B$ meson:
\begin{eqnarray}
\langle \bar B|T^{\mu\nu}|\bar B\rangle&=&\sum_{X_q} 
\frac{\langle\bar B|J^{\dagger\mu}(0)|X_q\rangle\langle X_q |J^\nu(0)|\bar B\rangle}{-M_B+E_X+q^0-i\epsilon}(2\pi)^3\delta^3(\overrightarrow{P}_X+\overrightarrow{q})\nonumber\\
&-&\sum_{X_{bb\bar q}} 
\frac{\langle\bar B|J^\nu(0)|X_{bb\bar q}\rangle\langle X_{bb\bar q} |J^{\dagger\mu}(0)|\bar B\rangle}{M_B-E_X+q^0+i\epsilon}(2\pi)^3\delta^3(\overrightarrow{P}_X-\overrightarrow{q}).
\end{eqnarray}
Taking the imaginary part, we find using equation (\ref{equation_Impart})
\begin{eqnarray}
\label{equation_Wmnappx}
&&W^{\mu\nu}=\frac{1}{\pi}{\rm Im}\frac{\langle \bar B|T^{\mu\nu}|\bar B\rangle}{2M_B}\nonumber
\\&=&
\frac {1}{2M_B}\sum_{X_q} 
\langle\bar B|J^{\dagger\mu}(0)|X_q\rangle\langle X_q |J^\nu(0)|\bar B\rangle\cdot(2\pi)^3\delta^4(P_B-P_X-q)\nonumber\\
&+&\frac {1}{2M_B}\sum_{X_{bb\bar q}} 
\langle\bar B|J^\nu(0)|X_{bb\bar q}\rangle\langle X_{bb\bar q} |J^{\dagger\mu}(0)|\bar B\rangle\cdot(2\pi)^3\delta^4(P_B-P_X+q) 
\end{eqnarray}
For $B$ decays the second sum does not contribute, since the energy of $X_{bb\bar q}$ is always large than $M_B$.

Using this expression for the hadronic tensor, it is easy to prove the following symmetry of the hadronic tensor:
\begin{equation}
\label{equation_Wstar}
W^{\mu\nu}=\left(W^{\nu\mu}\right)^*. 
\end{equation}
Using the definition of the Hermitian conjugate, we have for each term in the first sum in (\ref{equation_Wmnappx}):
\begin{equation}
\left(\langle\bar B|J^{\dagger\nu}|X_q\rangle\langle X_q |J^\mu|\bar B\rangle\right)^*=
\langle X_q|J^{\nu}|\bar B\rangle\langle \bar B |J^{\dagger\mu}|X_q\rangle=
\langle\bar B|J^{\dagger\mu}|X_q\rangle\langle X_q |J^\nu|\bar B\rangle
\end{equation}
A similar relation holds for the second sum in (\ref{equation_Wmnappx}). Combining them gives us  
(\ref{equation_Wstar}).

\chapter{Perturbative Expressions}

\label{apx:Sudakovs}

\section{Anomalous dimensions}

Here we list the known perturbative expansions of the $\beta$-function
and relevant anomalous dimensions. We work in the $\overline{\rm MS}$
scheme and define
\begin{eqnarray}
   \beta(\alpha_s)
   &=& \frac{d\alpha_s(\mu)}{d\ln \mu} 
    = -2\alpha_s \sum\limits_{n=0}^\infty \beta_n 
    \left( \frac{\alpha_s}{4\pi} \right)^{n+1} , \nonumber\\
   \Gamma_{\rm cusp}(\alpha_s) &=& \sum\limits_{n=0}^\infty \Gamma_n 
    \left( \frac{\alpha_s}{4\pi} \right)^{n+1} , \qquad
   \gamma'(\alpha_s) = \sum\limits_{n=0}^\infty \gamma'_n 
    \left( \frac{\alpha_s}{4\pi} \right)^{n+1} ,
\end{eqnarray}
as the expansion coefficients for the $\beta$-function, the leading-order 
SCET current anomalous dimension, and the cusp anomalous
dimension. To three-loop order, the $\beta$-function reads \cite{Tarasov:au} 
\begin{eqnarray}
   \beta_0 &=& \frac{11}{3}\,C_A - \frac23\,n_f \,, \qquad
    \beta_1 = \frac{34}{3}\,C_A^2 - \frac{10}{3}\,C_A\, n_f - 2C_F\,n_f \,, 
    \nonumber \\
   \beta_2 &=& \frac{2857}{54}\,C_A^3
    + \left( C_F^2 - \frac{205}{18}\,C_F C_A
    - \frac{1415}{54}\,C_A^2 \right) n_f
    + \left( \frac{11}{9}\,C_F + \frac{79}{54}\,C_A \right) n_f^2 \,,\nonumber \\
\end{eqnarray}
where $n_f=4$ is the number of light flavors, $C_A=3$ and $C_F=4/3$. The 
three-loop expression for the cusp anomalous dimension has
recently been obtained in \cite{Moch:2004pa}. The coefficients read
\begin{eqnarray}
   \Gamma_0 &=& 4C_F \,, \qquad
    \Gamma_1 = 8C_F \left[ \left( \frac{67}{18} - \frac{\pi^2}{6} \right) C_A 
    - \frac59\,n_f \right] , \nonumber\\
   \Gamma_2 &=& 16C_F \bigg[ \left(\frac{245}{24} - \frac{67\pi^2}{54}
    + \frac{11\pi^4}{180} + \frac{11}{6}\,\zeta_3 \right) C_A^2
    + \left( - \frac{209}{108} + \frac{5\pi^2}{27} - \frac73\,\zeta_3
    \right) C_A\,n_f \nonumber\\
   &&\mbox{}+ \left( - \frac{55}{24} + 2\zeta_3 \right) C_F\,n_f
    - \frac{1}{27}\,n_f^2 \bigg] \,.
\end{eqnarray}
The SCET anomalous dimension $\gamma$ is explicitly known only to
one-loop order. However, the two-loop coefficient can be extracted by
noting that $\gamma$ is related to the axial-gauge anomalous dimension
in deep inelastic scattering \cite{Neubert:2004dd}. The result is
\begin{eqnarray}
   \gamma_0' &=& -5 C_F \,,\nonumber \\
   \gamma_1' &=& -8 C_F \bigg[ \left(\frac{3}{16} - \frac{\pi^2}{4}
    + 3\zeta_3 \right) C_F
    + \left( \frac{1549}{432} + \frac{7\pi^2}{48} - \frac{11}{4}\,\zeta_3
    \right)\,C_A\nonumber\\
    &&- \left( \frac{125}{216} + \frac{\pi^2}{24} \right) n_f \bigg] \,.
\end{eqnarray}

\section{Evolution factor}

The exact expression for the evolution factor reads
\begin{equation}\label{eq:U1}
   \ln U(\mu_h,\mu_i) = 2 S_\Gamma(\mu_h,\mu_i)  
   - 2a_\Gamma(\mu_h,\mu_i) \,\ln \frac{m_b}{\mu_h} 
   - 2a_{\gamma'}(\mu_h,\mu_i) \,,
\end{equation}
where the functions of the right-hand side are solutions to the 
renormalization-group equations
\begin{eqnarray}\label{eq:aGammaETC}
   \frac{d}{d\ln\mu} S_\Gamma(\nu,\mu)
   &=& - \Gamma_{\rm cusp}(\alpha_s(\mu))\,\ln\frac{\mu}{\nu} \,, \nonumber\\
   \frac{d}{d\ln\mu} a_\Gamma(\nu,\mu)
   &=& - \Gamma_{\rm cusp}(\alpha_s(\mu)) \,, \qquad
   \frac{d}{d\ln\mu} a_{\gamma'}(\nu,\mu) = - \gamma'(\alpha_s(\mu)) \,,
\end{eqnarray}
with boundary conditions $S(\nu,\mu)=0$ etc.\ at $\mu=\nu$.
These equations can be integrated using that 
$d/d\ln\mu=\beta(\alpha_s)\,d/d\alpha_s$. The solutions are
\begin{equation}
   S_\Gamma(\nu,\mu)
   = -\int\limits_{\alpha_s(\nu)}^{\alpha_s(\mu)}\!\!d\alpha\,
    \frac{\Gamma_{\rm cusp}(\alpha)}{\beta(\alpha)} 
    \int\limits_{\alpha_s(\nu)}^{\alpha}\frac{d\alpha'}{\beta(\alpha')}
    \,, \qquad
   a_\Gamma(\nu,\mu)
   = -\int\limits_{\alpha_s(\nu)}^{\alpha_s(\mu)}\!\!d\alpha\,
    \frac{\Gamma_{\rm cusp}(\alpha)}{\beta(\alpha)} \,,
\end{equation}
and similarly for $a_{\gamma'}$. 

Next, we give explicit results for the Sudakov exponent $S_\Gamma$ and
the functions $a_\Gamma$ and $a_\gamma$ in (\ref{eq:U1}) at next-to-leading 
order in renormalization-group improved perturbation theory. We obtain
\begin{equation}
   a_\Gamma(\nu,\mu) = \frac{\Gamma_0}{2\beta_0}
   \left[ \ln\frac{\alpha_s(\mu)}{\alpha_s(\nu)}
   + \left( \frac{\Gamma_1}{\Gamma_0} - \frac{\beta_1}{\beta_0} \right)
   \frac{\alpha_s(\mu)-\alpha_s(\nu)}{4\pi} + \dots \right] ,
\end{equation}
and similarly for $a_\gamma$. The next-to-leading order expressions for the 
Sudakov exponent $S_\Gamma$ contains the three-loop coefficients $\beta_2$ 
and $\Gamma_2$. With $r=\alpha_s(\mu)/\alpha_s(\nu)$, it reads
\begin{eqnarray}
   S_\Gamma(\nu,\mu)
   &=& \frac{\Gamma_0}{4\beta_0^2}\,\Bigg\{
    \frac{4\pi}{\alpha_s(\nu)} \left( 1-\frac1r-\ln r \right)
    + \left( \frac{\Gamma_1}{\Gamma_0} - \frac{\beta_1}{\beta_0} \right)
    (1-r+\ln r) + \frac{\beta_1}{2\beta_0} \ln^2 r \nonumber\\
  &&\mbox{}+ \frac{\alpha_s(\nu)}{4\pi} \Bigg[ \left( 
   \frac{\beta_1\Gamma_1}{\beta_0\Gamma_0}-\frac{\beta_2}{\beta_0}
   \right) (1-r+r\ln r) + \left( 
   \frac{\beta_1^2}{\beta_0^2}-\frac{\beta_2}{\beta_0}\right)
   (1-r) \ln r \nonumber\\
  &&\hspace{1cm}\mbox{}- \left( \frac{\beta_1^2}{\beta_0^2}
   - \frac{\beta_2}{\beta_0}-\frac{\beta_1\Gamma_1}{\beta_0\Gamma_0}
   + \frac{\Gamma_2}{\Gamma_0} \right) \frac{(1-r)^2}{2}
   \Bigg] + \dots \Bigg\} \,.
\end{eqnarray}
The next-to-leading-logarithmic evolution factor $U(\mu_h,\mu_i)$
can be obtained by combining the above expressions according to (\ref{eq:U1}) 
and expanding out terms of order $\alpha_s$.

\chapter{Partially integrated decay rates}
\label{appendix:B}

With the exception of the combined cut on the lepton energy $E_l$ and
the hadronic quantity $s_H^{\rm max}$ studied in
Section~\ref{sec:babarcut}, all other partial rates investigated in our 
analysis can be derived by first integrating the triple differential decay 
rate (\ref{eq:tripleRate}) over the lepton energy $E_l\ge E_0$ 
and $P_-\le P_-^{\rm max}$ analytically, where the quantity 
$P_-^{\rm max}$ (and in principle even $E_0$) may depend on the value of 
$P_+$. The remaining integration over $P_+$ is then performed numerically. 
In such a situation, we need to evaluate the partially integrated decay rate
\begin{equation}\label{PSints}
   \frac{d\Gamma_u}{dP_+}
   = \int_{P_+}^{P_-^{\rm max}}\!dP_-
   \int_{P_+}^{{\rm min}(P_-,M_B-2E_0)}\!dP_l\,
   \frac{d^3\Gamma_u}{dP_l\,dP_-\,dP_+} \,.
\end{equation}
Changing variables from $P_-$ to
$y$ defined in (\ref{ydef}), the constraint $P_-\le P_-^{\rm max}$
translates into the integration domain $0\le y\le y_{\rm max}$, where in 
analogy to (\ref{ydef}) we define
\begin{equation}
   y_{\rm max} = \frac{P_-^{\rm max}-P_+}{M_B-P_+} \,, \qquad
   y_0 = \frac{P_l^{\rm max}-P_+}{M_B-P_+} = 1 - \frac{2 E_0}{M_B-P_+} \,.
\end{equation}
From the phase-space relation (\ref{equation_psl}) it follows
that a cut on the lepton energy has no effect if $y_0\ge y_{\rm max}$. 
The result of performing the integrations in (\ref{PSints}) can be written as
\begin{equation}
   \frac{d\Gamma_u(y_{\rm max},y_0)}{dP_+}
   = \left\{ \begin{array}{lll}
   \Gamma_u^A(y_{\rm max}) &;& \quad y_{\rm max} \le y_0 \,, \\
   \Gamma_u^A(y_0) + \Gamma_u^B &;& \quad y_{\rm max} > y_0 \,,
   \end{array} \right.
\end{equation}
where 
\begin{eqnarray}
   \Gamma_u^A(y_i)
   &=& \frac{G_F^2|V_{ub}|^2}{96\pi^3}\,(M_B-P_+)^5\,U(\mu_h,\mu_i)\nonumber\\
    &&\times\int_0^{y_i}\!dy\,y^{2-2a_\Gamma} \left[ (3-2y)\,\F_1
    + 6(1-y)\,\F_2 + y\,\F_3 \right] , \nonumber\\
   \Gamma_u^B
   &=& \frac{G_F^2|V_{ub}|^2}{96\pi^3}\,(M_B-P_+)^5\,U(\mu_h,\mu_i)
    \int_{y_0}^{y_{\rm max}}\!dy\,y^{-2a_\Gamma} y_0 \nonumber\\
   &&\times \left[ \left( 6y(1+y_0) - 6y^2
    - y_0(3+2y_0) \right) \F_1\right.\nonumber\\
    &&\left.+ 6y(1-y)\,\F_2
    + y_0(3y-2y_0)\,\F_3 \right] .
\end{eqnarray}
When the kinematical power corrections in (\ref{eq:fullDFN}) are 
expanded as in (\ref{eq:kinNLO}) and (\ref{eq:kinNNLO}), the resulting
integrals over $y$ can be expressed in terms of the master functions
$I_n(b,z)$ given in eq.~(86) of \cite{Bosch:2004th}. The resulting expressions 
are used to obtain the numbers in the various tables in 
Section~\ref{sec:analysio}.

We now list the values of $y_0$ and $y_{\rm max}$ for the different cuts 
studied in Section~\ref{sec:analysio}. Whenever a cut $E_l\ge E_0$ on the 
charged-lepton energy is applied, we have
\begin{equation}
    y_0 = 1 - \frac{2E_0}{M_B-P_+} \,.
\end{equation}
For an additional cut $P_+\le\Delta_P$, we have $y_{\rm max}=1$ and 
$0\le P_+\le\mbox{min}(\Delta_P,M_B-2E_0)$. 
For a cut on hadronic invariant mass, $M_X\le M_0$, we have
\begin{equation}
   y_{\rm max} = \frac{\mbox{min}(M_B,M_0^2/P_+) - P_+}{M_B - P_+}
\end{equation}
and $0\le P_+\le\mbox{min}(M_0,M_B-2E_0)$.
For a cut on leptonic invariant mass, $q^2\ge q_0^2$, we have
\begin{equation}
   y_{\rm max} = 1 - \frac{q_0^2}{(M_B - P_+)^2}
\end{equation}
and $0\le P_+\le\mbox{min}(M_B-q_0,M_B-2E_0)$. 
Finally, for the combined $M_X$--$q^2$ cut we take the minimum of the previous 
two $y_{\rm max}$ values.


\begin{thebibliography}{9}


\bibitem{Manohar:1993qn}
  A.~V.~Manohar and M.~B.~Wise,
  Phys.\ Rev.\ D {\bf 49}, 1310 (1994)
  [hep-ph/9308246].

\bibitem{DeFazio:1999sv}
  F.~De Fazio and M.~Neubert,
  JHEP {\bf 9906}, 017 (1999)
  [hep-ph/9905351].  
  
\bibitem{Chen:2001fj}
  S.~Chen {\it et al.}  [CLEO Collaboration],
  Phys.\ Rev.\ Lett.\  {\bf 87}, 251807 (2001)
  [hep-ex/0108032].  
  
\bibitem{Koppenburg:2004fz}
  P.~Koppenburg {\it et al.}  [Belle Collaboration],
  Phys.\ Rev.\ Lett.\  {\bf 93}, 061803 (2004)
  [hep-ex/0403004].
  
\bibitem{Aubert:2005cu}
  B.~Aubert {\it et al.}  [BABAR Collaboration],
  Phys.\ Rev.\ D {\bf 72}, 052004 (2005)
  [hep-ex/0508004].
  
\bibitem{Peskin:1995ev}
  M.~E.~Peskin and D.~V.~Schroeder,
  ``An Introduction to quantum field theory,'' Reading, USA: Addison-Wesley (1995)

\bibitem{Buras:1998ra}
  A.~J.~Buras,
  [hep-ph/9806471].
  

\bibitem{Gel'fand}
  I.M. Gelʹfand and G.E. Shilov
  "Generalized functions", New York, USA: Academic Press (1964)  

\bibitem{Jezabek:1988ja}
  M.~Jezabek and J.~H.~Kuhn,
  Nucl.\ Phys.\ B {\bf 320}, 20 (1989).    

\bibitem{Chetyrkin:1996vx}
  K.~G.~Chetyrkin, M.~Misiak and M.~Munz,
  Phys.\ Lett.\ B {\bf 400}, 206 (1997)
  [Erratum-ibid.\ B {\bf 425}, 414 (1998)]
  [hep-ph/9612313].
                                                            
\bibitem{Neubert:1993mb}
  M.~Neubert,
  Phys.\ Rept.\  {\bf 245}, 259 (1994)
  [hep-ph/9306320].

\bibitem{Manohar:2000dt}
   A.~V.~Manohar and M.~B.~Wise,
  ``Heavy quark physics,''
  Camb.\ Monogr.\ Part.\ Phys.\ Nucl.\ Phys.\ Cosmol.\  {\bf 10}, 1 (2000).

\bibitem{Blok:1993va}
  B.~Blok, L.~Koyrakh, M.~A.~Shifman and A.~I.~Vainshtein,
  Phys.\ Rev.\ D {\bf 49}, 3356 (1994)
  [Erratum-ibid.\ D {\bf 50}, 3572 (1994)]
  [hep-ph/9307247].

\bibitem{Chay:1990da}
J.~Chay, H.~Georgi and B.~Grinstein,
Phys.\ Lett.\ B {\bf 247}, 399 (1990).
  
\bibitem{Neubert:1993ch}
M. Neubert,
Phys.\ Rev.\ D {\bf 49}, 3392 (1994)
[hep-ph/9311325];

\bibitem{Neubert:1993um}
M.~Neubert,
Phys.\ Rev.\ D {\bf 49}, 4623 (1994)
[hep-ph/9312311].

\bibitem{Bigi:1993ex}
I.~I.~Y.~Bigi, M.~A.~Shifman, N.~G.~Uraltsev and A.~I.~Vainshtein,
Int.\ J.\ Mod.\ Phys.\ A {\bf 9}, 2467 (1994)
[hep-ph/9312359].

\bibitem{Korchemsky:1994jb}
G.~P.~Korchemsky and G.~Sterman,
Phys.\ Lett.\ B {\bf 340}, 96 (1994)
[hep-ph/9407344].

\bibitem{Falk:1993dh}
  A.~F.~Falk, M.~E.~Luke and M.~J.~Savage,
  Phys.\ Rev.\ D {\bf 49}, 3367 (1994)
  [hep-ph/9308288].

\bibitem{Bauer:2000ew}
C.~W.~Bauer, S.~Fleming and M.~E.~Luke,
Phys.\ Rev.\ D {\bf 63}, 014006 (2001)
[hep-ph/0005275].

\bibitem{Bauer:2000yr}
C.~W.~Bauer, S.~Fleming, D.~Pirjol and I.~W.~Stewart,
Phys.\ Rev.\ D {\bf 63}, 114020 (2001)
[hep-ph/0011336].

\bibitem{Bauer:2001yt}
C.~W.~Bauer, D.~Pirjol and I.~W.~Stewart,
Phys.\ Rev.\ D {\bf 65}, 054022 (2002)
[hep-ph/0109045].

\bibitem{Beneke:2002ph}
M.~Beneke, A.~P.~Chapovsky, M.~Diehl and T.~Feldmann,
Nucl.\ Phys.\ B {\bf 643}, 431 (2002)
[hep-ph/0206152];

\bibitem{Beneke:2002ni}
M.~Beneke and T.~Feldmann,
Phys.\ Lett.\ B {\bf 553}, 267 (2003)
[hep-ph/0211358].




\bibitem{Akhoury:1995fp}
R.~Akhoury and I.~Z.~Rothstein,
Phys.\ Rev.\ D {\bf 54}, 2349 (1996)
[hep-ph/9512303].

\bibitem{Bosch:2003fc}
S.~W.~Bosch, R.~J.~Hill, B.~O.~Lange and M.~Neubert,
Phys.\ Rev.\ D {\bf 67}, 094014 (2003)
[hep-ph/0301123].

\bibitem{Korchemsky:wg}
G.~P.~Korchemsky and A.~V.~Radyushkin,
Nucl.\ Phys.\ B {\bf 283}, 342 (1987);

\bibitem{Korchemskaya:1992je}
I.~A.~Korchemskaya and G.~P.~Korchemsky,
Phys.\ Lett.\ B {\bf 287}, 169 (1992).

\bibitem{Lange:2003ff}
B.~O.~Lange and M.~Neubert,
Phys.\ Rev.\ Lett.\  {\bf 91}, 102001 (2003)
[hep-ph/0303082].

\bibitem{Grozin:1996pq}
A.~G.~Grozin and M.~Neubert,
Phys.\ Rev.\ D {\bf 55}, 272 (1997)
[hep-ph/9607366].


\bibitem{Mannel:1994pm}
T.~Mannel and M.~Neubert,
Phys.\ Rev.\ D {\bf 50}, 2037 (1994)
[hep-ph/9402288].

\bibitem{Kagan:1998ym}
A.~L.~Kagan and M.~Neubert,
Eur.\ Phys.\ J.\ C {\bf 7}, 5 (1999)
[hep-ph/9805303].

\bibitem{Bigi:2002qq}
I.~Bigi and N.~Uraltsev,
Int.\ J.\ Mod.\ Phys.\ A {\bf 17}, 4709 (2002)
[hep-ph/0202175].

\bibitem{Mannel:1999gs}
T.~Mannel and S.~Recksiegel,
Phys.\ Rev.\ D {\bf 60}, 114040 (1999)
[hep-ph/9904475].

\bibitem{Mannel:2000aj}
T.~Mannel and S.~Recksiegel,
Phys.\ Rev.\ D {\bf 63}, 094011 (2001)
[hep-ph/0009268].

\bibitem{Falk:1992fm}
A.~F.~Falk, M.~Neubert and M.~E.~Luke,
Nucl.\ Phys.\ B {\bf 388}, 363 (1992)
[hep-ph/9204229].

\bibitem{Bigi:1994em}
I.~I.~Y.~Bigi, M.~A.~Shifman, N.~G.~Uraltsev and A.~I.~Vainshtein,
Phys.\ Rev.\ D {\bf 50}, 2234 (1994)
[hep-ph/9402360].

\bibitem{Beneke:1994sw}
M.~Beneke and V.~M.~Braun,
Nucl.\ Phys.\ B {\bf 426}, 301 (1994)
[hep-ph/9402364].

\bibitem{Bigi:1996si}
I.~I.~Y.~Bigi, M.~A.~Shifman, N.~Uraltsev and A.~I.~Vainshtein,
Phys.\ Rev.\ D {\bf 56}, 4017 (1997)
[hep-ph/9704245].

\bibitem{Beneke:1998rk}
M.~Beneke,
Phys.\ Lett.\ B {\bf 434}, 115 (1998)
[hep-ph/9804241].

\bibitem{Bigi:1997fj}
I.~I.~Y.~Bigi, M.~A.~Shifman and N.~Uraltsev,
Ann.\ Rev.\ Nucl.\ Part.\ Sci.\  {\bf 47}, 591 (1997)
[hep-ph/9703290].

\bibitem{Benson:2003kp}
D.~Benson, I.~I.~Bigi, T.~Mannel and N.~Uraltsev,
Nucl.\ Phys.\ B {\bf 665}, 367 (2003)
[hep-ph/0302262].

\bibitem{Martinelli:1995zw}
G.~Martinelli, M.~Neubert and C.~T.~Sachrajda,
Nucl.\ Phys.\ B {\bf 461}, 238 (1996)
[hep-ph/9504217];

\bibitem{Neubert:1996zy}
M.~Neubert,
Phys.\ Lett.\ B {\bf 393}, 110 (1997)
[hep-ph/9610471].



\bibitem{Becher:2003qh}
T.~Becher, R.~J.~Hill and M.~Neubert,
Phys.\ Rev.\ D {\bf 69}, 054017 (2004)
[hep-ph/0308122].

\bibitem{Hill:2002vw}
R.~J.~Hill and M.~Neubert,
Nucl.\ Phys.\ B {\bf 657}, 229 (2003)
[hep-ph/0211018].

\bibitem{Hill:2004if}
R.~J.~Hill, T.~Becher, S.~J.~Lee and M.~Neubert,
JHEP {\bf 0407}, 081 (2004)
[hep-ph/0404217].

\bibitem{Pirjol:2002km}
D.~Pirjol and I.~W.~Stewart,
Phys.\ Rev.\ D {\bf 67}, 094005 (2003)
[Erratum-ibid.\ D {\bf 69}, 019903 (2004)]
[hep-ph/0211251].

\bibitem{Lange:2003pk}
B.~O.~Lange and M.~Neubert,
  Nucl.\ Phys.\ B {\bf 690}, 249 (2004)
  [Erratum-ibid.\ B {\bf 723}, 201 (2005)]
  [hep-ph/0311345].

\bibitem{Neubert:2004dd}
 M.~Neubert,
  Eur.\ Phys.\ J.\ C {\bf 40}, 165 (2005)
  [hep-ph/0408179].


\bibitem{Bauer:2001mh}
C.~W.~Bauer, M.~E.~Luke and T.~Mannel,
Phys.\ Rev.\ D {\bf 68}, 094001 (2003)
[hep-ph/0102089].

\bibitem{Bauer:2002yu}
C.~W.~Bauer, M.~Luke and T.~Mannel,
Phys.\ Lett.\ B {\bf 543}, 261 (2002)
[hep-ph/0205150].

\bibitem{Luke:1990eg}
M.~E.~Luke,
Phys.\ Lett.\ B {\bf 252}, 447 (1990).

\bibitem{Falk:1992wt}
A.~F.~Falk and M.~Neubert,
Phys.\ Rev.\ D {\bf 47}, 2965 (1993)
[hep-ph/9209268].

\bibitem{Neubert:1996we}
M.~Neubert and C.~T.~Sachrajda,
Nucl.\ Phys.\ B {\bf 483}, 339 (1997)
[hep-ph/9603202].

\bibitem{Bigi:1994wa}
I.~I.~Y.~Bigi, B.~Blok, M.~A.~Shifman, N.~Uraltsev and A.~I.~Vainshtein,
hep-ph/9401298;

\bibitem{Bigi:1995jr}
I.~I.~Y.~Bigi,
hep-ph/9508408.

\bibitem{DiPierro:1998ty}
M.~Di Pierro and C.~T.~Sachrajda,
Nucl.\ Phys.\ B {\bf 534}, 373 (1998)
[hep-lat/9805028].

\bibitem{Becirevic:2001fy}
D.~Becirevic,
[hep-ph/0110124].

\bibitem{Baek:1998vk}
M.~S.~Baek, J.~Lee, C.~Liu and H.~S.~Song,
Phys.\ Rev.\ D {\bf 57}, 4091 (1998)
[hep-ph/9709386].



\bibitem{Voloshin:2001xi}
M.~B.~Voloshin,
Phys.\ Lett.\ B {\bf 515}, 74 (2001)
[hep-ph/0106040].






\bibitem{Bauer:2003pi}
C.~W.~Bauer and A.~V.~Manohar,
  Phys.\ Rev.\ D {\bf 70}, 034024 (2004)
  [hep-ph/0312109].

\bibitem{Bosch:2004th}
S.~W.~Bosch, B.~O.~Lange, M.~Neubert and G.~Paz,
Nucl.\ Phys.\ B {\bf 699}, 335 (2004)
[hep-ph/0402094].

\bibitem{Gardi:2004ia}
  E.~Gardi,
  JHEP {\bf 0404}, 049 (2004)
  [hep-ph/0403249].

\bibitem{Tackmann:2005ub}
  F.~J.~Tackmann,
  Phys.\ Rev.\ D {\bf 72}, 034036 (2005)
  [hep-ph/0503095].

\bibitem{Bosch:2004cb}
  S.~W.~Bosch, M.~Neubert and G.~Paz,
  JHEP {\bf 0411}, 073 (2004)
  [hep-ph/0409115].

\bibitem{Leibovich:1999xf}
A.~K.~Leibovich, I.~Low and I.~Z.~Rothstein,
Phys.\ Rev.\ D {\bf 61}, 053006 (2000)
[hep-ph/9909404].

\bibitem{Leibovich:2000ey}
  A.~K.~Leibovich, I.~Low and I.~Z.~Rothstein,
  Phys.\ Lett.\ B {\bf 486}, 86 (2000)
  [hep-ph/0005124].

\bibitem{Neubert:2001sk}
M.~Neubert,
Phys.\ Lett.\ B {\bf 513}, 88 (2001)
[hep-ph/0104280].


\bibitem{Greub:1996tg}
  C.~Greub, T.~Hurth and D.~Wyler,
  Phys.\ Rev.\ D {\bf 54}, 3350 (1996)
  [hep-ph/9603404].

\bibitem{Ali:1995bi}
  A.~Ali and C.~Greub,
  Phys.\ Lett.\ B {\bf 361}, 146 (1995)
  [hep-ph/9506374].


\bibitem{Becher:2003kh}
  T.~Becher, R.~J.~Hill, B.~O.~Lange and M.~Neubert,
  Phys.\ Rev.\ D {\bf 69}, 034013 (2004)
  [hep-ph/0309227].

\bibitem{Leibovich:2002ys}
A.~K.~Leibovich, Z.~Ligeti and M.~B.~Wise,
Phys.\ Lett.\ B {\bf 539}, 242 (2002)
[hep-ph/0205148].

\bibitem{Neubert:2002yx}
M.~Neubert,
Phys.\ Lett.\ B {\bf 543}, 269 (2002)
[hep-ph/0207002].



\bibitem{Lee:2004ja}
  K.~S.~M.~Lee and I.~W.~Stewart,
  Nucl.\ Phys.\ B {\bf 721}, 325 (2005)
  [hep-ph/0409045].

\bibitem{Beneke:2004in}
  M.~Beneke, F.~Campanario, T.~Mannel and B.~D.~Pecjak,
  JHEP {\bf 0506}, 071 (2005)
  [hep-ph/0411395].

\bibitem{Uraltsev:1999rr}
  N.~Uraltsev,
  Int.\ J.\ Mod.\ Phys.\ A {\bf 14}, 4641 (1999)
  [hep-ph/9905520].

\bibitem{Bigi:1993bh}
  I.~I.~Y.~Bigi and N.~G.~Uraltsev,
  Nucl.\ Phys.\ B {\bf 423}, 33 (1994)
  [hep-ph/9310285].



\bibitem{TomsThesis}
  T.~O.~Meyer, 
  ``Limits on weak annihilation in inclusive charmless semileptonic $B$ 
  decays'', Ph.D.\ Thesis, Cornell University (2005).


\bibitem{Neubert:2004cu}
  M.~Neubert,
  Eur.\ Phys.\ J.\ C {\bf 44}, 205 (2005)
  [hep-ph/0411027].

\bibitem{Aubert:2004aw}
  B.~Aubert {\it et al.}  [BaBar Collaboration],
  Phys.\ Rev.\ Lett.\  {\bf 93}, 011803 (2004)
  [hep-ex/0404017].


\bibitem{Hoang:1998hm}
  A.~H.~Hoang, Z.~Ligeti and A.~V.~Manohar,
  Phys.\ Rev.\ D {\bf 59}, 074017 (1999)
  [hep-ph/9811239].

\bibitem{Neubert:2004sp}
  M.~Neubert,
  Phys.\ Lett.\ B {\bf 612}, 13 (2005)
  [hep-ph/0412241].

\bibitem{vanRitbergen:1999gs}
  T.~van Ritbergen,
  Phys.\ Lett.\ B {\bf 454}, 353 (1999)
  [hep-ph/9903226].
  
\bibitem{Bosch:2004bt}
S.~W.~Bosch, B.~O.~Lange, M.~Neubert and G.~Paz,
Phys.\ Rev.\ Lett.\  {\bf 93}, 221801 (2004)
[hep-ph/0403223].

\bibitem{Falk:1997gj}
A.~F.~Falk, Z.~Ligeti and M.~B.~Wise,
Phys.\ Lett.\ B {\bf 406}, 225 (1997)
[hep-ph/9705235].

\bibitem{Bigi:1997dn}
I.~I.~Y.~Bigi, R.~D.~Dikeman and N.~Uraltsev,
Eur.\ Phys.\ J.\ C {\bf 4}, 453 (1998)
[hep-ph/9706520].

\bibitem{Bauer:2000xf}
  C.~W.~Bauer, Z.~Ligeti and M.~E.~Luke,
  Phys.\ Lett.\ B {\bf 479}, 395 (2000)
  [hep-ph/0002161].
  
\bibitem{Bauer:2001rc}
C.~W.~Bauer, Z.~Ligeti and M.~E.~Luke,
Phys.\ Rev.\ D {\bf 64}, 113004 (2001)
[hep-ph/0107074].

\bibitem{Aubert:2004tw}
  B.~Aubert {\it et al.}  [BaBar Collaboration],
  [hep-ex/0408045].

\bibitem{Neubert:2000ch}
  M.~Neubert,
  JHEP {\bf 0007}, 022 (2000)
  [hep-ph/0006068].
  
\bibitem{Neubert:2001ib}
T.~Becher and M.~Neubert,
Phys.\ Lett.\ B {\bf 535}, 127 (2002)
[hep-ph/0105217].


\bibitem{Aglietti:2002md}
U.~Aglietti, M.~Ciuchini and P.~Gambino,
Nucl.\ Phys.\ B {\bf 637}, 427 (2002)
[hep-ph/0204140].

\bibitem{Hoang:2005pj}
A.~H.~Hoang, Z.~Ligeti and M.~Luke,
Phys.\ Rev.\ D {\bf 71}, 093007 (2005)
[hep-ph/0502134].


\bibitem{Lange:2005yw}
  B.~O.~Lange, M.~Neubert and G.~Paz,
  Phys.\ Rev.\ D {\bf 72}, 073006 (2005)
  [hep-ph/0504071].

\bibitem{Neubert:2005nt}
M.~Neubert,
  Phys.\ Rev.\ D {\bf 72}, 074025 (2005)
  [hep-ph/0506245].
  
\bibitem{Becher:2006qw}
  T.~Becher and M.~Neubert,
  Phys.\ Lett.\ B {\bf 637}, 251 (2006)
  [hep-ph/0603140].
  
\bibitem{Moch:2004pa}
  S.~Moch, J.~A.~M.~Vermaseren and A.~Vogt,
  Nucl.\ Phys.\ B {\bf 688}, 101 (2004)
  [hep-ph/0403192].

\bibitem{Burrell:2003cf}
C.~N.~Burrell, M.~E.~Luke and A.~R.~Williamson,
Phys.\ Rev.\ D {\bf 69}, 074015 (2004)
[hep-ph/0312366].
  
\bibitem{Chay:2005ck}
J.~Chay, C.~Kim and A.~K.~Leibovich,
Phys.\ Rev.\ D {\bf 72}, 014010 (2005)
[hep-ph/0505030].

\bibitem{Gamiz:2004ar}
E.~Gamiz, M.~Jamin, A.~Pich, J.~Prades and F.~Schwab,
Phys.\ Rev.\ Lett.\  {\bf 94}, 011803 (2005)
[hep-ph/0408044].

\bibitem{Aubin:2004ck}
C.~Aubin {\it et al.}  [HPQCD Collaboration],
Phys.\ Rev.\ D {\bf 70}, 031504 (2004)
[hep-lat/0405022].

\bibitem{Brodsky:1982gc}
S.~J.~Brodsky, G.~P.~Lepage and P.~B.~Mackenzie,
Phys.\ Rev.\ D {\bf 28}, 228 (1983).

\bibitem{Melnikov:2005bx}
K.~Melnikov and A.~Mitov,
Phys.\ Lett.\ B {\bf 620}, 69 (2005)
[hep-ph/0505097].

\bibitem{Ligeti:1999ea}
Z.~Ligeti, M.~E.~Luke, A.~V.~Manohar and M.~B.~Wise,
Phys.\ Rev.\ D {\bf 60}, 034019 (1999)
[hep-ph/9903305].


   
\bibitem{Aubert:2005im}
  B.~Aubert {\it et al.}  [BABAR Collaboration],
  Phys.\ Rev.\ Lett.\  {\bf 95}, 111801 (2005)
  [hep-ex/0506036].
      
\bibitem{Aubert:2005hb}
  B.~Aubert {\it et al.}  [BABAR Collaboration],
  [hep-ex/0507017].
               
\bibitem{Aubert:2005mg}
  B.~Aubert {\it et al.}  [BABAR Collaboration],
  Phys.\ Rev.\ D {\bf 73}, 012006 (2006)
  [hep-ex/0509040].


\bibitem{Limosani:2005pi}
  A.~Limosani {\it et al.}  [Belle Collaboration],
  Phys.\ Lett.\ B {\bf 621}, 28 (2005)
  [hep-ex/0504046].
  
\bibitem{Bizjak:2005hn}
  I.~Bizjak {\it et al.}  [Belle Collaboration],
  Phys.\ Rev.\ Lett.\  {\bf 95}, 241801 (2005)
  [hep-ex/0505088].

\bibitem{Becher:2005pd}
  T.~Becher and M.~Neubert,
  Phys.\ Lett.\ B {\bf 633}, 739 (2006)
  [hep-ph/0512208].

\bibitem{Blokland:2005uk}
  I.~Blokland, A.~Czarnecki, M.~Misiak, M.~Slusarczyk and F.~Tkachov,
  Phys.\ Rev.\ D {\bf 72}, 033014 (2005)
  [hep-ph/0506055].


\bibitem{Tarasov:au}
  O.~V.~Tarasov, A.~A.~Vladimirov and A.~Y.~Zharkov,
  Phys.\ Lett.\ B {\bf 93}, 429 (1980).



\end{thebibliography}
\end{document}